\documentclass[final,3p]{elsarticle}
\usepackage{amssymb} % The amssymb package provides various useful mathematical symbols
\usepackage{amsthm}% The amsthm package provides extended theorem environments
\usepackage{graphicx,subfigure}% Include figure files
\usepackage{dcolumn}% Align table columns on decimal point
\usepackage{bm}% bold math
\usepackage{cancel}
\usepackage{float}
\usepackage{epstopdf}
\usepackage{hyperref}
\hypersetup{colorlinks = true}

\def\nuebar{\bar{\nu}_e}
\def\picwidth1{90mm}

\journal{Physics Reports}

\begin{document}

\begin{frontmatter}



\title{Theoretical Antineutrino Detection, Direction and Ranging at Long Distances}


\author{Glenn R. Jocher}
\ead{glenn.jocher@gmail.com}
\address{Integrity Applications Incorporated, 15020 Conference Center Drive, Chantilly, VA, 20151 USA}

\author{Daniel A. Bondy}
\ead{dbondy@integrity-apps.com}
\address{Integrity Applications Incorporated, 15020 Conference Center Drive, Chantilly, VA, 20151 USA}

\author{Brian M. Dobbs}
%\ead{bdobbs@integrity-apps.com}
\ead{Brian.M.Dobbs.ctr@nga.mil}
\address{Integrity Applications Incorporated, 15020 Conference Center Drive, Chantilly, VA, 20151 USA}

\author{Stephen T. Dye}
\ead{sdye@phys.hawaii.edu}
\address{College of Natural Sciences\\Hawaii Pacific University, Kaneohe, HI 96744 USA}
\address{Department of Physics and Astronomy\\University of Hawaii, Honolulu, HI, 96822 USA}

\author{James A. Georges III}
%\ead{jgeorges@integrity-apps.com}
\ead{James.A.Georges.ctr@nga.mil}
\address{Integrity Applications Incorporated, 15020 Conference Center Drive, Chantilly, VA, 20151 USA}

\author{John G. Learned\corref{cor1}}
\ead{jgl@phys.hawaii.edu}
%\email{Corresponding author: jgl@phys.hawaii.edu}
\address{Department of Physics and Astronomy\\University of Hawaii, Honolulu, HI, 96822 USA}

\author{Christopher L. Mulliss}
%\ead{cmulliss@integrity-apps.com}
\ead{Christopher.L.Mulliss.ctr@nga.mil}
\address{Integrity Applications Incorporated, 15020 Conference Center Drive, Chantilly, VA, 20151 USA}

\author{Shawn Usman}
\ead{Shawn.Usman@nga.mil}
\address{InnoVision Basic and Applied Research Office, Sensor Geopositioning Center\\National Geospatial-Intelligence Agency, 7500 GEOINT Dr., Springfield, VA, 22150 USA}

\cortext[cor1]{Corresponding author}

\begin{abstract}
In this paper we introduce the concept of what we call ``NUDAR" (NeUtrino Direction and Ranging), making the point that measurements of the observed energy and direction vectors can be employed to passively deduce the exact three-dimensional location and thermal power of  geophysical and anthropogenic neutrino sources from even a single detector. Earlier studies have presented the challenges of long-range detection, dominated by the unavoidable inverse-square falloff in neutrinos, which force the use of kiloton scale detectors beyond a few kilometers. Earlier work has also presented the case for multiple detectors, and has reviewed the background challenges. We present the most precise background estimates to date, all handled in full three dimensions, as functions of depth and geographical location.  For the present calculations, we consider a hypothetical 138 kiloton detector which can be transported to an ocean site and deployed to an operational depth. We present a Bayesian estimation framework to incorporate any \textit{a priori} knowledge of the reactor that we are trying to detect, as well as the estimated uncertainty in the background and the oscillation parameters. Most importantly, we fully employ the knowledge of the reactor spectrum and the distance-dependent effects of neutrino oscillations on such spectra. The latter, in particular, makes possible determination of range from one location, given adequate signal statistics. Further, we explore the rich potential of improving detection with even modest improvements in individual neutrino direction determination. We conclude that a 300 MW$_\mathrm{th}$ reactor can indeed be geo-located, and its operating power estimated with one or two detectors in the hundred kiloton class at ranges out to a few hundred kilometers.  We note that such detectors would have natural and non-interfering utility for scientific studies of geo-neutrinos, neutrino oscillations, and astrophysical neutrinos.  This motivates the development of cost effective methods of constructing and deploying such next generation detectors.
\end{abstract}

\begin{keyword}
antineutrino \sep neutrino \sep geo-neutrino \sep reactor \sep geo-reactor \sep oscillation
\end{keyword}

\end{frontmatter}

\newpage
\tableofcontents
\newpage

\section{Introduction}
\label{Introduction}
In this paper we present a careful discussion of scientific and applied experiments based on the measurement of electron antineutrinos from distant sources (tens to hundreds of km) in a $10^{34}$ proton class detector. Because of the well-known ghostly nature of the neutrinos, we focus on large detectors and strive to employ all information potentially available, including our best present knowledge of all background processes. One cannot escape the small neutrino cross-sections or the inverse-square falloff in neutrino flux with range, which requires the detector size to grow with the square of the range. One also cannot escape the increasing importance of the background at long distances as the source signal becomes a relatively small component of the total signal.  Since some (dominant) background sources depend on cosmic rays which come into the atmosphere, one is driven to place the larger detectors deeper underground (or under water) at longer distances.

\subsection{Types of neutrinos}
Since the original observation of neutrinos by Cowan and Reines in the mid 1950’s \cite{reines_1956}, operating in close proximity to nuclear reactors, people have speculated on the possible detection of reactors from longer ranges.  Note that we focus here on the detection of electron antineutrinos in the energy range of about 1 to 11 MeV.  The flux of neutrinos from reactors is relatively small compared to antineutrinos ($<1\%$).  Additionally, flux from the sun is composed almost entirely of neutrinos and not antineutrinos.  Experiments have set strong limits on the flux of antineutrinos from the sun \cite{kamland_2005}.  The neutrino flux from radioactive decays throughout the Earth is composed almost entirely of electron antineutrinos \cite{Krauss1}. Hence we can speak of the electron antineutrinos from reactors and the Earth (geo-neutrinos), without ambiguity, as neutrinos.

A useful complication (as we shall see) is that the flux of electron antineutrinos consists of varying fractions of all three types of antineutrinos (electron, muon and tauon) when observed from distances beyond a few kilometers.  For the energies and the detection mechanism under consideration here, however, only the electron type is detectable due to the masses of the muon and tauon. All three types of neutrinos may be sensed with a coherent neutrino scattering detector \cite{Drukier}, but such devices will only be useful at small distances ($<$1km) from a reactor and will be dominated by solar neutrinos at larger distances. % \cite{Collar_2009}.

\subsection{Detection mechanism: inverse beta decay}
Remarkably, the detection mechanism for reactor neutrinos remains exactly the same as in the original Reines-Cowan discovery studies: inverse beta decay (IBD), detected in a scintillating medium with surrounding optical sensors.  In the charge exchange process, the electron antineutrino becomes a positron and the target free proton becomes a neutron.  The positron receives energy proportional to the incoming neutrino energy, less the threshold of 1.8 MeV for the reaction, and the positron scatters most often nearly perpendicular to the incoming direction.  The neutron, being much heavier than the positron, acquires little velocity and rolls forward with tens of keV of kinetic energy while the positron may have several MeV.  One may think of the positron as keeping the kinetic energy of the neutrino and the neutron as inheriting the momentum.

The neutron slows to thermal velocities and gets captured within microseconds, as compared to the positron stopping and annihilating with an electron (usually after forming positronium) within nanoseconds.  The net signature of the IBD reaction is two spatially correlated depositions of energy, resulting from positron energy loss and annihilation and from nuclear de-excitation after neutron capture. These are detected by optical sensors as flashes of light in a scintillating liquid medium, two pulses occurring close in space (less than 1 m) and time (few microseconds) with the second pulse having a known intensity.  This distinctive double hit signature distinguishes this reaction from the many processes which produce only one flash (such as from solar neutrinos and many types of radioactivity). Large detectors such as the 1000 ton Kamioka Liquid-scintillator Anti-Neutrino Detector (KamLAND) \cite{KamLAND_collaboration} and Borexino \cite{Borexino_2010} have demonstrated clean extraction of a few IBD events per month from a single pulse background rate more than a million times greater.  We discuss this process in more detail later in the discussion of neutrino fluxes and cross sections in Section \ref{Estimation theory}.

\subsection{Electron antineutrino production in reactors}
The source of the neutrinos in the nuclear fire of the reactor originates in beta decay, which is simply neutron decay to a less massive proton, an electron (the beta particle) and the generally unseen electron antineutrino.  Nuclear reactors operate by fission chain reactions, wherein the fission of a heavy nucleus such as U-235 becomes two lighter but unstable elements, freeing typically 7 neutrons, and liberating approximately 200 MeV in total energy \cite{kopeikin_2004}.  Most of the resulting nuclei are well away from the nuclear stable valley and are neutron rich, resulting in many beta decays.  The calculation of the flux of neutrinos involves, as one may imagine, substantial nuclear physics and we use the results of existing calculations herein.  It should not be a surprise that the spectrum may be approximated by an exponentially falling flux with neutrino energy.

In reality though, both the neutrino flux and energy spectrum from a reactor are not constant in time. They depend upon the design of the reactor, its operation, and the reactor fuel mix. (The measured flux at the detector is calculable to within a few percent given the source information.) There is also a well -known predicted difference in the neutrino spectra between uranium and plutonium, with roughly a 5\% slope difference above 4 MeV. This ``burn-up" effect has been observed by the SONGS1 collaboration, operating cubic meter scale detectors about 20 m from a power reactor \cite{songs_2008}. This effect has been the source of much interest as to the possibility of directly detecting the fuel mix and observing the enrichment of Pu over time as the U is consumed (see Nieto \textit{et al.} \cite{Nieto}). However, since the fuel is always a mixture of U and Pu, the statistical requirements for spectral  determination are on the order of tens of thousands of events, and something not as yet achieved by detectors more than 20m from a fission source \cite{songs_2008}.  Thus, for longer ranges, it is reasonable to assume a reactor energy spectrum which is constant in time.

Thermal power output from a reactor (typically three times the electrical power) tracks the neutrino flux to within about one percent.  Since this is also about the precision to which power output can be measured at reactors, and of the order of the accuracy of the flux calculations, we can calculate the expected neutrino flux from a reactor facility, which usually reports power output (typically daily or more often) to about 3\% including all sources of error.  See Lasserre, \textit{et al.} \cite{lasserre_2010} for more discussion of this topic. Generally, the reactor can be treated as a point source of neutrino emission for ranges beyond tens of meters since most of the emission occurs within about a one meter radius of the center. 

The inverse beta decay cross section for $\nuebar$ upon free protons is basic to weak interactions, and is well studied. It is proportional to the square of the visible neutrino energy in this energy regime, and on the scale of $10^{-42}\mathrm{cm}^2$ (see Equation \ref{JNBeq_19}). The target and detection medium are usually the same for this purpose, typically a long organic molecule with roughly twice as many hydrogen nuclei as carbon nuclei. Interestingly, for many materials the density of hydrogen in chemical form is not far from that of liquid hydrogen. Hence one cannot do much to increase the free proton density. The protons bound in heavier nuclei do not help, since it requires many MeV of energy to get them out of the nucleus, hence only free protons count.

\subsection{Detectors}
With IBD energy depositions in the range of a few MeV, the optical signal in large detectors (tons) will not be terribly large.  For economic reasons the configuration of such instruments has generally been of a large volume of liquid surrounded by optical sensors, generally photomultipliers.  Smaller detectors, such as employed in laboratory-scale double beta decay and direct dark matter searches, utilize a variety of techniques such as crystals and solid state detectors.  The price of these options, however, limits their practical use to detectors on the scale of tons.  At the kiloton scale, the only viable options at present are organic fluids or water.   Fortunately, these liquids may be ``doped" with a variety of materials in small concentration to enhance light output and spectral match to the light sensors.

We do not perform a detailed investigation of possible liquid scintillators for this study, but rather we assume an organic liquid such as those employed in the KamLAND and Borexino detectors, and that proposed for the 50 kiloton LENA detector \cite{lena_2005}. Of course, purified water costs far less but it produces roughly 30 times less light (the organic liquid producing light from scintillation and the water only from the Cherenkov radiation).  For comparison, the KamLAND scintillation detector produces signals of about 250 photoelectrons per million electron volts of ionization (PE/MeV) whereas the water Cherenkov Super-Kamiokande \cite{superk_2006} produces about 8-10 PE/MeV.  Thus water should have additives to increase light production, and such is under active study at present.  Therefore, we  conservatively assume organic liquids in this study.

Neutron absorption depends upon the choice of medium as well. In water and organic liquids the IBD-produced neutron generally wanders for a hundred or more microseconds before capture by a hydrogen nucleus. The deuterium formed de-excites with emission of a 2.2 MeV gamma ray.  The addition of an absorber with a larger neutron capture cross section and a larger de-excitation energy improves the measurement resolution by causing the neutron capture to be closer in time and space to the neutrino interaction.  A favorite material considered for addition to water at present is gadolinium \cite{Yeh}, which is inexpensive, has water soluble salts and an emission of order 8 MeV.

\subsection{Background Sources}
There are several types of background sources which can cause false neutrino signatures.  Fortunately, the main discriminant for IBD is the spatial coincidence (within a meter or so) and the temporal coincidence (within a few microseconds) between the prompt (positron) and delayed (neutron capture) signals, along with the prompt signal providing a known energy release.  Background sources which may fake neutrino signatures fall into two categories, one which simulates both prompt and delayed signals, and one where the prompt and delayed signals are of separate uncorrelated origin but randomly occur in near coincidence.  These latter events, the so called ``accidentals", are generally restricted to the lower end of the energy range since the frequency of single noise hits increases at lower energies.  The sources of background can be from internal residual radioactivity, cosmic rays (only muons are important at depths greater than a few meters) passing through the detector, cosmic rays outside the detector which make products such as fast neutrons which enter the detector \cite{kamland_2010}, and finally from radioactivity outside the detector.  The external radioactivity of concern is largely gamma rays since alphas and electrons do not travel far.

Of course, there are neutrinos which come from other reactors around the world, there are neutrinos from radioactive decays within the Earth, there are cosmic ray neutrinos, and there are potentially neutrinos from beyond the Earth. We deal with these sources one at a time, here qualitatively and later numerically:

\begin{enumerate}
\item {\bf Accidental Coincidences.}
It is difficult to make general statements about the rate of random coincidences, since they depend entirely upon environmental and detector construction details.  However, based upon the experience of the two extant large neutrino instruments KamLAND and Borexino, we know that this rate can be kept to nearly ignorable levels.  Much of the source of single pulses will originate in the detector structure and environs, so there is a clear benefit to larger detectors with smaller surface-to-volume ratios.

\item {\bf Detector Internal Radioactivity.}
With care in detector preparation, the neutrino community has learned to fabricate detectors with negligible levels of internal radioactivity.  The U and Th decays chains present a continuing concern, and radon presents a dangerous and subtle pollutant. However, the technology to reach extremely low levels for internal radioactivity has been well demonstrated in the last several decades.

\item {\bf Detector External Radioactivity.}
External radioactivity, such as from $^{40}$K only causes problems near the detector surface.  Two meters (water equivalent) thickness of shielding is generally enough to reduce it significantly.  Such a background is easily revealed by the clustering of reconstructed event locations that occurs near the surface.

\item {\bf Penetrating Cosmic Ray Muons.}
Muons originating from cosmic ray interactions high in the atmosphere (typically 20 km above) penetrate to the deepest mines, having been observed two miles underground and under water.  Since the incoming cosmic ray spectrum falls steeply with energy ($\sim E^{-2.7}$) the energy of muons falls as well, and the flux falls steeply with depth. A strong constraint on detector depth arises simply from having a rate of muons so high that the electronics become saturated if the depth is too shallow. We deal with this quantitatively in this study.  To give some sense of scale, current large detectors need to be under more than about 2000 meters of water equivalent (MWE) to have low enough rates to reduce electronics dead time to negligible levels. These events are due to long lived $^9$Li and $^8$He isotopes, and are generally referred to as ``cosmogenic" background (see Section \ref{Non-neutrino background}).

\item {\bf External Cosmic Ray Interaction Products.}
Fast neutrons from muon interactions in the detector surroundings pose a similar concern to entering external radioactivity products.  These neutrons can scatter elastically from protons and then be absorbed, mimicking the prompt and delayed signals of a true IBD event.  Protection against such comes from possible shielding with muon detectors in the outer layers.  But in the case of large detectors, again, these events are confined to the near surface and are discernible. 

\item {\bf Geo-neutrinos.} 
The study of geo-neutrinos and their origins is an interesting goal in itself, which we do not discuss at any length in this paper.  We carefully consider geo-neutrinos because they represent the largest background for applied experiments (such as those related to reactors).  The uranium and thorium decay chains have energies extending up to 3.27 MeV.  Hence, one common way to reduce geo-neutrino background is simply to reject events with visible energies above some threshold, but this approach can discard nearly half of the desired (and never too strong) reactor signal.  The rates of the geo-neutrino events are low in current detectors, only recently being reported for the first time by KamLAND \cite{kamland_2005} and Borexino \cite{Borexino_2010}. 

If direction resolution improves in the future, then event reconstruction (discussed below) can potentially identify geo-neutrinos events by isolating those events arriving from below the horizon.  This would dramatically improve the signal-to-noise ratio for applied experiments.  In such experiments, geo-neutrino data can be harvested on a non-interfering basis to provide excellent parallel science.  In this paper we utilize all the data, employing all possible information and a Bayesian approach to include the geo-neutrino data in applied experiments.

\item {\bf Atmospheric Cosmic Ray Produced Neutrinos.}
Fortunately for this subject, most of the cosmic rays are of much higher energy, in the hundreds of MeV range with average around a GeV.  These events have a low enough rate not to be a problem and are well separated in detected energy.  With large enough detectors, the study of these events themselves will become a useful and non-interfering byproduct. See Honda \textit{et al.} \cite{Honda 2011} for more information on atmospheric neutrinos.

\item {\bf Extraterrestrial Neutrinos.}
The search for extraterrestrial neutrinos, such as those from ancient supernovae and from gamma ray bursts, has been a grand goal for many in the neutrino field for decades.  The only occurrence of such a detection was during the Supernova 1987A, observing a total of 19 interactions between the Irvine-Michigan-Brookhaven (IMB) detector \cite{IMB 1987} and Kamioka \cite{Hirata} detectors. Nearby supernovae in our galaxy are rare (a few per century) and thus pose no problem for neutrino experiments, but provide a potentially wonderful opportunity for parallel science.

\item {\bf Other Reactors.}
In Figure \ref{worldIAEA} we present a map of the calculated flux of neutrinos from power and research reactors around the world.  One cannot currently distinguish a neutrino's reactor-of-origin on an event by event basis since all nuclear reactors share similar spectra (to a few percent) and current detectors have nearly isotropic direction resolution.  Statistically however, neutrino oscillation modifies the reactor spectra and reduces the flux (measured over a wide range of energy) at great distances.  If one can achieve sufficient angular measurement resolution, then one may reject events based on arrival directions.

There have been a number of suggestions of a natural nuclear reactor, at the Earth's core \cite{Herndon_1996} or perhaps at the inner core boundary \cite{Rusov_2007} or at the core-mantle boundary \cite{deMeijer_2008}.  While geologists generally reject such models, they have not yet been ruled out at the level below 3 TW$_\mathrm{th}$ \cite{Borexino_2010}.  Such would provide a significant, and detectable, background for a very large detector located far from any nearby reactors. 
\end{enumerate}

\subsection{Other physics and astrophysics}

The list of possible scientific experiments that can be accomplished with the sort of detector we are discussing is quite impressive, both in the fields of elementary particle physics and in astrophysics.  Study of these topics will generate considerable interest in the scientific community and will result in many theses, publications, and in the training of a cadre of professionals in these fields.

The physics and astrophysics study list includes the following:

\begin{enumerate}
\item Long baseline studies using a neutrino beam from an accelerator
\item Proton decay searches, particularly in the Kaon decay modes, not well studied by SuperKamiokande
\item Search for supernova burst neutrinos, most likely from our galaxy
\item Search for relic neutrinos summed from all past type II supernovae
\item Atmospheric neutrinos, moving beyond SuperKamiokande
\item Solar neutrinos, at least careful monitoring of solar output versus time
\item Reactor neutrino properties, depending upon the proximity and power of reactors
\item Geo-neutrinos both as object of study and background, depending upon location and as discussed below
\item Pion decay at rest neutrinos from a possible portable low energy accelerator (the DAEDALUS idea \cite{Conrad_2010})
\item Possible experiments employing radioactive sources for short range searches for or studies of sterile neutrinos
%\item As yet undefined experiments associated with the recent reports of superluminal neutrinos \cite{MINOS_2007} \cite{Adam_2011}
\end{enumerate}

These are already discussed in the specific framework of the LENA ``whitepaper" for a 50 kiloton liquid scintillation detector in Europe \cite{lena_2005}, and we shall not elaborate further upon them here.  First, it should be noted that all of these are low duty factor studies which carry on without interference and in parallel with applied experiments... they are essentially ``free".  Some, such as those involving a neutrino beam from an accelerator will depend upon placement.  Others, such as employing a radioactive source may require access to the detector.  For the most part, the important searches for proton decay, studies of atmospheric neutrinos, and for rare astrophysical phenomena will go along unimpeded wherever the detector is placed.  The presence of such detectors would serve to engage a large scientific community.

In the following we make explicit and conservative models of background sources, in three dimensions, and we test the capabilities of the simulated detectors in a variety of situations via simulation.  We construct a hypothetical detector in Section \ref{Antineutrino Detector Model}, which we call TREND (Test REactor antiNeutrino Detector), and explicitly define its properties, resolution and the expected background.  In Section \ref{Geospatial Model}, we construct a thorough geospatial model including the variation of all background sources with location and depth.  In Section \ref{Estimation theory}, we present our optimal estimator theory and apply it.  In Section \ref{Oscillation parameter estimation results}, we discuss the synergistic determination of neutrino properties in terms of neutrino oscillation parameters.  Finally in Section \ref{Reactor geolocation results}, we explore the applied experiment of geolocating a reactor in several different detector configurations\footnote{The choice of locations was designed to match those studied previously by a French group \cite{lasserre_2010}, to explore different geographical constraints on detector placements, and to sample different background environments.}. We finish with a summary of our findings and we highlight the areas deserving further research.

\section{Antineutrino Detector Model}
\label{Antineutrino Detector Model}
The Detector Model determines an arbitrary detector's neutrino measurement resolution (both angle and energy) via Monte Carlo (MC) simulations.  It also determines the energy-dependent efficiencies associated with various candidate selection criteria. The Detector Model is composed of three distinct parts: a simulation of particles (and their energy depositions) produced by inverse beta decay in liquid scintillator, a model for photon production and propagation, and a detector hardware model for the photon detection process. The Detector Model has been validated two ways, against CHOOZ experimental \textit{angular} resolution and against KamLAND experimental \textit{energy} resolution.

\begin{figure}[!htbp]
\centering
\includegraphics[width=.75\linewidth]{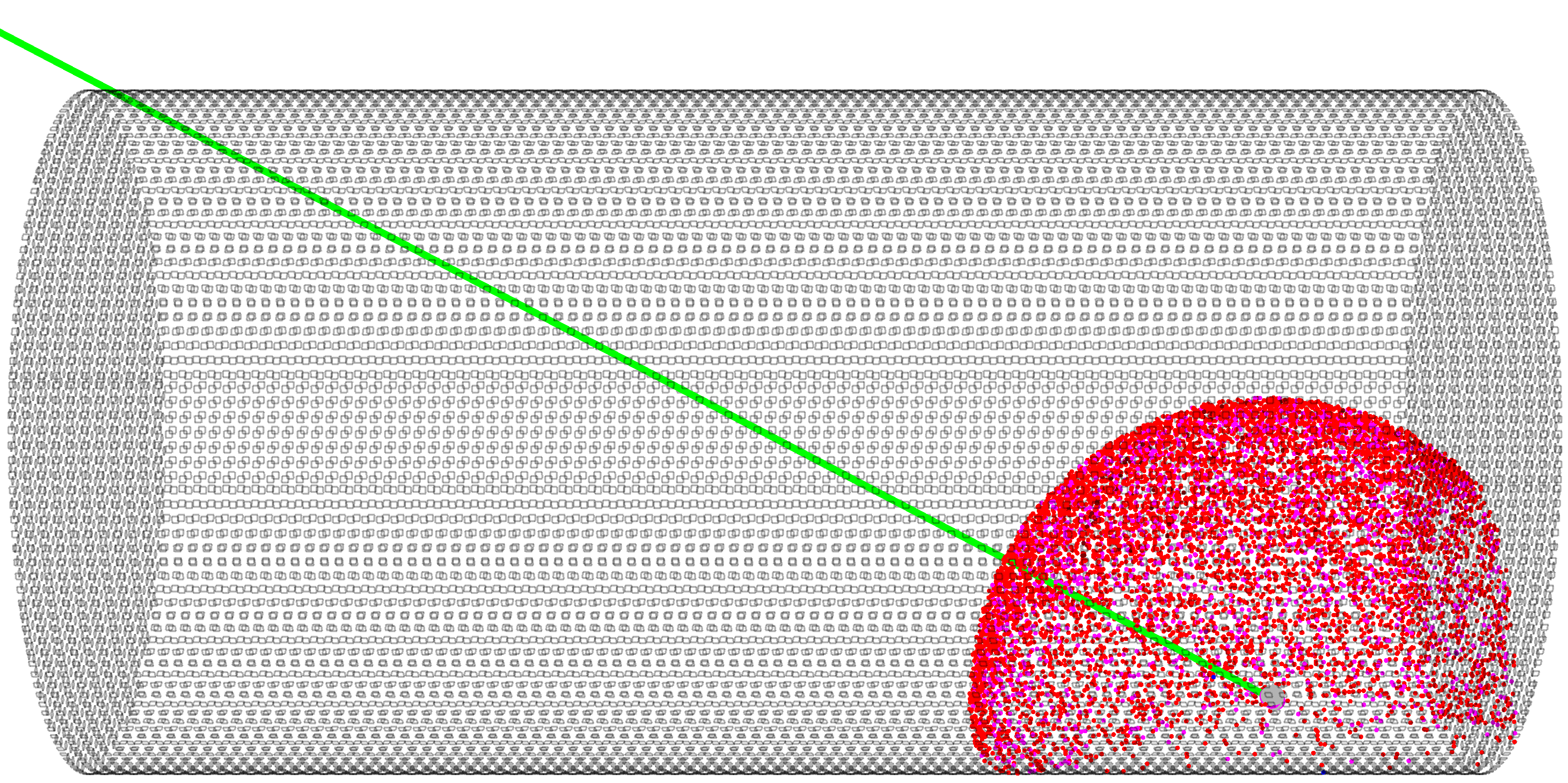}
\caption{$E_{\nuebar}=5$MeV randomly placed inverse beta decay event within TREND, about 100 ns after the event. The event vertex occurred about 4m from the detector's lower cylinder wall. $\nuebar$ trajectory in green, photons in red, PMT outlines in grey. Cylinder is 96m long and 46m in diameter.  The 1 m thick outer region of purified water for Cherenkov detection of cosmic muons is not shown for purposes of visual clarity.}
\label{TRENDdetector}
\end{figure}

Determination of a detector's measurement resolution is a precursor to the placement of the detector within the Geospatial Model discussed in Section \ref{Geospatial Model}. The resulting energy measurement resolution consists of an energy-dependent table of standard deviations which correspond to the energy measurement uncertainty probability density function (pdf). This uncertainty pdf is assumed Gaussian.  The resulting angle measurement resolution consists of an energy-dependent table of angular signal-to-noise ratio (SNR) values.  The angular SNR is defined as the mean displacement between the prompt and delayed signal divided by the direction vector uncertainty per Cartesian axis.  A detector's measurement resolution is employed in two ways by the Geospatial Model. First, it is used to add the correct amount of energy and angle measurement noise to the measurements of detectors placed in the Geospatial Model. Second, the detector resolution is passed to the maximum \textit{a posteriori} (MAP) estimator discussed in Section \ref{Estimation theory} where it is used to account for the impact of measurement noise on parameter estimation.  

\subsection{Inverse beta decay model}
\label{Inverse beta decay model}

The Detector Model begins with output from a GEANT \cite{GEANT} (Version 4.9.1 patch 3) simulation.  The GEANT simulation provides the locations, timestamps, kinetic energy and energy deposition of all of the particles and daughter particles of the positron and neutron produced by the inverse beta decay event.  A large, homogeneous volume of linear alkyl benzene (LAB) scintillation fluid was chosen as the target medium.  Positrons and neutrons were directly introduced into GEANT simulations in a manner consistent with the statistical properties described by Vogel \cite{vogel_1999} while enforcing the appropriate conservation laws (e.g. positron-neutron momentum along all three dimensions). 

The GEANT dataset used in this paper consists of 10,000 sample IBD events at 10 different discrete neutrino energies from 2MeV through 11MeV. These 100,000 sample events were used in the Detector Model to characterize the measurement resolutions of TREND, CHOOZ and KamLAND, assuming random placement of events within each detector.  It is important to note that the Geospatial Model does not directly utilize these 100,000 sample events.  Instead, it uses the energy-dependent measurement resolution that was derived from these 100,000 sample events.  This allows statistical outliers (a real possibility) to occur in the Geospatial Model even if such outliers were not included in the 100,000 sample events.

\subsection{Photon propagation model}
\label{Photon propagation model}
The Photon Model (written in MATLAB \cite{MATLAB}) accepts GEANT output and generate MC photons via scintillation and Cherenkov radiation \cite{Cerenkov}. The model assumes an amalgamation of parameters derived from various sources to model photon scintillation, Cherenkov radiation, attenuation and re-emission. The scintillation spectrum was defined by Eljen Technology's EJ-254 1\% boron-doped plastic scintillator datasheet \cite{EJ-254}. The scintillation spectrum has a peak at 421 nm, a scintillation yield of 9200 photons per MeV (deposited), and a scintillation decay constant of 2.2 ns (fast component only). Photo-multiplier tube (PMT) quantum efficiency (QE) curves were modeled upon a notional absorption spectra which is peaked at 425 nm and extends down to 200 nm in the ultraviolet region.

Oleg Perevozchikov's thesis \cite{Oleg_2009} was used to define an energy-dependent re-emission efficiency curve, scaled to about 20\% efficiency at 425 nm. The wavelength-dependent attenuation length curve was taken from Lightfoot \cite{Lightfoot 2004} and scaled to give a scintillation-weighted mean of slightly over 20 m, per Lasserre's example \cite{lasserre_2010}. The refractive index was defined by the Sellmeier equation shown in Equation \ref{sellmeier}, scaled to provide a scintillation-weighted mean index of refraction of about $n=1.58$.

\begin{equation}
\label{sellmeier}
n(\lambda) = 1+\frac{B_1\lambda^2}{\lambda^2-C_1} + \frac{B_2\lambda^2}{\lambda^2-C_2} + \frac{B_3\lambda^2}{\lambda^2-C_3}
\end{equation}

Here $B_1=1.44$, $C_1=81.76\mathrm{nm}^2$, $B_2=B_3=0$, and $C_2=C_3=0\mathrm{nm}^2$.

\begin{figure}[!htbp]
\centering
\includegraphics[width=\linewidth]{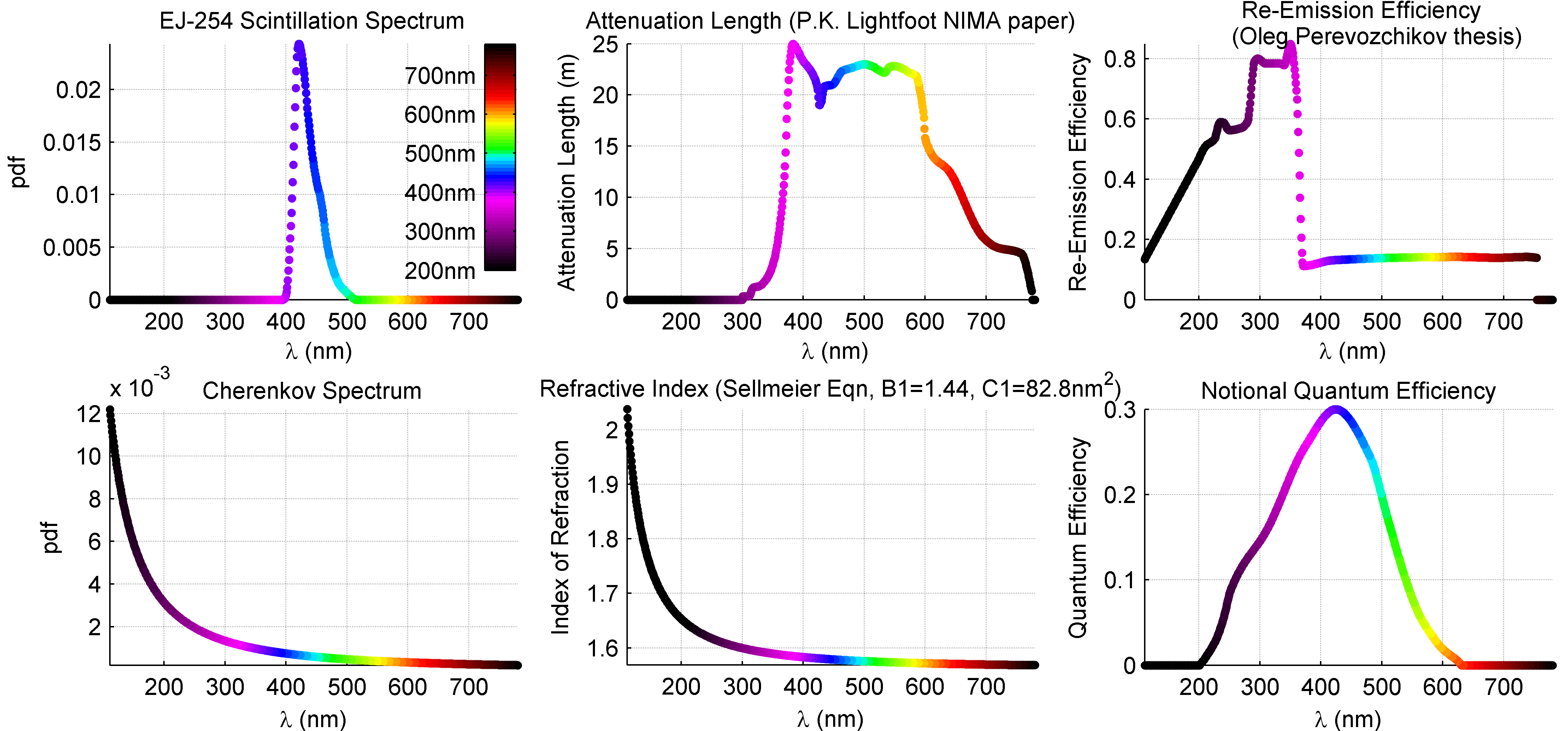}
\caption{Photon model optical properties. For our emission spectrum we used Eljen Technology's EJ-254 1\% boron-doped LS datasheet \cite{EJ-254}. Oleg Perevozchikov's thesis \cite{Oleg_2009} defines our energy-dependent re-emission efficiency curve shape, scaled to 20\% at 425nm. Attenuation length curve derived from Lightfoot \cite{Lightfoot 2004} and scaled according to Lasserre \textit{et al.} \cite{lasserre_2010}. Sellmeier equation fitted to refractive index curve to achieve a scintillation wavelength-weighted mean of about $n=1.58$. Notional PMT QE curve set to peak at 425nm and survive down to 200nm.}
\label{spectra1}
\end{figure}

All of the optical properties referenced above are displayed in Figure \ref{spectra1}. Figure \ref{spectra2} shows the photons created in one MC run compared to the optical properties described above (and shown in Figure \ref{spectra1}) for a neutrino inverse beta decay event centered within TREND.

Poisson statistics were applied according to Equation \ref{dE_to_photon0} to create discrete integer photon counts from GEANT's continuous energy deposition values. In Equation \ref{dE_to_photon0}, $N$ is the number of photons produced at a single energy deposition point.

\begin{equation}
\label{dE_to_photon0}
N = \mathcal{P}\left(\bar{N}\right)
\end{equation}

$N$ is a Poisson random number, sampled from the mean photon count, $\bar{N}$, defined in Equation \ref{dE_to_photon1}

\begin{equation}
\label{dE_to_photon1}
\bar{N} = y(kB)dE
\end{equation}
where $dE$ is the energy deposition given by GEANT (in MeV), $y$ is the scintillation yield (in photons/MeV) and $kB$ is the ``Birk's quenching factor\footnote{True quenching factor values vary by scintillator and are a function of particle ionization state, kinetic energy, and mass energy. These values are typically experimentally determined, but notional values were assumed for simplicity, independent of energy.}", a ratio of visible energy released to total energy released. In the case of light ionizing particles such as electrons and positrons a value of $kB=1.00$ was assumed. In the case of heavy ionizing particles such as protons a value of $kB=0.10$ was assumed.

Charged particles exceeding the local speed of light in the medium produce Cherenkov radiation. Equation \ref{cherenkov1} defines the scaled Cherenkov emission spectrum.

\begin{equation}
\label{cherenkov1}
f(\lambda) = \left\{ 
\begin{array}{l l}
    \beta\geq\frac{1}{n(\lambda)} & \quad 2\pi\alpha q^2 \mu \left(\frac{1-(\beta n(\lambda))^{-2}}{\lambda^2}\right) \times 10^6\\
    \beta<\frac{1}{n(\lambda)} & \quad 0\\
\end{array} \right.
\end{equation}

Here, $\beta=\frac{|\vec{v}|}{c}$, is the ratio of the particle speed through the fluid to the speed of light in a vacuum, $n(\lambda)$ is the unitless fluid index of refraction at wavelength $\lambda$ (in nanometers), $\mu$ is the relative permeability of the fluid (assumed to be 0.999992), $q$ is the particle charge in units of elementary charge, and the fine structure constant $\alpha=1/137.036$.

\begin{figure}[!htbp]
\centering
\includegraphics[width=\linewidth]{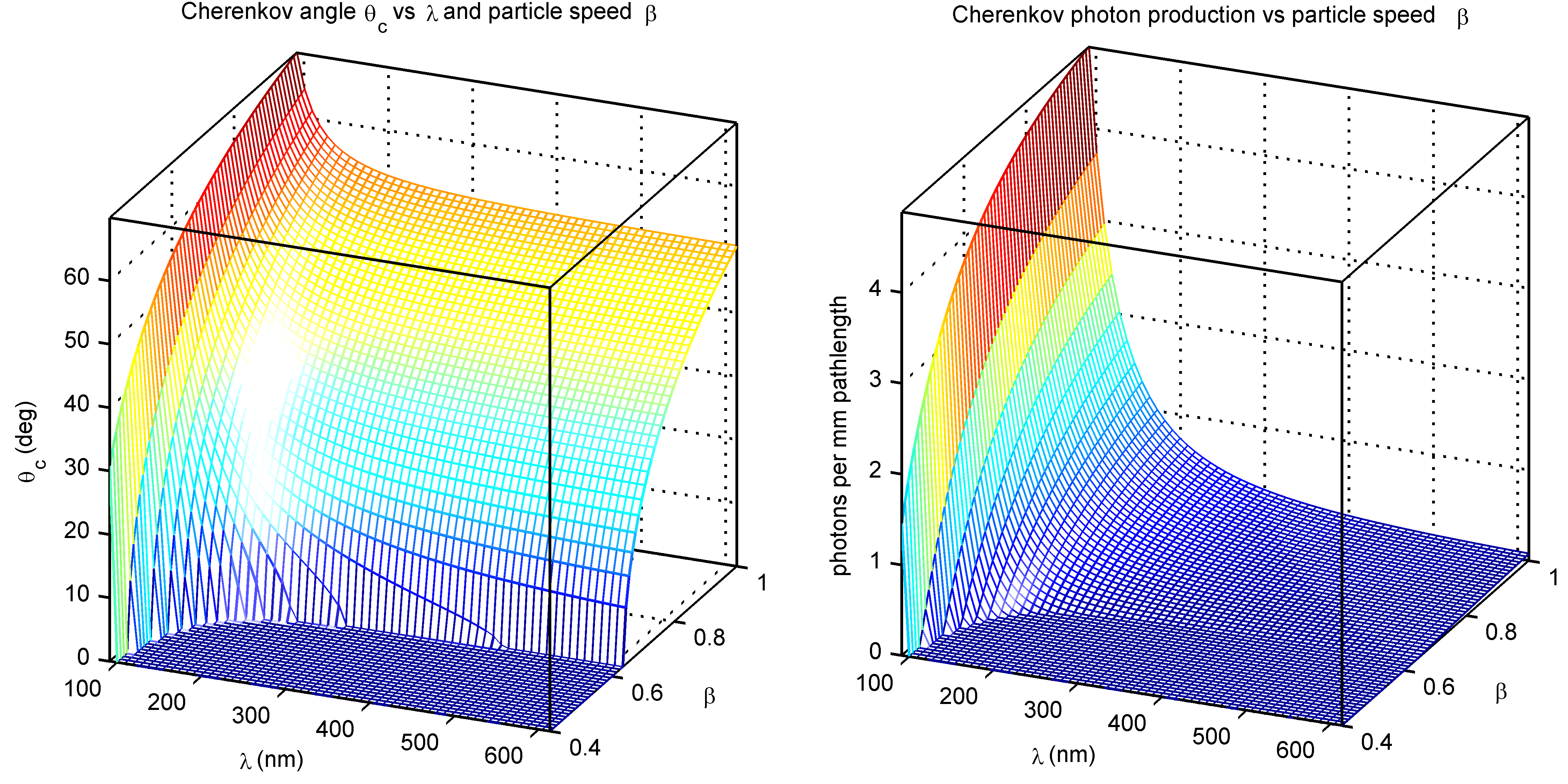}
\caption{On the right panel, Equation \ref{cherenkov1} is evaluated: Cherenkov photon production per mm of particle path-length as a function of photon wavelength $\lambda$ and parent particle speed $\beta$. On the left panel, Equation \ref{lorentz1} is evaluated: Cherenkov angles $\theta_c$ as a function of wavelength $\lambda$ and parent particle speed $\beta$.}
\label{cherenkov1fig}
\end{figure}

\begin{figure}[!htbp]
\centering
\includegraphics[width=\linewidth]{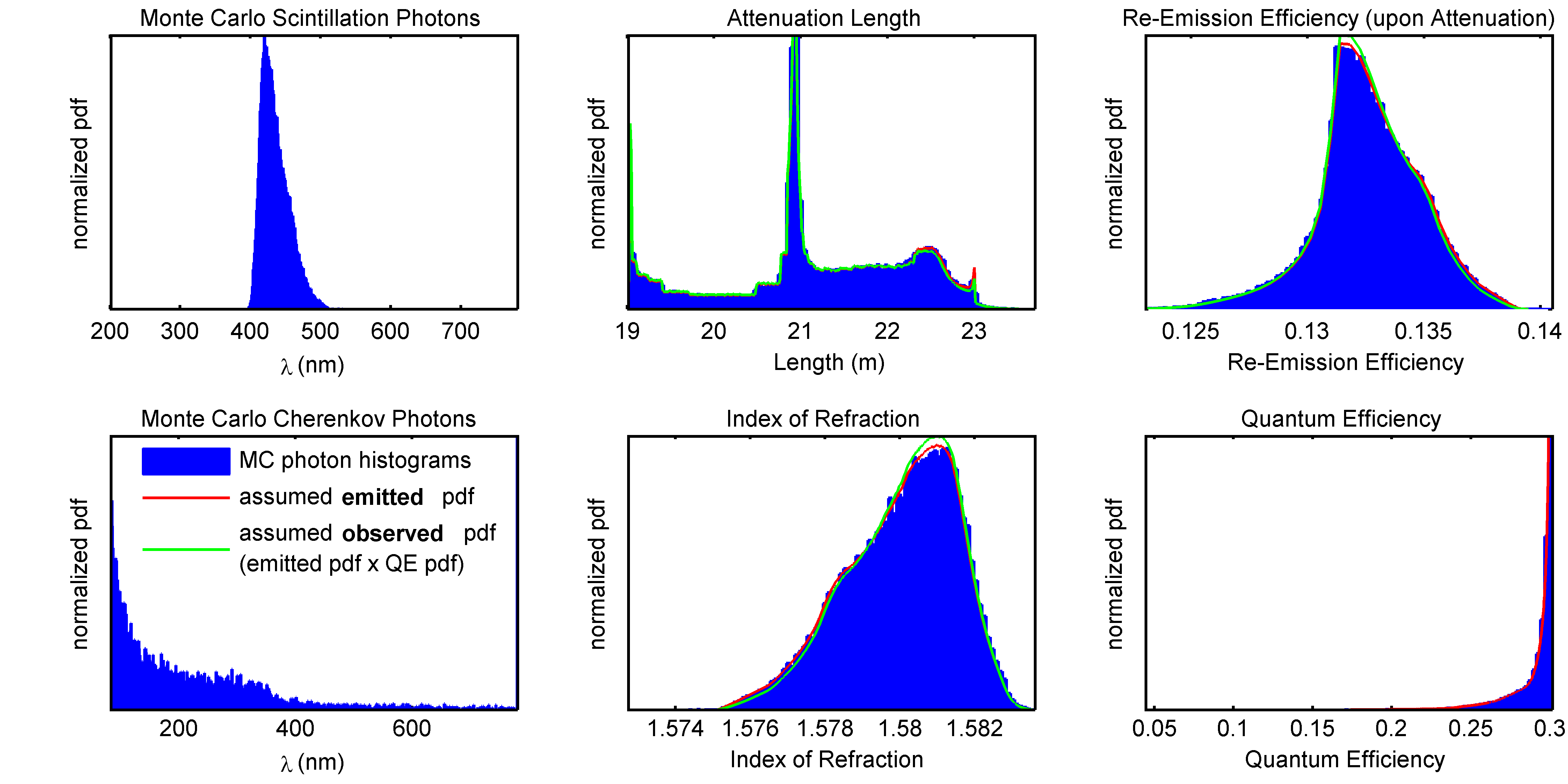}
\caption{Photon histograms from one TREND MC IBD event, using LS optical properties from Figure \ref{spectra1}. Blue histograms represent  MC photons. Red lines are estimator-assumed \textit{emission} spectra. Green lines are the estimator-assumed \textit{observed} spectra, whose $\lambda$-weighted mean values are used (since PMTs are ``colorblind") to estimate the prompt and delayed ``bookends" (positions and energies). Deviation between the emitted and observed spectra are caused by operation of the QE curve on all observed photons.}
\label{spectra2}
\end{figure}

The units of $f(\lambda)$ are the mean number of Cherenkov photons produced per mm of particle travel distance per nanometer of photon wavelength. Figure \ref{cherenkov1fig} plots this value across a typical $\lambda$-$\beta$ domain. 

Determination of the (Poisson mean) number of Cherenkov photons produced per mm from $\lambda_1$ to $\lambda_2$ requires an integration of Equation \ref{cherenkov1} across the $\lambda_1$ to $\lambda_2$ spectrum:

\begin{equation}
\label{cherenkov2}
\bar{N} = \int\limits_{\lambda_1}^{\lambda_2}f(\lambda)d\lambda
\end{equation}

Cherenkov radiation ``cone" half-angles $\theta_c$ were assumed to be continuously variable over the portion of each charged particle's trajectory where $\beta \geq \frac{1}{n(\lambda)}$. These angles, $\theta_c$, defined in Equation \ref{lorentz1} and shown in Figure \ref{cherenkov1fig}, are a function of a Cherenkov photon's wavelength $\lambda$ (in nm) and parent particle's speed $\beta$.

\begin{equation}
\label{lorentz1}
\theta_c(\lambda) = \arccos\left(\frac{1}{\beta n(\lambda)}\right)
\end{equation}

The parent particle speed $\beta$ may be defined as a function of its kinetic energy KE and rest mass $m_0$ by Equation \ref{lorentz2}.

\begin{equation}
\label{lorentz2}
\beta = \frac{\sqrt{\mathrm{KE}^2+2m_0c^2\mathrm{KE}}}{\mathrm{KE}+m_0c^2}
\end{equation}

A single-exponential time delay (2.2 ns) was applied to all scintillation photon emissions, and photon ``start" times were randomly sampled from this exponential distribution. We expect effects of a second slower exponential to be small and ignore them in the present simulation.
%For the TREND Detector Model, Eljen Technology's EJ-254 \cite{EJ-254} 1\% boron-doped LS scintillator decay time constant was assumed.
Photon attenuation and re-emission was also modeled. Wavelength dependencies including attenuation length, re-emission fraction, refractive index and QE were defined for each individual photon as a function of each photon's wavelength\footnote{No two photons share the same exact wavelength (or the same exact wavelength-dependent optical properties) as the scintillation spectrum is treated as continuous; i.e. it is randomly sampled using a linear interpolant between defined table lookup points.}.

Photons which are absorbed and then re-emitted are probabilistically assigned a new wavelength using a re-emission spectrum which is different than the original scintillation spectrum.  For each such photon, the re-emission spectrum is scaled to ensure that energy conservation is not violated (i.e. the new wavelength is longer than its predecessor). Photon \textit{re-emission} directions are assumed isotropic for all re-emitted photons regardless of their properties or source of origin. This causes photons to gradually lose directional ``information"\footnote{The term ``information" here denotes any information which may contribute to a mathematical reconstruction of an event's location and $dE$.} over time as they repeated absorptions and re-emission.  This is especially true of high energy Cherenkov photons (see Figure \ref{spectra1}) because of the shorter attenuation lengths in the ultraviolet/blue region of the spectrum.

\subsection{Detector hardware model}
Any arbitrary detector design can be selected for use with the Inverse Beta Decay Model (described in Section \ref{Inverse beta decay model}) and the Photon Model (described in Section \ref{Photon propagation model}) to quickly test different detector designs of varying volume on the same GEANT dataset.  The Detector Hardware Model consists primarily of a physical description of the size and shape of the scintillating volume and its enclosure(s), the size and shape of the photo-detectors, the optical coverage due to photo-detectors, and the wavelength-dependent QE curve for the photo-detectors.
%This versatile MC engine has successfully simulated detectors from several liters to several hundred kilotons fiducial volume. 
The physical structure of TREND (the detector selected for this study) was largely adopted from the Secret Neutrino Interaction Finder (SNIF) detector proposed by Lasserre \textit{et al.} \cite{lasserre_2010}.  We are, however, proposing to operate TREND quite differently.  We propose a different set of selection criteria, to operate without a low energy cut, to utilize energy and angle measurements, to use a different estimation technique, and to operate at deeper depths.  In future work, we hope to further optimize TREND using the flexibility of the Detector Hardware Model.

\begin{figure}[!htbp]
\centering
\includegraphics[width=\linewidth]{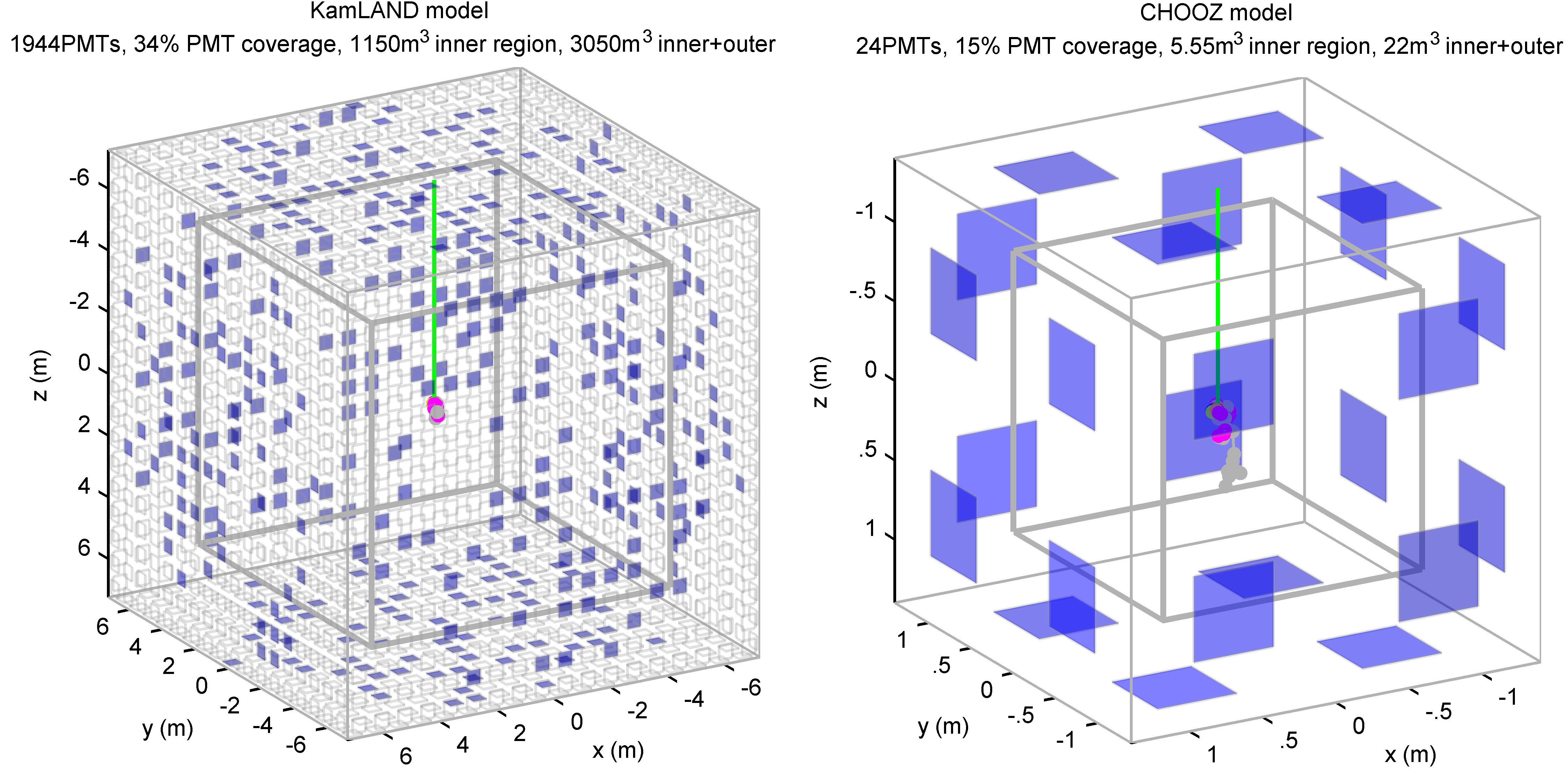}
\caption{CHOOZ and KamLAND models as created by the Detector Model. Each detector (modeled as a cube rather than a sphere for simplicity) shows the same GEANT IBD event at detector center. Purple PMTs (square ``pixels" lying on the outer region walls) indicate active voltage jumps from photon detection (the images were taken about 500 ns after $\nuebar$ annihilation). Note separate inner regions (\textit{thick} grey lines) and outer regions (\textit{thin} grey lines).}
\label{CHOOZ_and_KamLAND}
\end{figure}

In addition to TREND, Regions I and II of CHOOZ were also modeled for angular measurement resolution validation, as well as KamLAND for energy measurement resolution validation. Our Detector Model incarnations of these two real world detectors are shown in Figure \ref{CHOOZ_and_KamLAND}. As evidenced in Figure \ref{CHOOZ_and_KamLAND}, PMTs were modeled as flat squares lying tangent to the detector wall. PMT temporal response functions and wavelength-dependent QE curves were imported from manufacturer's datasheets. Rayleigh scattering, and photon reflection at the LS/PMT boundary are not modeled\footnote{Concerns over the omission of Rayleigh scattering and photon reflection at the LS/PMT boundary may be allayed by the positive validation results for CHOOZ and KamLAND presented in Section \ref{Validation results}.}. 

For simplicity both CHOOZ and KamLAND were modeled as cube-shaped in our Detector Model, though we did model TREND as a cylinder. We notice almost no adverse effects in our validations due to simplifying the detector spherical geometries into a cube shapes.

\subsection{Validation results}
\label{Validation results}

To establish confidence in the measurement resolution values that the Detector Model predicts for TREND, two separate validation tests were carried out.  The first against experimental KamLAND energy resolution \cite{KamLAND_collaboration} (Section \ref{KamLAND energy resolution validation}), and the second against experimental CHOOZ direction vector resolution \cite{chooz_2003} (Section \ref{CHOOZ direction resolution validation}).

\subsubsection{KamLAND energy resolution validation}
\label{KamLAND energy resolution validation}

The Kamioka Liquid-scintillator Anti-Neutrino Detector (KamLAND) is located in the Kamioka mine, which is approximately 1 km ($\sim$2700 MWE) below the summit of Mt. Ikenoyama, Gifu prefecture, Japan (36.43$^\circ$ N, 137.31$^\circ$ E). The presently operating detector measures electron antineutrinos from nuclear reactors (Eguchi \textit{et al.} \cite{kamland_eguchi_2003}) and the earth (Araki \textit{et al.} \cite{Araki_Nature}), using $\sim$1000 tons of ultra-pure scintillating liquid monitored by $\sim$1900 PMTs providing $\sim$34\% photo-coverage. The detector realizes an energy resolution of
$6.5\%/\sqrt{E_\mathrm{vis}[\mathrm{MeV}]}$ (Abe \textit{et al.} \cite{kamland_2010}).

KamLAND was simulated with a refractive index of 1.44, a (scintillation wavelength-weighted) mean attenuation length of 8.46m, a LS exponential decay constant of 6ns, a yield of 9200 photons/MeV, (scintillation wavelength-weighted) mean quantum efficiency of about 29\%, a (scintillation wavelength-weighted) mean re-emission efficiency of about 13\%, a 1150m$^3$ inner region and a 3050m$^3$ outer region with 324 PMT's per face (1944 total). A fixed quenching factor of 0.1 was assumed for heavy ionization particles, and the scintillation spectrum was lifted from our EJ-254 datasheet \cite{EJ-254}. Each PMT was assumed capable of 1 ns timing resolution, with no flat-fielding effects. Each PMT was modeled as a square, 0.47m per side, separated from its neighbor by 0.81 m (center-to-center), providing 34\% detector surface area coverage. An earlier alternate configuration is also modeled with 22\% PMT coverage. This earlier configuration uses identical fluid properties as above but has 225 PMT's per face (1350 total). Each PMT in this configuration was modeled as a square, 0.45m per side, separated from its neighbors by 0.97 m (center-to-center), providing 22\% total surface area coverage.

KamLAND energy validation results are shown in Table \ref{table:KAMLAND}. Note that these validation results pertain specifically to center-detector point-sources of energy, \textit{not} IBD sources. A full comparison between IBD energy resolution and point-source energy resolution, as well as center vs off-center event resolution, is presented in Section \ref{TREND energy and direction resolution MC results}. 

\begin{figure}[!htbp]
\centering
\includegraphics[width=.5\linewidth]{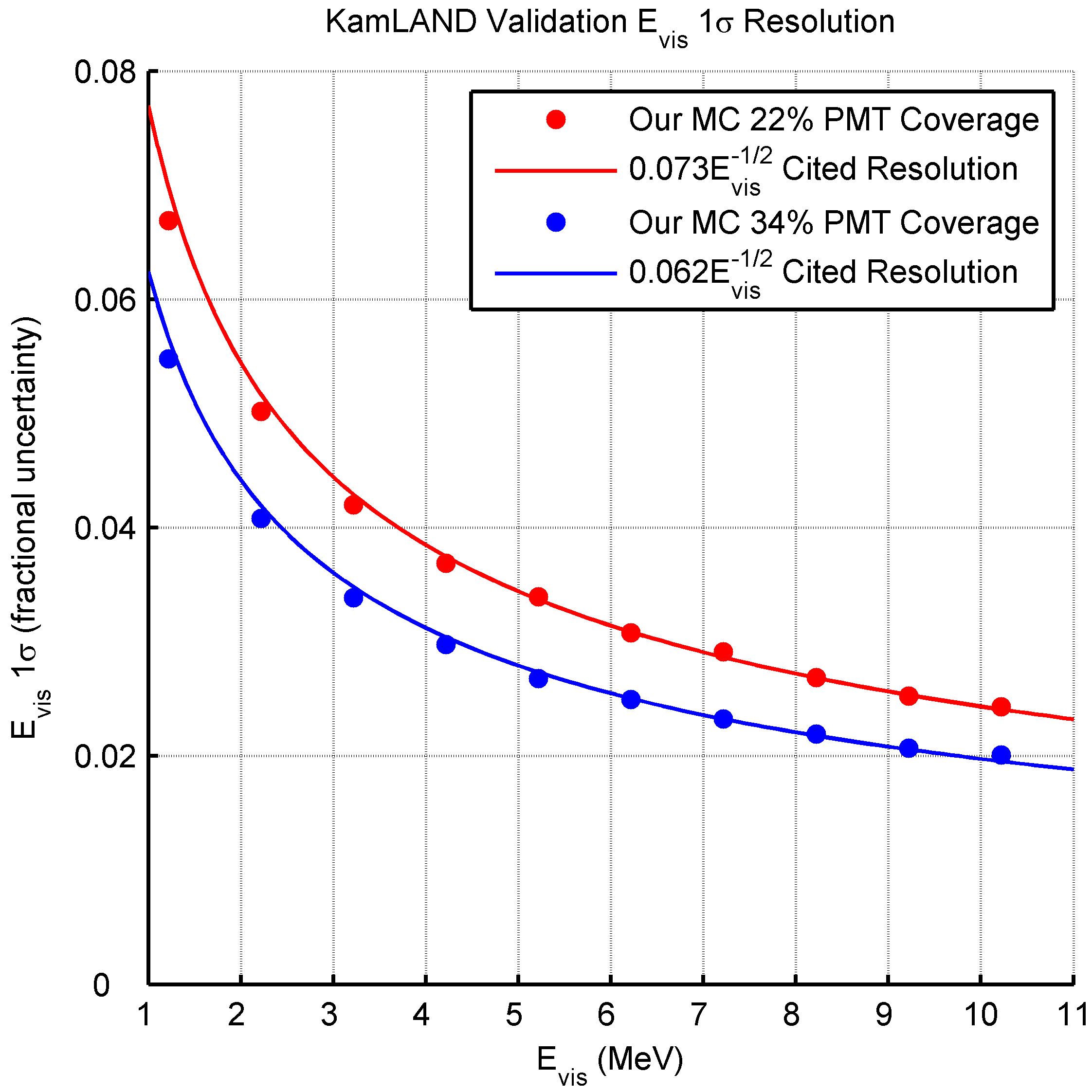}
\caption{KamLAND validation results \textit{for center-detector point sources of energy}. Our Detector Model MC results (red and blue \textit{points}) are shown vs  KamLAND \cite{KamLAND_collaboration} cited values (red and blue \textit{solid lines}). Blue results show 34\% PMT coverage validation, red results show 22\% PMT coverage validation.}
\label{KamLANDvalidation}
\end{figure}

The full comparison over energy can be seen in Figure \ref{KamLANDvalidation}. Our Detector Model produced very similar resolution values across the energy spectrum when compared to actual KamLAND experimental data, both before and after the KamLAND PMT additions of 2003. This provides us with high confidence  in the $\left. 8.9\% \right|_{E_\mathrm{vis}=1\mathrm{MeV}}$ energy resolution that our Detector Model produced for TREND in Table \ref{table:TREND2}. 

\begin{table} [!htbp]
\fontsize{8}{10}\selectfont
\begin{center}
\begin{tabular}{l|l|l}
\hline
\textbf{Comparison Metric} 				&\textbf{Cited \cite{KamLAND_collaboration}} &\textbf{Our MC}\\
\hline Energy Resolution $1\sigma$ 	&$6.2\%/\sqrt{E_\mathrm{vis}[\mathrm{MeV}]}$ 		&$6.24\%/\sqrt{E_\mathrm{vis}[\mathrm{MeV}]}$\\
(34\% PMT coverage) & &(82.4\% candidates)\\
\hline Energy Resolution $1\sigma$ 	&$7.3\%/\sqrt{E_\mathrm{vis}[\mathrm{MeV}]}$ 		&$7.69\%/\sqrt{E_\mathrm{vis}[\mathrm{MeV}]}$\\
(22\% PMT coverage) & &(82.0\% candidates)\\
\end{tabular}
\caption{KamLAND $\nuebar$ visible-energy resolution validation results \textit{for center-detector point sources of energy}. Values in right column represent nonlinear least squares fits to our KamLAND MC energy estimate errors. We compared the KamLAND detector before and after the 2003 additional PMT installation cited by \cite{KamLAND_collaboration}: ``The central detector PMT array was upgraded on February 27, 2003 by commissioning 554 20-inch PMTs, increasing the photo-cathode coverage from 22\% to 34\% and improving the energy resolution from $7.3\%/\sqrt{E_\mathrm{vis}[\mathrm{MeV}]}$ to $6.2\%/\sqrt{E_\mathrm{vis}[\mathrm{MeV}]}$." Our KamLAND ``candidates" met all the criteria specified for CHOOZ candidates in Table \ref{table:CHOOZ}.}
\label{table:KAMLAND}
\end{center}
\end{table}

\subsubsection{CHOOZ direction resolution validation}
\label{CHOOZ direction resolution validation}

The CHOOZ detector was located $\sim$1 km from the 4.25 GW$_\mathrm{th}$ nuclear reactor near the village Chooz, along the River Meuse in northern France. To reduce the cosmic ray muon flux the detector was placed underground with an overburden of $\sim$300 MWE. The active target of the detector consisted of 5 tons of Gd-loaded scintillating liquid. Scintillation light was collected by $\sim$200 PMTs, achieving $\sim$15\% photo-coverage and an energy resolution of $9\%/\sqrt{E_\mathrm{vis}[\mathrm{MeV}]}$ (Apollonio \textit{et al.} \cite{chooz_2003}). This detector measured the direction of reactor antineutrinos with a 1$\sigma$ resolution of 18$^\circ$.

CHOOZ was simulated with a refractive index of 1.47, a (scintillation wavelength-weighted) mean attenuation length of 3.4m, a LS exponential decay constant of 7ns, a yield of 9200 photons/MeV, (scintillation wavelength-weighted) mean quantum efficiency of about 29\%, a (scintillation wavelength-weighted) mean re-emission efficiency of about 13\%, a 5.6m$^3$ inner region and a 22m$^3$ outer region with 6 PMT's per face (24 total). A fixed quenching factor of 0.1 was assumed for heavy ionization particles, and the scintillation spectrum was lifted from our EJ-254 datasheet \cite{EJ-254}. Each PMT was assumed capable of 8 ns timing resolution, with no flat-fielding effects. Each PMT was modeled as a square, 0.54m per side, separated from its neighbor by 1.40 m (center-to-center), providing 15\% detector surface area coverage.

With CHOOZ, as with TREND, the direction of the incoming neutrino is calculated by forming a vector between the estimated location of neutron capture (the delayed signal) and the estimated location of the positron annihilation (the prompt signal).  Because the neutron can bounce many times before its capture, the neutrino directional information recorded for each IBD event is typically very poor using this method.  Because of this CHOOZ (and others) often quote their direction resolution as a spherical 1$\sigma$ uncertainty after a fixed large number of IBD events.  In this work, we introduce a metric called ``angular SNR" which quantifies the directional information contained per single IBD event.  The angular SNR (SNR$\approx$.07 in CHOOZ), defined in Equations \ref{JNBeq_23a} - \ref{JNBeq_23c}, is the mean positron-neutron displacement ($\sim$15mm in CHOOZ) divided by the norm ($\sim$210mm) of the combined neutron ($\sim$200mm) and positron ($\sim$60mm) location uncertainties.

CHOOZ direction validation results are shown in Table \ref{table:CHOOZ}. The Detector Model values compared well to the CHOOZ experimental values across all the different comparison metrics, including candidate selection cuts, mean neutron displacement, neutron position resolution, and the primary metric of interest, direction vector SNR. This successful validation establishes grounds for confidence in the 0.045 SNR direction vector resolution our Detector Model produced for TREND in Table \ref{table:TREND2}

\begin{table} [!htbp]
\fontsize{8}{10}\selectfont
\begin{center}
\begin{tabular}{l|l|l}
\hline
\textbf{Comparison Metric} 						& \textbf{Cited \cite{chooz_2003}} 				& \textbf{Our MC}\\
\hline positron energy cut  					& 97.8\% 										& 99.9\% \\
\hline positron-geode distance cut  			& 99.9\% 										& 99.7\% \\
\hline neutron capture cut  					& 84.6\% 										& 82.5\% \\
\hline capture energy containment cut  			& 94.6\% 										& 94.7\% \\
\hline neutron-geode distance cut   			& 99.5\% 										& 97.1\% \\
\hline neutron delay cut   						& 93.7\% 										& 99.8\% \\
\hline positron-neutron distance cut   			& 98.4\% 										& 99.9\% \\
\hline neutron multiplicity cut   				& 97.4\% 										& 100\%* \\
\hline combined cut   							& 69.8\% 										& 75.1\% \\
\hline Mean Neutron Displacement  				& 19mm (17mm sim)								& 16.6mm \\ 
\hline Neutron Position Resolution $1\sigma$  	& 190.0mm 										& 181.9mm \\
\hline Angular $1\sigma$, $n$=2500 events		& 18.0$^\mathrm{o}$ (19.0$^\mathrm{o}$ sim) 	& 18.9$^\mathrm{o}$ \\
\hline SNR  									& 0.100 (.091 sim) 								& 0.091 \\
\end{tabular}
\caption{CHOOZ direction resolution validation results. Energy dependent values weighted to a 1km range reactor spectrum. *Neutron multiplicity cut not implemented because our GEANT dataset was comprised of single-neutron inverse beta decay events only. CHOOZ own simulation results \cite{chooz_2003} shown in parenthesis next to CHOOZ experimental results \cite{chooz_2003}.}
\label{table:CHOOZ}
\end{center}
\end{table}

\subsubsection{TREND energy and direction resolution MC results}
\label{TREND energy and direction resolution MC results}
\begin{figure}[!htbp]
\centering
\includegraphics[width=1\linewidth]{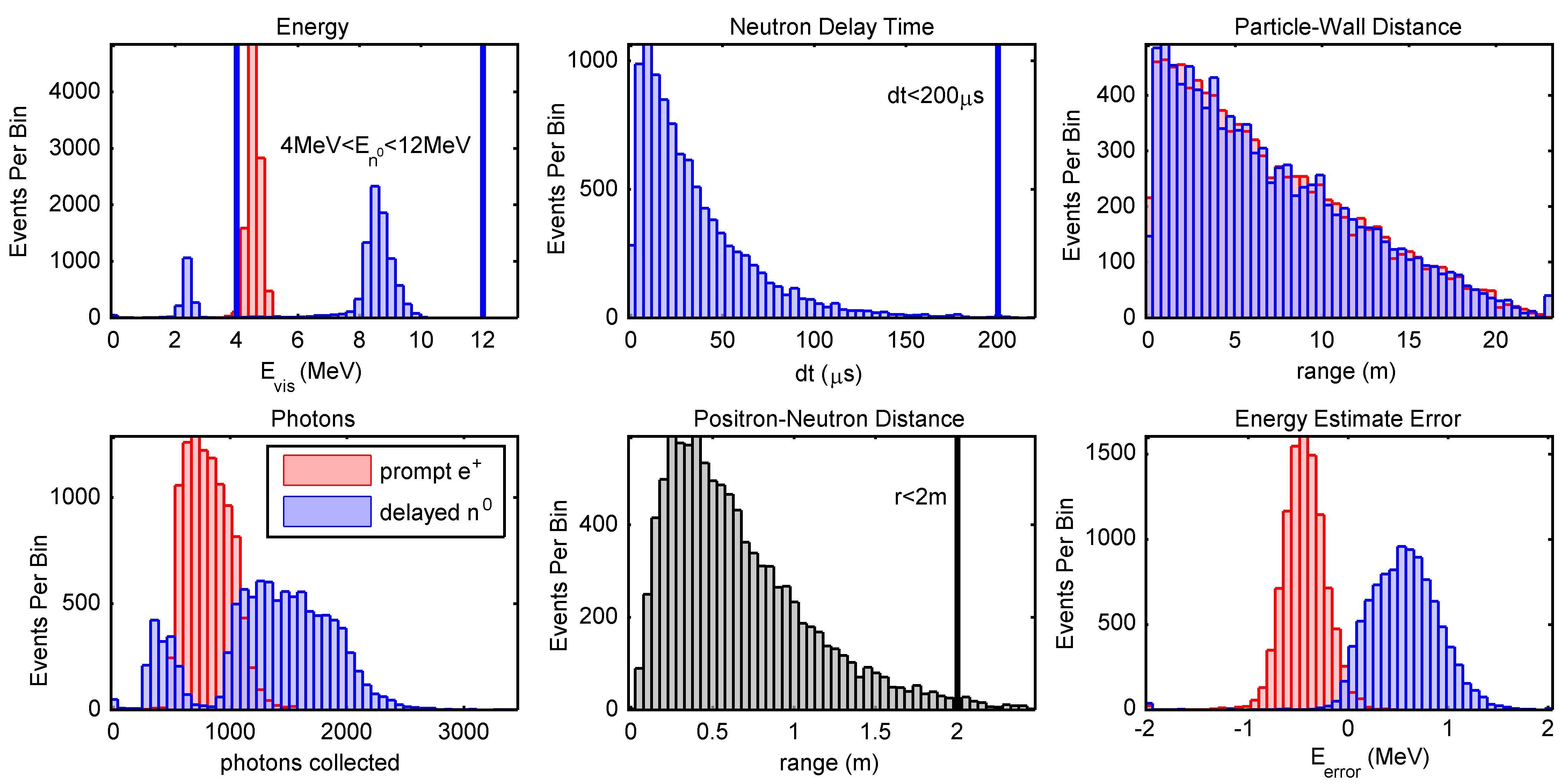}
\caption{$E_{\nuebar}=5$MeV MC statistics, 10,000 $\nuebar$ events. Prompt ``bookends" in red, delayed ``bookends" in blue. Select candidate criteria are also shown in this figure: the 200$\mu$s time cut (99.8\% candidates), 2m range cut (96.5\% candidates), and 4MeV$<E^{n^0}_\mathrm{vis}<$12MeV neutron energy cut (83.3\% candidates).}
\label{MCstats}
\end{figure}

Once validated against CHOOZ and KamLAND experimental results, the Detector Model was tasked with determining the energy and direction resolution of the proposed TREND detector. TREND ``candidate" events were selected to meet all the CHOOZ candidate criteria specified in Table \ref{table:CHOOZ} with the exception of three TREND-specific modifications:

\begin{enumerate}
\item Neutron delay cut increased from 100$\mu$s in CHOOZ to 200$\mu$s in TREND.
\item Positron energy cut increased from $E_\mathrm{vis}^{e^+}<$8MeV in CHOOZ to $E_\mathrm{vis}^{e^+}<$12MeV in TREND.
\item Positron-neutron distance cut increased from 1m in CHOOZ to 2m in TREND.
\end{enumerate}
We believe the time and distance increases above were warranted by the poor position resolution and vastly larger size of TREND compared to CHOOZ. The positron energy cut increase was implemented to capture reactor antineutrinos to the very end of their detectable energy spectrum (near 11MeV). Figure \ref{MCstats} shows how increasing the positron-neutron distance cut from 1m to 2m lets us accept a much larger percentage of neutrinos as candidates, while increasing the neutron delay cut from 100$\mu$s to 200$\mu$s has a similar, but smaller, effect.

For TREND, the scintillator optical properties assumed by the Photon Model were defined by the various sources cited in Section \ref{Photon propagation model}. The modeled TREND detector (shown intercepting a $\nuebar$ in Figure \ref{TRENDdetector}) is a cylinder 96m high with a radius of 23m, comprising 16906 PMTs of 0.2m$^2$ area each to provide almost 20\% optical coverage. The fiducial mass of 138,000 metric tons brings the free proton count to $\sim 10^{34}p^+$. Table \ref{table:TREND1} provides a complete list of parameter inputs used by the Detector Model to construct TREND, including the candidate cut criteria used to produce the statistical results presented in Table \ref{table:TREND2}.  

\begin{figure}[!htbp]
\centering
\includegraphics[width=\linewidth]{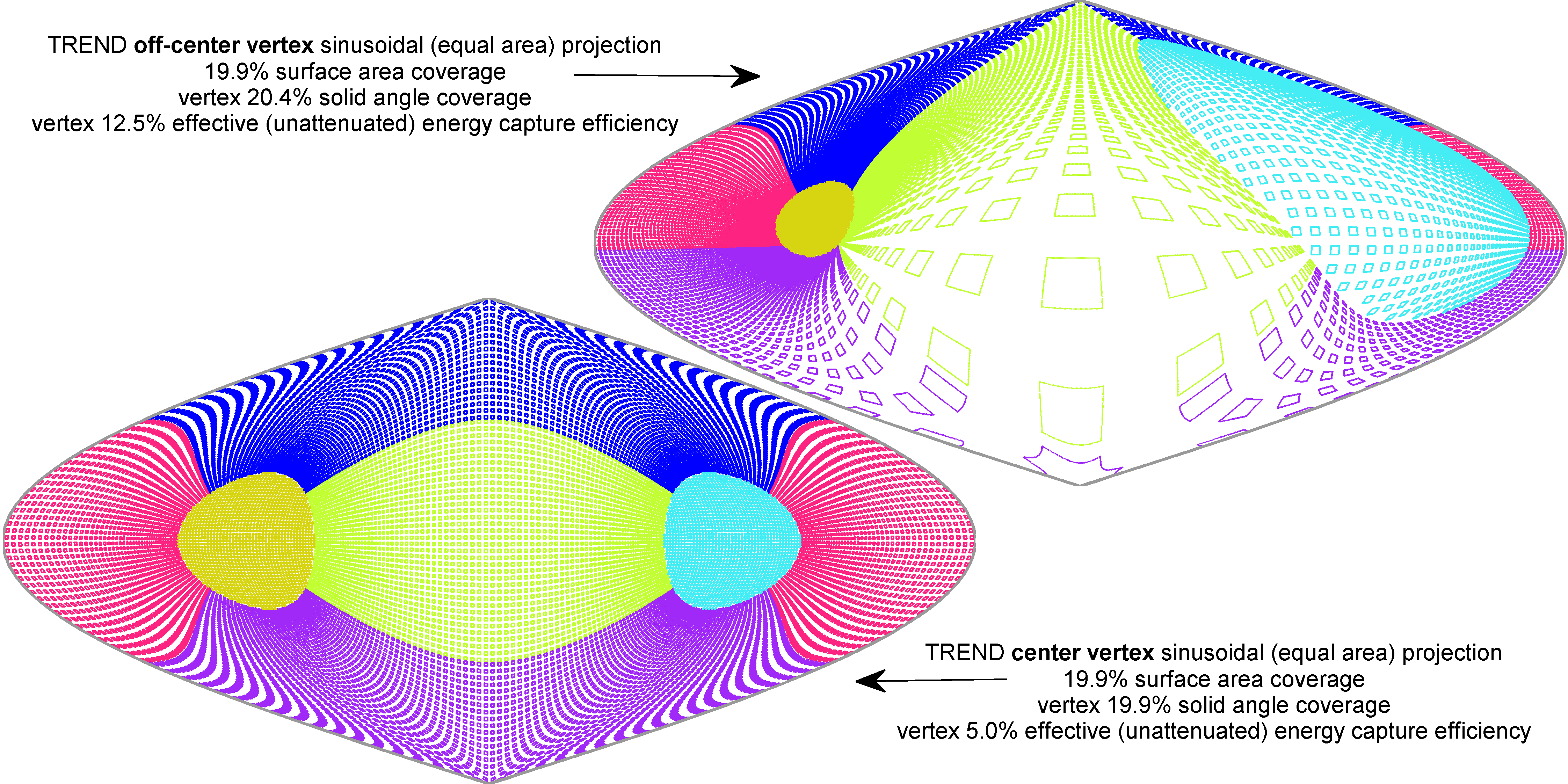}
\caption{2D PMT solid angle projections for center vertex and off-center vertex cases. The off-center vertex case in the upper plot  corresponds to the IBD event shown in Figure \ref{TRENDdetector}. Different colors indicate six different PMT regions: front (green, 3456 PMTs), top (blue, 3456 PMTs), bottom (purple, 3456 PMTs), back (pink, 3456 PMTs), right circular cap (aqua, 1541 PMTs) and left circular cap (yellow, 1541 PMTs). The effective energy capture efficiency values shown include attenuation effects inside the detector. The mean attenuation length modeled for our LS was 21.1m. TREND has a fiducial-volume radius of 23m and a height of 96.5m, so photon attenuation plays a significant role, harming detector-center events more than detector-edge events.}
\label{TRENDsinusoidal}
\end{figure}

Figure \ref{TRENDsinusoidal} shows all 16906 square-shaped PMTs in the TREND detector model, as seen from the perspective of an observer placed at the $\nuebar$ vertex shown in the Figure \ref{TRENDdetector} IBD event. Summing over the angular area within the pixels shown in Figure \ref{TRENDsinusoidal} can provide a useful measure of the \textit{solid angle} coverage from the IBD vertex. The solid angle coverage is naturally a function of the simpler TREND \textit{surface area}\footnote{Note that even though the TREND PMT \textit{surface area} coverage is about 19.9\%, the PMT \textit{solid angle} coverage will vary slightly depending on the observer's location within the TREND detector. As an example, solid angle coverage from the $\nuebar$ vertex near the detector wall shown in Figure \ref{TRENDdetector} is about 20.4\%, while the solid angle coverage as seen by an observer exactly in the center of the detector is 19.9\%, more closely matching the surface area coverage.} coverage (19.9\%), but is a more accurate measure of how much energy the detector can expect to capture from a single IBD event.

A third very useful metric, the \textit{energy capture efficiency}, or ECE, is also shown in Figure \ref{TRENDsinusoidal}. The ECE includes attenuation effects from the fluid, and can predict the total fraction of energy that a neutrino will deposit on the detector PMTs. In this respect it is a much more useful metric than surface area coverage or even solid angle coverage. The ECE will naturally vary as a function of an event's location within the detector. Comparing the ECEs of the two vertices shown in Figure \ref{TRENDsinusoidal}, we can see that a detector-center event in TREND deposits only about 5\% of its energy at the actual PMT surfaces, while the example event near the detector wall deposits much more energy at the PMT surfaces, about 12.5\%. This indicates to us that off-center events are actually preferred to detector-center events, as they will yield greater photon collection per their greater ECE, and they should provide us with consequentially better energy and position resolution.

100,000 events were used for the TREND MCs, 10,000 each at $\nuebar$ energies of 2meV through 11MeV. Some example histograms produced from the 10,000 5MeV MCs can be seen in Figure \ref{MCstats}. Application of select candidate ``cuts" are also shown in Figure \ref{MCstats}, including the neutron delay cut, neutron energy cut and positron-neutron distance cut.

\begin{table} [!htbp]
\fontsize{8}{10}\selectfont
\begin{center}
\begin{tabular}{l|l}
\hline
\textbf{TREND Parameter}				& \textbf{Value}\\
\hline Cylinder Radius 					& 23.0m \\
\hline Cylinder Height  				& 96.5m \\
\hline Volume    						& $\sim$160,000m$^3$ \\
\hline Mass 	 						& 138kT \\
\hline Protons   						& $10^{34}\mathrm{p}^+$ \\
\hline Attempted PMTs					& 17000 \\
\hline Achieved PMTs					& 16906 \\
\hline Optical Coverage    				& 19.9\% \\
\hline PMT Time Resolution     			& $<1$ns \\
\hline PMT Area     					& .203m$^2$ \\
\hline PMT Shape    					& flat square \\
\hline Mean PMT QE						& 23.1\% \\
\hline Peak Absorption Wavelength  		& 425nm \\
\hline GEANT Scintillator     			& 0.5\% Gd doped LS\\
\hline Peak Emission Wavelength  		& 421nm \\
\hline Yield  							& 9200photons/MeV \\
\hline Scintillator Decay Constant  	& 2.20ns \\
\hline Mean Refractive Index    		& 1.58 \\
\hline Mean Attenuation Length 			& 21.1m \\
\hline Mean Re-Emission Fraction		& 13.2\% \\
\hline Measured Cherenkov Fraction      & $\sim 2\%$ \\
\hline Positron energy cut				& $E^{e^+}_\mathrm{vis}<12$MeV\\
\hline Positron-wall distance cut		& $||x_{e^+}-x_\mathrm{wall}||>30$cm\\
\hline Neutron capture cut  			& $12$MeV$>E^{n^0}_\mathrm{vis}>4$MeV\\
\hline Capture energy containment cut   & $E^{n^0}_\mathrm{vis}<6$MeV\\
\hline Neutron-wall distance cut		& $||x_{n^0}-x_\mathrm{wall}||>30$cm\\
\hline Neutron delay cut				& $t_{n^0}-t_{e^+}<200$ns\\
\hline Positron-neutron distance cut    & $||x_{e^+}-x_{n^0}||<2$m\\
\hline Neutron multiplicity cut 		& num$\left(n^0\right)=1$\\
\end{tabular}
\caption{Assumed TREND characteristics used in TREND detector MCs. All candidate cut percentages shown are energy weighted mean values for a 1km range reactor spectra.}
\label{table:TREND1}
\end{center}
\end{table}

\begin{table} [!htbp]
\fontsize{8}{10}\selectfont
\begin{center}
\begin{tabular}{l|l}
\hline
\textbf{Metric} 											& \textbf{TREND MC Results}\\
\hline Positron energy cut									& 99.9\%\\
\hline Positron-geode distance cut							& 97.1\%\\
\hline Neutron capture cut  								& 83.3\%\\
\hline Capture energy containment cut       				& 98.8\%\\
\hline Neutron-geode distance cut							& 97.8\%\\
\hline Neutron delay cut									& 99.8\%\\
\hline Positron-neutron distance cut        				& 96.5\%\\
\hline Neutron multiplicity cut 							& 100.0\%\\
\hline Combined candidate cut								& 78.5\% 		 \\
\hline Mean Neutron Displacement  							& $20.7$mm 	 	 \\ 
\hline Positron Annihilation Point Resolution $1\sigma$  	& $374.5$mm  	 \\
\hline Neutron Capture Point Resolution $1\sigma$  			& $243.7$mm 	 \\
\hline Direction Vector Resolution $1\sigma$  				& $461.3$mm  	 \\
\hline Angular Direction $1\sigma$, n=2500 events			& 37.0$^\mathrm{o}$ 		 \\
\hline SNR  												& 0.045			 \\
\hline Energy Resolution $1\sigma$  						& $\left. 8.9\% \right|_{E_\mathrm{vis}=1\mathrm{MeV}}$ 	\\
\end{tabular}
\caption{TREND MC results, energy dependent values weighted to a 1km range reactor spectrum. Candidate events used in these statistics meet all the CHOOZ candidate criteria specified in Table \ref{table:CHOOZ} with three TREND specific modifications: neutron delay cut (200$\mu$s from 100$\mu$s), positron energy cut ($E_\mathrm{vis}^{e^+}<12$MeV from $E_\mathrm{vis}^{e^+}<8$MeV), and positron-neutron distance cut (2m from 1m).}
\label{table:TREND2}
\end{center}
\end{table}

\begin{figure}[!htbp]
\centering
\includegraphics[width=\linewidth]{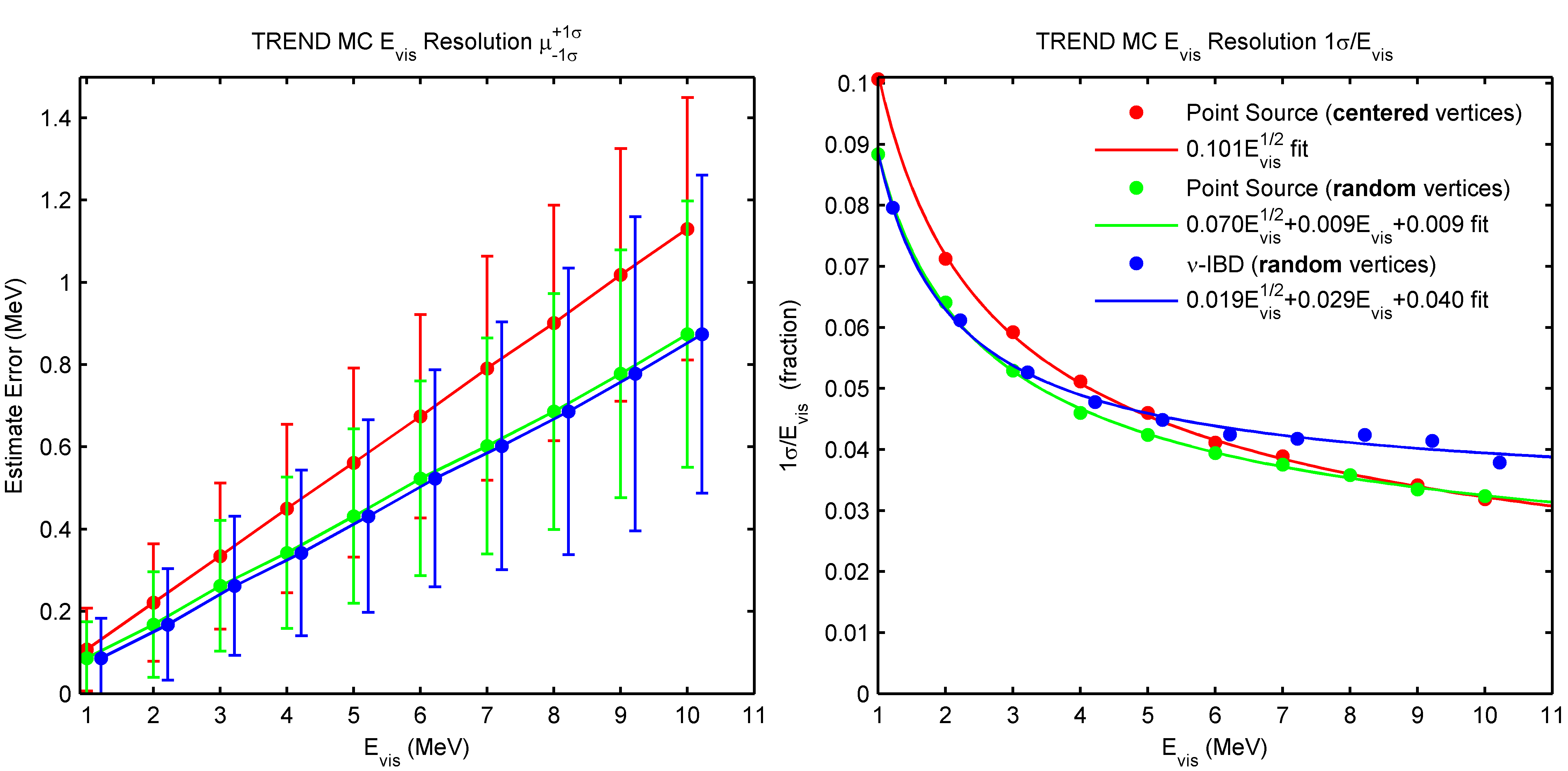}
\caption{TREND Detector energy resolution results. Three energy resolutions are shown: 1. Center-detector point-source resolution (red), 2. Random-location point-source resolution (green), and 3. Random-location $\nuebar$ energy resolution (blue). $\nuebar$ statistics were harvested from candidate events out of 10,000 possible GEANT neutrino IBD events at integer energy levels of 2MeV through 11MeV (100,000 total events). CHOOZ maximum likelihood estimator (MLE) \cite{chooz_2003} was used to estimate energies and locations within TREND Detector Model for all three types of sources. Energy estimate \textit{bias} can be seen in left plot, useful for detector calibration.}
\label{TRENDenergyMCresults}
\end{figure}

The mean TREND energy resolution based on 100,000 randomly placed inverse beta decay events (shown in Figure \ref{TRENDenergyMCresults}) was found to be about $\left. 8.9\% \right|_{E_\mathrm{vis}=1\mathrm{MeV}}$ at E$_\mathrm{vis}$=1MeV. This \textit{random-vertex IBD energy resolution} is not directly comparable to the KamLAND \textit{center-detector point-source energy resolution}, though we \textit{can} directly compare center-detector point-source energy resolutions. In TREND we find a center-detector point-source energy resolution of 10.1\%, about 50\% worse than the cited KamLAND energy resolution of $7.3\%/\sqrt{E_\mathrm{vis}[\mathrm{MeV}]}$. TREND surface area coverage is similar to KamLAND, however in TREND fewer photons reach the detector walls due to the inevitably larger travel distances mandated by a 138kT detector, which the increased attenuation length in TREND apparently does not completely make up for.

Figure \ref{TRENDenergyMCresults} shows the complete TREND energy resolution picture, including measurement bias on the left panel. Estimation biases in this context usually arise from a mismatch between the truth model and the estimator assumptions; in our case this may be caused by several real-world effects occurring in the Detector Model which are absent from the CHOOZ MLE, such as photon re-emission. Such biases would scale almost linearly with $\nuebar$ energy; and indeed in Figure \ref{TRENDenergyMCresults} we do observe bias to be a linear function of $\nuebar$ energy.

Determination of measurement bias in neutrino physics is useful for detector calibration to remove such biases from future energy measurements. Accordingly, in this paper we assume that all such biases have been removed from energy measurements and we concern ourselves solely with measurement \textit{noise}. Three energy resolutions are shown in Figure \ref{TRENDenergyMCresults}: 1. Center-detector point-source resolution (red), 2. Random-location point-source resolution (green), and 3. Random-location $\nuebar$ energy resolution (blue). This figure substantiates our claim that detector-centered events have worse resolution (at least for low-energy events) in very large detectors. 

Figure \ref{TRENDenergyMCresults} also allows us to directly compare IBD energy resolution with point-source energy resolution. Interestingly, we see that IBD energy resolution and point-source energy resolutions are comparable up until about E$_\mathrm{vis}$=3MeV, at which point they begin to diverge as the IBD energy resolution suffers to a greater degree than the point-source energy resolution. We believe this deviation in resolution at the higher $\nuebar$ energies to be a product of several factors:

\begin{enumerate}
\item \textbf{Cherenkov radiation.} Cherenkov radiation is directional in nature (produced in significant quantities by the positron streak and any ``knockoff" electrons), and not suitably compensated for by the simple CHOOZ MLE estimator, which assumes isotropic radiation.
\item \textbf{Extended source vs point source.} IBD events are, naturally, extended sources of radiation, a conglomeration of many smaller point-sources of radiation produced by the IBD positron streak, gammas, and neutron-proton collisions, among others. This extended-source nature violates the point-source assumption of the simple CHOOZ MLE estimator.
\item \textbf{Prompt gamma energy floor.} A portion of an IBD event's energy is not a function of $\nuebar$ energy, but rather a constant value (an energy ``floor") of energy release produced by two prompt gammas created upon positron annihilation. Each gamma releases a fixed 511keV of energy, and this radiation is not a function of the original antineutrino energy, as the simple $kE^{1/2}$ fit equation seeks to provide. Thus we would not expect IBD events to conform to the standard energy resolution fit equation either. This is the reason that we have fit our IBD MC events with some additional constants rather than trying to apply the usual $kE^{1/2}$ fit.
\end{enumerate}

Direction vector resolution for TREND was found to be about SNR=0.045 (weighted for the neutrino energy spectrum of a reactor at a distance of 1km), or 21mm/461mm, from a mean neutron capture point resolution of 243mm and a mean positron annihilation point resolution of 374mm, which combine to form a reconstruction vector resolution of 461mm per cartesian axis. 

Note that the reconstruction vector resolution 461mm $1\sigma$ noise value includes additive noise incurred by neutron random-walk natural to inverse beta decay, and is not simply the norm of the 374mm positron CE $1\sigma$ and 243mm neutron CE $1\sigma$ values\footnote{A simple norm of the 374mm positron CE $1\sigma$ and 243mm neutron CE $1\sigma$ values would produce a false smaller reconstruction vector resolution $1\sigma$ of only 446m.}.

\section{Geospatial Model}
\label{Geospatial Model}
The Geospatial Model calculates the expected rate of detections within the scintillator volume ($\bar{n}$) from all modeled sources at a specified detector location.  This rate includes neutrinos from reactors and the Earth, as well as non-neutrino background which survives all selection criteria applied, and accounts for the modeled energy-dependent detector efficiency and veto-related duty cycle losses.  The model also computes the elevation-azimuth-energy measurement probability density function (pdf) for each detector, \textit{smeared} by the modeled measurement resolution. For simplicity, the measurement resolution found for neutrino events were applied to all detections, including those arising from non-neutrino background.  The total number of detections in each detector, $n_z$, was randomly determined according to Poisson statistics.  The resulting noisy measurements were randomly selected from the appropriate measurement pdf and fed to the estimator detailed in Section \ref{Estimation theory} for parameter estimation.

This process is repeated one thousand times to construct an \textit{a posteriori} parameter space pdf.  The pdf defines the location and thermal power observability of an unknown neutrino source or, in the case of oscillation parameter estimation, the oscillation parameter observability. MC results for an unknown neutrino source are presented in Section \ref{Reactor geolocation results} and oscillation parameter estimation results are presented in Section \ref{Oscillation parameter estimation results}.

The Geospatial Model operates on a three dimensional representation of the Earth.  It uses the National Oceanic and Atmospheric Administration (NOAA) Earth TOPOgraphical 1 (ETOPO1) ``ice" data \cite{ETOPO1} to represent the land/ice surface and the ocean bathymetric data, relative to the World Geodetic System 84 (WGS84) ellipsoid. The ETOPO1 dataset consists of 233 million tiles in a 10800 by 21600 matrix, providing global elevation resolution of 1 arc-minute. For the oceans, the ``surface" was taken to be the zero-tide elevation determined via the National Geospatial-Intelligence Agency (NGA) Earth Gravity Model 2008 \cite{EGM2008} (EGM2008).

Several different coordinate systems are employed by the Geospatial Model. Detections are measured in the local North-East-Down (NED) reference frame.  This reference frame is different for each detector in a given scenario. For estimating the location of an unknown neutrino source, the parameter space is expressed in the WGS84 Earth centered, Earth-fixed reference frame.  This reference frame can be expressed in (Latitude, Longitude, Height) or in a Cartesian coordinate system that rotates with the Earth.  When the unknown source is constrained to the local terrain height, the  parameter space reduces to just two local dimensions, WGS84 Latitude and Longitude.  The standard Direction Cosine Matrix (DCM) defined in Equation \ref{ned2ecef} is used to rotate between the NED and ECEF reference frames. Note that this rotation depends on the detector's geodetic latitude $\phi$ and longitude $\lambda$.

\begin{eqnarray}
\label{ned2ecef}
C_{NED_d}^{ECEF} = \left[ \begin{array}{ccc}-\cos\lambda \sin\phi &-\sin\phi \sin\lambda &\cos\phi	\\
-\sin\lambda &         \cos\lambda &0		\\
-\cos\phi \cos\lambda &-\cos\phi \sin\lambda &-\sin\phi\end{array} \right]
\end{eqnarray}

\subsection{Reactor Neutrinos}

Neutrinos are detected in TREND by the inverse beta decay (IBD) reaction, where a $\nuebar$ interacts with a free proton in the liquid scintillator (\ref{ibd}). 

\begin{equation}
\label{ibd}
\nuebar + p \rightarrow e^+ + n
\end{equation}

\begin{figure}[!htbp]
\centering
\includegraphics[width=.5\linewidth]{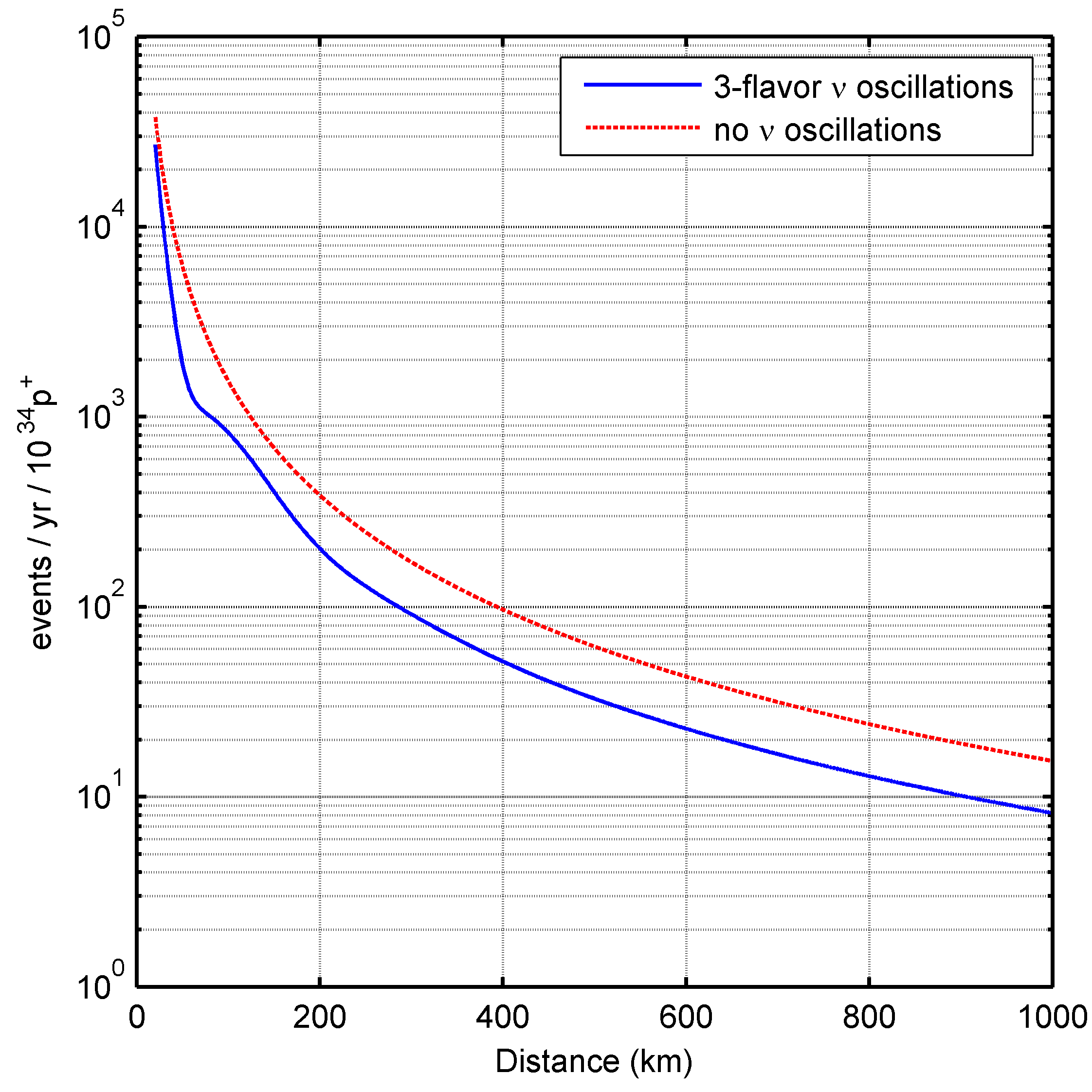}
\caption{Number of predicted $\nuebar$ inverse beta decay events observed over 1 year (100\% duty cycle) by a $10^{34}p^+$ detector, as a function of distance from a 300 MW$_\mathrm{th}$ reactor. This assumes 100\% efficiency for both the detector and the reactor, perfect measurements of neutrino energy, no energy cut, and no background contribution.}
\label{number_of_events}
\end{figure}

In this study, the reactor energy spectrum is approximated as an exponential fall-off with respect to a 2nd order polynomial in neutrino energy (see Equation \ref{JNBeq_18} in Section \ref{Estimation theory}).  The scaling of the total neutrino flux assumes approximately 200 MeV and 6 neutrinos per fission, with an average of two  neutrinos per fission created above the inverse beta decay energy threshold of $E_{\nuebar}\approx 1.8$MeV. These assumptions yield $1.872 \times 10^{20}$ detectable neutrinos per GW$_\mathrm{th}$ per second emitted isotropically from a reactor.  This rate is comparable to that produced by a typical pressurized water reactor at the beginning of its fuel cycle \cite{bernstein_2009}. The TREND yearly predicted $\nuebar$ inverse beta decay observation rate for a 300MW$_\mathrm{th}$ source is shown in Figure \ref{number_of_events} as a function of source range.

For the longer ranges and smaller reactors in this study, it would be difficult to extract much information about fuel mix from the spectrum.  Thus long range detection of reactors should focus upon detection, location, and estimation of the time-averaged power output.  If there is sufficient signal, it might be possible to detect interruptions in the reactor operation over time scales of months and crudely monitor long term power output.

\begin{figure}[!htbp]
\includegraphics[width=\linewidth]{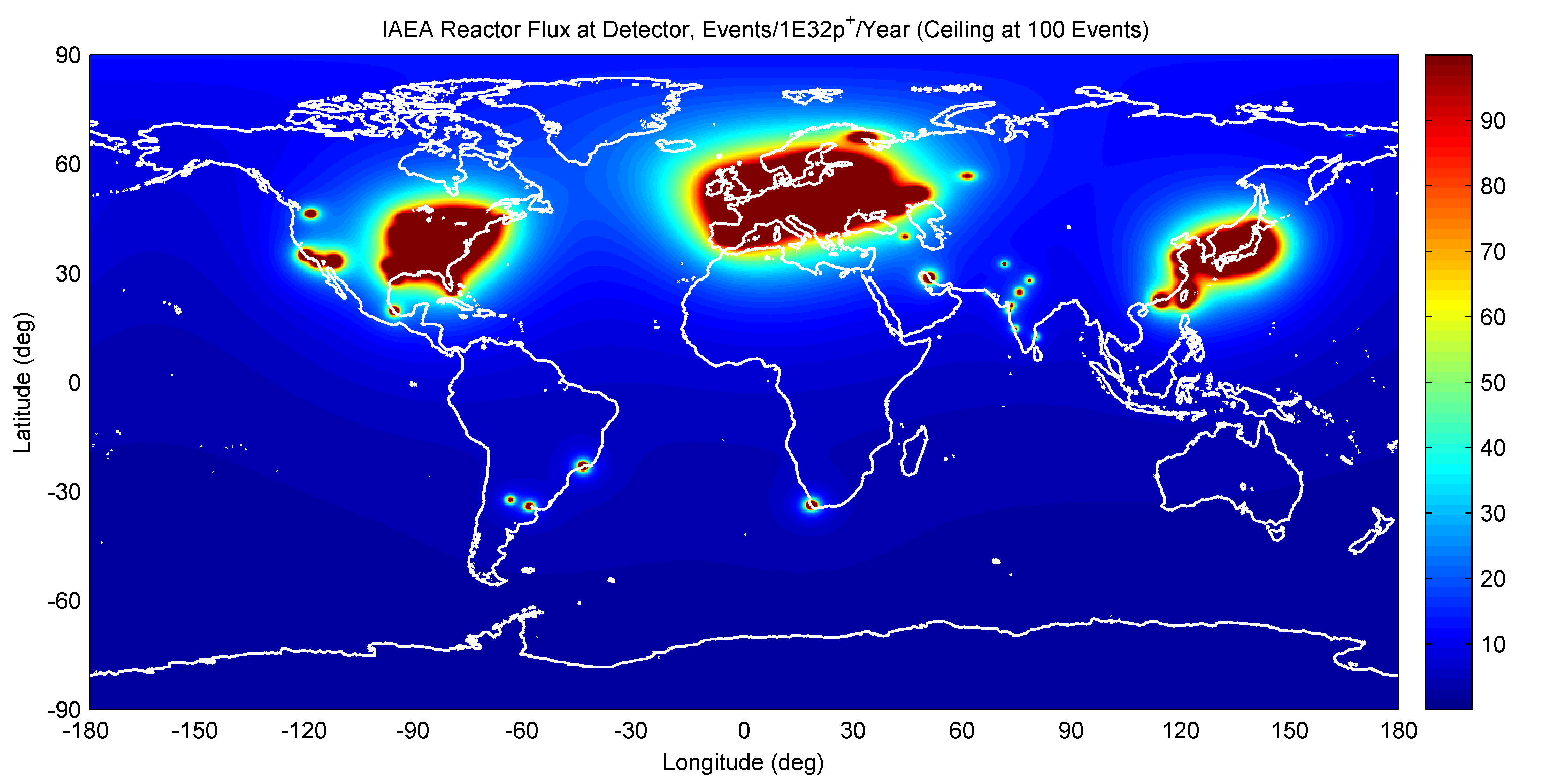}
\caption{IAEA known reactor background for a $10^{32}\mathrm{p}^+$ detector, saturated at 100 events per year.}
\label{worldIAEA}
\end{figure}

The 1999 International Atomic Energy Agency (IAEA) data on nuclear power reactors \cite{IAEA1999} was used to model the location and power of known reactors.  The data include 436 active reactor cores distributed among 206 locations (total 1063GW$_{\mathrm{th}}$) and 35 reactor cores (total 93GW$_{\mathrm{th}}$) under construction. The reported electrical power was converted to thermal power assuming an efficiency of 33\%, regardless of the reactor design.  A reactor duty cycle of 80\% was assumed and applied in the Geospatial Model. The neutrino event rate per 10$^{32}$ protons per year (assuming 100\% efficiency) due solely to these active reactor cores can be seen in Figure \ref{worldIAEA}.

\subsection{Geo-neutrinos}
The underlying interior structure and composition of the Earth is, in some regards, still poorly understood. The concentration and distribution of radio-isotopes, whose decay chains are capable of producing significant neutrino flux, does not escape this uncertainty. Therefore modeling of the distribution, energy spectra, and total flux of geo-neutrinos remains a challenging task on its own. The Geospatial Model does not attempt to be a hi-fidelity model for geo-neutrino research, but rather it provides for a practical representation of this complex and significant background neutrino source.

The modeling of this flux was broken down into the mantle and the crust. The Earth's core was assumed to have no significant contribution to neutrino flux (i.e. no geo-reactor) in this analysis. For the mantle (radii from 3480km to 6291km), the spherically symmetric density profile in the Preliminary Reference Earth Model \cite{PREM} (PREM) was used.  For elemental abundances, the two-layer stratified model suggested by Fiorentini \textit{et al.} \cite{fiorentini} was used.  For the mantle above a depth of 670 km, the elemental abundances were 6.5 ppb, 17.3 ppb, and 78 ppm for $^{238}$U, $^{232}$Th, and $^{40}$K respectively.  Below this, the abundances were 13.2 ppb, 52 ppb, and 160 ppm, respectively.  The mantle isotope abundances were assumed to be constant with values of approximately 99.3\%, 100\%, and 0.01\% for $^{238}$U, $^{232}$Th, and $^{40}$K respectively.

\begin{figure}[htbp]
\includegraphics[width=\linewidth]{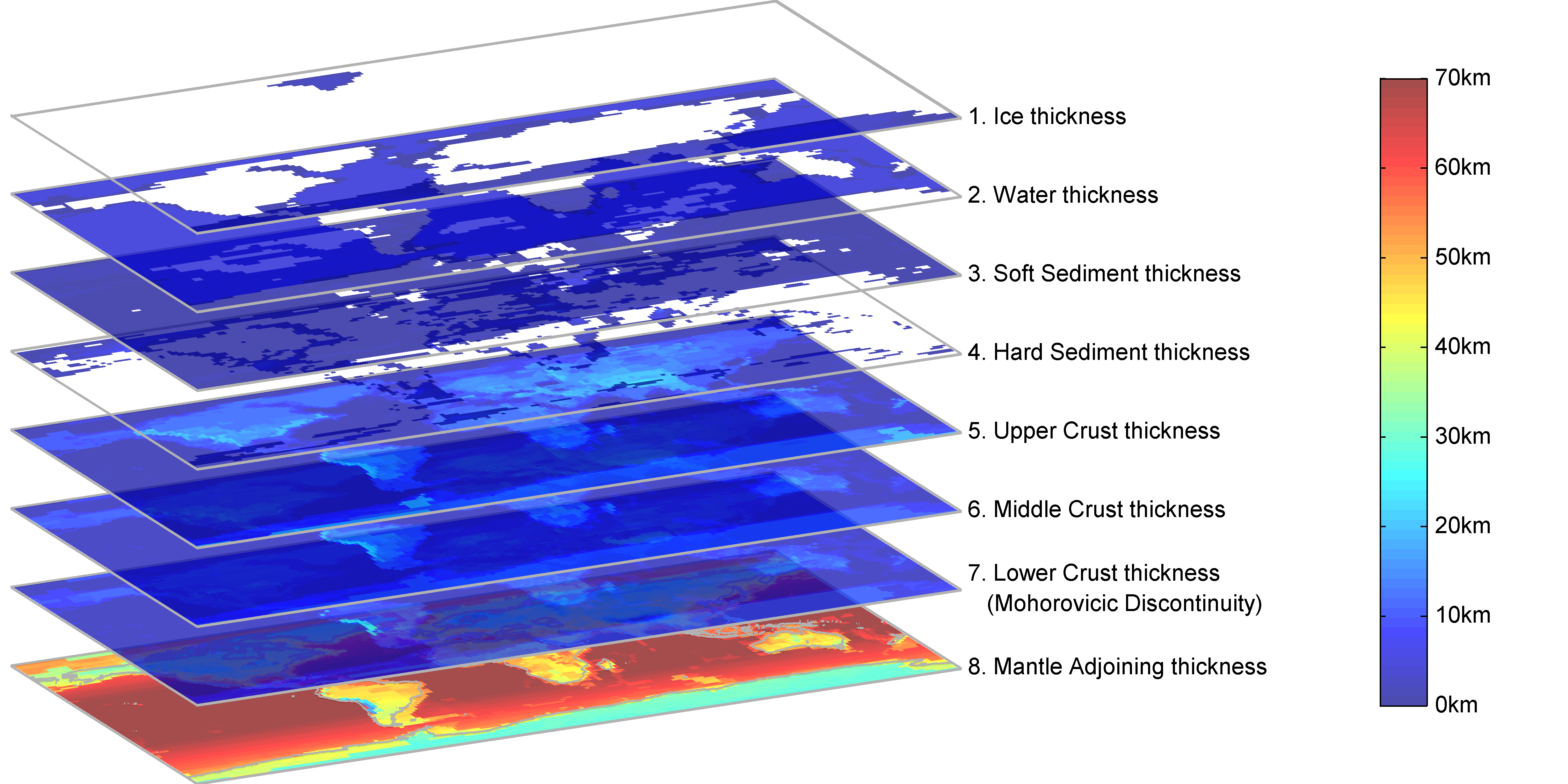}
\caption{CRUST 2.0 + Mantle Adjoining Layer thickness, in km. The Mantle Adjoining Layer connects the ellipsoidal 7th CRUST 2.0 layer, the ``lower crust (Mohorovicic discontinuity)", to the spherical upper mantle. Each 8th layer tile's thickness is defined by the gap between the spherical mantle below it and the floor of the 7th layer tile above it. Zero-thickness tiles are omitted from this figure.}
\label{crust_thickness}
\end{figure}

The Geospatial Model uses CRUST 2.0 \cite{crust2.0} to describe the Earth's crust.  This model consists of seven layers (to which we add an 8th) of $2^\circ \times 2^\circ$ tiles beginning at the ETOPO1 surface and typically descending to depths of 30km-70km. The lowest points of the 7th layer tiles range from 6302km Earth radius at the poles to 6366km Earth radius near the equator.  A spherical mantle is assumed in the Geospatial Model, and seamlessly joined to the CRUST 2.0 by introducing an 8th ``Mantle Adjoining" layer with properties identical to the upper mantle.  Figure \ref{crust_thickness} shows how the thickness of the CRUST 2.0 and Mantle Adjoining Layer tiles varies across the Earth. Typical CRUST 2.0 tiles are a few km thick, while some layers (like 1. Ice) are primarily made up of zero-thickness tiles across wide swaths of the Earth's surface.  Our Mantle Adjoining Layer tiles are considerably thicker than the CRUST 2.0 tiles. The thinnest Mantle Adjoining Layer tile, at about 11km thick, can be found beneath the Himalayas in Nepal, while the thickest, at 76km thick, is found right at the equator. These Mantle Adjoining tiles create a seamless Earth geo-neutrino source model with no air gaps.

The eight layers we assume are:
\begin{enumerate}
\item Ice
\item Water
\item Soft Sediment
\item Hard Sediment
\item Upper Crust
\item Middle Crust
\item Lower Crust (Mohorovicic Discontinuity)
\item Mantle Adjoining Layer (not part of CRUST 2.0)
\end{enumerate}

\begin{figure}[htbp]
\includegraphics[width=\linewidth]{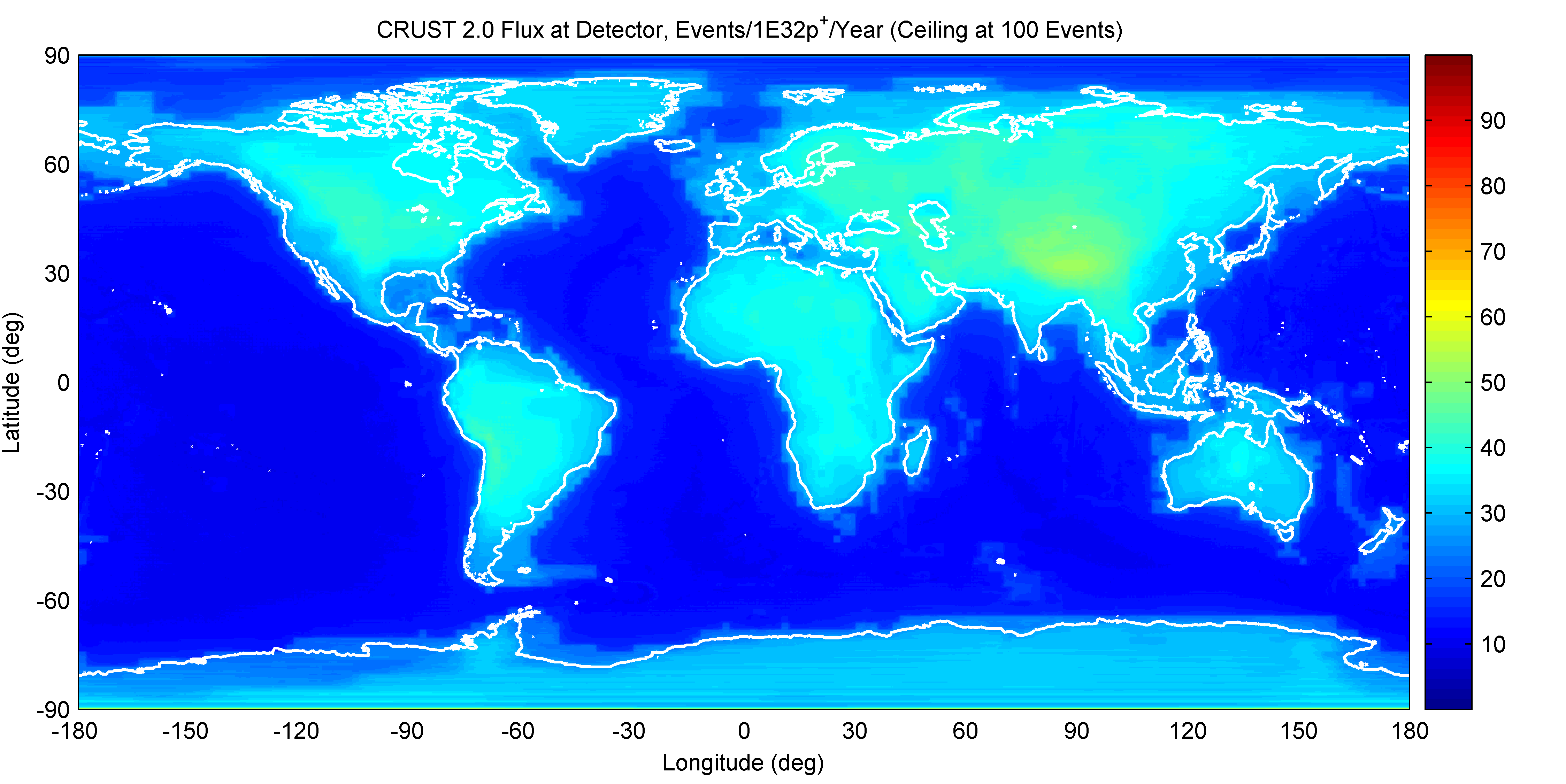}
\caption{Crust background for a $10^{32}\mathrm{p}^+$ detector, saturated at 100 events per year.}
\label{worldcrust}
\end{figure}

\begin{figure}[htbp]
\includegraphics[width=\linewidth]{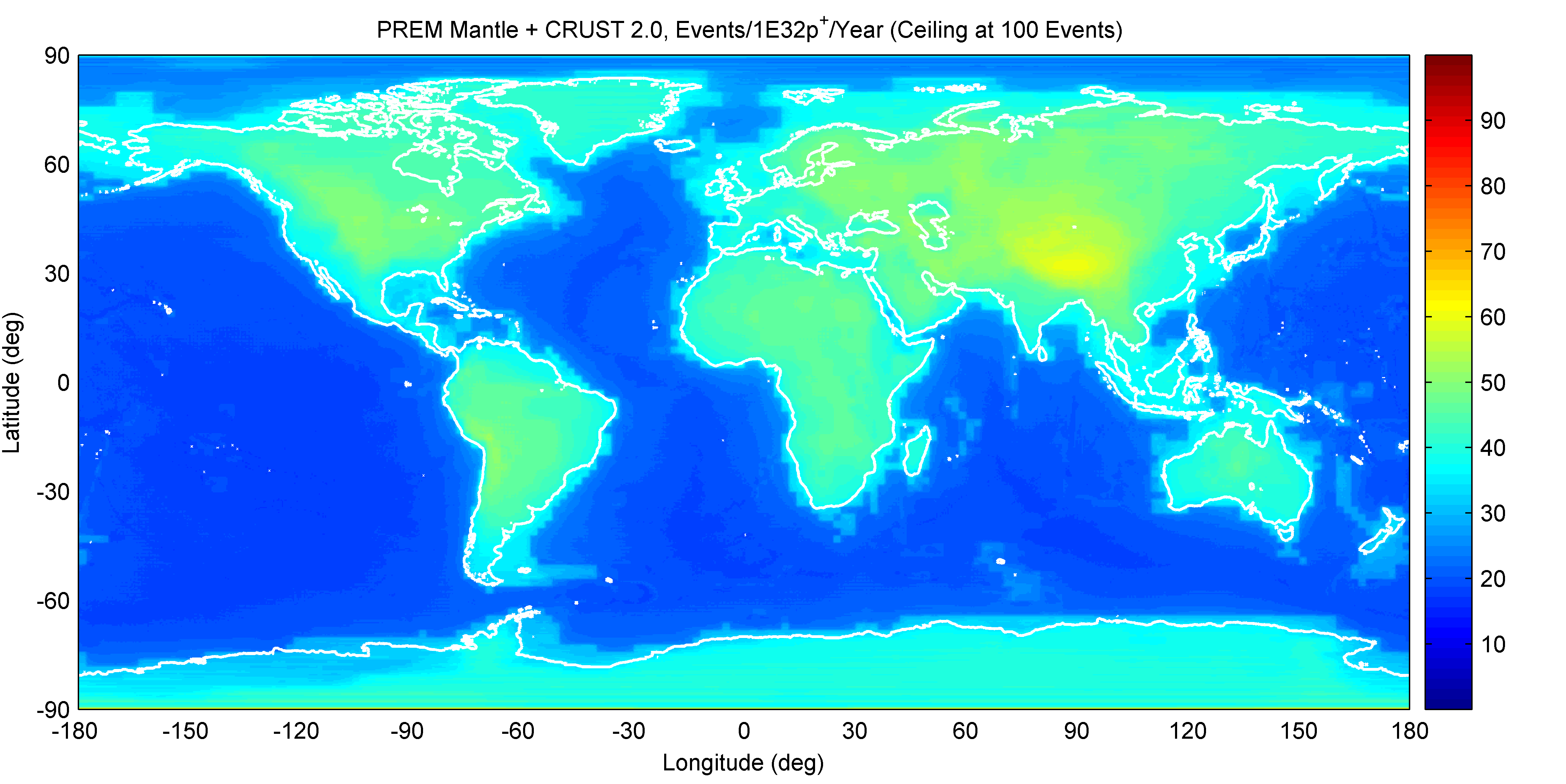}
\caption{Crust+Mantle background for a $10^{32}\mathrm{p}^+$ detector, saturated at 100 events per year.}
\label{worldcrustmantle}
\end{figure}

\begin{figure}[htbp]
\includegraphics[width=\linewidth]{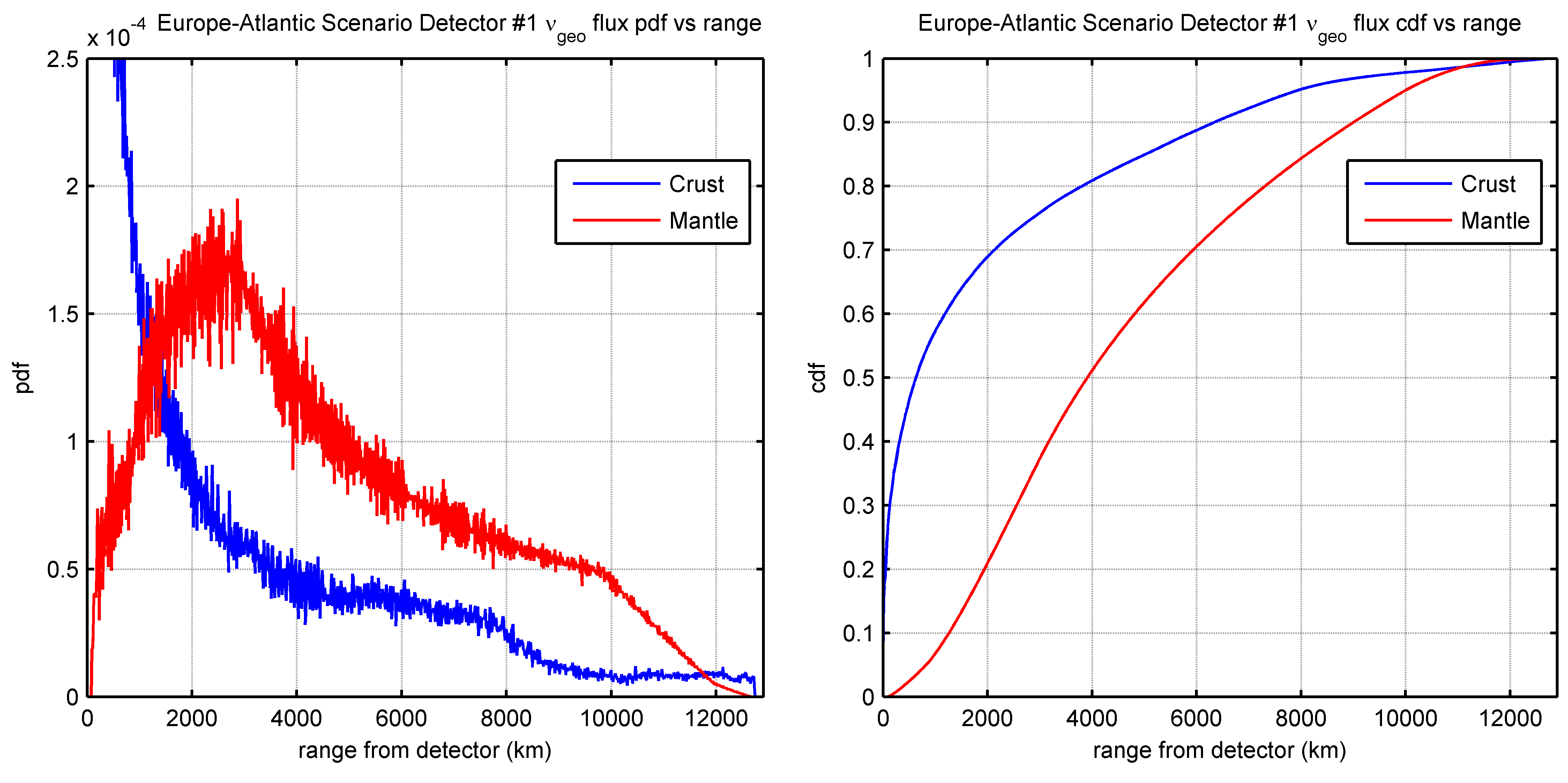}
\caption{Random Crust and Mantle geo-neutrino flux vs range from a TREND detector off the coast of Europe-Atlantic. Refer to Section \ref{High reactor background (Europe-Atlantic)} for more information concerning this detector's placement and its environment.}
\label{geopdf}
\end{figure}

\begin{figure}[htbp]
\includegraphics[width=\linewidth]{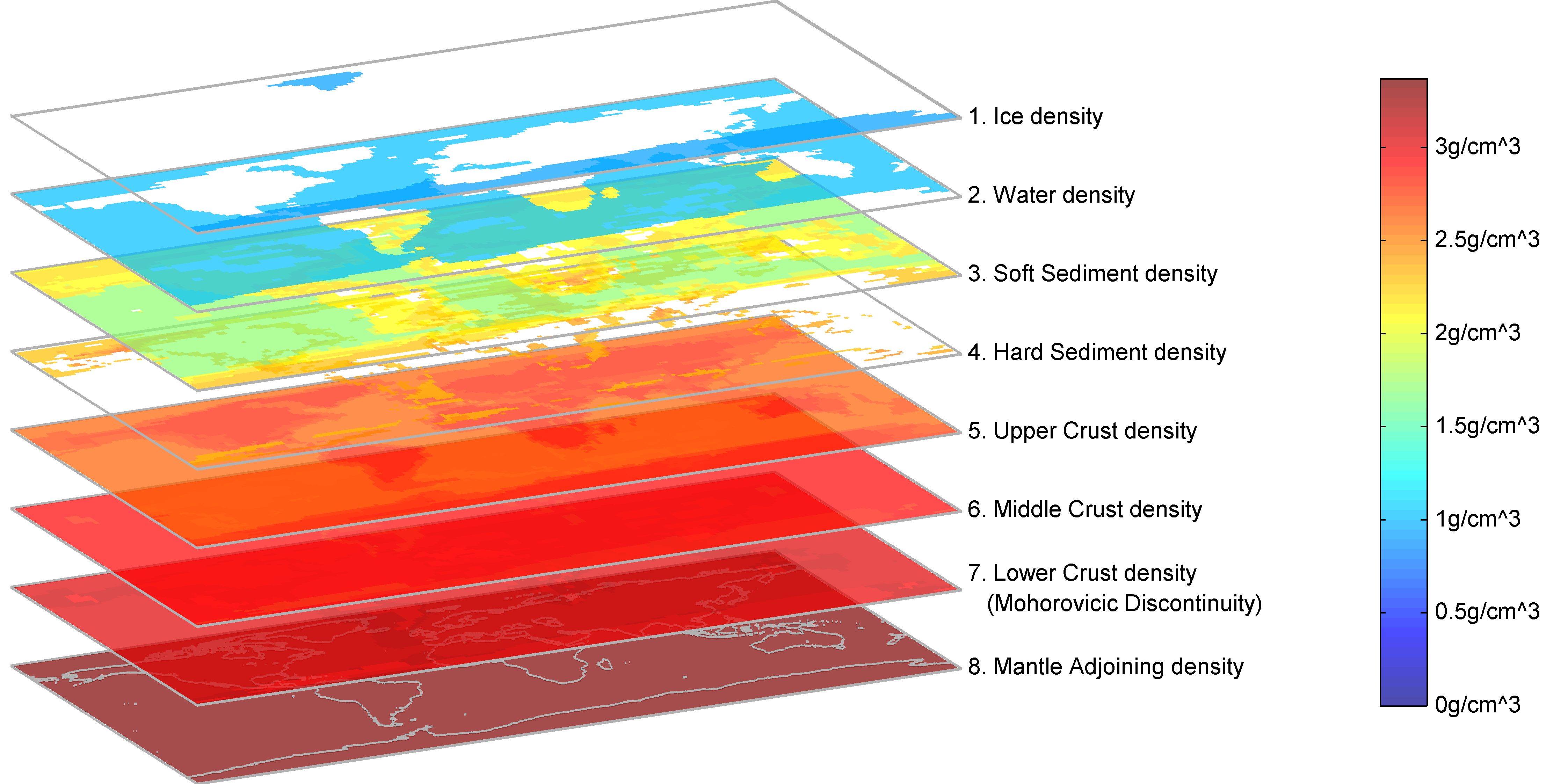}
\caption{CRUST 2.0 + Mantle Adjoining Layer density, in g/cm$^3$. Density is seen to generally increase as depth increases. The top ice layer naturally shows the lowest density, slightly under g/cm$^3$, while the bottom Mantle Adjoining Layer density is about 3.4g/cm$^3$. Zero-thickness tiles are omitted from this figure.}
\label{crust_density}
\end{figure}

\begin{figure}[htbp]
\includegraphics[width=\linewidth]{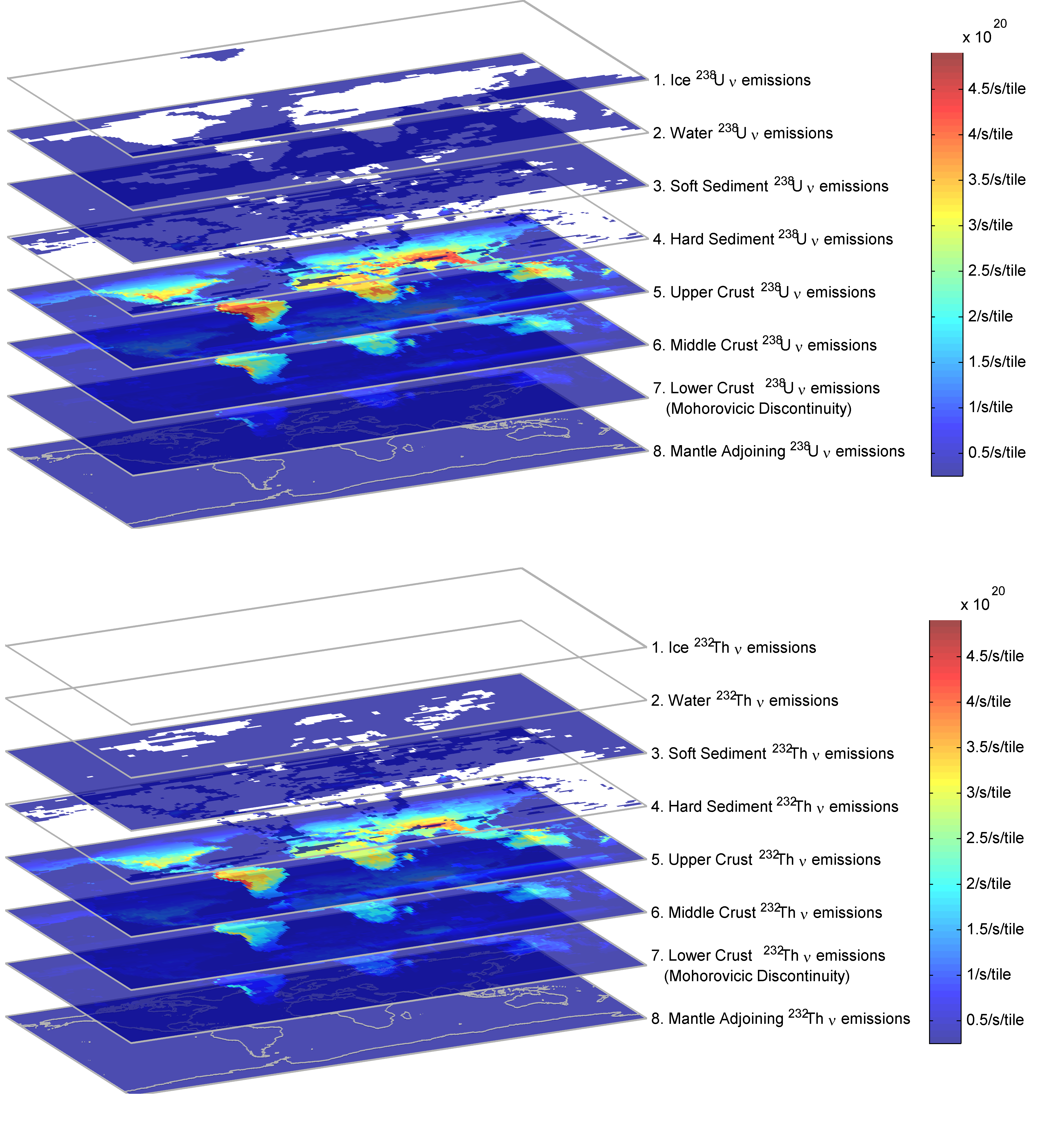}
\caption{CRUST 2.0 + Mantle Adjoining Layer emissions ($^{238}$U in upper, $^{232}$Th in lower), in $\nuebar$ per second per tile. This figure combines density and thickness information from CRUST 2.0 \cite{crust2.0} and PREM \cite{PREM} with elemental abundances from Mantovani \textit{et al.} \cite{mantovani_2004}, (along with $^{232}$Th half life of $20.3 \times 10^{9}$ years and multiplicity of 4$\nuebar$ per $^{232}$Th decay, and $^{238}$U half life of $6.45 \times 10^{9}$ years and multiplicity of 6$\nuebar$ per $^{238}$U decay) to compute the total $\nuebar$ emissions from a tile per second isotropically. Layers 5 and 6, the Upper Crust and Middle Crust emit the most neutrinos due to their high elemental abundance \cite{mantovani_2004}, despite having lower density and thickness (as seen in Figures \ref{crust_density} and \ref{crust_thickness}) than many other layers.}
\label{thorium_emissions}
\end{figure}

Each crust tile has a defined location, thickness (shown in Figure \ref{crust_thickness}), and density (shown in Figure \ref{crust_density}). Elemental abundances for each layer were taken from Mantovani \textit{et al.} \cite{mantovani_2004}. Using the CRUST 2.0 densities and volumes, the Mantovani \textit{et al.} elemental abundances for the appropriate layer, and the same constant isotope abundances assumed for the mantle, the total mass of each isotope of interest can be calculated for each tile.  The relationship between total abundance and neutrino flux is determined from isotope half-life and multiplicity (the number of neutrinos emitted per decay).
The complete neutrino emissions calculated for each tile in each crust layer is shown in Figure \ref{thorium_emissions} for $^{238}$U and $^{232}$Th.
%, and in Figure \ref{thorium_emissions} for $^{232}$Th.

The neutrino energy spectra for the isotopes considered were taken from Enomoto \cite{Enomoto_Sanshiro}.  The flux and energy spectrum, together, fully define the neutrino source signal (assumed to be isotropic) for each tile and for each isotope.  In Section \ref{Optimal MAP estimator} these source spectra are combined with the detector cross section, defined in Equation \ref{JNBeq_19} and shown in Figure \ref{twoXsections}, to create the observed geo-neutrino spectrum for each tile and isotope.

The computation burden of modeling the crust geo-neutrino signal, even at this level of fidelity, is quite significant.  For each detector in a scenario, there are 129,600 (90x180x8) observed energy spectra for each isotope.   Each spectra must be distorted by neutrino oscillations with the appropriate range.  Additionally, the crust tiles near a given detector are too large (about 200km across at the equator) to be adequately modeled as point sources.  Because of this fact, a ``smart" discrete integration process was developed to recursively subdivide the nearest tiles into progressively smaller sub-tiles until the contribution of each is below a threshold of 0.1 events per detector-year. With the large detector sizes in this study, the creation of hundreds of thousands of additional sub-tiles to replace the nearest high-flux crust tiles is required.  Figure \ref{fig4_smartIntegration2} is a Google Earth overlay showing the uppermost layer of these sub-tiles following smart integration after TREND placement within the Geospatial Model.

The process described for the crust must be repeated for the mantle in order to compute the total geo-neutrino signal at a given detector.  The mantle is much easier since spherical symmetry is assumed here and because the ``smart" discrete integration process is rarely needed. 
The Geospatial Model can predict the geo-neutrino flux for detector placements at locations around the Earth. The mantle component will vary neutrino flux only with detector depth, however the crust component will vary neutrino flux with detector latitude, longitude \textit{and} depth. Figure \ref{worldcrust} shows the observed crust-only measurement rate calculated for a $10^{32}\mathrm{p}^+$ detector located anywhere in the world and Figure \ref{worldcrustmantle} shows the combined crust+mantle geo-neutrino flux on the same scale.  Figure \ref{geopdf} shows MC geo-neutrino flux contributions separately for the crust and mantle as a function of range away from a detector placed off the southern coast of Europe-Atlantic\footnote{The Europe-Atlantic scenario is one of four used for  reactor searches in this paper, and Detector \#1 shown in Figure \ref{geopdf} is discussed in detail in Section \ref{High reactor background (Europe-Atlantic)}.}.

%\subsubsection{Geo-reactor}
%GEOREACTOR DISCUSSION HERE.

\subsection{Non-neutrino background}
\label{Non-neutrino background}

Non-neutrino background can be categorized into two types. The first type is one where a single complex event mimics both the prompt and delayed signals. The second type is one where the prompt and delayed signals are of separate uncorrelated origin, the so called ``accidental" background. Accidental background may be caused by internal residual radioactivity, cosmic rays (only muons are important at depths greater than a few meters) passing through the detector, cosmic rays outside the detector which make products such as fast neutrons which enter the detector, and finally radioactivity outside the detector, largely gamma rays being the concern (alphas and electrons do not travel far). The three specific components making up the total non-neutrino background considered in this study are:

\begin{enumerate}
\item \textbf{Accidentals.} Accidental prompt and delayed signals are caused by two separate uncorrelated sources, and occur close enough in time and space to fool a typical inverse beta decay filter.
\item \textbf{Fast Neutrons.} A fast neutron may mimic an inverse beta decay prompt signal during its random walk process, recoiling off free protons in the liquid scintillator, and then when finally captured on a dopant (such as Gd) it may produce a signal indistinguishable from a normal inverse beta decay delayed signal.
\item \textbf{Cosmogenic $^9$Li/$^8$He.} Cosmogenic isotopes produced by showering muons fall into this category, and are most problematic due to the long lifetimes of $^8$He and $^9$Li isotopes in the detector, which may trick a time- or space-based coincidence filter designed to detect inverse beta decay events.
\end{enumerate}

Fortunately, inverse beta decay is a very distinctive process that involves a spatial coincidence on the order of a meter and the temporal coincidence on the order of 10 $\mu$s between the prompt (positron) and delayed (neutron capture) signals, followed by a neutron capture providing a known energy release. Thus, a ``delayed coincidence filter" can be used in the data processing to dramatically reduce the number of background events mistaken for neutrinos.  This and other processing filters reduce the non-neutrino background, usually at the expense of reducing the effective fiducial volume, reducing the duty cycle, and/or decreasing the detection efficiency for true neutrino sources. 

In order to estimate the count rate of the non-neutrino background in the detector, a series of scaling relationships were established based upon KamLAND and Borexino background measurements.  Table \ref{table:backgrounds} shows a summary of these estimated rates, along with the predicted reactor neutrino and geo-neutrino background rates.  The cosmogenic $^9$Li/$^8$He and fast neutron count rates and uncertainties were carefully scaled from KamLAND \cite{koichi_2007} and the accidental count rate and uncertainty from Borexino \cite{Borexino_2010}.  The geo-neutrino and reactor neutrino background count rates were computed as described in Section \ref{Geospatial Model} for one of the scenarios.  The background count rates are given for the entire fiducial volume (no fiducial cuts) and for one year of live-time.  The actual background counts expected over one year must include reductions caused by the muon-related and cosmogenic-related downtimes, which are very sensitive to detector depth.   Note that the geo-neutrino signal does not disappear entirely with the $E_\mathrm{vis}\geq 2.6$MeV energy cut due to blurring from the energy measurement resolution.

\begin{table*}
\fontsize{8}{10}\selectfont
\begin{center}
\begin{tabular}{l|cc|ccc}
\hline
 & \textbf{KamLAND} & \textbf{Borexino} & 
\multicolumn{3}{c}{\textbf{TREND extrapolated from}} \\
 & & & \multicolumn{3}{c}{\textbf{KamLAND} \&  \textbf{Borexino}}  \\
\hline
Flat eq. depth & 2,050~m & 3,050~m &  \multicolumn{3}{c}{3,500~m} \\
\hline
Scintillator & C$_{11.4}$H$_{21.6}$  &C$_9$H$_{12}$ &  
\multicolumn{3}{c}{C$_{16}$H$_{30}$} \\
H/m$^3$ & $6.60\;10^{28}$& $5.30\;10^{28}$&  \multicolumn{3}{c}{$6.24\;10^{28}$} \\
C/m$^3$ &$3.35\;10^{28}$& $3.97\;10^{28}$ &  \multicolumn{3}{c}{$3.79\;10^{28}$} \\
density & 0.78 & 0.88 &  \multicolumn{3}{c}{0.86} \\
\hline
Mass (tons) & 912& 278 &  \multicolumn{3}{c}{138,000} \\
Volume (m$^3$) & 1170& 316 &  \multicolumn{3}{c}{160,000} \\
Radius (m) & 6.5& 4.25 &  \multicolumn{3}{c}{23} \\
Cyl. Length (m) &  --- & --- & \multicolumn{3}{c}{96.5}  \\
$\mu-$Section (cm$^2$) &  $1.3\;10^{6}$& $0.57\;10^{6}$ &  
\multicolumn{3}{c}{$4.4\;10^{7}$} \\
\hline
$\mu-$Flux (cm$^{-2}$s$^{-1}$) & $1.6\;10^{-7}$& $0.3\;10^{-7}$ &  
\multicolumn{3}{c}{$1.4\;10^{-8}$} \\
$\mu-$Energy (MeV) &219& 276 &  \multicolumn{3}{c}{295} \\
$\mu-$Rate (s$^{-1}$) & $2.13\;10^{-1}$ & $1.6\;10^{-2}$ &  \multicolumn{3}{c}{$6.2\;10^{-1}$} \\
$\mu-$DT (200~$\mu$s) & $4\;10^{-5}$ & $0.3\;10^{-5}$ &  \multicolumn{3}{c}{$12\;10^{- 5}$} \\
Co-DT (1500~ms) & $1.0\;10^{-1}$& $7.5\;10^{-3}$ &  \multicolumn{3}{c}{$1.5\;10^{-1}$} \\
\hline
Exposure (H.y) & $2.44\;10^{32}$ & $6.02\;10^{30}$ & \multicolumn{3}{c}{$10^{34}$} \\
\hline
$E_\mathrm{vis}$ threshold [MeV] & 0.9 & 1 & 0.9 & 1 & 2.6\\
\hline
Accidental Rate & 80.5$\pm$0.1 & 0.080$\pm$0.001 & 277$\pm$3 & 183$\pm$2 & 0.3$\pm$0.003 \\
%Uncertainty &  0.1&	0.001	&  4 &	2 \\
$^9$Li/$^8$He Rate &13.6$\pm$1.0 &0.03$\pm$0.02 & 20$\pm$1 & 20$\pm$1 & 18$\pm$1\\
%Uncertainty & 1&0.02&12.2&	71.9\\
Fast n Rate & $<$9 (4.5$\pm$2.6)& $<$0.05 (0.025$\pm$0.014)  &31$\pm$18 & 31$\pm$18 & 26$\pm$15\\
%Uncertainty & 9& 0.025	&17& 9\\
Geo-$\nu$ & 69.7 & 2.5$\pm$0.2 & 2,111 & 2,088 & 18 \\
%Uncertainty & ---   & 0.2 &  ---    & 332  \\
\hline
Known Reactor-$\nu$ & --- & --- & 499 & 499 & 352 \\
%Uncertainty & 51  & 0.3 &  2,080    & 498  \\
\hline
\end{tabular}
\caption{Breakdown of the expected background count rates for TREND. 
The $\mu-$DT and Co-DT are the estimates of muon and cosmogenic induced downtime (DT). 
The flat equivalent depths and muon fluxes are taken from \cite{mei_2006}.}
\label{table:backgrounds}
\end{center}
\end{table*}

The accidental rate was modeled as having two components: one that is uniform within the fiducial volume (dictated by the radiopurity of the LS) and another that is an exponential function of the (closest) distance from the inner-most structure.   The scale length of the exponential component (0.3~m) and its magnitude relative to the uniform component at the edge of the structure (1,400) were inferred from KamLAND measurements of the accidental rate using various fiducial cuts \cite{miletic_2009}.  The shape of the exponential component was assumed to hold in both Borexino and TREND, with the absolute scaling being determined by a fit to the Borexino measurements.  For the TREND detector, there is not a significant fiducial volume cut, so the predicted accidental rate includes regions of increased rate (per unit volume) near the edges of the cylinder.

The fast neutron rate was modeled as an exponential function of the (closest) distance from the outer structure that might produce fast neutrons without having them be tagged to a passing muon.  The exponential curve was defined such that 90\% of the rate (per unit volume) comes from within 3 m of the ``edge''.  This is somewhat similar to the assumption made by Lasserre {\it et al} \cite{lasserre_2010} that ``all'' of the fast neutron rate comes uniformly from within 3 m of the ``edge''.  This corresponds to a scale length of 1.3 m for the exponential.   The shape of the exponential was assumed to hold in both KamLAND and TREND, with the absolute scaling being determined by a fit to the KamLAND measurements.  For the TREND detector, this ``edge'' was considered to be the cylinder itself. 

In constructing Table \ref{table:backgrounds}, several significant discrepancies with the background calculations of Lasserre \textit{et al.} \cite{lasserre_2010} were found.  The first discrepancy involved the calculation of the expected cosmogenic $^9$Li/$^8$He count rate.  Lasserre \textit{et al.} scaled the expected cosmogenic $^9$Li/$^8$He count rate from KamLAND without accounting for the fact that KamLAND uses a different veto time (2s) than the $\sim$600ms time proposed for SNIF.  This lead Lasserre \textit{et al.} to significantly underestimate the expected cosmogenic count rate.  This, in turn, contributed to their conclusion that SNIF can be deployed at depths as 
shallow as 1500m.  In order to reduced the count rate back to an acceptable level, the veto time was increased from 600ms in SNIF to 1500ms in TREND.

\begin{figure}[htbp]
\centering
\includegraphics[width=\linewidth]{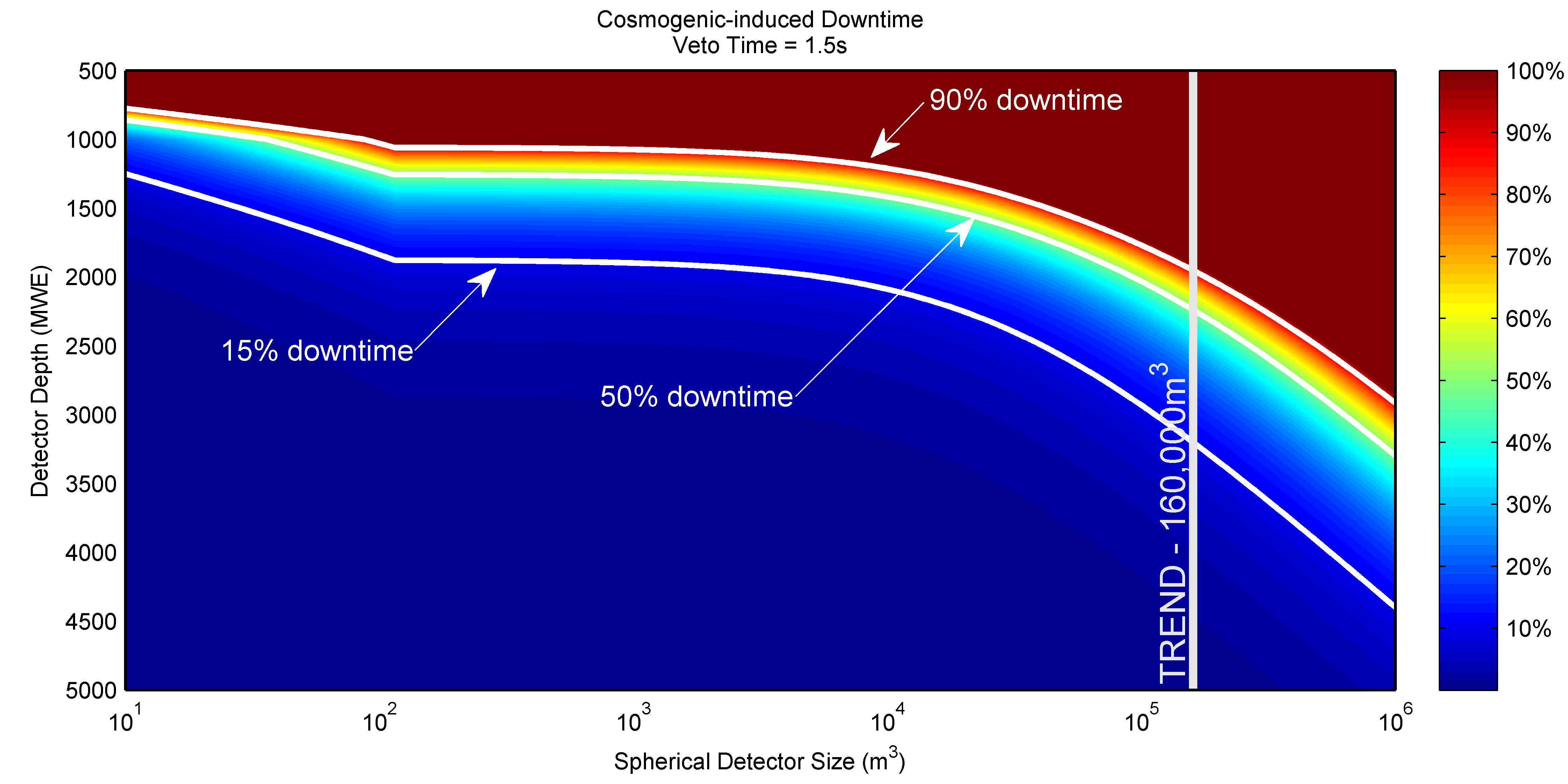}
\caption{Downtime due to Cosmogenic $^9$Li/$^8$He veto as a function of detector depth in meters of water equivalent and detector volume.  The dashed white lines represent downtimes of 15\%, 50\%, and 90\%.   The solid white line indicates the 160,000 m$^3$ fiducial volume of TREND.  The discontinuity below 113 m$^3$ occurs because the detector becomes too small to use the r=3m veto cylinder; the entire volume is vetoed even for well-reconstructed muon tracks.  Note: this figure assumes spherical detector geometry, that 0.17\% of muons are showering per meter of average muon path length (regardless of depth), and that the entire volume is vetoed for showering muons.  The calculations for TREND use the actual cylindrical geometry and a conservative estimate of 15\% showering muons; this yields a larger downtime than implied by this figure (e.g. 15\% downtime at a depth of $\sim$3,500m).}
\label{snifdt}
\end{figure}

The second discrepancy involved the calculation of downtime related to cosmogenic $^9$Li/$^8$He veto.  In their calculations, Lasserre \textit{et al.} vetoed  all muons with a 3m radius cylinder through the detector along the track; they never veto the entire fiducial volume for their proposed 600ms veto time. This did not impose a penalty, in terms of reduced duty cycle, for the large unsegmented design.  Based on KamLAND, approximately 1.5\% of muons passing through the detector were showering\cite{koichi_2007} for an average muon path length of approximately 8.7m.  Scaling this to the average muon path length computed for TREND ($\sim$37m) yields a prediction that 6.4\% of the muons are showering.  This prediction is highly uncertain and the scaling ignores any dependence on the average muon energy with depth.  To be conservative, we assume 15\% of the muons passing through TREND are showering (regardless of depth) and subject to veto of the entire fiducial volume.  These assumptions caused our estimate of the cosmogenic downtime to be significantly larger.  Figure \ref{snifdt} shows the computed cosmogenic downtime percentage as a function of depth (for a veto time of 1500ms).  This analysis indicates that a depth of approximately 3500m is required to reduce the cosmogenic downtime to values less than 15\%.  A deeper ocean site increases the stand-off distances involved, making the TREND concept even more challenging in practice.  This study will show, however, that TREND can still be effective using the estimator presented in this paper.

The third discrepancy involved the calculation of the expected accidental count rate. Lasserre \textit{et al.}
scaled the accidental count rate from Borexino without accounting for the fact that Borexino uses a different delayed coincidence filter 
($\tau$=1280 $\mu$s, $\Delta$R=1 m) than the filter they proposed ($\tau$=30 $\mu$s, $\Delta$V=1 m$^{3}$).  This lead 
Lasserre \textit{et al.} to significantly overestimate the expected accidental count rate.
While this is good news for the technical feasibility of TREND, Lasserre \textit{et al.} correctly point out that a concept as futuristic as 
TREND can safely expect the higher level of Borexino's radiopurity due to technological improvements over the next 30 years.
Based on detailed modeling of the TREND detector design, a modified delayed coincidence filter is proposed here ($\tau$=200 $\mu$s, $\Delta$R=2 m) to balance the competing desires to minimize the expected accidental count rate while 
preserving a relatively high neutrino detection efficiency.

\begin{figure}[htbp]
\centering
\includegraphics[width=.5\linewidth]{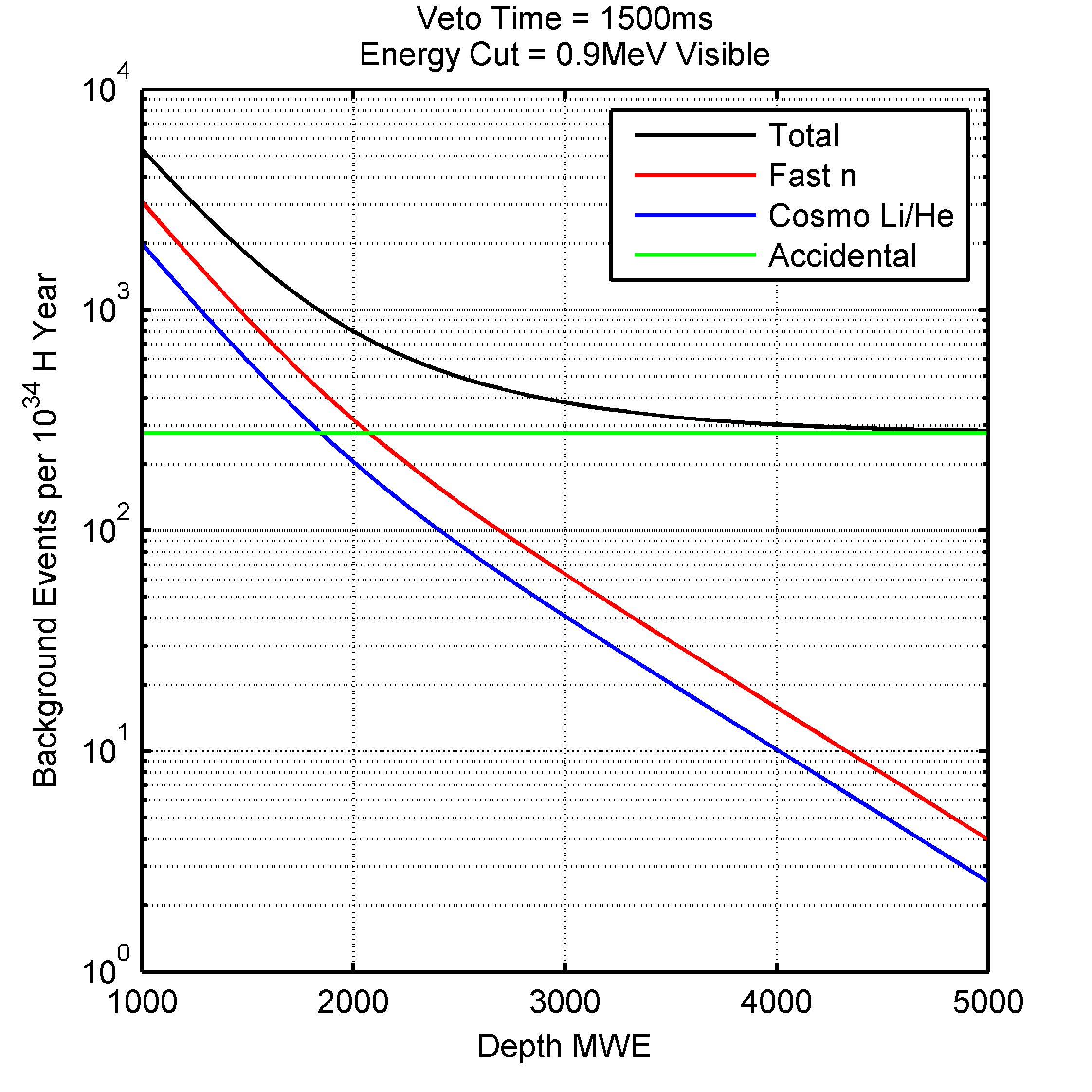}
\caption{\label{snifbkg} Non-neutrino background rate as a function of the detector depth in meters of water equivalent for an exposure of $10^{34}\mathrm{p}^+$-year. The limitation line for accidentals depends upon the detailed detector design, as it will be largely due to radiation from outside the detector or the detector structure and components. This limitation may be considerably lower in practice for TREND, and in any event is largely confined to the peripheral region of the detector.  The same is true of the entering fast neutron background.}
\end{figure}

Figure \ref{snifbkg} shows the non-neutrino background as a function depth.  It is interesting to note that the muon-related non-neutrino background no longer dominates for depths greater than 2000m.  What appears to be preventing TREND from operating at a shallow depth (2000m - 2500m) is the high cosmogenic downtimes (100\% - 70\%) incurred primarily by the vetoing of showering muons in such a large unsegmented detector.

There is a well-known seasonal variation in the muon flux on the surface and at depth due, primarily, to seasonal variations in the atmospheric temperature profile \cite{ambrosio_1997}.  Ambrosio \textit{et al.} found a 2\% seasonal variation with the MACRO detector located at the San Grasso underground laboratory in Italy. Furthermore, the relative magnitude of this variation is expected to increase with the average muon energy and, thus, with increasing depth.  For simplicity, this paper assumes that the (Poisson mean) muon flux, and its contribution to the non-neutrino background, is constant in time regardless of the detector depth.  In actual operation, the expected temporal variation in muon flux during an experiment can be accounted for if it is deemed necessary.

For simplicity an isotropic angular distribution is assumed for each non-neutrino background.  Thus, unlike reactor and geo-neutrino background, the non-neutrino background have the same energy spectrum in all directions seen by a detector.  Even if this assumption is not strictly true, the non-neutrino background very likely have a slowly and smoothly varying angular distribution.   The nearly isotropic directional measurement resolution of the detector further reduces the importance of modeling the angular distribution of non-neutrino background with high fidelity.

\begin{figure}[!htbp]
\centering
\includegraphics[width=90mm]{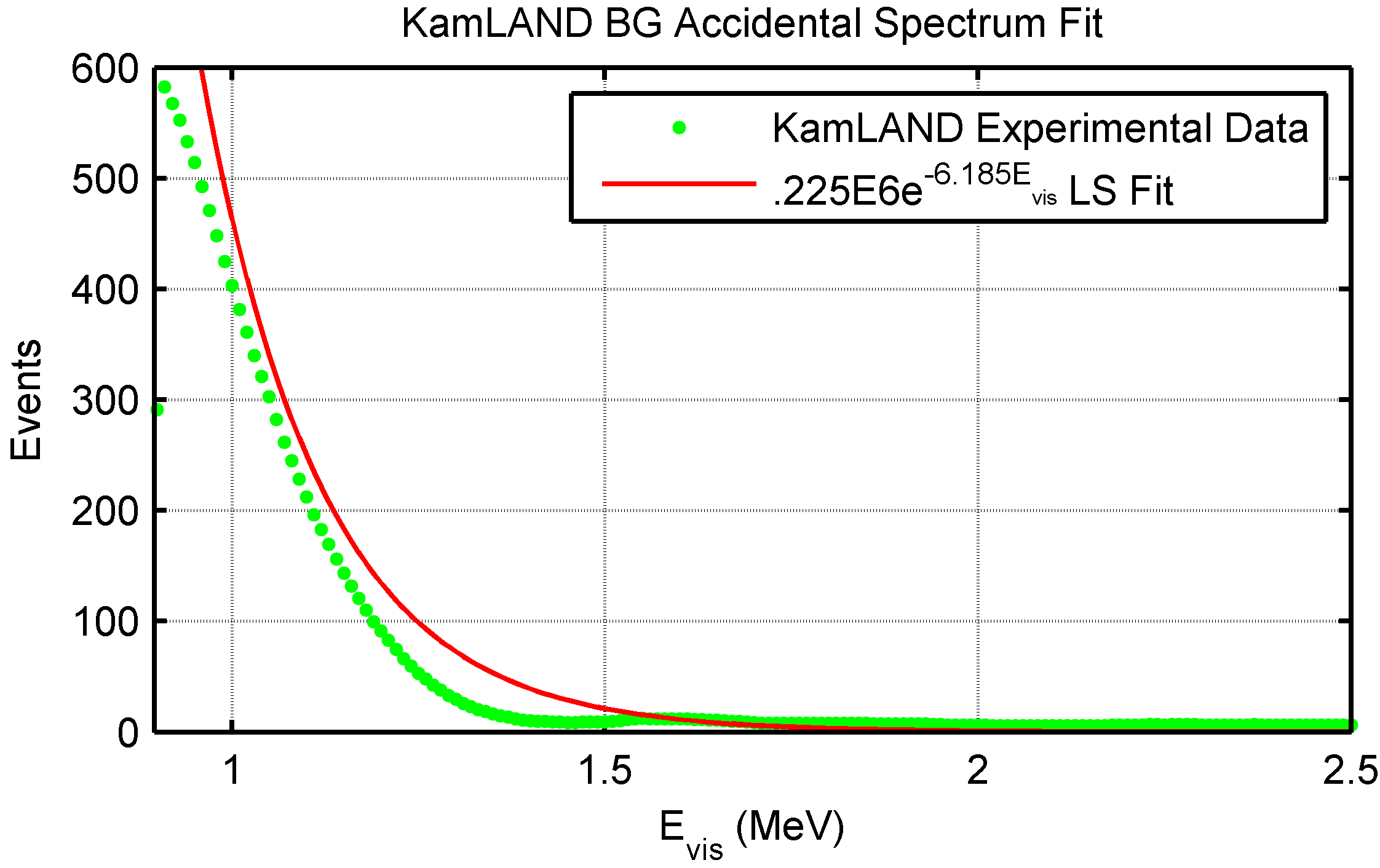}
\centering
\caption{Exponential least squares fit to experimental KamLAND data from \cite{KamLAND_accidentals}. Fit was performed using KamLAND data from $E_\mathrm{vis}=0.9$MeV to $E_\mathrm{vis}=1.7$MeV, then applied over the wider energy spectrum seen in this figure.}
\label{accidentalfit}
\end{figure}

\begin{figure}[!htbp]
\centering
\includegraphics[width=\linewidth]{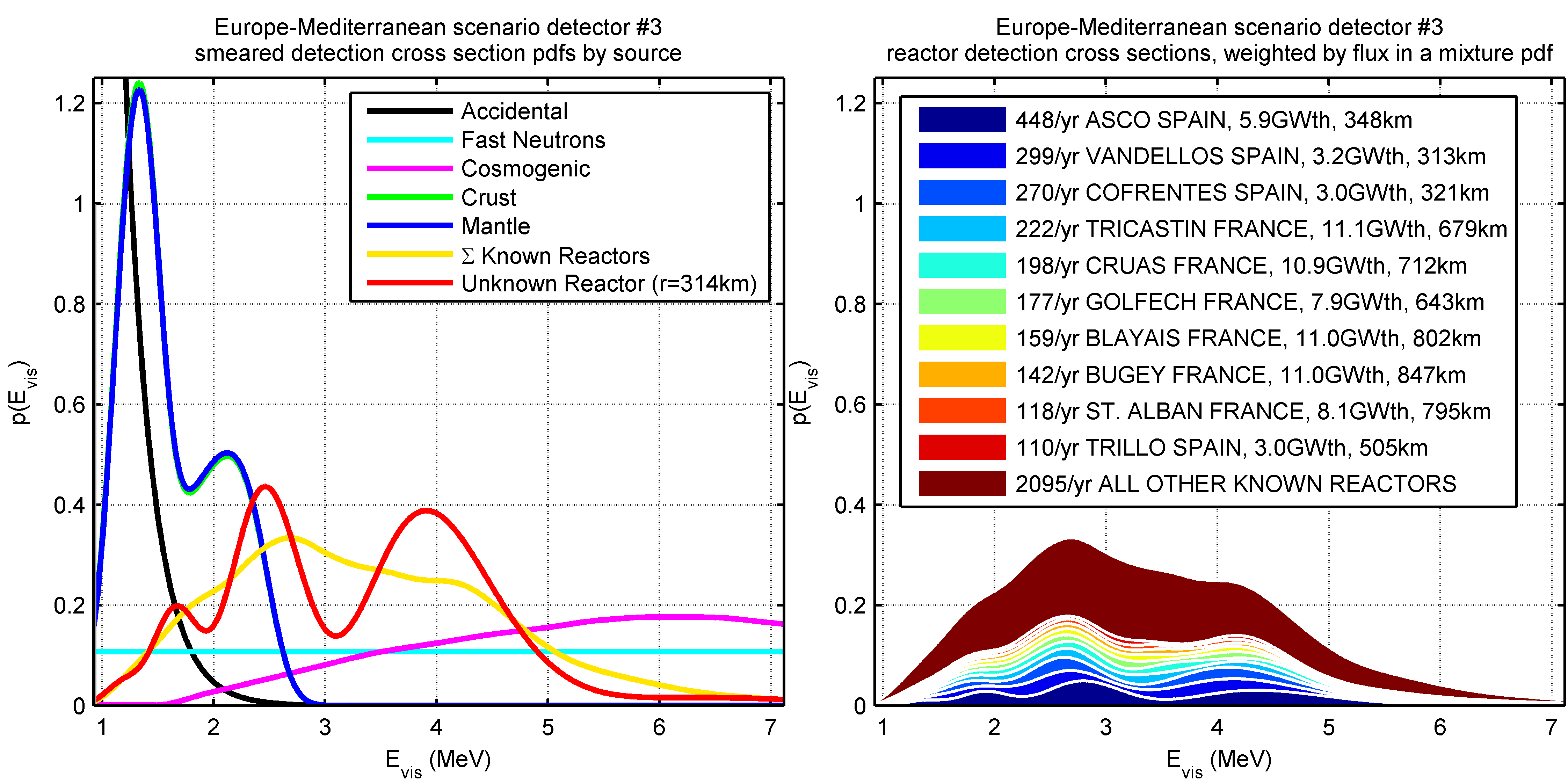}
\caption{Smeared energy spectrum of each measurement source type in $\theta_k$ as seen by TREND detector \#3 in our Europe-Mediterranean unknown-reactor scenario (refer to Section \ref{High reactor background with nearby reactors (Europe-Mediterranean)} for more information concerning this detector's placement and its environment). The known-reactor spectrum is a mixture distribution ($Z=w_n\sum\limits_{n} H(\theta_n,d)+v(d)$) of all the individual known-reactor spectra from around the world, which tend to ``wash out" the effect of oscillations from each source when summed together. See Abe {\it et. al.} \cite{kamland_2010} for more information concerning cosmogenic background energy spectra.}
\label{epdf}
\end{figure}

The non-neutrino spectra were all taken from empirical observations.  For fast neutron background and cosmogenic background the energy spectra defined by Mention \textit{et al.} \cite{mention_2007} were used. For accidentals the energy spectrum was obtained by fitting an exponential distribution (see Equation \ref{kamland_exp_fit}) to KamLAND data \cite{KamLAND_accidentals}.  Figure \ref{accidentalfit} shows the fit.  Being based on observations from other detectors, these spectra likely have their own energy measurement resolution and energy-dependent efficiencies applied to them.  For this study, the observed spectra were simply assumed to represent the source spectra already smeared by our modeled detector energy resolution.  Thus the non-neutrino spectra are somewhat notional for this detector.  This approach is not perfect, but it was deemed more than sufficient for this study.  The three non-neutrino energy pdfs can be seen in normalized form in Figure \ref{epdf} alongside four neutrino source categories: known reactors, unknown reactor, crust and mantle.

\begin{equation}
\label{kamland_exp_fit}
p(E_\mathrm{vis}) = \left( 2.25 \times 10^5 \right) e^{-6.185E_\mathrm{vis}}
\end{equation}

\subsection{Uncertainty model}
The estimator presented in Section \ref{Estimation theory} naturally incorporates the expected count rate for background sources, similar to the approach taken by Lasserre \textit{et al.} \cite{lasserre_2010}.  This work additionally allows the mean count rate for each background source, the mean free proton count of each detector, and the neutrino oscillation parameters to have realistic Gaussian-distributed systematic uncertainty.  The inclusion of these real-world uncertainties addresses many questions raised by simpler assumptions and increases MC fidelity.  While this widens the uncertainty contours about any maximum \textit{a posteriori} estimate, it acknowledges the real uncertainties involved in accurate parameter estimation and leads to a more accurate \textit{a posteriori} uncertainty interpretation about the maximum \textit{a posteriori} estimates. 

The \textit{a priori} systematic uncertainties used by the Geospatial Model are shown in Table \ref{table:background_uncertainty}.  The uncertainty model assumes that the energy spectra and angular distributions for the various flux sources have no uncertainty to their intrinsic shapes.  All of the uncertainty for a particular point source, or for a class of point sources, is assumed to be located in the mean flux rate itself.
For flux originating from discrete sources (crust with CRUST 2.0 tiles, known reactors), there may be a systematic uncertainty that affects the entire class ($\sigma_\mathrm{syst_{class}}$) as well as a systematic uncertainty which is independent for each source in the class ($\sigma_\mathrm{syst_{source}}$).  For continuous background sources (non-neutrinos), where $\sigma_\mathrm{syst_{source}}$ = 0\%, the ``number of sources" column indicates the number of discrete sample points used to approximate the continuous background in the Geospatial Model.
It is important to note that the background uncertainties described here are distinct from the statistical uncertainty that naturally arises from Poisson noise.

The assumption that the uncertainty can be restricted to the mean flux rate seems quite reasonable for several reasons.  We assume that careful modeling and extensive calibration runs would be performed on TREND during its design and construction to quantify the spectral character of non-neutrino background surviving the various processing filters. For known-reactor flux and geo-neutrino flux, the spectral shape of the intrinsic signal is relatively well understood (to within a few percent). The noted sterile neutrino uncertainty in Table 6 refers to the present controversy referred to as the Reactor Neutrino Anomaly \cite{Mention_2011}.  Reanalysis of old data and calculation presently point to a deficit in near reactor counting rates as compared to calculated reactor neutrino fluxes, of the order of $6.3+/-2.7\%$.  This plus other hints may point towards a small level oscillation coupling from normal neutrinos into sterile neutrinos.  The mass difference determined range is implied to be a few meters, such that the oscillations are in equilibrium for almost all near-reactor experiments.  Detectors located further away, as is the case for all situation considered in this paper, the result of the putative sterile neutrinos would be a few percent reduction in the expected flux, and some rather small shift in oscillation parameters.  At present there are also some peculiarities in the spectra of the experiments (Daya Bay, Double Chooz and RENO) measuring $\theta_{13}$ from about 2 km away from reactors and reported at conferences. This may be pointing to problems with the calculations of reactor borne neutrino spectra.  Of course it is reasonable to expect that this anomaly will be resolved by the time the long range detectors discussed herein will be built, and will require in any event only minor adjustment of mixing parameters, but will change no conclusions about our results.

Finally, small uncertainties in the spectral shapes are smoothed out by the energy measurement noise and the poor directional resolution of TREND\footnote{``Poor" directional resolution of TREND defined as $\sim$0.045 SNR per Section \ref{TREND energy and direction resolution MC results}.}.

\begin{table*}
\fontsize{8}{10}\selectfont
\centering
\begin{tabular}{l|c|l|l}
\hline
\textbf{Source} & \textbf{Uncertainty} & \textbf{Number of Sources} & \textbf{Comment} \\
&$\pm 1\sigma_\mathrm{syst_{class}} \pm 1\sigma_\mathrm{syst_{source}}$ & & \\
\hline
Known reactor flux 		&$\pm 2.0\% \pm 3.4\%$ & 471 reactor cores				&does not include sterile $\nu$ uncertainty\\
\hline
Crust flux 				&$\pm 20.0\% \pm 8.0\%$ & $\sim 129600$ tiles 			&100\% correlated across $^{238}$U and $^{232}$Th\\
\hline
Mantle flux 			&$\pm 50.0\% \pm 0\%$ & $\sim 50396$ tiles 				&100\% correlated across $^{238}$U and $^{232}$Th\\
\hline
Cosmogenic $^9$Li/$^8$He&$\pm 3.3\% \pm 0\%$ & 4584/detector 					&KamLAND \cite{kamland_2008} extrapolation\\
\hline
Fast neutron flux 		&$\pm 10.0\% \pm 0\%$ & 4584/detector 					&KamLAND \cite{kamland_2008} (capped at 10.0\%)\\
\hline
Accidental 				&$\pm 1.3\% \pm 0\%$ & 4584/detector 					&Borexino \cite{koichi_2007,Borexino_2010} isopurity extrapolation\\
\hline
Detector proton count	&$\pm 0.0\% \pm 2.4\%$ & 1/detector 					&i.e. fiducial volume uncertainty\\
\hline
$\Delta m^2_{12}$		&$^{+2.4\%}_{-2.8\%} \pm 0\%$ & $\sim 180467$ 			&per Fogli \textit{et al.} \cite{fogli_2012}\\
\hline
$\Delta m^2_{13}$		&$^{+3.4\%}_{-2.6\%} \pm 0\%$ & $\sim 180467$ 			&per Fogli \textit{et al.} \cite{fogli_2012}\\
\hline
$\sin^2\theta_{12}$		&$^{+5.2\%}_{-5.7\%} \pm 0\%$ & $\sim 180467$ 			&per Fogli \textit{et al.} \cite{fogli_2012}\\
\hline
$\sin^2\theta_{13}$		&$^{+10.0\%}_{-9.8\%} \pm 0\%$ & $\sim 180467$ 		&per Fogli \textit{et al.} \cite{fogli_2012}\\
\end{tabular}
\caption{Breakdown of the \textit{a priori} systematic uncertainties assumed by the Geospatial Model.  These uncertainties are assumed zero-mean normally distributed, and are each randomly sampled once per Geospatial Model MC.}
\label{table:background_uncertainty}
\end{table*}

Significant effort was spent on creating a realistic uncertainty model, populated with reasonable values.  The uncertainty model also reflects the long distance, multi-detector nature of the scenarios in this study by allowing for the fact that the detected signal is not dominated by one detector or by one source.  The systematic uncertainties for the non-neutrino background were generally taken from KamLAND and Borexino measurements.  The one exception was fast neutrons, whose highly uncertain rate was based upon a measured upper limit.  It was assumed that, for a futuristic detector such as TREND, measurement methods would exist to quantify this to 10\%.  The 3.4\% systematic uncertainty of each known reactor core corresponds to the combined reactor-related system uncertainty from S. Abe \textit{et al.}\cite{kamland_2008}.  The 2.4\% systematic uncertainty of each detector corresponds to the combined detector-related system uncertainty from S. Abe \textit{et al.}\cite{kamland_2008}.  Finally, oscillation parameter uncertainty was defined per Fogli \textit{et al.} \cite{fogli_2012} and applied to all neutrino sources present in the Geospatial Model, including 471 reactor cores, all crust tiles and the mantle.

\section{Estimation theory}
\label{Estimation theory}
A Bayesian maximum \textit{a posteriori} (MAP) estimator is presented here, suitable for a wide variety of neutrino experiments. The estimator utilizes all the available measurements and exploits all available prior information such as the measurement noise (detector hardware), uncertainty in the background signal, and constraints on the parameters to be estimated. The optimal\footnote{Optimal is defined here as using all readily available measurements (count rate, energy, angle, no energy cut) and marginalizing over (or co-estimating) all parameters with significant uncertainty.} estimator is presented, as well as two tractable suboptimal approximations appropriate for real world implementation with finite computing resources.  A few example applications include the refinement of neutrino oscillation parameters, the characterization of the geo-neutrino background, or placing an upper limit on the power of a geo-reactor. 

In this paper we test the estimator on two separate applications. First, it is used to refine $\nuebar$ oscillation parameters and to reduce systematic uncertainty related to our \textit{a priori} knowledge of backgrounds (e.g. geo-neutrinos and reactor neutrinos). Second, it is used to geolocate an unknown reactor. Both of these applications assume that four TREND detectors are operating simultaneously and that they combine their detections for the estimation process.

\subsection{Optimal MAP estimator}
\label{Optimal MAP estimator}
In the Bayesian approach, one starts with the \textit{a priori} probability density function of the parameter space, $p(\theta)$, from which one can obtain its \textit{a posteriori} pdf:

\begin{equation}
\label{JNBeq_1}
p(\theta |Z)=\frac{p(Z|\theta)p(\theta)}{p(Z)}=\frac{1}{c}p(Z|\theta)p(\theta)
\end{equation}
where $c=p(Z)$ acts as a normalization constant\footnote{In the Bayesian approach, $p(Z)$ is a constant scalar value (sometimes referred to as the `marginal likelihood') equal to an integral across the entire measurement space $Z$.It is deterministic in nature since it is not evaluated at any specific measurement. Since it is not a function of the parameter space $\theta$ and does not vary as a function of the random measurements, it follows that the {\it a posteriori} probability may be expressed as proportional to the \textit{numerator only} of Bayes' equation.}, which, since it does not depend on $\theta$ can be considered irrelevant for maximization \cite{Bar-Shalom}. $Z$ is a stacked vector of $n_z$ measurements. $p(\theta)$ is the Bayesian prior which describes our \textit{a priori} knowledge of the parameter space $\theta$. $p(Z|\theta)$ is the conditional probability of $Z$ given $\theta$, commonly referred to as the \textit{likelihood function} or, more simply, as the \textit{likelihood}.

The measurement vector $Z$, corrupted by a detector-dependent zero-mean white noise $v(d)$, is a function of the parameter vector $\theta$ and the detector $d$. In a linear system, $H$ would be a Jacobian of partial derivatives expressing the relationship between $Z$ and $\theta$: $Z=H\theta+v$. In our nonlinear system, $H$ is a function which translates the parameter vector $\theta$ into the smeared measurement space $Z$:

\begin{equation}
\label{JNBeq_2}
Z=H(\theta,d)+v(d)
\end{equation}

The parameter vector $\theta$ is composed of a stacked vector of $n_u$ unknown sources (e.g. reactors) to be estimated ($\theta_u$), $n_k$ known sources ($\theta_k$), and various physical constants contained in $\theta_c$. $\theta_k$ contains the known background sources: geo-neutrinos, known-reactor neutrinos, accidentals, fast neutrons, and cosmogenics. $\theta_c$ contains all the physical constants employed by the estimator, such as the solar mixing angle and the constants contained within PREM used for geo-neutrino flux calculations.

\begin{eqnarray}
\label{JNBeq_3}
\theta &=& \left[ \begin{array}{c}
\theta_u
\\\theta_k
\\\theta_c
\end{array}\right]\label{JNBeq_3.0.1}\\
\theta_u &=& \left[ \begin{array}{c}
\theta_{u _{1}}
\\\vdots
\\\theta_{u_{n_u}}
\end{array}\right]\label{JNBeq_3.0.2}\\
\theta_k &=& \left[ \begin{array}{c}
\theta_{k_{1}} 
\\\vdots
\\\theta_{k_{n_k}}
\end{array}\right]\label{JNBeq_3.0.3}
\end{eqnarray}

To simplify the problem, we assume each unknown reactor in $\theta_u$ can be accurately modeled as a point source which emits neutrinos isotropically, and that all the unknown reactors are constrained to lie on the Earth's surface. Following these assumptions each unknown reactor in $\theta_u$ can now be defined by just four elements: geodetic latitude $\phi$, geodetic longitude $\lambda$, thermal power output $p$ (i.e. flux), and zero-range energy spectrum $s$\footnote{Note that the zero-range spectrum $s$ is a function of energy, $s(E)$, but for simplicity it is assumed that all sources have one of a finite number of well-defined energy spectra.  Thus $s$ is treated as a discrete parameter.}. For $n_u$ unknown reactors:

\begin{equation}
\label{JNBeq_3.1}
\theta_u = 
\left[ \begin{array}{cccc}
\phi_1 & \lambda_1 & p_1&s_1\\ 
\vdots &\vdots &\vdots &\vdots \\
\phi_{n_{u}} & \lambda_{n_{u}} & p_{n_{u}} &s_{n_{u}}
\end{array}\right]
\end{equation}

We further simplify the problem by assuming that only one unknown source exists in the vicinity of our detector. Thus $n_u=1$ and $\theta_u$ reduces to one row. For this 1x4 row vector $\theta_u$, we apply a four-dimensional uniform (but bounded) \textit{a priori} probability $U(\theta_u)$.  This effectively constrains the unknown source to exist within a certain latitude-longitude-flux-spectrum domain with equal \textit{a priori} probability at every point. $p(\theta_u)$ need not be uniform, however, or even linear. Any informative \textit{a priori} probability may be used for $p(\theta_u)$, but it must be \textit{properly normalized} to reflect a 100\% confidence that the unknown source exists somewhere within the full bounds of the four-dimensional parameter space under consideration. Note that behavior of the estimator when the unknown source is not, in reality, contained within $p(\theta_u)$ is examined later on in Section \ref{No reactor}.

\begin{figure}[htbp]
\centering
\includegraphics[width=\linewidth]{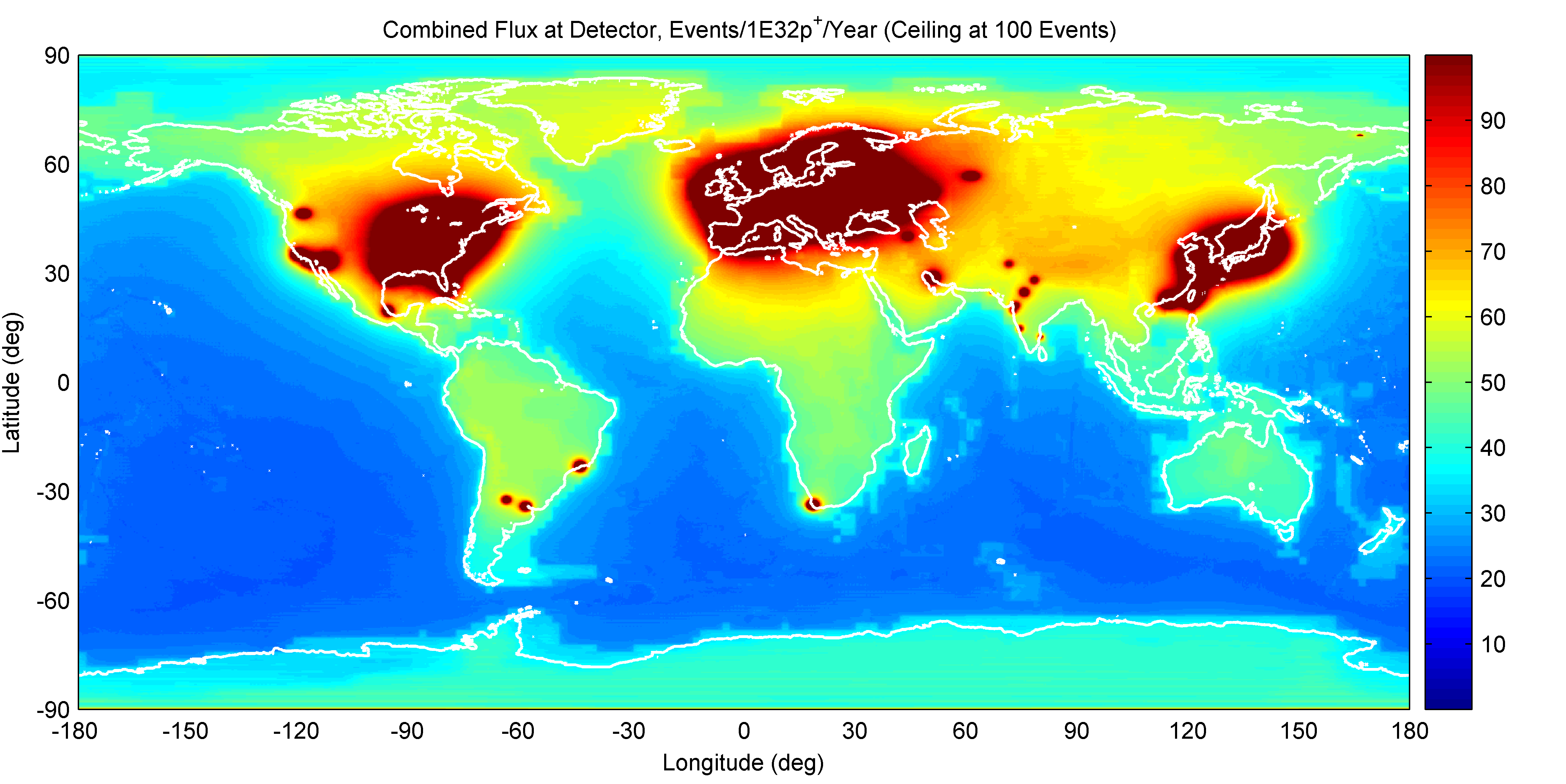}
\caption{Background $\nuebar$ flux expected at varying locations worldwide from the $\nuebar$ sources in $\theta_k$. Mantle flux computed  from PREM \cite{PREM} and Mantovani \textit{et al.} \cite{Mantovani_2004}. Crust data computed from CRUST 2.0 model and Mantovani \textit{et al.} \cite{Mantovani_2004}. Reactor flux computed from publicly available IAEA data \cite{IAEA1999}.}
\label{fig3_worldBackground}
\end{figure}

$\theta_k$ represents the known background sources expected at the detector location, independent of the unknown sources in $\theta_u$.  $^{232}$Th and $^{238}$U isotope geo-neutrino sources in the mantle and crust are considered here, though potassium $^{40}$K isotopes are ignored as their neutrino energies are below the inverse beta decay energy measurement threshold of about $E_{\nuebar}\geq 1.8$MeV. The combined flux from all neutrino sources for a $10^{32}p^+$ detector are shown in Figure \ref{fig3_worldBackground}.

We have eight known categories of background in $\theta_k$:

\begin{equation}
\label{JNBeq_4}
\theta_k = \left[ \begin{array}{c}
\theta_{\mathrm{IAEA}}
\\ \theta_{\mathrm{crust_{^{238}U}}}
\\ \theta_{\mathrm{crust_{^{232}Th}}}
\\ \theta_{\mathrm{mantle_{^{238}U}}}
\\ \theta_{\mathrm{mantle_{^{232}Th}}}
\\ \theta_{\mathrm{accidental}}
\\ \theta_{\mathrm{fast\ }n^0}
\\ \theta_{\mathrm{cosmogenic}}
\end{array}\right]
\end{equation}

%\begin{figure}[htbp]
%\centering
%\includegraphics[width=.5\linewidth]{./figures/fig1_eventRate}
%\caption{Example measurement count breakdown among the various contributors to the measurement vector Z.}
%\label{fig1_eventRate}
%\end{figure}

Nearly all the neutrino source locations in $\theta_k$ are most naturally expressed in geodetic, or ECEF reference frames, though the non-neutrino background may be more naturally expressed in a local spherical coordinate system with respect to the detector center-point. Some coordinate system transformations may need to be defined if sources are defined with respect to different origins, or expressed in different reference frames\footnote{Note that these low level coordinate system transformations are implicitly assumed in all the equations in this paper.}.

Some sources in $\theta_k$ are assumed to be point sources, such as the known reactors in $\theta_{\mathrm{IAEA}}$. Others are clearly volume sources, such as the mantle and crust $^{238}$U and $^{232}$Th isotopes. These volume sources can be treated as integrations of point sources, where each point in the discrete integral would occupy a row in $\theta_k$. Discrete volume integral step sizes must be carefully chosen to ensure accurate approximation to a continuous volume integral, especially when dealing with volumes in the immediate vicinity of the detector such as the Earth's oceans and crust.

\begin{figure}[htbp]
\centering
\includegraphics[width=1.00\linewidth]{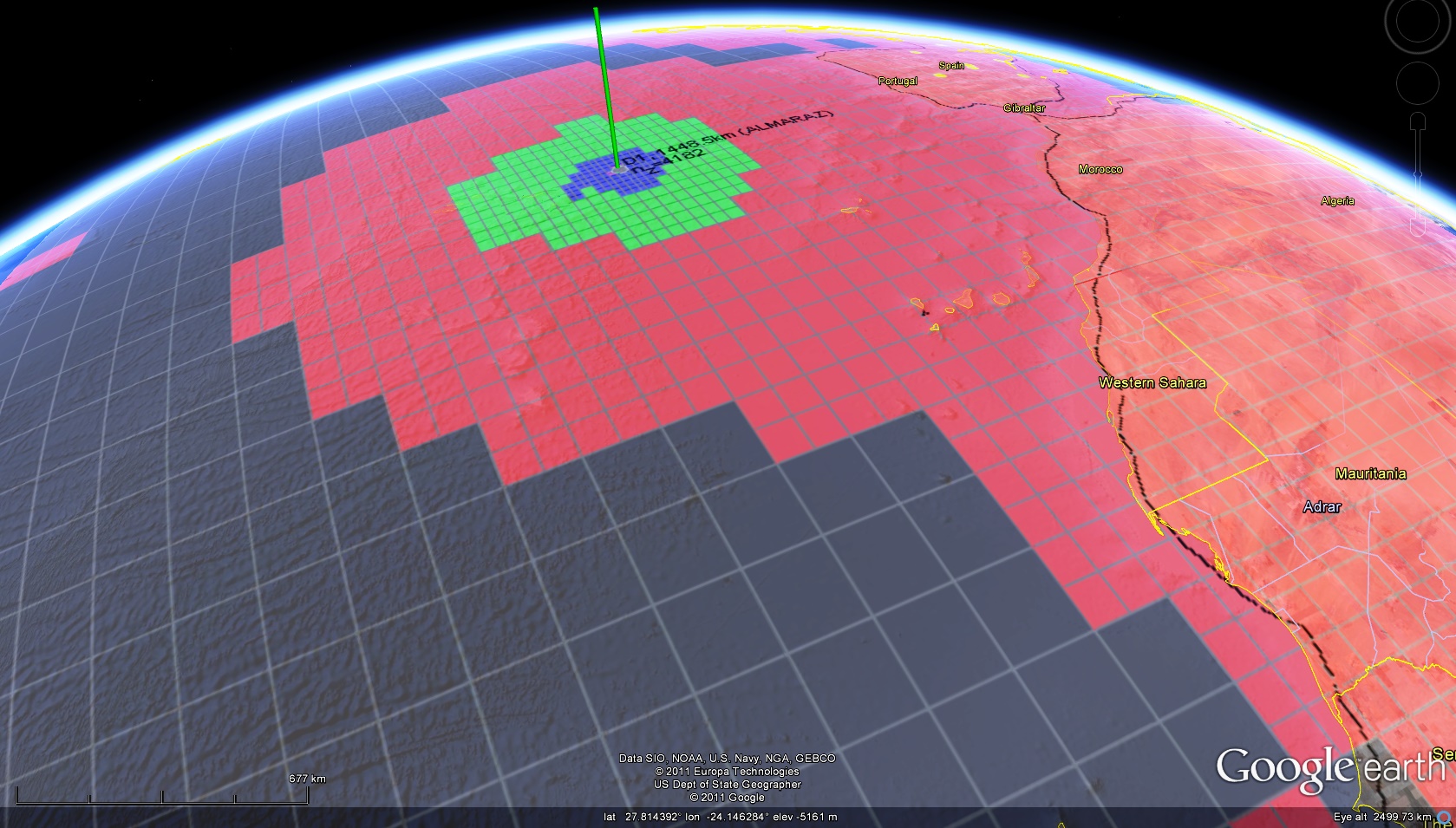}
\caption{Example ``smart" spherical integration of the Earth's upper crust, a volume source. Gray squares are original CRUST 2.0 Upper Crust tiles. Each grey tile is $2^\circ \times 2^\circ$. Red tiles are $1^\circ \times 1^\circ$, green tiles are $0.5^\circ \times 0.5^\circ$ and so on, up to a 30th generation tile, at $\left(4 \times 0.5^{-30} \right )^\circ \times \left(4 \times 0.5^{-30} \right )^\circ$. Note that smart discrete integration is performed in three dimensions simultaneously, though the height axis is not shown in this figure.}
\label{fig4_smartIntegration2}
\end{figure}

To avoid excessive computational burden caused by small integral step-sizes over large volumes, a recursive ``smart" integration technique may be implemented, varying local integral step-size efficiently while retaining high model fidelity. Large volume sources are recursively broken into groups of smaller volume sources until a predetermined flux threshold is no longer exceeded by each resultant discrete integral point. In our tests we found a threshold of 0.1 events/detector/year per discrete point source to offer an excellent compromise between discrete integral accuracy and computational burden. Figure \ref{fig4_smartIntegration2} shows one such variable step-size crust integration for a detector submersed in the waters of the North Atlantic.

It is natural for the expected flux of known sources in $\theta_k$ to contain some degree of systematic uncertainty. An example of this would be the large systematic uncertainty in the expected flux of geo-neutrinos produced from the mantle.  We assume here that any systematic uncertainty in $\theta_k$ can be fully (and accurately) defined by an \textit{a priori} probability, $p(\theta_k )$. Uncertainty in many elements of $\theta_c$, such as the solar mixing angle uncertainty, is also expected as well.   We, therefore, also assume that a well-defined prior probability $p(\theta_c)$ can describe these uncertainties.

There are two mutually exclusive options\footnote{Approximation is a third option, used in our suboptimal approximations in Section \ref{Applied suboptimal MAP estimator}.} for optimally dealing with such parameter space uncertainty: marginalization or estimation. In marginalization, the parameter is treated as a nuisance parameter, and a weighted integration through the parameter dimension is performed. This results in a marginal probability which is no longer conditional on the uncertain parameter:

\begin{equation}
\label{JNBeq_5}
p(x)=\int\limits_{-\infty }^{\infty }p(x,i)p(i)di
\end{equation}
where $i$ is the uncertain parameter, $p(i)$ is the Bayesian prior of $i$, $p(x,i)$ is the conditional likelihood and $p(x)$ is the marginal likelihood. Note that $p(x)$ is independent of $i$ following marginalization. This method can be extrapolated to correlated multidimensional uncertainties through the use of higher dimensional integrals, albeit at a much high computational cost. It may be more easily applied to independent (zero-covariance) multidimensional uncertainties via the product of several single dimensional integrals.

The second option when dealing with uncertain nuisance parameters in $\theta_k$ or $\theta_c$ is to co-estimate more accurate values for them alongside the original parameters of interest, $\theta_u$ in this case. Assuming at least partial observability of the uncertain nuisance parameters from the measurements, $Z$, this option may deliver superior results to marginalization.

There is no correct answer for deciding when to marginalize or optimize, the choice depends largely on the problem at hand and the computational resources available. Marginalization usually expends more computational resources as it requires more objective function evaluations. In general, information is lost during marginalization as a higher dimensional \textit{a priori} pdf collapses into a lower dimensional \textit{a priori} pdf. This results in flatter \textit{a posteriori} pdf with smaller gradients and larger confidence bounds about the maximum \textit{a posteriori} estimate. For these reasons this method should only be applied when there is quite high uncertainty in the ``known" parameter value, and poor prospects for estimating a better value (i.e. low parameter observability). Marginalization may also be appealing when faced with higher degrees of freedom, nonlinear or multimodal distributions, or deteriorated \textit{a posteriori} parameter space topology inhibiting reliable optimizer performance.

So, recognizing the need to marginalize some of the uncertainty dimensions while optimizing others, we update Bayes' equation with the requisite multidimensional integral through the uncertain nuisance dimensions we have chosen to marginalize:

\begin{equation}
\label{JNBeq_6}
p(\theta|Z)=\frac{1}{c}\int\limits_{-\infty }^{\infty }\cdots\int\limits_{-\infty }^{\infty }p(Z|\theta)p(\theta)d\theta_1\cdots\theta_n
\end{equation}

\noindent Now, the estimation problem can be defined as the maximization of Bayes' posterior probability density $p(\theta|Z)$ over the parameters in $\theta$ that remain following marginalization of the nuisance parameters:

\begin{equation}
\label{JNBeq_7}
\mathrm{arg}_\theta \mathrm{max}~p(\theta|Z)
\end{equation}
In other words, we want to create a maximum \textit{a posteriori} (MAP) estimator to find a value for $\theta$ which maximizes $p(\theta|Z)$ conditional on $p(\theta)$, post-marginalization of the nuisance elements of $\theta$.\footnote{Note that while this paper primarily covers estimation of oscillation parameters and unknown nuclear reactors that may exist in $\theta_u$, the estimator is not limited to these applications. Any uncertain element in $\theta$, including the neutrino physics constants in $\theta_c$, such as possible sterile neutrino mixing angles and mass differences, may be estimated by allowing an optimizer to travel along those additional dimensions in $\theta_c$ rather than marginalizing them.}

Prior to optimization (\textit{maximization} of the likelihood function, or alternatively, \textit{minimization} of the negative log-likelihood function), each element of the Equation \ref{JNBeq_7} must be defined. Bayes' normalization constant, $p(Z)$, or $c$ is addressed first. $c$ is intended to normalize Bayes' \textit{a posteriori} pdf into a proper\footnote{A ``proper" pdf integrates to unity, (i.e. cdf=1).} pdf, however since $c$ is irrelevant during optimization, it does not require definition and is ignored from here onward.

$p(\theta_u)$ has been defined to be a 4D uniform distribution, $p(\theta_u)=U(\theta_u)$, across our latitude-longitude-power-spectrum parameter space.  \textit{A priori} pdfs for our other parameters have been assumed\footnote{The exact form that these uncertainties take is not important for description of the estimator, though every effort should be make to accurately represent them when these equations are implemented in an experiment.} for $p(\theta_k )$ and $p(\theta_c)$ as well. The only remaining term in Bayes' equation we have not yet explicitly defined is the likelihood function, $p(Z|\theta)$.

We begin by examining the measurement vector $Z$. $Z$ is composed of energy measurements ($Z_E$) in the first column and direction vector measurements ($Z_V$) in the last three columns.

\begin{equation}
\label{JNBeq_8}
Z = \left[\begin{array}{cc}Z_E&Z_V \end{array}\right],\;\;\;\;\;Z_E = \left[\begin{array}{c}E_1\\\vdots\\E_{n_{Z}} \end{array}\right],\;\;\;\;\;Z_{V} = \left[\begin{array}{ccc}V_1^x&V_1^y &V_1^z\\\vdots&\vdots&\vdots\\V_{n_{Z}}^x&V_{n_{Z}}^y&V_{n_{Z}}^z\end{array}\right]
\end{equation}

\noindent $Z_V$ is simply the Cartesian vector which connects the estimated IBD \textit{delayed} center-of-energy (CE) to the \textit{prompt} CE, following a neutrino IBD event within a liquid scintillator detector. In our Detector Model (Section \ref{Antineutrino Detector Model}), prompt and delayed CEs are separately estimated by two variants of the CHOOZ \cite{chooz_2003} maximum likelihood estimator (MLE) for each event\footnote{$Z_V$ is expressed here in the ECEF reference frame, though $Z_V$ is naturally first recorded in the local detector reference frame before rotation into the ECEF frame. This is a natural prerequisite to combining direction vector measurements from multiple detectors with different orientations in the MAP estimator. Units of $Z_V$ assumed in this paper are mm.}.

\begin{equation}
\label{JNBeq_9}
Z_V = \widehat{CE}_{\mathrm{delayed}}-\widehat{CE}_{\mathrm{prompt}} = \left[\begin{array}{ccc}\hat{x}^{CE}_{1} &\hat{y}^{CE}_{1}&\hat{z}^{CE}_{1}\\\vdots&\vdots&\vdots\\\hat{x}^{CE}_{n_{Z}} &\hat{y}^{CE}_{n_{Z}}&\hat{z}^{CE}_{n_{Z}}\end{array}\right]_{\mathrm{delayed}} - \left[\begin{array}{ccc}\hat{x}^{CE}_{1} &\hat{y}^{CE}_{1}&\hat{z}^{CE}_{1}\\\vdots&\vdots&\vdots\\\hat{x}^{CE}_{n_{Z}} &\hat{y}^{CE}_{n_{Z}}&\hat{z}^{CE}_{n_{Z}}\end{array}\right]_{\mathrm{prompt}}
\end{equation}

The two CHOOZ MLEs co-estimate the visible energy as well as position of both CEs. The prompt visible-energy estimate,  $\widehat{E}_{\mathrm{prompt}}$ is highly correlated with the actual neutrino energy and is converted to an estimate of neutrino energy via Equation \ref{JNBeq_9.1}.  The MAP estimator has been designed to exploit these neutrino energy estimates because the detected neutrino energy spectrum encodes significant information about the range and spectrum of the contributing sources.  Thus we retain the corrected CHOOZ MLE prompt energy estimate, $Z_E$ in units of $\mathrm{MeV}$, as an estimate of the neutrino energy.  Note that the delayed visible-energy estimate is used in the candidate selection cut, but not directly in the MAP estimator.

\begin{equation}
\label{JNBeq_9.1}
Z_E = \left[\begin{array}{c}\widehat{E}_1\\\vdots\\\widehat{E}_{n_{Z}}\end{array}\right]_{\mathrm{prompt}}+(E_{n^{o}})-(E_{p^{+}})-(E_{e^{+}})
\approx\left[\begin{array}{c}\widehat{E}_1\\\vdots\\\widehat{E}_{n_{Z}}\end{array}\right]_{\mathrm{prompt}}+0.782\mathrm{MeV}
\end{equation}

The measurements $Z$ natively reside within the $\left[ \begin{array}{cccc} E & V_x &  V_y & V_z \end{array}\right]$ measurement space, and each individual measurement may be thought of as a random sample of this continuous 4D measurement space (effectively a 4D pdf). The MAP estimator exploits only 3 degrees of freedom (energy, azimuth of direction, and elevation of direction) from each measurement; it does not utilize the CE-to-CE distance. One may thus normalize the direction vectors and convert them to spherical elevation and azimuth angles, ``throwing away" the unneeded CE-to-CE distance information in the process (assuming this event has already been passed candidate selection cut).

\begin{figure}[htbp]
\centering
\includegraphics[width=\linewidth]{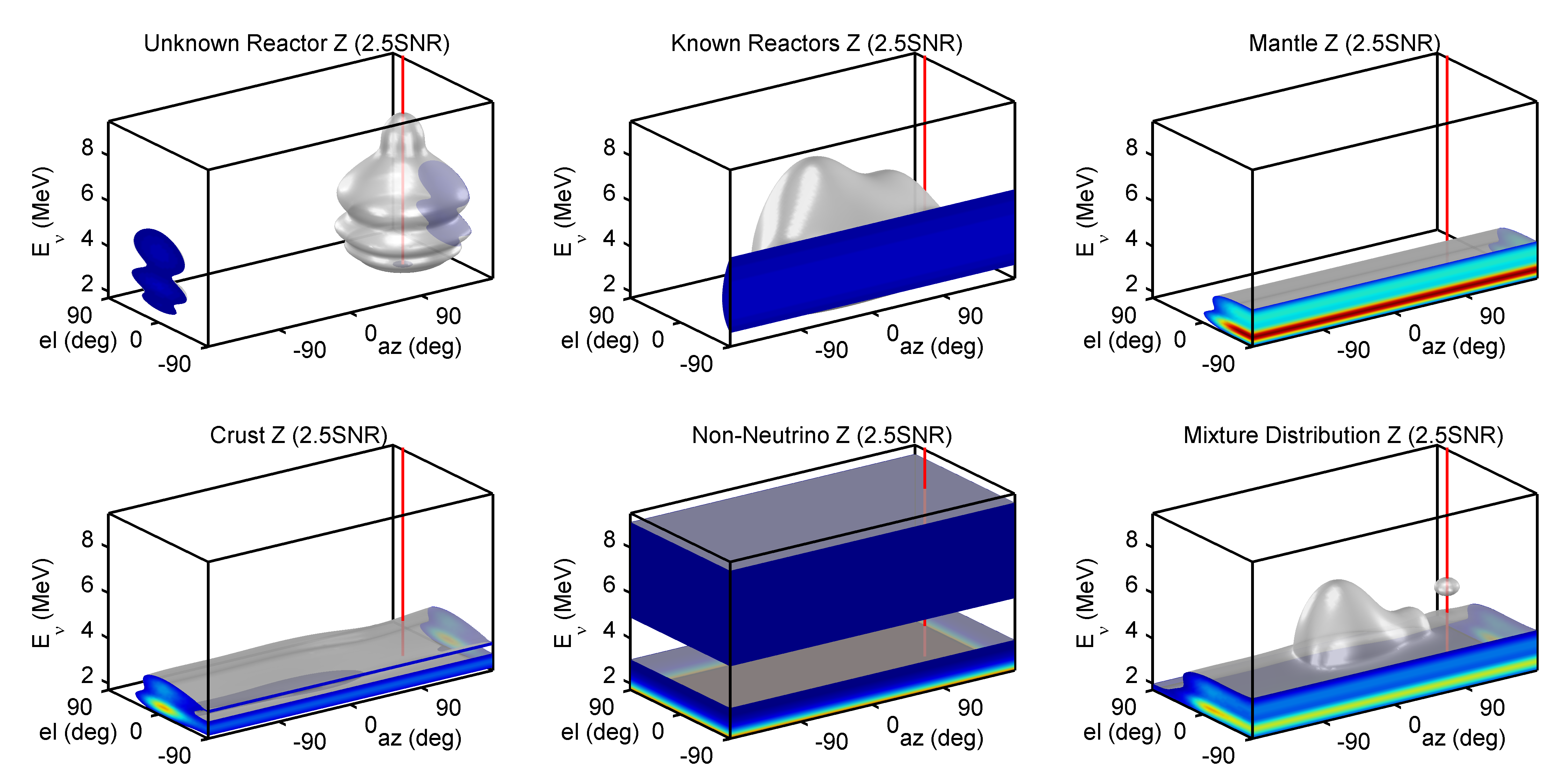}
\caption{Europe-Mediterranean scenario TREND detector \#1 measurement space (showing 2.5 angular SNR), a mixture distribution ($Z=w_n\sum\limits_{n} H(\theta_n,d)+v(d)$) of measurement spaces defined by $\theta_u$ and $\theta_k$. 90\% probability confidence isosurface is shown in silver; red line represents true direction of unknown reactor. More information concerning this detector within the Europe-Mediterranean unknown-reactor scenario can be found in Section \ref{High reactor background with nearby reactors (Europe-Mediterranean)}.}
\label{SNR3_3D}
\end{figure}

\begin{figure}[htbp]
\centering
\includegraphics[width=\linewidth]{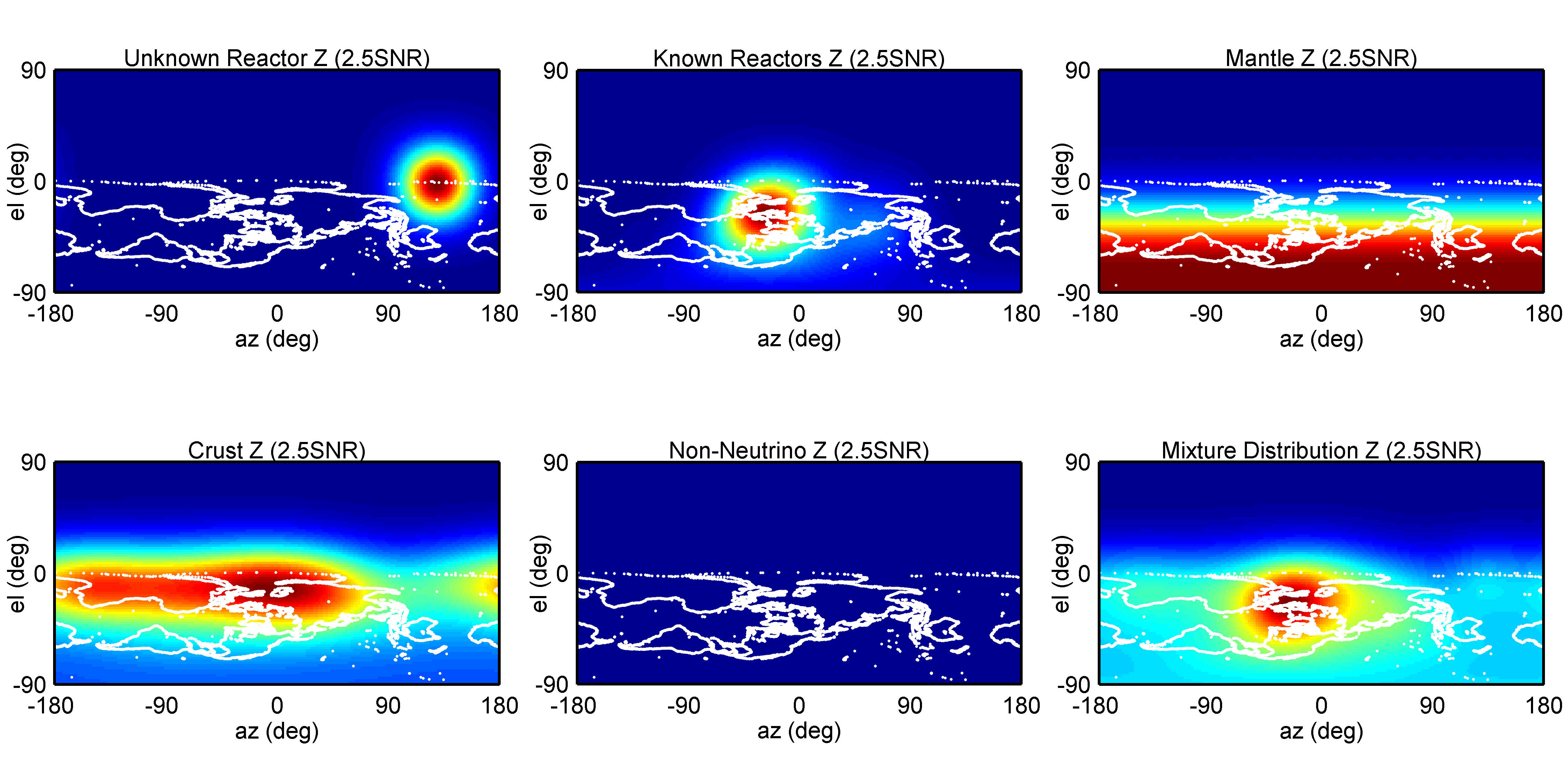}
\caption{Europe-Mediterranean scenario TREND detector \#1 2D elevation-azimuth \textit{marginal} measurement space following integration (across  energy) of Figure \ref{SNR3_3D} 3D measurement spaces. Continent outlines are shown in white. More information concerning this detector within the Europe-Mediterranean unknown-reactor scenario can be found in Section \ref{High reactor background with nearby reactors (Europe-Mediterranean)}.}
\label{SNR3_2D}
\end{figure}

\begin{figure}[htbp]
\centering
\includegraphics[width=\linewidth]{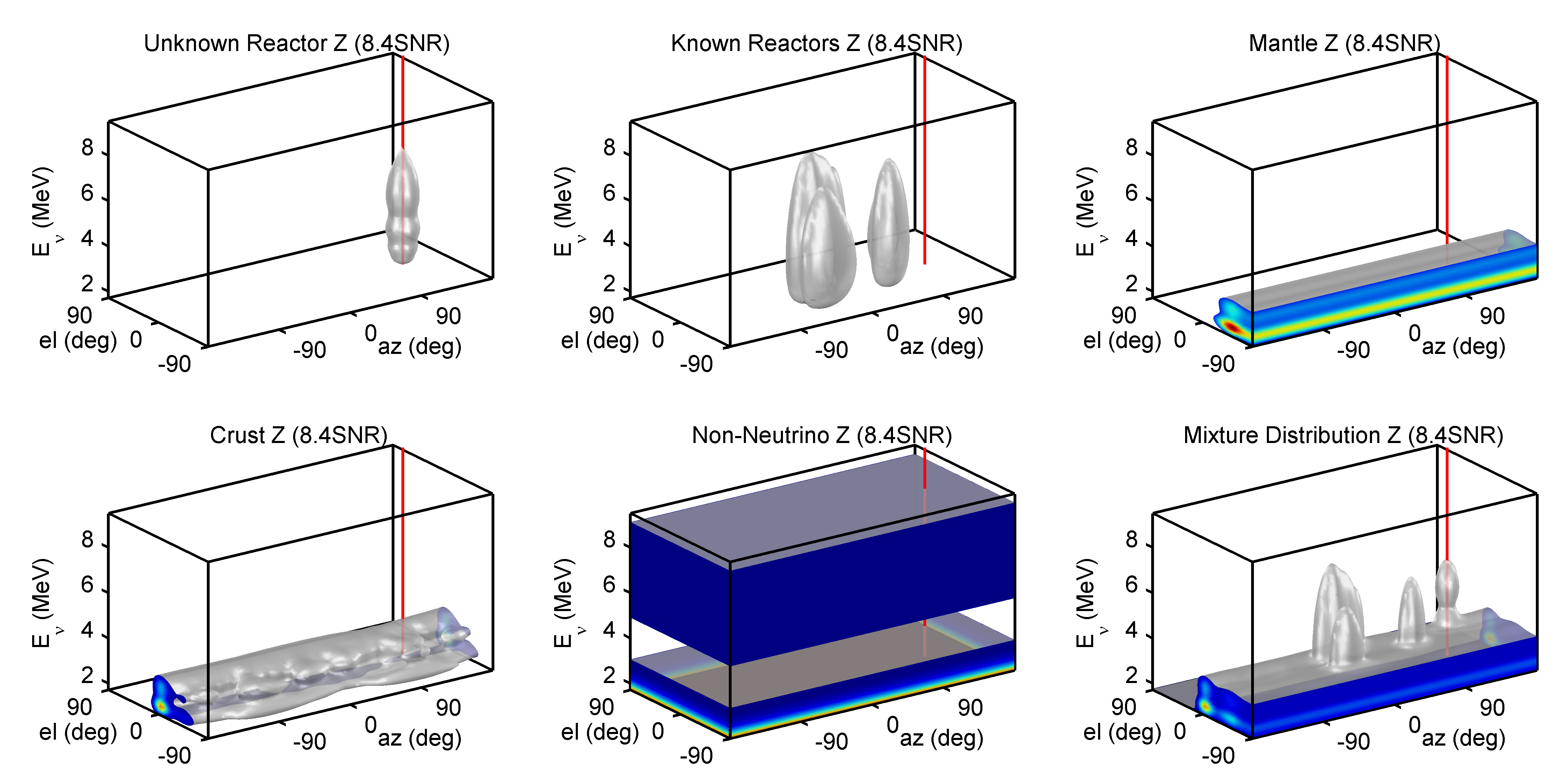}
\caption{Europe-Mediterranean scenario TREND detector \#1 measurement space (showing 8.4 angular SNR), a mixture distribution ($Z=w_n\sum\limits_{n} H(\theta_n,d)+v(d)$) of measurement spaces defined by $\theta_u$ and $\theta_k$. 90\% probability confidence isosurface is shown in silver; red line represents true direction of unknown reactor. More information concerning this detector within the Europe-Mediterranean unknown-reactor scenario can be found in Section \ref{High reactor background with nearby reactors (Europe-Mediterranean)}.}
\label{SNR9_3D}
\end{figure}

\begin{figure}[htbp]
\centering
\includegraphics[width=\linewidth]{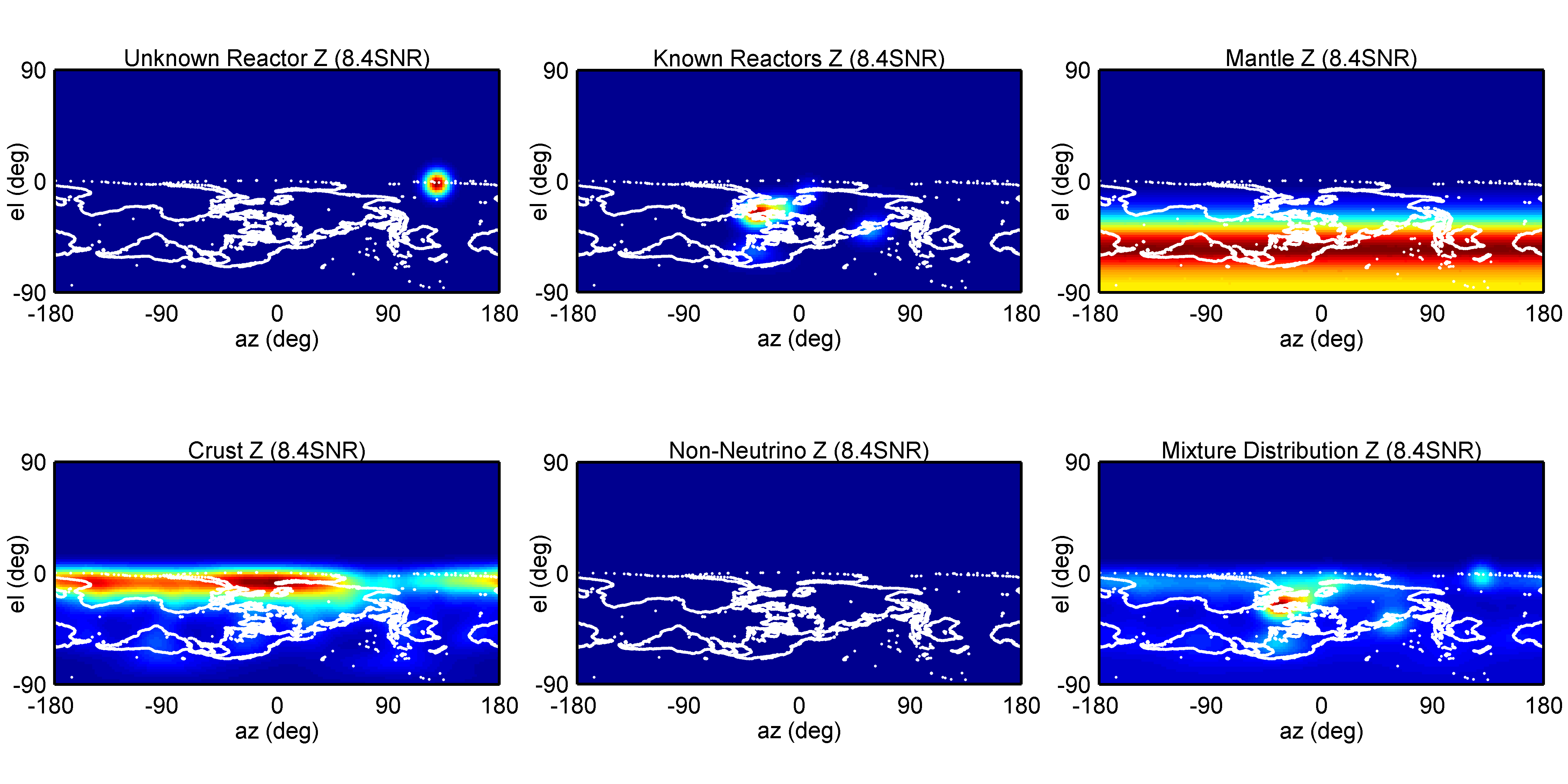}
\caption{Europe-Mediterranean scenario TREND detector \#1 2D elevation-azimuth \textit{marginal} measurement space following integration (across  energy) of Figure \ref{SNR9_3D} 3D measurement spaces. Continent outlines are shown in white. More information concerning this detector within the Europe-Mediterranean unknown-reactor scenario can be found in Section \ref{High reactor background with nearby reactors (Europe-Mediterranean)}.}
\label{SNR9_2D}
\end{figure}

Examples of such 3D measurement spaces are shown in Figures \ref{SNR3_3D} to \ref{SNR9_2D} for each source category (unknown reactor, known reactors, crust, mantle and non-neutrinos) as well as for the cumulative signal. Figures \ref{SNR3_3D} and \ref{SNR9_3D} compare the full 3D measurement space at two different (highly exaggerated) levels of detector angular resolution. Figures \ref{SNR3_2D} and \ref{SNR9_2D} show representations of Figures \ref{SNR3_3D} and \ref{SNR9_3D} respectively where the energy dimensional has been fully integrated (or \textit{marginalized}), resulting in a marginal 2D elevation-azimuth measurement space. The unique directional characteristics of each background source can be seen in these figures, including the isotropic nature of the non-neutrino background signal we have chosen to assume.

Before defining $p(Z|\theta)$ we address a somewhat unique aspect of this problem: information is contained not just within the measured energies and direction vectors, but also in the number of measurements: $n_Z$, defined by the row count of measurement vector $Z$. This aspect of the problem demands application of Poisson statistics within the likelihood $p(Z|\theta)$. Thus a new measurement count likelihood $p(n_Z|\theta)$ is combined with the energy and directional likelihood $p(Z_E,Z_V |\theta)$ to get a more explicit definition of the overall likelihood $p(Z|\theta)$:

\begin{equation}
\label{JNBeq_10}
p(Z|\theta)=p(n_Z|\theta)p\left(Z_E, Z_V|\theta\right)
\end{equation}
where the measurement count likelihood $p(n_Z|\theta$) is simply a Poisson pdf ($\mathcal{P}$):

\begin{equation}
\label{JNBeq_11a}
p\left(n_Z|\theta\right)=\mathcal{P}\left (n_Z, \bar{n}\right)
\end{equation}
\begin{equation}
\label{JNBeq_11b}
\bar{n}=\sum\limits_{j=1}^{n_u}\bar{n}\left (\theta_{u}^j\right)+\sum\limits_{l=1}^{n_k}\bar{n}\left (\theta_{k}^l\right)
\end{equation}
The quantity $n_Z$, a scalar integer, is the observed number of candidate measurements in the detector, $\bar{n}(\theta_k^l)$ is the expected number of candidate measurements at the detector from known source $\theta_k^l$, $n_k$ is the number of known sources, $\bar{n}(\theta_u^j)$ is the expected number of candidate measurements from unknown source $\theta_u^j$, and $n_u$ is the number of unknown sources.

The second component of the likelihood, $p(Z_E,Z_V |\theta)$, is the conditional likelihood of all the energy and direction measurements given $\theta$. Considering all the measurements to be independent of each other, the likelihood of all $n_Z$ measurements equals the product of the likelihoods of each of the measurements:

\begin{equation}
\label{JNBeq_12}
p\left(Z_E, Z_V|\theta\right)=\prod\limits_{i=1}^{n_Z}p\left(Z_E^i,Z_V^i|\theta \right)
\end{equation}

The energy and direction likelihood for one single measurement $i$, $p(Z_E^i,Z_V^i |\theta)$, corrupted by \textit{true} energy-dependent white measurement noise $v_E (E^i,d)$, but assumed (in the estimator) corrupted by \textit{measured} energy-dependent white measurement noise $\sigma(Z_E^i,d)$, equals the sum of a convex\footnote{A convex combination is a weighted sum of a finite number of points where all coefficients are non-negative and sum up to unity.} combination of likelihoods representing a finite set of possible sources, commonly known as a mixture distribution.  Stated in words, the mixture distribution simply states that each measurement must come from some source included in the model (be it a known source or an unknown source that we are trying to estimate).

\begin{equation}
\label{JNBeq_13}
p\left(Z_E^i,Z_V^i|\theta \right)=\sum\limits_{j=1}^{n_u}w_jp\left(Z_E^i,Z_V^i|\theta_u^j,\theta_c\right)+\sum\limits_{l=1}^{n_k}w_lp\left(Z_E^i,Z_V^i|\theta_k^l, \theta_c\right)
\end{equation}

\noindent The mixture weight $w_j$ for unknown source $j$, $\theta_u^j$, is defined as the number of expected measurements from source $\theta_u^j$ divided by the total number of measurements expected from all sources in $\theta$, known and unknown:

\begin{equation}
w_j = \frac{\bar{n}\left(\theta_u^j\right)}{\sum\limits_{j}\bar{n}\left(\theta_u^j\right)+\sum\limits_{l}\bar{n}\left (\theta_k^l\right)}
\label{JNBeq_13.1.1}
\end{equation}

\noindent The mixture weight $w_l$ for the known source $l$, $\theta_k^l$, is correspondingly defined:

\begin{equation}
w_l = \frac{\bar{n}\left(\theta_k^l\right)}{\sum\limits_{j}\bar{n}\left(\theta_u^j\right)+\sum\limits_{l}\bar{n}\left(\theta_k^l\right)}
\label{JNBeq_13.1.2}
\end{equation}
where $\bar{n}(\theta_u^j)$ is just the cumulative distribution function (cdf) of the numerator inside the integral in Equation \ref{JNBeq_15}, $\mathrm{cdf}_d(\theta_u^j)$, defined in Equation \ref{JNBeq_16} below for detector \textit{d}. Thus $\bar{n}(\theta_u^j)=\mathrm{cdf}_d(\theta_u^j)$ and $\bar{n}(\theta_k^l)=\mathrm{cdf}_d(\theta_k^l)$.

The likelihood of a single measurement $i$ given a single \textit{unknown} source $\theta_u^j$\footnote{Note that the process is nearly identical when defining the likelihood of a single measurement $i$ given a single \textit{known} source $\theta_k^l$} can now be defined. The likelihood of energy and direction measurement $i$ given unknown source $\theta_u^j$ and constants $ \theta_c,$ prior to weighting by $w_j$ can finally be split into independent energy and direction likelihoods:

\begin{equation}
\label{JNBeq_14}
p\left(Z_E^i,Z_V^i|\theta_u^j, \theta_c\right) = p\left (Z_E^i|\theta_u^j, \theta_c\right)p\left (Z_V^i|\theta_u^j, \theta_c\right)
\end{equation}
allowing each of these independent likelihoods to be defined in turn. First the energy likelihood is defined: the likelihood of energy measurement $i$ given unknown source $\theta_u^j$ and constants $\theta_c$, prior to weighting by $ w_j:$

\begin{equation}
\label{JNBeq_15}
p\left(Z_E^i|\theta_u^j, \theta_c\right)=\left. \int\limits_{1.8\mathrm{MeV}}^{11\mathrm{MeV}}\frac{s_{\theta_{u}^{j}}{(E)s_d(E)f_ {\theta_c}\left(E,r_d^j\right)}}{\mathrm{cdf}_d\left (\theta_u^j\right)}\mathcal{N}\left (Z_E^i, \sigma_d^i\right)dE \right|_{E=E_{\bar{\nu}}}
\end{equation}
where the $\mathrm{cdf}_d (\theta_u^j)$ merely normalizes the unsmeared energy spectrum in the numerator into a proper pdf:

\begin{equation}
\label{JNBeq_16}
\mathrm{cdf}_d\left(\theta_u^j\right)= \left. \int\limits_{1.8\mathrm{MeV}}^{11\mathrm{MeV}}s_{\theta_{u}^{j}}{(E)s_d(E)f_{\theta_c}\left (E,r_d^j\right)}dE \right|_{E=E_{\bar{\nu}}}
\end{equation}

Matter oscillation effects are assumed to be negligible and, as such, we ignore the Mikheyev\textendash Smirnov\textendash Wolfenstein (MSW) effect \cite{Wolfenstein}. Following this simplifying assumption, the $\nuebar$ survival fraction $f_{\theta_c}\left (E,r_d^j\right)$ used in Equations \ref{JNBeq_15} and \ref{JNBeq_16} can now be defined for neutrino sources.  At a given range $r$ (in meters) and neutrino energy $E$ (in MeV), $f_{\theta_c}\left (E,r_d^j\right)$ is defined by:

\begin{eqnarray}
\label{JNBeq_17_0}
f\left(E,r\right)=1-\cos^4\theta_{13}\sin^2 2\theta_{12}\sin^2\frac{1.27\Delta m^2_{12} r}{E}\nonumber\\
-\sin^2 2\theta_{13}\cos^2 \theta_{12}\sin^2\frac{1.27\Delta m^2_{23} r}{E}\nonumber\\
-\sin^2 2\theta_{13}\sin^2 \theta_{12}\sin^2\frac{1.27\Delta m^2_{31} r}{E}
\end{eqnarray}
where $\Delta m_{23}^2 = \Delta m^2_{13}-\Delta m^2_{12}$. We've assumed the following nominal values for the neutrino mixing angles and mass constants contained in $\theta_c$ per Fogli \textit{et al.} \cite{fogli_2012}:

\begin{eqnarray} 
\theta_{12}=0.566 \nonumber\\
\theta_{13}=0.232 \nonumber\\
\sin^2 2\theta_{12}=0.856 \nonumber\\
\sin^2 2\theta_{13}=0.156 \nonumber\\
\Delta m^2_{12}=8.00\times 10^{-5} eV^2\nonumber\\
\Delta m^2_{13}=2.50\times 10^{-3} eV^2\nonumber\\
\Delta m^2_{23}=2.42\times 10^{-3} eV^2\nonumber
\end{eqnarray}

\begin{figure}[htbp]
\centering
\includegraphics[width=1\linewidth]{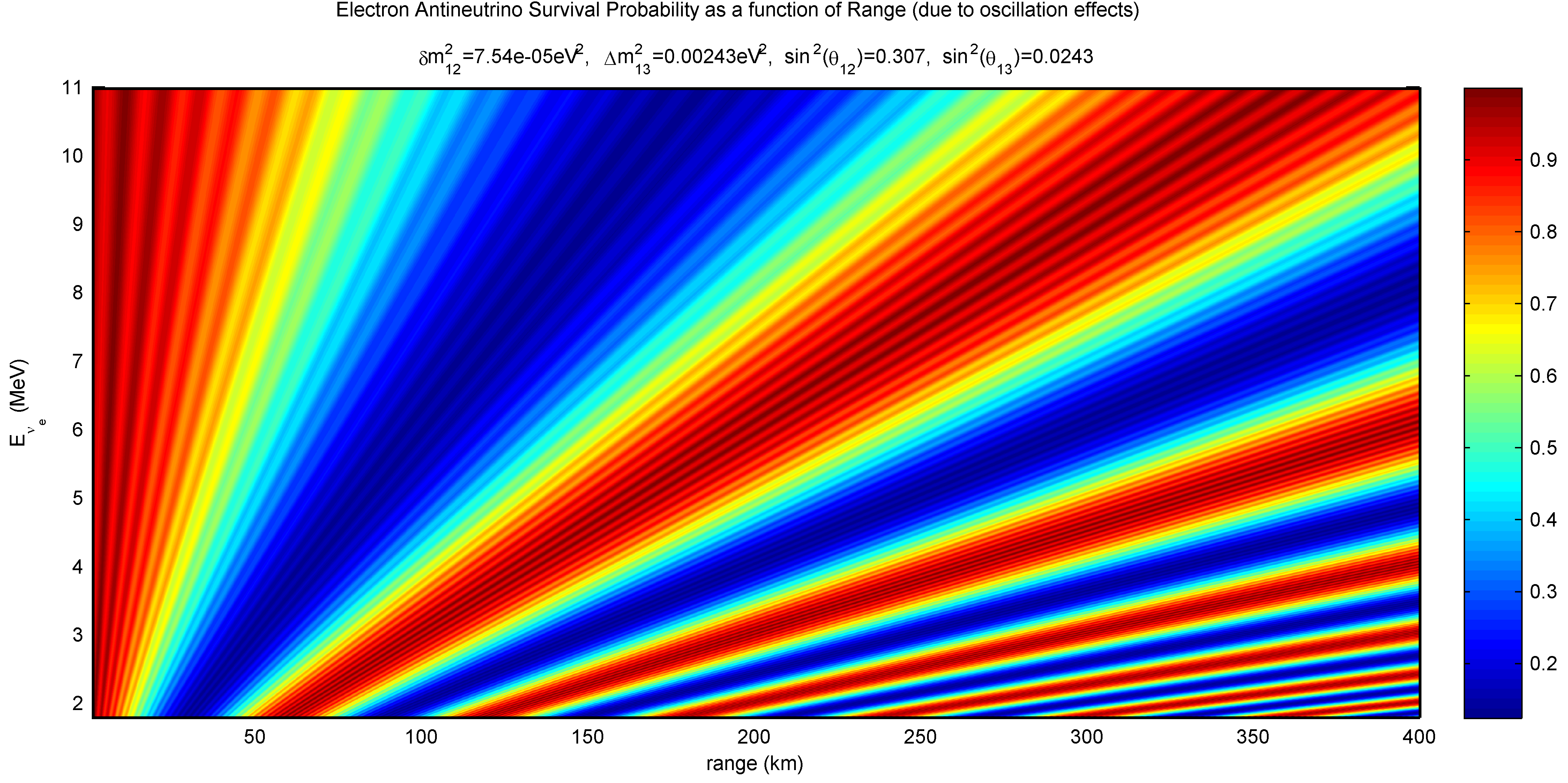}
\caption{$\nuebar$ survival fraction versus range, produced by the evaluation of Equation \ref{JNBeq_17}. Note that, as reflected in this figure, Equation \ref{JNBeq_17} does not evaluate the solid angle as a function of range. This task is performed by the mixture weights in Equation \ref{JNBeq_13}.}
\label{oscFractionSurf}
\end{figure}

After applying the above constants, Equation \ref{JNBeq_17_0} can be approximated by Equation \ref{JNBeq_17} which is evaluated in Figure \ref{oscFractionSurf}:

\begin{equation}
\label{JNBeq_17}
f\left(E,r\right)\approx 0.532+0.029\cos\left(\frac{6.147r}{E}\right)+\ 0.071\cos\left(\frac{6.350r}{E}\right)+\ 0.368\cos\left(\frac{0.203r}{E}\right)
\end{equation}

In Equations \ref{JNBeq_15} and \ref{JNBeq_16}, $s_{\theta_u^j}$ is the assumed zero-range neutrino energy spectrum from unknown source $j$ in the unknown parameter vector $\theta_u$. For a reactor, $s_{\theta_u^j}$ may be approximated by:

\begin{equation}
\label{JNBeq_18}
s_{\theta_u^j}\left(E\right)\left[\frac{1}{\mathrm{sec} \times \mathrm{GW_{th}}} \right]\approx 1.407 \times \left(1.872 \times 10^{20} \right)\left(\frac{1}{3} \right)
\times e^{-(0.3125E+0.25)^{2}}
\end{equation}

\begin{figure}[htbp]
\centering
\includegraphics[width=.5\linewidth]{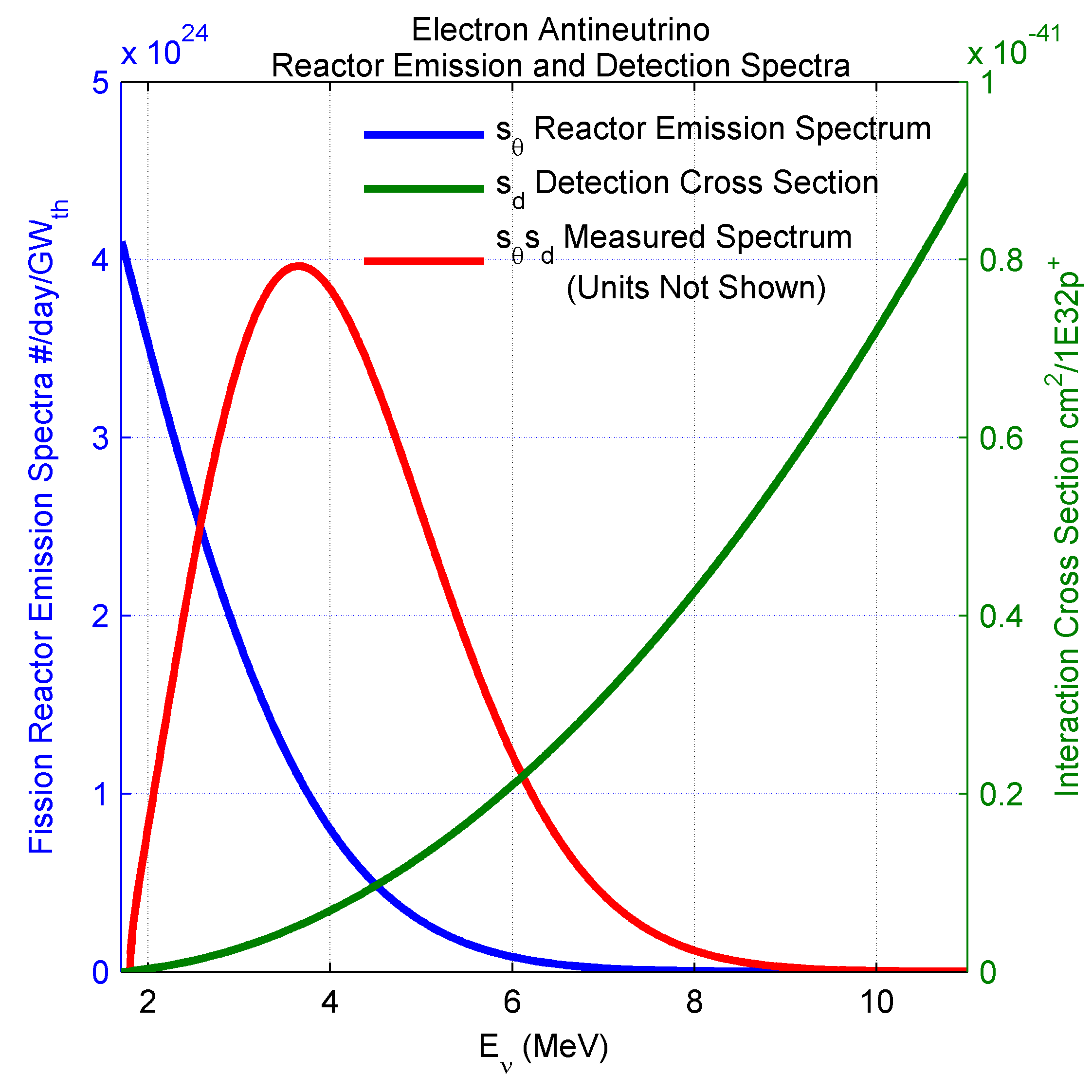}
\caption{$\nuebar$ reactor emission and detection spectra. The measured spectrum is the product of the emitted and the detected spectra, and represents the inverse beta decay $\bar{\nu}_e$ spectrum that the detector should expect to measure from a nearby nuclear power plant. The detection spectra should be independent of the $\bar{\nu}_e$ source, and applies to geo-neutrinos as well as reactor neutrinos.}
\label{twoXsections}
\end{figure}

Equation \ref{JNBeq_18}, evaluated in Figure \ref{twoXsections}, uses the 2$^{nd}$ order exponential flux model first proposed by Vogel and Engel \cite{vogel 1989} with updated model coefficients.  It assumes 200MeV per fission and 6 antineutrinos of all energies per fission to get $1.872 \times 10^{20}$ antineutrinos per GW$_{\mathrm{th}}$ per second, per Bernstein \textit{et al.} \cite{bernstein_2009}. The fraction $\frac{1}{3}$ represents the proportion of neutrinos produced above $E_{\bar{\nu}}\geq 1.8$MeV, per Gil-Botella \cite{Gil-Botella}.

The detector IBD cross section $s_d$ is shown in Figure \ref{twoXsections}; it is defined by the constants in $\theta_c,$ and approximated by:

\begin{equation}
\label{JNBeq_19} 
s_d(E)\left[\frac{\mathrm{cm}^2}{10^{32}\mathrm{p}^+} \right ]\approx (0.095E-0.123)\times \sqrt{(E-1.293)^2-0.261} \times 10^{-42}
\end{equation}

In Equation \ref{JNBeq_15}, $\mathcal{N}(Z_E^i,\sigma_d^i)$ describes the assumed energy measurement uncertainty about energy measurement $Z_E^i$. This uncertainty is approximated by a normal distribution with mean $Z_E^i$ and variance $\sigma_{d^i}^2,$ where $\sigma_d^i$ is the $1\sigma$ standard deviation attached to energy measurement $i$, interpolated from previously compiled MC results using GEANT and MATLAB. Our TREND energy MC results, as well as a least squares fit to the results, are shown in Section \ref{TREND energy and direction resolution MC results}.

\begin{figure}[htbp]
\centering
\includegraphics[width=\linewidth]{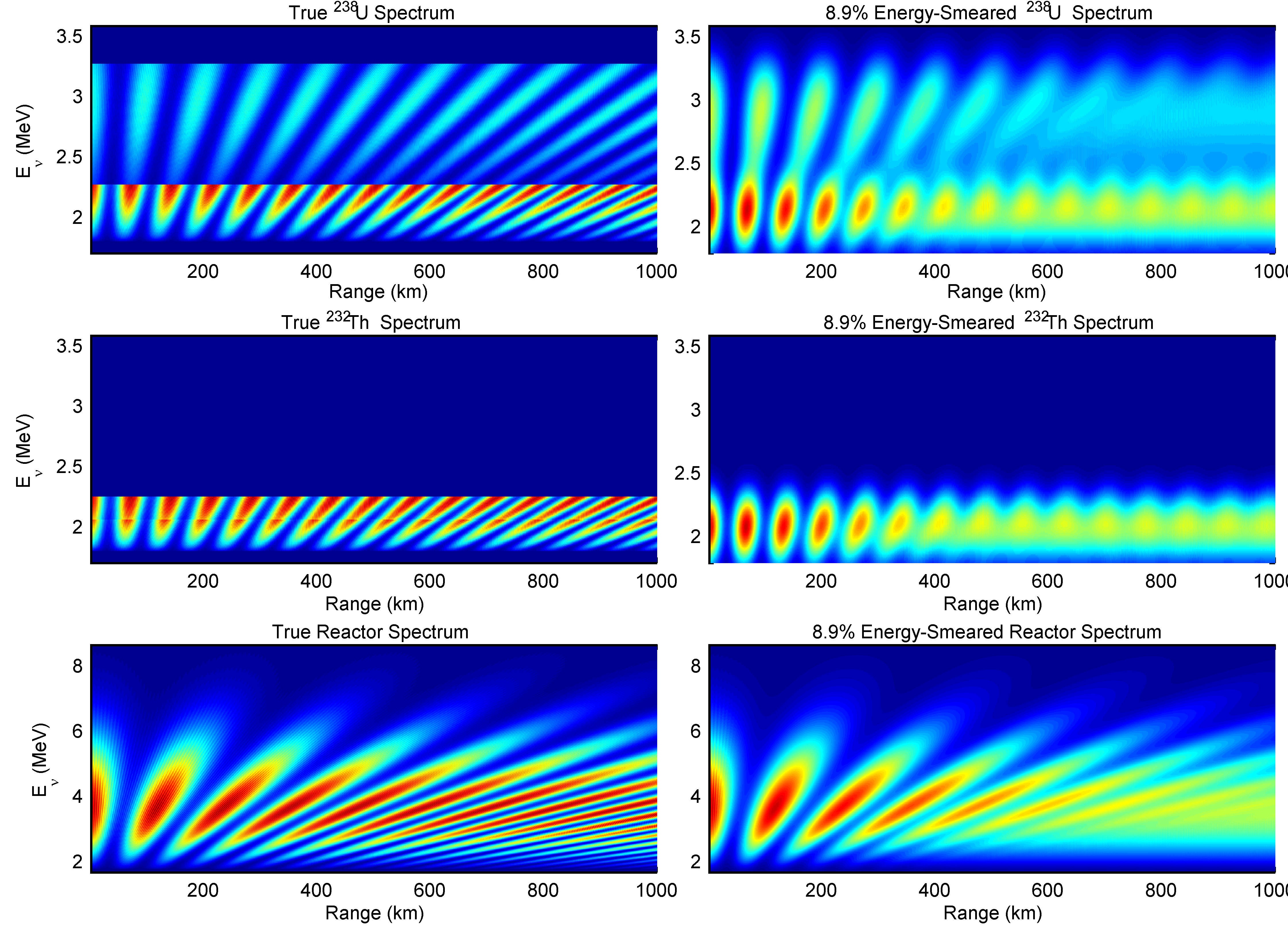}
\caption{Comparison of true and energy-smeared $\nuebar$ spectra vs range for reactors, $^{238}$U and $^{232}$Th, without solid-angle compensation.}
\label{3spectra}
\end{figure}

Each energy measurement carries a different energy measurement noise, with higher energy measurements being associated with higher absolute energy measurement uncertainty (though less fractional uncertainty). This mandates an energy-dependent ``smear" of the expected energy measurement space, shown in Figure \ref{3spectra}. Note that higher energies in Figure \ref{3spectra} can be observed to be smeared to a greater degree than lower energy rows.

In the absence of energy measurement noise, the true energy likelihood would be:

\begin{equation}
\label{JNBeq_20}
p\left(Z_E^i|\theta_u^j,\theta_c\right)=\left. \frac{s_{\theta_{u}^j}(E)s_d(E)f_{\theta_c}\left(E,r_d^j \right )}{\mathrm{cdf}_d\left(\theta_u^j \right )} \right|_{E=Z_E^i}
\end{equation}

\noindent This unsmeared energy spectrum, dependent only on source spectrum and detector range, can be seen on the left side of Figure \ref{3spectra} for the three different antineutrino sources. Conditioning Equation \ref{JNBeq_20} above on the uncertainty in the energy measurements, we get our original smeared Equation \ref{JNBeq_15}, shown on the right side of Figure \ref{3spectra}. It is this smeared energy measurement which is used to evaluate the likelihood of noisy energy measurements $Z_E$. 

Once the likelihood of energy measurement $i$ given \textit{unknown sources} ($\theta_u$) has been evaluated, the process is repeated identically for all the \textit{known} sources ($\theta_k$). The likelihood of the same energy measurement $i$ given known source $\theta_k^l$ and constants $\theta_c$, prior to weighting by $w_l$ is:
\begin{equation}
\label{JNBeq_21}
p\left(Z_E^i|\theta_k^l, \theta_c\right)= \left. 
\int\limits_{1.8\mathrm{MeV}}^{11\mathrm{MeV}}\frac{s_{\theta_{k}^{l}}{(E)s_d(E)f_{\theta_c}\left(E,r_d^l\right)}}{\mathrm{cdf}_d\left(\theta_k^l\right)}\mathcal{N}\left(Z_E^i, \sigma_d^i\right)dE \right|_{E=E_{\bar{\nu}}}
\end{equation}
where $s_{\theta_{k}^{l}}$ now represents any number of possible known-source spectra. This zero-range spectrum may represent geo-neutrinos emitted by $^{238}$U or $^{232}$Th decay. It may also represent the same zero-range nuclear reactor energy spectrum as $s_{\theta_{u}^{j}}$, or it may represent one of the three non-neutrino energy spectra the detector will encounter: accidentals, fast neutrons and cosmogenics. Naturally the oscillation survival fraction function $f_{\theta_c}\left(E,r_d^l\right)=1.0$ for all non-neutrino sources, as they do not partake in neutrino oscillations.

The energy likelihoods have now been explicitly defined. The \textit{direction vector} likelihood is expressed in Equation \ref{JNBeq_22}: the likelihood of direction measurement $i$ given unknown source $\theta_u^j$ and constants $\theta_c$, prior to weighting by $w_j$. This equation was derived specifically for this application and represents an analytical expression for the angular pdf of a point source in terms of the angular signal-to-noise ratio (SNR). Equation \ref{JNBeq_22} replaces other commonly-used directional distributions such as the Von Mises-Fisher\footnote{In directional statistics, the Von Mises-Fisher distribution uses a ``mean direction" and ``concentration factor" to form a probability distribution defined on an $n$-dimensional sphere. When n=2, the distribution reduces to the Von-Mises distribution on a circle.} distribution, and has been numerically verified via MC methods.

\begin{eqnarray}
\label{JNBeq_22}
p\left(Z_V^i|\theta_u^j, \theta_c\right)=\frac{1}{4}\left(\mathrm{SNR}^2_d\left(Z_E^i \right )\cos^2\phi_d\left(Z_V^i, \theta_u^j \right ) +1\right )\mathrm{erfc}\left(\frac{-\mathrm{SNR}_d\left(Z_E^i \right )}{\sqrt{2}}\mathrm{\ cos}\phi_d\left(Z_V^i, \theta_u^j \right ) \right )\nonumber \\
\times\left(e^{-\frac{1}{2}\mathrm{SNR}^2_d\left(Z_E^i \right) \sin^2\phi_d\left(Z_V^i, \theta_u^j\right)}\right)+\left(\frac{2\mathrm{SNR}_d\left(Z_E^i \right )}{\sqrt{2\pi}}\mathrm{\ cos}\phi_d\left(Z_V^i, \theta_u^j \right ) \right ) e^{-\frac{1}{2}\mathrm{SNR}^2_d\left(Z_E^i \right )}
\end{eqnarray}
where $\phi_d (Z_V^i,\theta_u^j )$ is the angle between the measured direction vector $Z_V^i$ and the unit vector starting at detector $d$ and terminating at source $\theta_u^j$.

The assumed SNR in Equation \ref{JNBeq_22}, $\mathrm{SNR}_d (Z_E^i )$, is a function of measured neutrino energy $Z_E^i$ and detector $d$ hardware: 

\begin{equation}
\label{JNBeq_23a}
\mathrm{SNR}_d(E) = \frac{\mu_{d_{MC}}(E)}{\sigma_x^{d_{MC}}(E)}
\end{equation}

where $\mu_{d_{MC}}(E)$ is the mean forward-scatter range of the reconstructed positron-neutron vectors $V_{xyz}^{d_{MC}}(E)$ (in TREND $\mu_{d_{MC}}(4\mathrm{MeV})\approx 16$mm):

\begin{equation}
\label{JNBeq_23c}
\mu_{d_{MC}}(E) = \left \| V_{xyz}^{d_{MC}}(E) \right \|
\end{equation}

and $\sigma_x^{d_{MC}}(E)$ is the Cartesian uncertainty about each of the 3 axes of the mean reconstructed positron-neutron vector (in TREND $\sigma_x^{d_{MC}}(4\mathrm{MeV})\approx 460$mm). MC results show that vector uncertainty will be nearly equal about each of the 3 cartesian axes: x, y and z:

\begin{equation}
\label{JNBeq_23b}
\sigma_x^{d_{MC}}\approx\sigma_y^{d_{MC}}\approx\sigma_z^{d_{MC}}
\end{equation}

Both the mean vector range $\mu$ and the mean vector uncertainty $\sigma$ are functions of detector $d$'s scintillator and hardware and the original neutrino energy $E$, and should ideally be determined via MC methods using GEANT.

\begin{figure}[htbp]
\centering
\includegraphics[width=\linewidth]{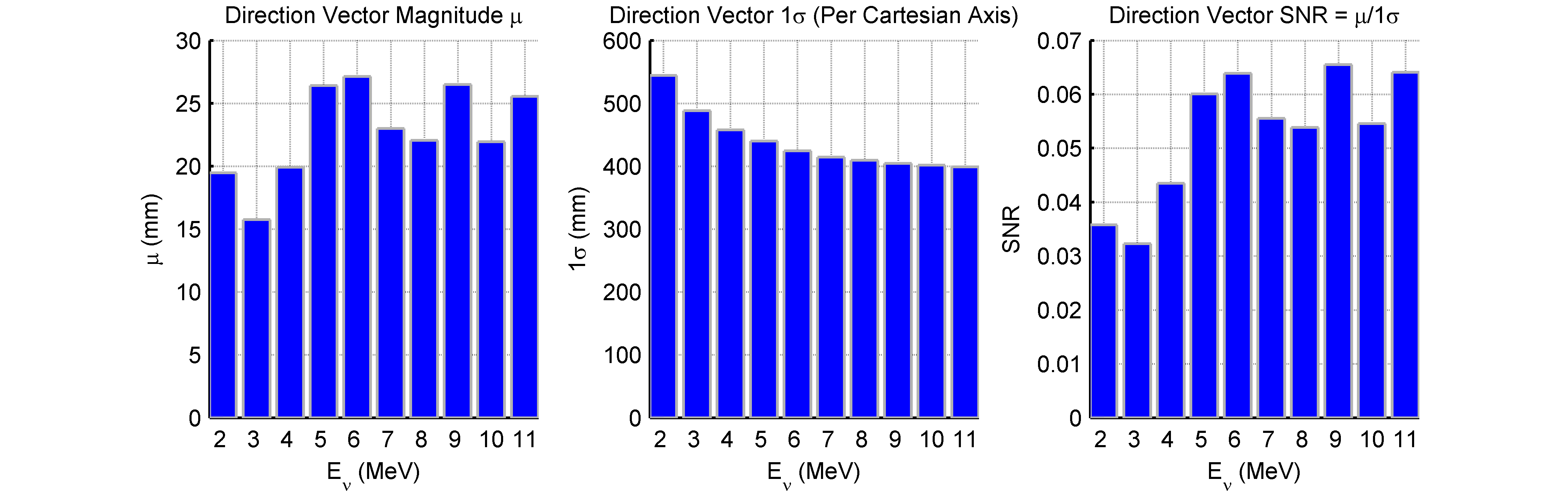}
\caption{TREND MC direction vector resolution as a function of $E_{\nuebar}$. Mean reconstruction vector resolution of 461mm $1\sigma$ per axis includes additive noise incurred by neutron random-walk natural to inverse beta decay, and is not simply the norm of the 374mm positron CE $1\sigma$ and 243mm neutron CE $1\sigma$ values. See Section \ref{TREND energy and direction resolution MC results} for more details.}
\label{SNR1}
\end{figure}

\begin{figure}[htbp]
\centering
\includegraphics[width=\linewidth]{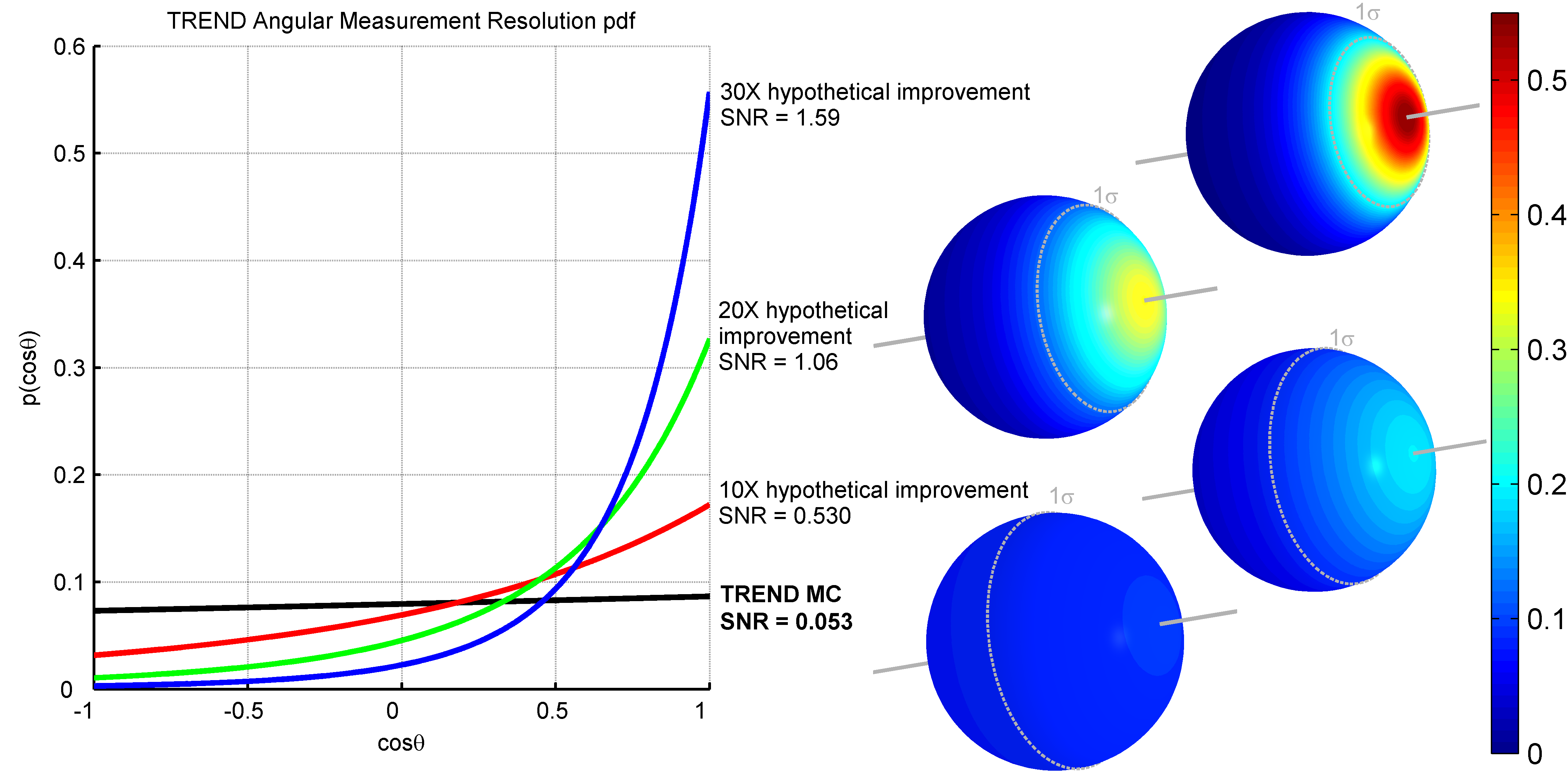}
\caption{Directional probability plot produced by evaluation of Equation \ref{JNBeq_22} for the TREND MC-derived mean angular SNR, as well as hypothetical improvements in TREND SNR by factors of 10X, 20X, and 30X.}
\label{SNR2}
\end{figure}

Appropriate $\mathrm{SNR}_d (E) $ values for TREND were determined via MC simulation of neutrino IBD events within the modeled detector at various neutrino energy levels. Figure \ref{SNR1} shows TREND MC angle SNR results, and Figure \ref{SNR2} shows the application of Equation \ref{JNBeq_22} to the TREND MC baseline SNR of about 0.05, as well as hypothetical improvements in direction vector resolution by factors of 10X, 20X, and 30X.


Once the likelihood of direction measurement $i$ given all the \textit{unknown} sources ($\theta_u$) has been evaluated, the process is repeated identically for all the \textit{known} sources ($\theta_k.$) The corresponding likelihood of the \textit{same direction measurement} $i$ given known source $\theta_k^l$ and constants $\theta_c$, prior to weighting by $w_l$ is:
\begin{eqnarray}
\label{JNBeq_24} 
p\left(Z_V^i|\theta_k^l, \theta_c\right)=\frac{1}{4}\left(\mathrm{SNR}^2_d\left(Z_E^i \right )\cos^2\phi_d\left(Z_V^i, \theta_k^l \right ) +1\right )\mathrm{erfc}\left(\frac{-\mathrm{SNR}_d\left(Z_E^i \right )}{\sqrt{2}}\mathrm{\ cos}\phi_d\left(Z_V^i, \theta_k^l \right ) \right )\nonumber\\
\times \left(e^{-\frac{1}{2}\mathrm{SNR}^2_d\left(Z_E^i \right ) \sin^2\phi_d\left(Z_V^i, \theta_k^l\right)}\right)+\left(\frac{2\mathrm{SNR}_d\left(Z_E^i \right )}{\sqrt{2\pi}}\mathrm{\ cos}\phi_d\left(Z_V^i, \theta_k^l \right ) \right ) e^{-\frac{1}{2}\mathrm{SNR}^2_d\left(Z_E^i \right )}
\end{eqnarray}

Thus the complete direction likelihood has been explicitly defined and, with it, the entire likelihood $p(Z|\theta)$. Substituting Equation \ref{JNBeq_10} into Equation \ref{JNBeq_6}, and taking the product through all detectors $1$ through $n_d$, we get our complete Bayesian posterior probability:

\begin{eqnarray}
\label{JNBeq_25}
p\left(\theta|Z \right )= \frac{1}{c}\int\limits_{-\infty }^{\infty }\cdots \int\limits_{-\infty }^{\infty }\prod\limits_{d=1}^{n_d}p(n_Z|\theta)p\left(Z_E, Z_V|\theta\right)p(\theta)d\theta_1\cdots d\theta_n
\end{eqnarray}

All that is left is to optimize a value for $\theta$ which maximizes the \textit{a posteriori} probability $p(\theta|Z)$. A MAP estimator, using the objective function defined in Equation \ref{JNBeq_25}, should be the optimal estimator for this task.

\subsection{Applied suboptimal MAP estimator}
\label{Applied suboptimal MAP estimator}
Unfortunately, the aforementioned optimal MAP estimator is likely not tractable due, primarily, to the multidimensional integration called for during marginalization of the nuisance parameters. To address this, we have developed two suboptimal approximations to Equation \ref{JNBeq_25} which attempt to strike a balance between accuracy and computational efficiency. 

The oscillation parameter approximation presented in Section \ref{Antineutrino oscillation parameter estimation} seeks to decrease the uncertainty about the four $\nuebar$ oscillation parameters (two mixing angles and two mass differences). It optimizes the four oscillation parameters as the primary parameters of interest, but it also carries six nuisance parameters into the optimization as well, which help define the flux at the detector from each of six main categories: known reactors, crust, mantle, accidentals, fast neutrons and cosmogenics. This approximation assumes current \textit{a priori} oscillation parameter observability levels per Fogli \textit{et al.} \cite{fogli_2012}. These values are shown in Table \ref{table:worldwide_op}.

The unknown-reactor approximation presented in Section \ref{Reactor geolocation}, on the other hand, seeks out the presence of possible unknown reactors within a defined search area. It optimizes along nine dimensions, three for each unknown reactor (only one unknown reactor is assumed here), and one for each of the six background flux categories. This approximation enjoys the use of \textit{a posteriori} oscillation uncertainty values found in our Section \ref{Oscillation parameter estimation results} (four-detector worldwide oscillation estimation MC runs). These \textit{a posteriori} oscillation uncertainties can be found in Table \ref{table:worldwide_op} also, and are much smaller than the \textit{a priori} values assumed by the oscillation parameter estimator approximation in Section \ref{Oscillation parameter estimation results}.

In addition to smaller oscillation parameter uncertainty, the unknown-reactor approximation presented in Section \ref{Reactor geolocation} also benefits from smaller background flux uncertainty, since it uses the \textit{a posteriori} flux uncertainties shown in Table \ref{table:worldwide_flux}, derived simultaneously with the oscillation results during the four-detector worldwide optimization performed in Section \ref{Oscillation parameter estimation results}.

\subsubsection{Reactor geolocation}
\label{Reactor geolocation}
We designate an unknown-reactor estimator suboptimal \textit{a posteriori} probability $\tilde{p}(\theta|Z)$, and assert that it is approximately equal to the optimal \textit{a posteriori} probability $p(\theta|Z)$:

\begin{equation}
\label{JNBeq_26}
\tilde{p}\left(\theta|Z \right )\approx p\left(\theta|Z \right )
\end{equation}

\noindent The central tenet of the approximation is that it assumes no uncertainty about any non-estimated nuisance parameter ($x$) in $\theta_k$ and $\theta_u$:

\begin{equation}
\label{JNBeq_27}
p(x) = \delta(x)
\end{equation}

\noindent This removes the marginalization requirements through the non-estimated nuisance dimensions, correspondingly reducing computational burden. We designate our new Bayesian prior approximation as $\tilde{p}(\theta)$, and assert that it is approximately equal to $p(\theta)$ given certain assumptions.

\begin{equation}
\label{JNBeq_28}
\tilde{p}(\theta) \approx p(\theta)
\end{equation}
where the assumptions are:

\begin{enumerate}
\item low variances in $p(\theta)$
\item symmetry in the topology of $p(\theta)$ about the expected mean
\item low correlation across dimensions of $p(\theta|Z)$
\end{enumerate}

\noindent These assumptions are naturally imperfect. Their validity will vary from one scenario to another, and also from one nuisance parameter to another. In general however, we recognize that the more valid the assumptions, the more similar $\tilde{p}(\theta|Z)$ will be to $p(\theta|Z)$. 

In order to achieve the most accurate approximation we seek to reduce the gap between the assumptions and the truth as much as possible. One way to do this is to estimate more accurate values for select uncertain parameters in $\theta_k$ and $\theta_c$ (rather than marginalizing). To do this we retain the full priors $p(\theta_k)$ and $p(\theta_c)$ for the select parameters we choose to estimate while assigning delta functions $\delta(\theta)$ for the rest. We can co-estimate these select nuisance parameters simultaneously with the unknown parameters in $\theta_u$.

Sensitive to the computational burden of the problem and our limited computer resources, we must carefully select the most effective parameters to optimize. Among the most important parameters to us are systematic correction factors (applied as gains in $\theta_c$) to our understanding of the mean background flux from each background flux category. We have six gains in $\theta_c$ which apply to each of the six main background categories: IAEA, crust, mantle, accidental, fast neutron, and cosmogenic:

\begin{equation}
\label{JNBeq_30}
\theta_c = \left[ \begin{array}{c}
\mathrm{}PREM
\\ \bar{\nu}_{\theta_{13}} 
\\ \vdots
\\ g_{\mathrm{IAEA}}
\\ g_{\mathrm{crust}}
\\ g_{\mathrm{mantle}}
\\ g_{\mathrm{accidental}}
\\ g_{\mathrm{fast\ n^0 }}
\\ g_{\mathrm{cosmogenic}}
\end{array} \right]
\end{equation}

Each flux gain ($g$) is valued at 1.0 by default to reflect current knowledge. We choose to co-estimate these gains in $\theta_c$ with the unknown parameters of interest in $\theta_u$, applying estimated correction gains ($\hat{g}$) to expected counts ($\bar{n}$) during optimization:

\begin{equation}
\label{JNBeq_31}
\bar{n}\left(\theta_k^l \right )g_l \to \bar{n}\left(\theta_k^l \right )\hat{g_l}
\end{equation}

Estimating these six gains ($g_l$) in $\theta_c$ will reduce their \textit{a posteriori} uncertainties, Var$\left(\theta_c\right)$, and help legitimize our first assumption: low variance in $p(\theta_c)$. We assert that $\theta_c$ \textit{a posteriori} uncertainty will be less than $\theta_c$ \textit{a priori} uncertainty following optimization:

\begin{equation}
\label{JNBeq_32}
\mathrm{Var}(\hat\theta_c) < \mathrm{Var}(\theta_c)
\end{equation}

Some elements of the parameter space are used repeatedly by the optimizer during low level calculations and, where appropriate, these low level calculations can be bundled into pre-computed ``lookup tables". Lookup tables not only reduce computation time, but they can also \textit{increase the fidelity of the approximation}.  This is because the marginalization of some uncertain parameters can be replaced by the appropriate smearing of specific pre-computed lookup tables.  This enables certain parameters to be marginalized at relatively little computation expense, even when using the suboptimal approximation to the estimator.  We have found that the largest return for this computational investment is in pre-computing a smeared neutrino survival fraction (Equation \ref{JNBeq_17_0}) to account for the uncertainty in the four oscillation parameters.

Assuming matter oscillation effects to be negligible, we ignore the MSW effect \cite{Wolfenstein} and define the \textit{oscillation-smeared} $\nuebar$ survival fraction $f(E,r)$ at a given range $r$ (in meters) and antineutrino energy $E$ (in MeV) to be:

\begin{equation}
\label{smear1}
f(E,r) = \int\limits_{-\infty}^{\infty} \int\limits_{-\infty}^{\infty} \int\limits_{-\infty}^{\infty} \int\limits_{-\infty}^{\infty}
f(E,r,\theta_{13},\Delta m^2_{13},\theta_{12},\Delta m^2_{12})
d \left( \Delta m^2_{13} \right) d \left( \sin^2 \theta_{13} \right) 
d \left( \Delta m^2_{12} \right) d \left( \sin^2 \theta_{12} \right) 
\end{equation}
where $f(E,r,\theta_{13},\Delta m^2_{13},\theta_{12},\Delta m^2_{12})$ is defined previously in Equation \ref{JNBeq_17_0}. Following integration (or \textit{marginalization}) through the four oscillation parameter dimensions as shown in Equation \ref{smear1}, $f(E,r)$ now completely captures all  oscillation-related uncertainty; it is no longer conditional on the exact values of any of the uncertain oscillation parameters.

For the \textit{a priori} oscillation uncertainties in Equation \ref{smear1} we assume modern values from Fogli \textit{et al.} \cite{fogli_2012}.  We assume that applied experiments, such as a search for an unknown reactor, occur only after the TREND detectors have been calibrated and deployed for a year to refine estimates of the oscillation parameters and to establish gain corrections for our knowledge of the background flux (see Section \ref{Oscillation parameter estimation results}).  To this end we assume the oscillation parameter search \textit{a posteriori} values from Table \ref{table:worldwide_op} in Section \ref{Oscillation parameter estimation results} to be the \textit{a priori} values for all unknown reactor scenarios in Section \ref{Reactor geolocation results}, both as truth in the Geospatial Model and as assumptions used by the estimator.

To the oscillation-smeared lookup table $f(E,r)$ we apply further energy measurement resolution smear per the energy resolution determined from our TREND detector MC runs detailed in Section \ref{Antineutrino Detector Model}: $\left. 8.9\% \right|_{E_\mathrm{vis}=1\mathrm{MeV}}$ for random placement of neutrino events within TREND\footnote{Center-detector vertices were found to have worse energy resolution than uniformly dispersed vertices, presumably because of their greater distance from the closest PMT. Point sources of photons were found to have better energy resolution than $\nuebar$ inverse beta decay prompt signals, generally a few percentage points better for both centered and randomly placed vertices.}. We utilize Equation \ref{JNBeq_21} for this task, creating energy \textit{and} oscillation smeared lookup tables for each of our three neutrino sources: reactors, and $^{238}$U and $^{232}$Th in the Earth's crust and mantle. Each of the three sources passes its respective source spectrum ($s_{\theta_{k}^{l}}$) into Equation \ref{JNBeq_21}. 

\begin{figure}[!htbp]
\includegraphics[width=\linewidth]{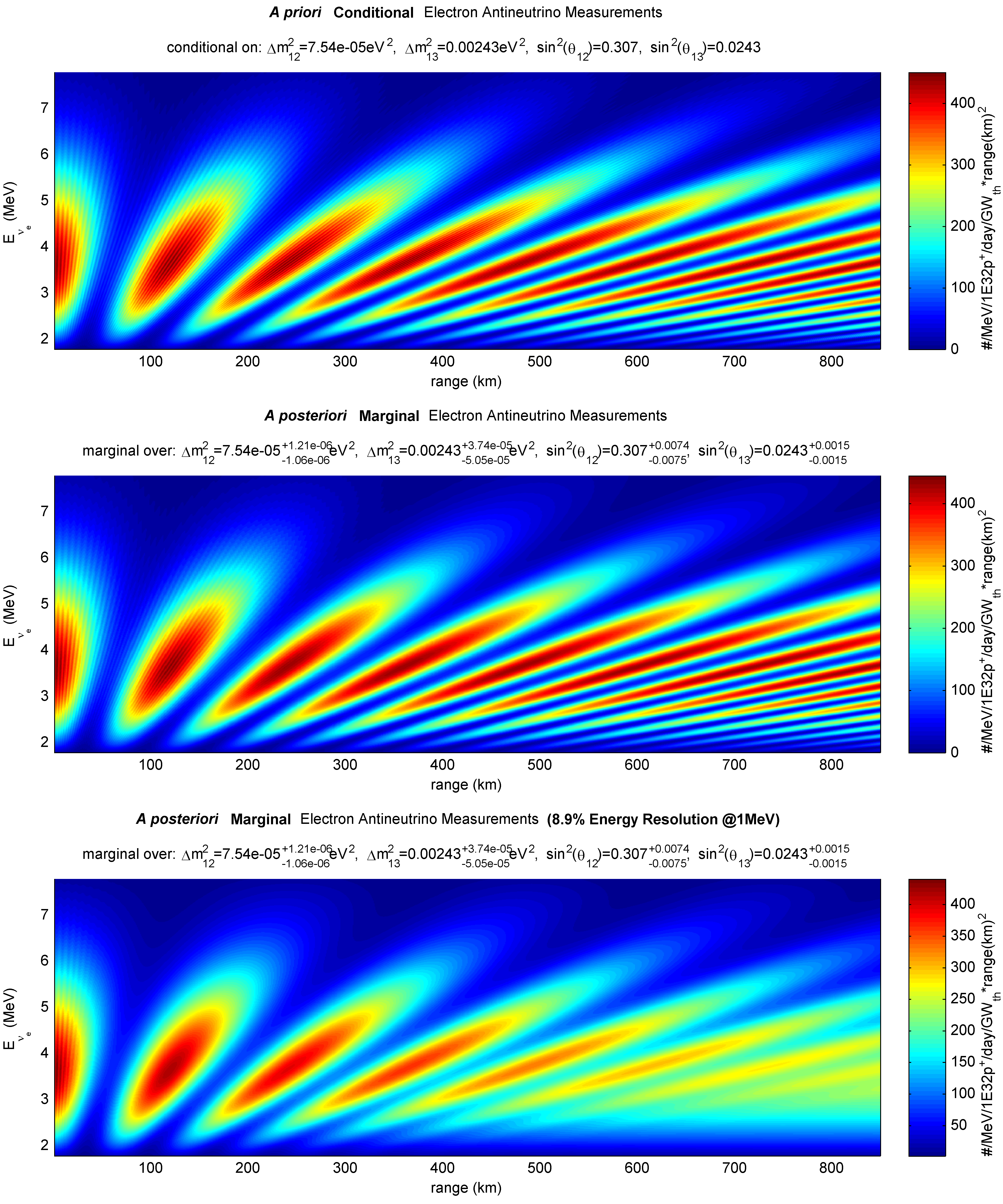}
\centering
\caption{Reactor spectra lookup tables showing \textit{a posteriori} oscillation uncertainty after a 4-TREND detector worldwide optimization simulation detailed in Section \ref{Oscillation parameter estimation results}. Note that the \textit{a posteriori} uncertainty shown here are significantly less than the \textit{a priori} uncertainty presented by Fogli \textit{et al.} \cite{fogli_2012}. The top subplot is the reactor oscillation spectrum before accounting for solid angle, conditional on the maximum likelihood values presented by Fogli \textit{et al.} \cite{fogli_2012}. The second subplot is the marginal spectrum produced by Equation \ref{smear1}, and shows the oscillation-smeared version of the top subplot. The bottom subplot is produced by applying Equation \ref{JNBeq_21} to Equation \ref{smear1}, smearing the middle subplot by TREND's energy measurement resolution of $\left. 8.9\% \right|_{E_\mathrm{vis}=1\mathrm{MeV}}$.}
\label{smearedfigure1}
\end{figure}

A subset\footnote{The data plotted in Figure \ref{smearedfigure1} spans only 0MeV-8MeV and 0km-1000km, though a lookup table in practice should span from 0MeV-11MeV and 0km-13000km, far enough to sufficiently encompass one Earth diameter so that a detector on one side of the Earth may access the lookup table to find the smeared spectrum of a reactor located across the opposite side of the Earth.} of the smeared \textit{reactor}\footnote{Note that corresponding $^{238}$U and $^{232}$Th lookup tables identical to the reactor lookup tables shown in Figure \ref{smearedfigure1} must also be created, though they are not shown in this paper. Smeared geo-neutrino lookup tables will actually be more useful than smeared reactor tables because a good smart integration consisting of millions of point sources of the Earth's crust and mantle will create many point sources at nearly every range in the lookup table out to one Earth diameter, but a reactor lookup table may only be used for a couple hundred detector-reactor ranges because of the limited numbers of known reactors, and the relatively small unknown-reactor search area, which is not likely to span the entire globe.} spectrum lookup table is shown in Figure \ref{smearedfigure1}. The top subplot is the reactor oscillation spectrum before accounting for solid angle, conditional on the maximum likelihood values presented by Fogli \textit{et al.} \cite{fogli_2012}. The second subplot is the marginal spectrum produced by Equation \ref{smear1}, and shows the oscillation-smeared version of the top subplot. The bottom subplot is produced by applying Equation \ref{JNBeq_21} to Equation \ref{smear1}, smearing the middle subplot by TREND's energy measurement resolution of $\left. 8.9\% \right|_{E_\mathrm{vis}=1\mathrm{MeV}}$.

The last approximation we make is to assume perfect knowledge of the unknown reactor energy spectrum $s^j_u$. This reduces our unknown parameter count in $\theta_u$ from four to three. We rationalize this change by noting the very small difference in reactor energy spectrum (shape) over the lifetime of its fuel, and our low signal statistics. In a higher signal-to-background environment the time-varying energy spectrum will be more observable. However, in our long-range ($>100$km) reactor scenarios, we expect very low observability of the true energy spectrum of the unknown reactor. This is due entirely to the low signal rate $n_Z<100$ and the high background rate $n_Z>1000$ typical for long-range unknown reactor scenarios.

Implementing the approximation per the stated changes to the optimal estimator, we now have nine dimensions through which we wish to optimize, and none which we wish to marginalize. The nine optimization dimensions are the unknown reactor latitude, longitude and power, and the six background flux correction gains:

\begin{equation}
\label{JNBeq_33}
\theta_{opt}= \left[\begin{array}{c}
\left[\begin{array}{c}\phi\\\lambda\\p\end{array}\right]_{\theta_u}
\\ 
\left[\begin{array}{c}g_{\mathrm{IAEA}}\\ g_{\mathrm{crust}}\\ g_{\mathrm{mantle}}\\ g_{\mathrm{accidental}}\\ g_{\mathrm{fast\ n^0}}\\ g_{\mathrm{cosmogenic}}\end{array}\right]\end{array}\right]
\end{equation}

\noindent Simplifying Equation \ref{JNBeq_25} to reflect the approximation expressed in Equation \ref{JNBeq_26} we get:

\begin{equation}
\label{JNBeq_34}
\tilde{p}\left(\theta|Z \right )=\frac{1}{c}\prod\limits_{d=1}^{n_d}p\left( n_Z|\theta\right)p\left(Z_E, Z_V|\theta \right )p(\theta)
\end{equation}

Equation \ref{JNBeq_34} assumes that the uncertainty in the number of measurements is independent from one detector to another. This assumption may be accurate following optimization and application of the six correction gains discussed above, or it may not be depending on the specific scenario. There is one final change we may apply to the approximation to quell any fears over this assumption, while keeping computational burden to a minimum.

Rather than treating the count probability as independent from detector to detector, we can treat it as a multidimensional correlated probability. There is no analytical expression for a correlated multivariate Poisson distribution, however at high counts ($>$30 events) a Poisson distribution approaches a Normal distribution, where the Normal variance is equal to its mean:

\begin{equation}
\label{JNBeq_35}
\mathcal{N}(n,n) = \lim_{n\to\infty}\mathcal{P}(n)
\end{equation}

We expect count rates well over 30 events for our TREND detectors, so we can exploit this characteristic to replace multiple independent Poisson likelihoods with a single multivariate normal likelihood:

\begin{equation}
\label{JNBeq_36.0.1}
\prod\limits_{i=1}^{n_d}\mathcal{P}\left(n_Z^i, \mu_i\right)= \mathcal{N}_{n_d}\left(n_Z; \mu,\Sigma \right )
\end{equation}

\noindent The new count probability, removed outside of the detector product sign, is now simply a multivariate normal likelihood:

\begin{equation}
\label{JNBeq_37.0.1}
\tilde{p}\left( \theta|Z\right)=\frac{1}{c}p\left(n_Z|\theta \right )\prod\limits_{d=1}^{n_d}p\left(Z_E, Z_V|\theta \right )p\left(\theta \right )
\end{equation}
\begin{equation}
\label{JNBeq_37.0.2}
p\left(n_Z|\theta \right ) = \mathcal{N}_{n_d}\left(x;\mu, \Sigma \right )
\end{equation}
\begin{equation}
\label{JNBeq_36.0.2}
\mathcal{N}_{n_d}\left(x; \mu,\Sigma \right)= \frac{1}{\left (2\pi\right)^{\frac{n_d}{2}}\sqrt{|\Sigma|}} e^{-\frac{1}{2}\left(x-\mu \right )'\Sigma^{-1}\left(x-\mu \right )}
\end{equation}
evaluated at the detector's measured counts vector:

\begin{equation}
\label{JNBeq_38}
x = \left[n_Z^1 \cdots n_Z^{n_d}\right ]
\end{equation}
where $\mu$ is a vector of all the detectors' expected measurement counts given $\theta$:

\begin{eqnarray}
\label{JNBeq_39}
&\mu = \left[\bar{n}_1 \cdots \bar{n}_{n_d}\right ]& \label{JNBeq_39.0.1}\\ 
&\bar{n}_i=\sum\limits_{j=1}^{n_u}\bar{n}_i\left(\theta_u^j \right )+\sum\limits_{l=1}^{n_k}\bar{n}_i\left(\theta_k^l \right )&\label{JNBeq_39.0.2}
\end{eqnarray}
and $\Sigma$ is the $n_d$ x $n_d$ count covariance matrix. We assume three separate (and independent) types of uncertainty in $\Sigma$, which we sum together to produce the full covariance matrix:

\begin{equation}
\label{JNBeq_40}
\Sigma = \Sigma_{\mathrm{Poisson}}+\Sigma_{\mathrm{flux}}+\Sigma_{\mathrm{detector}}
\end{equation}

{\bf 1. Poisson Uncertainty}. This represents the count uncertainty we normally associate with a Poisson distribution, where the distribution variance equals the mean:

\begin{equation}
\label{JNBeq_41}
\Sigma_{\mathrm{Poisson}} = \left[ \begin{array}{ccc}
\mu_1 & \cdots & 0\\ 
\vdots &\ddots &\vdots \\ 
0 & \cdots & \mu_n \end{array} \right]
\end{equation}

{\bf 2. Flux Uncertainty} ($\sigma^f$). Defined by $p(\theta_k )$, this represents the uncertainty in our knowledge of the background flux from each known category (excluding Poisson uncertainty and detector uncertainty), displayed in Table \ref{table:background_uncertainty}. This uncertainty is highly correlated among all point-sources within each category in $\theta_k$ (i.e. IAEA), though independent from one category to another (i.e. IAEA to  geo-neutrino). This uncertainty is highly correlated across detectors.

\begin{equation}
\label{JNBeq_42}
\Sigma_{\mathrm{flux}} = \left[ \begin{array}{ccc}
\sigma_{11}^f & \cdots & \sigma_{1n}^f\\ 
\vdots &\ddots &\vdots \\ 
\sigma_{n1}^f & \cdots & \sigma_{nn}^f \end{array} \right]
\end{equation}

\begin{equation}
\label{JNBeq_42.1}
\left(\sigma_{i}^f\right)^2=\left( \sigma_{i}^{f_{\mathrm{IAEA}}}\right)^2+\left( \sigma_{i}^{f_{\mathrm{geo}}}\right)^2+\left( \sigma_{i}^{f_{\mathrm{acc}}}\right)^2+\left( \sigma_{i}^{f_{\mathrm{fast\ n^0}}}\right)^2+\left( \sigma_{i}^{f_{\mathrm{cosmo}}}\right)^2
\end{equation}

\begin{equation}
\label{JNBeq_42.2}
\sigma_{i}^{f_A^2}=\sum\limits_{j=1}^{n_A}\sum\limits_{k=1}^{n_A}\mathrm{Cov}_i^A\left( \bar{n}_j, \bar{n}_k \right)
\end{equation}

The units of $\sigma^f$\ are measurements per unit time per detector. $\left( \sigma_{i}^{f_{A}}\right)^2$ is the variance in the mean count from source category $A$ at detector $i$ due to flux uncertainty only (excluding Poisson and detector uncertainty). 

$\mathrm{Cov}_i^A $ describes the flux uncertainty at detector $i$ between all points in source category $A$, not to be confused with $\Sigma_{\mathrm{flux}}$, which describes measurement count uncertainty between detectors due to flux uncertainties only (excluding Poisson and detector uncertainty).

$\mathrm{Cov}_i^A$ is an $n_A$ x $n_A$ matrix, with $n_A$ discrete point sources comprising source category $A$. Note that each point source may be a true point source or a sampled point source from a larger continuous source. In some volume source categories, such as the mantle, $n_A$ will depend on the step-size used during numerical integration.

{\bf 3. Detector uncertainty} ($\sigma^d$). This is incurred due to uncertain detector cross section, caused to a number of factors such as detector fiducial volume uncertainty. Our assumed value of 2.4\% is displayed in Table \ref{table:background_uncertainty}. This uncertainty is small but important, because it causes correlation in $\sigma^f$ between previously independent categories of flux. The simple act of measuring event counts using an uncertain detector cross section causes this correlation. \textit{We do not attempt to address this small incurred correlation between flux uncertainties in our estimator, due primarily to its assumed small magnitude}. $\sigma^d$ is independent from one detector to another. The units of $\sigma^d$ are measurements per unit time per detector.

\begin{equation}
\label{JNBeq_43}
\Sigma_{\mathrm{detector}} = \left[ \begin{array}{ccc}
\sigma_{11}^d+\left(\frac{\sigma_1^f\sigma_1^d}{\mu_1}\right)^2 & \cdots & 0\\
\vdots &\ddots &\vdots \\
0& \cdots & \sigma_{nn}^d+\left(\frac{\sigma_n^f\sigma_n^d}{\mu_n}\right)^2\end{array} \right]
\end{equation}

\begin{equation}
\label{JNBeq_43.1}
\sigma_i^d= f_i^d\mu_i
\end{equation}
where $f_i^d$ is the fractional $1{\sigma}$ detector uncertainty for detector $i$, with true and assumed values of $f_i^d=.024$ for all detectors in this paper.

$\Sigma$ may be attempted analytically from $\theta$ and $p(\theta)$ using Equation \ref{JNBeq_40} above or may, alternatively, be estimated numerical via MC methods using Equation \ref{JNBeq_44}:

\begin{equation}
\label{JNBeq_44}
\Sigma = \left[ \begin{array}{ccc}E\left[\left(Z_1-\mu_1\right)\left(Z_1-\mu_1\right)\right]& \cdots & E\left[\left(Z_1-\mu_1\right)\left(Z_n-\mu_n\right)\right]\\ 
\vdots &\ddots &\vdots \\ 
E\left[\left(Z_n-\mu_n\right)\left(Z_1-\mu_1\right)\right]& \cdots & E\left[\left(Z_n-\mu_n\right)\left(Z_n-\mu_n\right)\right]\end{array} \right]
\end{equation}

\begin{figure}[!htbp]
\includegraphics[width=\linewidth]{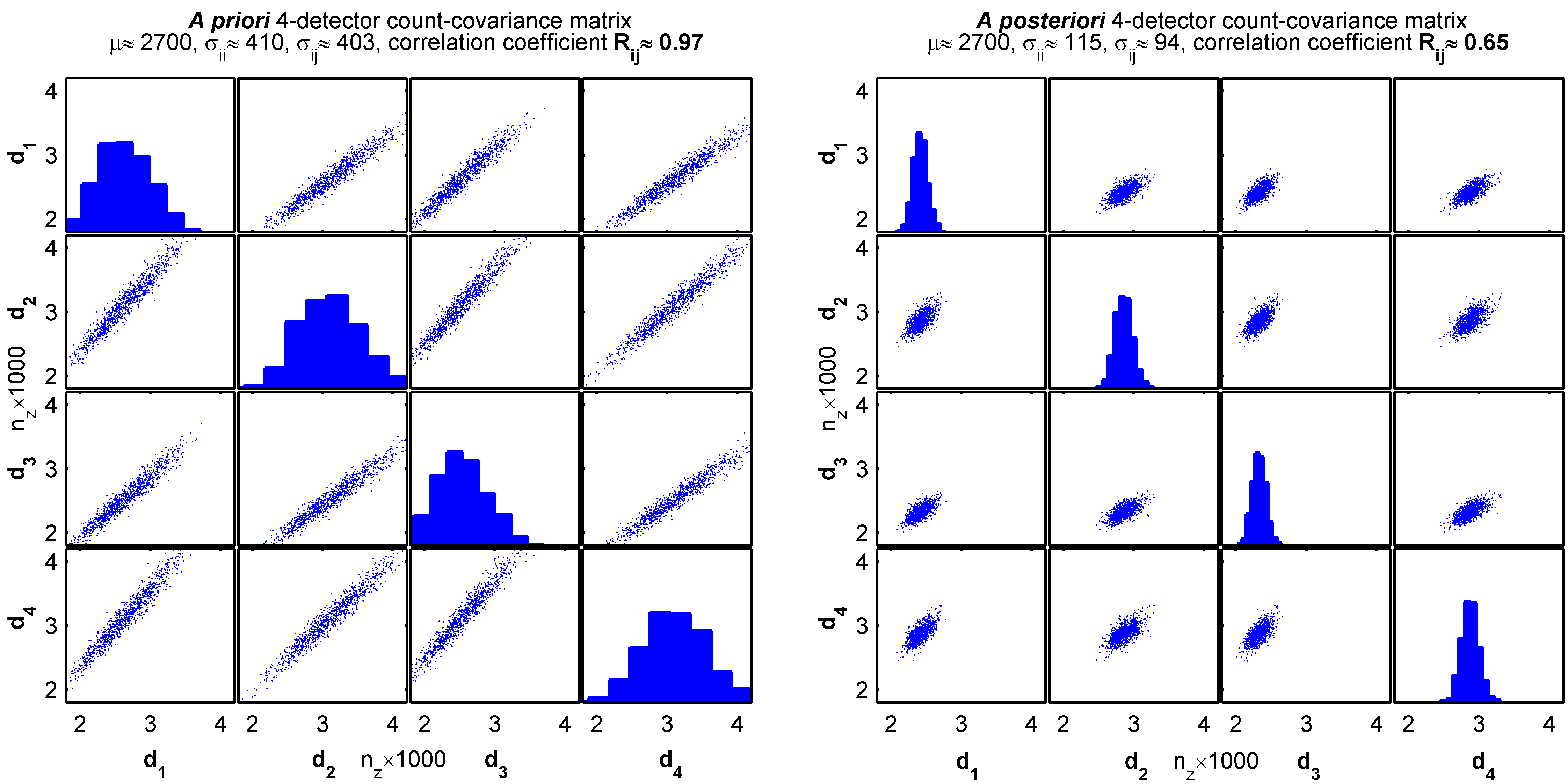}
\centering
\caption{Count error covariance matrices $\Sigma$ for our Europe-Mediterranean four-detector scenario. On the left, present day high  $\Sigma_\mathrm{flux}$ is assumed, reflecting today's poor knowledge of the true background flux levels. On the right, well known backgrounds are assumed and $\Sigma_\mathrm{flux}$ is substantially lower. These two background uncertainty levels correspond to the first and last rows of Table \ref{table:worldwide_flux} in Section \ref{Oscillation parameter estimation results}, respectively. The other two components of $\Sigma$, $\Sigma_\mathrm{Poisson}$ and $\Sigma_\mathrm{detector}$ do not vary between the two cases shown in this Figure. Note that since each detector is 100\% correlated with itself, the plots along the diagonals are used to display single-detector measurement count distributions.}
\label{d4corr}
\end{figure}

Figure \ref{d4corr} shows two such numerical estimates of the count error covariance matrix $\Sigma$ for our Europe-Mediterranean four-detector scenario. On the left it shows present day, high $\Sigma_\mathrm{flux}$, reflecting today's poor knowledge of the true background flux levels. On the right it assumes a hypothetical, well known $\Sigma_\mathrm{flux}$. These two background uncertainty levels correspond to the first and last rows of Table \ref{table:worldwide_flux} in Section \ref{Oscillation parameter estimation results}, respectively.\footnote{The other two components of $\Sigma$, $\Sigma_\mathrm{Poisson}$ and $\Sigma_\mathrm{detector}$ do not vary between the two cases shown in Figure \ref{d4corr}.} The correlation between detector count-errors can be seen to drop (but not disappear) as knowledge of the background improves, indicating that the (independent) Poisson and detector uncertainties $\Sigma_\mathrm{Poisson}$ and $\Sigma_\mathrm{detector}$ are playing a larger role in $\Sigma$ on the Figure \ref{d4corr} right-side plot than on the Figure \ref{d4corr} left-side plot.

Naturally, in the event that only one single detector is employed, the multi-detector correlation problem disappears. The multivariate normal distribution expressed in Equation \ref{JNBeq_36.0.2} reduces to the ubiquitous single-dimension normal distribution, where the variance equals the first element in the diagonal of the covariance matrix expressed in Equation \ref{JNBeq_40}.

Correlation is still present in the single detector problem by virtue of the flux uncertainties, which are all systematically affected by the detector uncertainty. This correlation is largely ignored as it is assumed small (2.4\% from Table \ref{table:background_uncertainty}) and directly addressing it in this estimator approximation would not yield significant benefits for the added complexity.

Thus the optimal Bayesian objective function has been presented, along with a suboptimal approximation appropriate for unknown reactor searches. The problem of correlation in detector measurement counts, due to possible systematic errors in our background knowledge has also been adequately addressed in this approximation. Different applications may merit different approximations but we believe the approximation presented here to strike a good compromise between accuracy and computational burden.  Section \ref{Reactor geolocation results} characterizes the performance of the suboptimal estimator approximation presented here as it applies to unknown-reactor searches at various spots around the world.

\subsubsection{Antineutrino oscillation parameter estimation}
\label{Antineutrino oscillation parameter estimation}
\begin{figure}[!htbp]
\includegraphics[width=\linewidth]{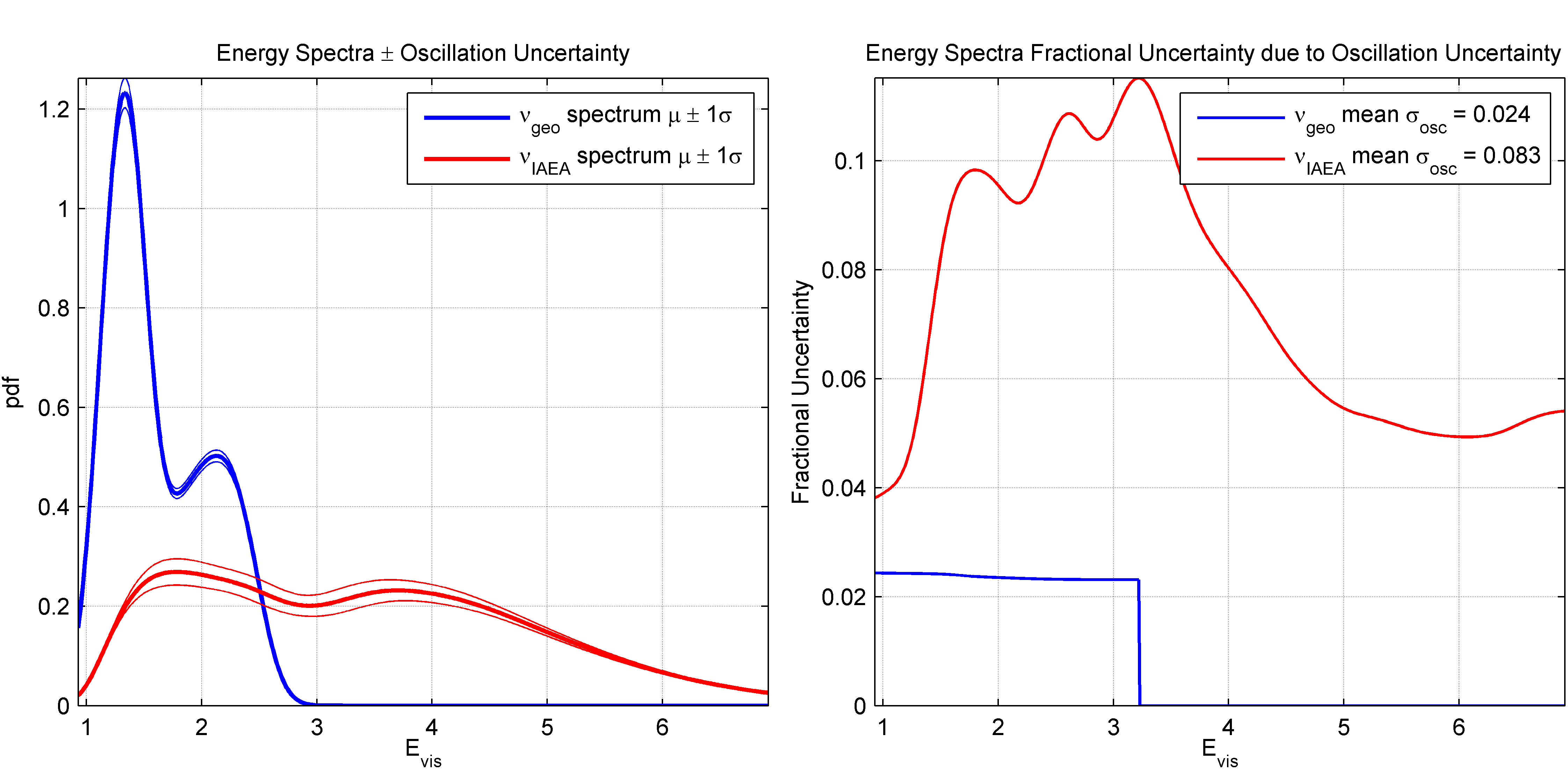}
\centering
\caption{Reactor and geo-neutrino energy-spectra uncertainty due solely to oscillation uncertainty. The higher fractional uncertainty (right plot) observed in the reactor spectrum indicates greater oscillation parameter observability from a reactor source than from geo-neutrino sources \textit{due to spectral shape  only}, irrespective of flux from either source. Naturally, increased flux from any source leads to increased oscillation parameter observability. However, this figure aims to illustrate the information content contained in the shape of the source spectrum itself rather than the source flux.}
\label{oscunc}
\end{figure}

\begin{figure}[!htbp]
\includegraphics[width=\linewidth]{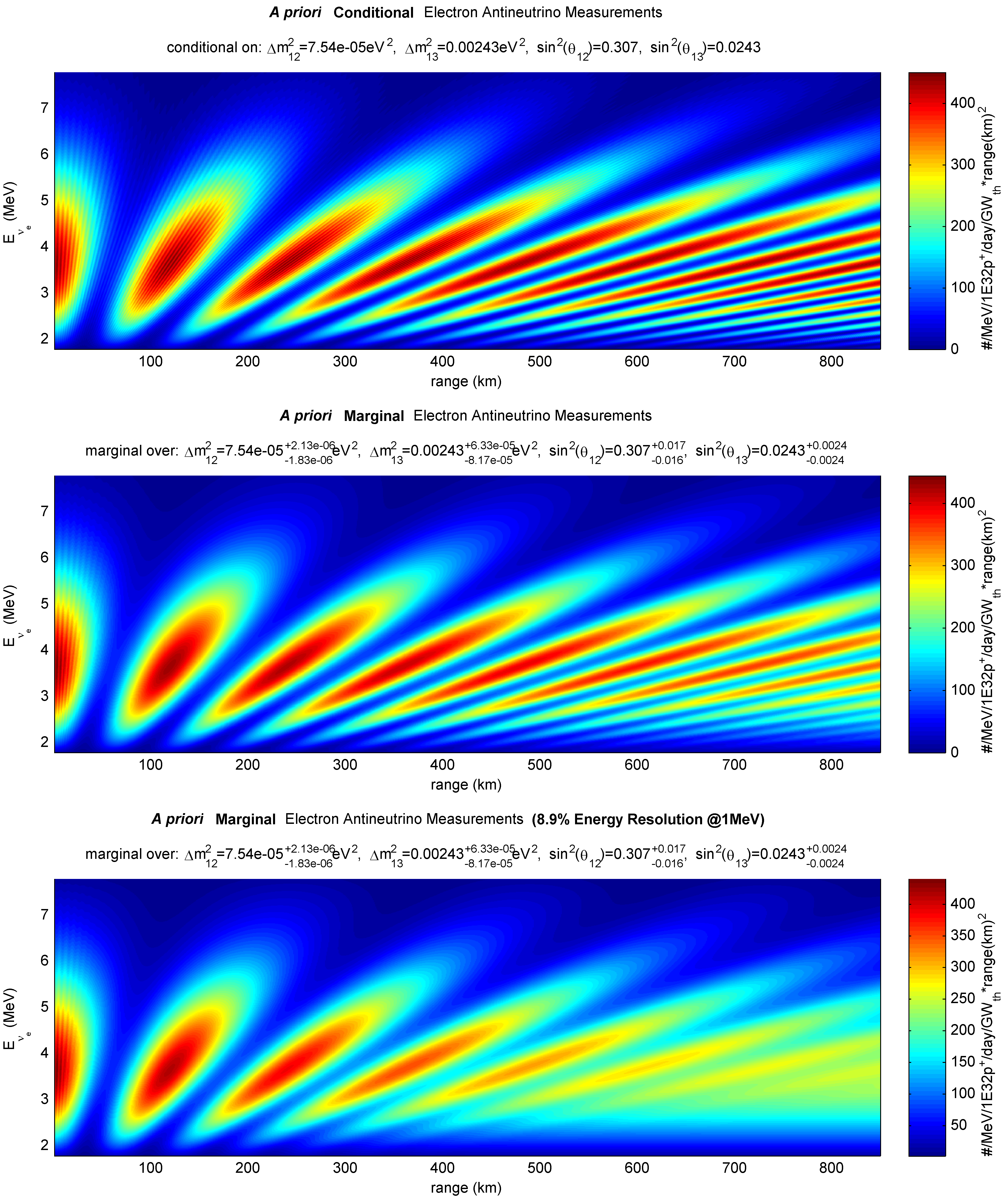}
\centering
\caption{Reactor spectra lookup tables showing \textit{a priori} oscillation uncertainty per Fogli \textit{et al.} \cite{fogli_2012}. The top subplot is the reactor oscillation spectrum before accounting for solid angle, conditional on the maximum likelihood values presented by Fogli \textit{et al.} \cite{fogli_2012}. The second subplot is the marginal spectrum produced by Equation \ref{smear1}, and shows the oscillation-smeared version of the top subplot. The bottom subplot is produced by applying Equation \ref{JNBeq_21} to Equation \ref{smear1}, smearing the middle subplot by TREND's energy measurement resolution of $\left. 8.9\% \right|_{E_\mathrm{vis}=1\mathrm{MeV}}$.}
\label{smearedfigure2}
\end{figure}

Oscillation parameter estimation proceeds almost exactly the same as the unknown-reactor estimation presented in Section \ref{Reactor geolocation} with a few minor differences:

\begin{enumerate}
\item \textbf{Optimized parameters.} The three unknown reactor parameters (latitude, longitude and power) in $\theta_u$ are removed and replaced by the four oscillation parameters: $\Delta m^2_{13}$, $\sin^2\theta_{13}$, $\Delta m^2_{12}$ and $\sin^2\theta_{12}$.
\item \textbf{Smeared lookup table.} The smeared lookup table built in Figure \ref{smearedfigure1} is replaced with an identical version which uses the modern day (worse) uncertainties expressed in Fogli \textit{et al.} \cite{fogli_2012}. This modern-day smeared lookup table is shown in Figure \ref{smearedfigure2}.
\item \textbf{Optimized vs Smeared \textit{sources}.} Computing neutrino oscillation fractions is a computationally burdensome task, especially for large numbers of point sources (at unique ranges). Thus it is not practical to precisely compute oscillation effects for all sources in the Geospatial Model simultaneously. Instead, when estimating oscillation parameters, we evaluate oscillation fractions only for the most prominent point sources, nuclear reactors, and use the smeared lookup table shown in Figure \ref{smearedfigure2} for our less influential geo-neutrino point sources.

\indent Thus Equation \ref{JNBeq_21}, the energy likelihood for known source $l$ at detector $d$, will be expressed in two forms in this approximation, one polling a lookup table for survival fraction values $f_{\theta_c}\left(E,r^l_d\right)$ and the other one computing survival fraction values $f(E,r^l_d,\theta_{13},\Delta m^2_{13},\theta_{12},\Delta m^2_{12})$ on the fly per Equation \ref{JNBeq_17_0} as the oscillation parameters ``travel" during optimization. 

\indent Equation \ref{JNBeq_21_1} thus applies to all geo-neutrino sources, and polls the lookup table in Figure \ref{smearedfigure2} for pre-smeared survival fraction values. The values returned by the lookup table $f_{\theta_c}\left(E,r^l_d\right)$ will not change as the optimizer travels around the $\theta_{13}$-$\Delta m^2_{13}$-$\theta_{12}$-$\Delta m^2_{12}$ space. Equation \ref{JNBeq_21_2}, on the other hand, computes survival fraction values $f(E,r^l_d,\theta_{13},\Delta m^2_{13},\theta_{12},\Delta m^2_{12})$ as a function of the exact current location of the optimizer within the continuous $\theta_{13}$-$\Delta m^2_{13}$-$\theta_{12}$-$\Delta m^2_{12}$ space. Equation \ref{JNBeq_21_2} is evaluated once per known reactor source (about 200 sources) during evaluation of the objective function, while the $f_{\theta_c}\left(E,r^l_d\right)$ lookup table expressed in Equation \ref{JNBeq_21_1} is evaluated hundreds of thousands of times (once per geo-neutrino point source) per objective function evaluation. Thus evaluation of the objective function is substantially sped up by decreasing evaluation of Equation \ref{JNBeq_21_2} from hundreds of thousands to simply hundreds per objective function evaluation.

\begin{equation}
\label{JNBeq_21_1}
p\left(Z_E^i|\theta_k^l, \theta_c\right)= \left.
\int\limits_{1.8\mathrm{MeV}}^{11\mathrm{MeV}}\frac{s_{\theta_{k}^{l}}{(E)s_d(E)
f_{\theta_c}\left(E,r^l_d\right)
}}{\mathrm{cdf}_d\left(\theta_k^l\right)}\mathcal{N}\left(Z_E^i, \sigma_d^i\right)dE \right|_{E=E_{\bar{\nu}}}
\end{equation}

\begin{equation}
\label{JNBeq_21_2}
p\left(Z_E^i|\theta_k^l, \theta_c\right)= \left.
\int\limits_{1.8\mathrm{MeV}}^{11\mathrm{MeV}}\frac{s_{\theta_{k}^{l}}{(E)s_d(E)
f(E,r^l_d,\theta_{13},\Delta m^2_{13},\theta_{12},\Delta m^2_{12})
}}{\mathrm{cdf}_d\left(\theta_k^l\right)}\mathcal{N}\left(Z_E^i, \sigma_d^i\right)dE \right|_{E=E_{\bar{\nu}}}
\end{equation}

\indent This approximation will naturally be more accurate in regions where geo-neutrinos comprise a smaller percentage of the total signal, the exact type of area where an oscillation parameter search will likely be conducted (near a reactor site). We accept this approximation for the following three reasons:

\begin{enumerate}
\item \textbf{Low geo-neutrino flux vs reactor-neutrino flux.} To increase signal statistics any oscillation parameter search must be undertaken in a high neutrino flux environment (i.e. near one or more known nuclear reactor sites). In this type of environment geo-neutrinos will comprise a small percentage of the total antineutrino flux, and they will thus have a correspondingly small effect on the likelihood function.
\item \textbf{Geo-neutrino flux uncertainty dominates oscillation uncertainty.} Current geo-neutrino flux uncertainties ($\sim 20$\%,  quantized in Table \ref{table:background_uncertainty}) are much higher than comparable reactor-neutrino flux uncertainties ($\sim 3$\%, also quantized in Table \ref{table:background_uncertainty}), making oscillation parameter estimation in high geo-neutrino flux areas less preferable than estimation in high reactor-flux environments.
\item \textbf{Geo-neutrinos (a volume-source) suffer from range smearing.} Due to the range smearing inherent in any volume source (as opposed to a point source), and because neutrinos oscillate as a function of range, \textit{observed} geo-neutrino spectra typically carry less information about oscillation parameters than their nuclear reactor counterparts.  This relative information loss\footnote{Information in the measurements is defined as the inverse of the variance of a parameter estimated from those measurements (also known as the Fisher information).} between reactor neutrinos and geo-neutrinos is loosely quantized in Figure \ref{oscunc} for a TREND detector 60km off the coast of Hamaoka, Japan. From Figure \ref{oscunc}, we can see that a perturbation in the true oscillation parameters will yield greater difference (across all energies) in the observed reactor spectrum than in the observed geo-neutrino spectrum. This condition will likely remain true even in locations where the geo-neutrino flux overshadows the nearby reactor-flux by factors as high as 2 to 3.
\end{enumerate}

\end{enumerate}

Section \ref{Oscillation parameter estimation results} characterizes the performance of the suboptimal estimator approximation presented here as it pertains to oscillation parameter estimation using four strategically placed underwater TREND detectors.

\section{Oscillation parameter estimation results}
\label{Oscillation parameter estimation results}

The best way to advance current knowledge about neutrino oscillations is to study neutrino emissions from a known point source. Luckily there are plenty such sources here on Earth in the form of civilian nuclear power reactors. Unfortunately, though, the TREND detectors are intended to operate at substantial depth to shield them from large levels of cosmogenics and muons. This constrains them to operate at certain distances from the coastline where the ocean is deep enough to provide sufficient shielding and, thus, this distances them from some of the most desirable locations around the world for neutrino oscillation parameter estimation.

\subsection{Estimator Cramer-Rao lower bound}

To determine an optimal placement strategy for oscillation parameter estimation using a single TREND detector, we numerically evaluated the oscillation parameter Cramer-Rao lower bound (CRLB) at all possible underwater locations worldwide. The CRLB defines (for a given estimator) a lower bound on the variance of the parameter estimates, essentially a floor which the variance of an \textit{efficient} estimator will approach (but may not pass) as the number of MC trials approach infinity. CRLB evaluation of the MAP estimator approximation is much less computationally intensive than comparable MC analysis, and is a useful substitute for MC study under certain circumstances. In our case, we found it very useful in conducting an exhaustive observability study (presented in this section) of the four $\nuebar$ oscillation parameters as a function of the placement of the TREND detectors (with the detector depth depending on the ocean depth at each location).

Just as the CRLB varies over detector location, it also varies as a function of measurement resolution. In this section we assume that all Cramer-Rao lower bounds are defined per our ``Count+Energy" measurement resolution. This resolution includes the measured \textit{counts} at the detector and the measured \textit{energies}, but not any \textit{direction vector} measurements. Thus the expectations called for in the later equations in this section are taken across the energy and count dimensions only, they omit the elevation and azimuth measurement dimensions.

The results of our Count+Energy CRLB exhaustive worldwide oscillation parameter observability study are shown in Figures \ref{crlb_dm12}, \ref{crlb_s2t12}, \ref{crlb_dm13} and \ref{crlb_s2t13} for $\Delta m^2_{12}$, $\sin^2\theta_{12}$, $\Delta m^2_{13}$ and $\sin^2\theta_{13}$ respectively.

Our CRLB was derived from Bar-Shalom's \cite{Bar-Shalom} multidimensional definition for \textit{non-random} (non-Bayesian) parameters, whereby the CRLB states that the covariance matrix of an unbiased estimator is bounded from below as follows:

\begin{equation}
\label{crlb1}
E \left[ \left[\theta(Z)-\theta_0\right]\left[\theta(Z)-\theta_0\right]'\right] \geq J^{-1}
\end{equation}
where the Fisher information matrix (FIM), $J$, is given by

\begin{equation}
\label{crlb2}
J = -E[\nabla_\theta \nabla'_\theta \textrm{ln} p(Z|\theta)]|_{\theta=\theta_0} = 
E\left[\left[\nabla_\theta \textrm{ln} p(Z|\theta)\right]\left[\nabla_\theta \textrm{ln} p(Z|\theta)\right]'\right]
|_{\theta=\theta_0}
\end{equation}
and $\theta_0$ is the true value of the parameter vector $\theta$, $p(Z|\theta)$ is the likelihood function, and ln$p(Z|\theta)$ is the log-likelihood. The ``nabla" (or ``del") operator denotes a gradient here, and thus $\nabla_\theta \textrm{ln} p(Z|\theta)$ is the (multidimensional) gradient vector of the log-likelihood function. The expectations ($E$) are taken through the measurement space $Z$ in both Equation \ref{crlb1} and Equation \ref{crlb2}. As mentioned before, for simplicity all CRLB evaluations in this section take the expectations ($E$) only through the ``Count+Energy" dimensions, and omit the elevation and azimuth dimensions of the measurement space $Z$.

Equation \ref{crlb2} shows the two forms of the FIM: one with the Hessian $\nabla_\theta \nabla'_\theta \textrm{ln} p(Z|\theta)$ of the log-likelihood function and the other with the dyad\footnote{A dyad is the product of a column vector with its transpose. All vector gradients referred to here are  expressed as column vectors, thus the dyad of an nx1 column vector will create an nxn matrix.} of its gradient. In this paper we evaluate the second form of the FIM, the expectation across the measurement space of the dyad of the parameter space gradient. We also update Bar-Shalom's non-random CRLB to support random (Bayesian) parameter spaces by substituting the likelihood function $p(Z|\theta)$ with the MAP estimator approximation\footnote{MAP estimator approximation for oscillation parameter estimation defined in Section \ref{Antineutrino oscillation parameter estimation}.} \textit{a posteriori} pdf $\tilde{p}(\theta|Z)=\tilde{p}(Z|\theta)\tilde{p}(\theta)c^{-1}$.

Due to the asymmetry seen in Fogli's oscillation parameter uncertainty \cite{fogli_2012}, the FIM is evaluated twice for each parameter space dimension, first using forward differences for numerical gradient $\nabla_\theta \textrm{ln} \tilde{p}(\theta|Z)$ in $J_u$ (upper FIM), and then a second time using backward differences for numerical gradient computation in $J_l$ (lower FIM). To arrive at a single FIM the corresponding upper and lower FIM pairs \textit{for each dimension} were combined using weighted means per Equation \ref{crlb3}. Note that in general each dimension has unique upper and lower weights, and thus the weightings shown in Equation \ref{crlb3} are vector element-by-element multiplication.

\begin{equation}
\label{crlb3}
J = w_l J_l + w_u J_u
\end{equation}
where the weights are

\begin{equation}
\label{crlb4}
w_l = \int\limits_{-\infty}^{0}p(\theta)d\theta
\end{equation}
\begin{equation}
\label{crlb5}
w_u = \int\limits_{0}^{\infty}p(\theta)d\theta
\end{equation}
for $\theta = \left[ \begin{array}{cccc} \Delta m^2_{12} & \sin^2\theta_{12} &  \Delta m^2_{13} &  \sin^2\theta_{13} \end{array}\right]'$. 

In the absence of knowledge of the true parameter vector $\theta_0$ (we do not know the true oscillation parameters after all), we have substituted Fogli's \cite{fogli_2012} maximum likelihood values, $\theta_{ML}$ (shown in Table \ref{table:worldwide_op}), again per Bar-Shalom's \cite{Bar-Shalom} guidance. 
Implementing all the changes discussed above to Bar-Shalom's original FIM and substituting Equations \ref{crlb2}, \ref{crlb4} and \ref{crlb5} into \ref{crlb3} we arrive at the complete FIM expression used to define all the Cramer-Rao lower bounds presented in this section:

\begin{eqnarray}
\label{crlb6}
J = \int\limits_{-\infty}^{0}p(\theta)d\theta
E \left[ \left[ \frac{\textrm{ln} \tilde{p}(\theta_{ML}-\delta\theta|Z) - \textrm{ln} \tilde{p}(\theta_{ML}|Z)}{\delta\theta}\right]
\left[ \frac{\textrm{ln} \tilde{p}(\theta_{ML}-\delta\theta|Z) - \textrm{ln} \tilde{p}(\theta_{ML}|Z)}{\delta\theta}\right]' \right]
+ \nonumber \\ 
\int\limits_{0}^{\infty}p(\theta)d\theta 
E \left[ \left[ \frac{\textrm{ln} \tilde{p}(\theta_{ML}+\delta\theta|Z) - \textrm{ln} \tilde{p}(\theta_{ML}|Z)}{\delta\theta}\right]
\left[ \frac{\textrm{ln} \tilde{p}(\theta_{ML}+\delta\theta|Z) - \textrm{ln} \tilde{p}(\theta_{ML}|Z)}{\delta\theta}\right]' \right]
\end{eqnarray}

\subsection{Cramer-Rao lower bound verification results}

\begin{figure}[!htbp]
\includegraphics[width=\linewidth]{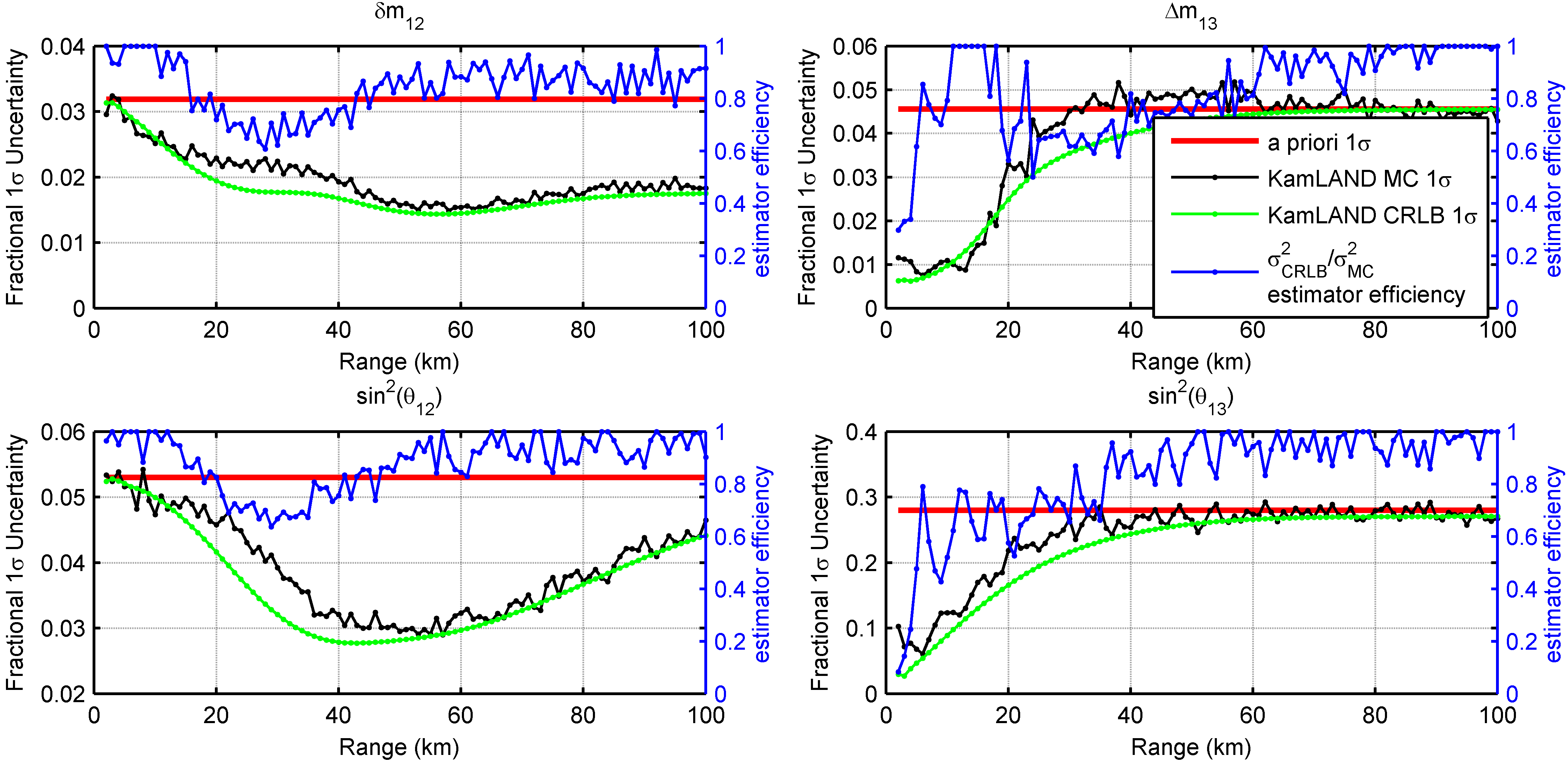}
\centering
\caption{CRLB verification against MC results. Range is in a south-east direction from Hamaoka nuclear reactor site in Japan. A constant shielding value of 2500m water overburden equivalent was assumed at each range to remove shielding as an independent variable in the results. Modified KamLAND style detector was used for CRLB and MC. \textit{A priori} uncertainty values are per Fogli \textit{et al.} \cite{fogli_2012}.}
\label{crlbverification}
\end{figure}

\begin{figure}[!htbp]
\includegraphics[width=\linewidth]{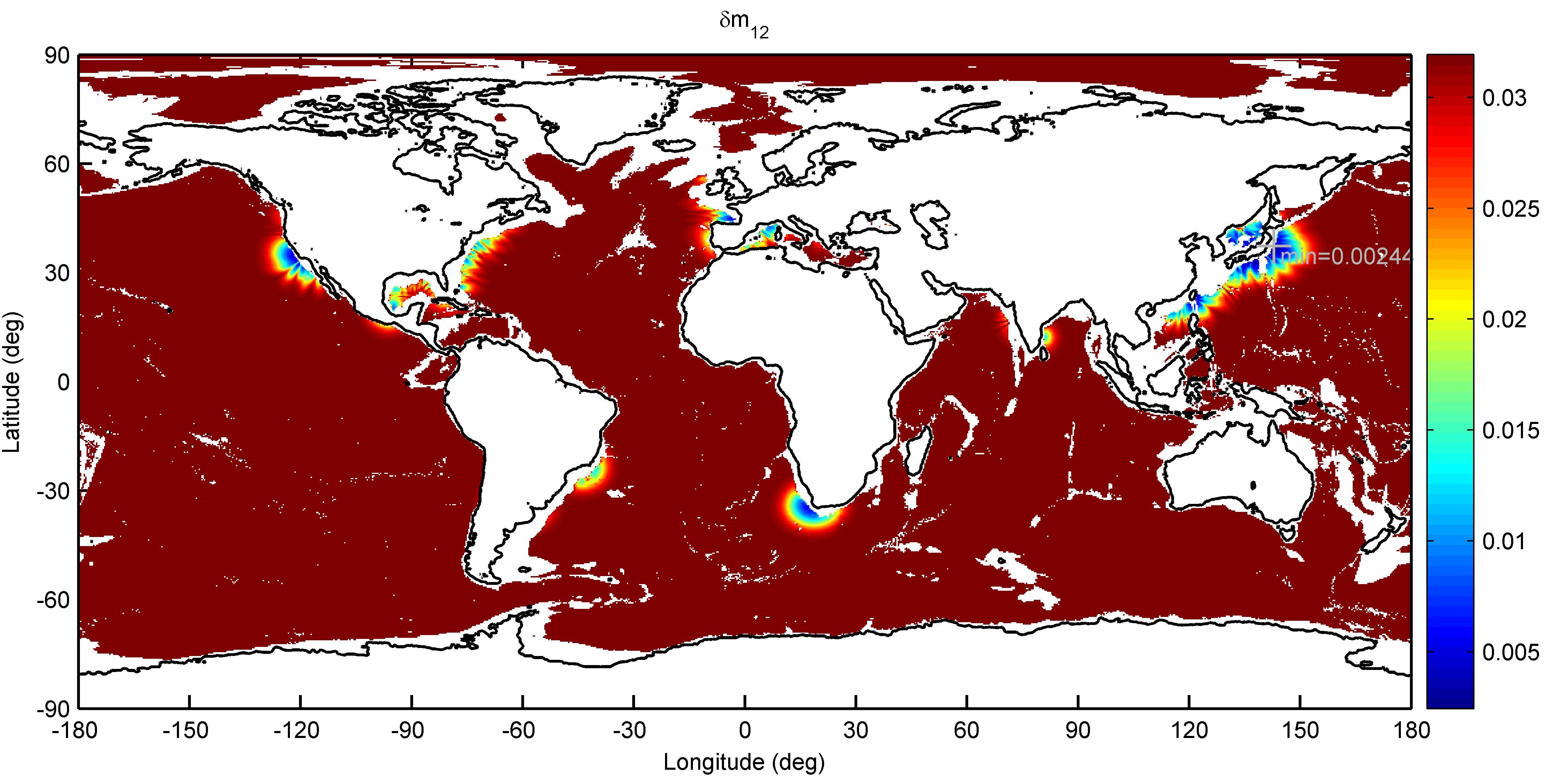}
\centering
\caption{MAP estimator approximation CRLB, $\Delta m^2_{12}$. Units are \textit{a posteriori} fractional uncertainty. The results show $\Delta m^2_{12}$ observability to be highest near coastal reactor sites. The most prominent locations seem to be near Japan and Taiwan, off South Africa by the Koeberg reactor, in the Bay of Biscay west of France, and off the west coast of the United States.}
\label{crlb_dm12}
\end{figure}

For CRLB verification, shown in Figure \ref{crlbverification} for all four oscillation parameters, we compared MC and CRLB results side by side for detector-reactor ranges up to 100km. 500 MC runs were performed at 99 discrete ranges from 2km to 100km. In our verification tests, we substituted a KamLAND-sized detector for TREND. This was done to reduce measurement count at the detector and, consequently, the computational burden associated with the MC evaluation (CRLB computational burden is unaffected by detector measurement count). 

For the verification site, the Japanese Hamaoka nuclear reactor site was selected due to its high thermal power output (14.5GW$_\mathrm{th}$), relative isolation, and coastal proximity. The reactor-detector range vector points south-east (away from Hamaoka) to avoid as many of the other nearby reactors as possible and the ``destructive interference" to the observed oscillations in the energy spectrum they might bring. For the CRLB verification tests, a constant shielding of 2500m water overburden equivalent was assumed at each range.  For the worldwide TREND CRLB results, however, the detector was assumed to lie on the ocean floor at each point on the world map.

\subsection{Optimal detector placement via CRLB results}

\begin{figure}[!htbp]
\includegraphics[width=\linewidth]{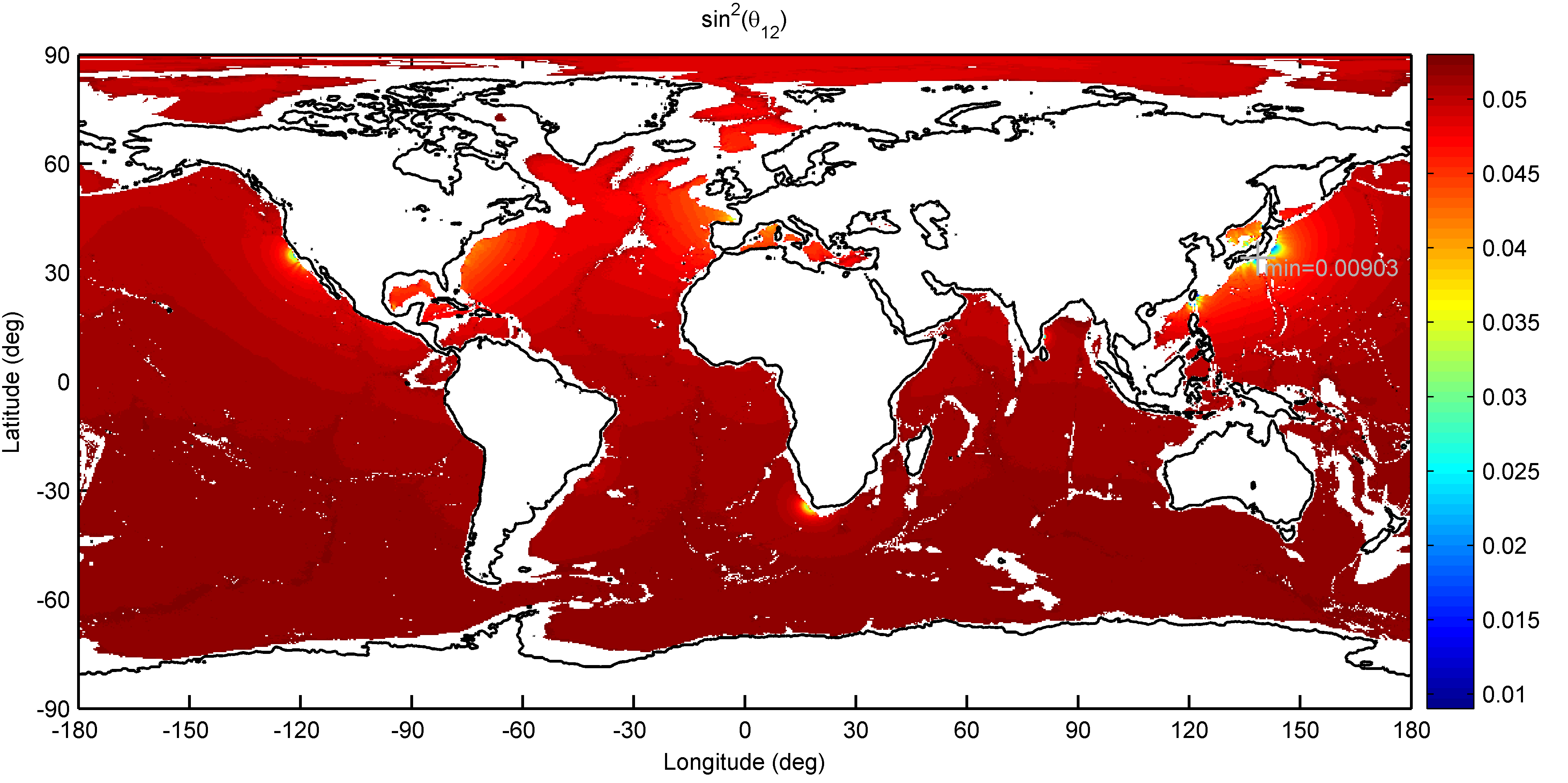}
\centering
\caption{MAP estimator approximation CRLB, $\sin^2\theta_{12}$. Units are \textit{a posteriori} fractional uncertainty. This figure indicates that $\sin^2\theta_{12}$ observability is substantially lower than $\Delta m^2_{12}$ observability at a given range. Curiously the $\sin^2\theta_{12}$ CRLB is showing remnant observability out to thousands of kilometers from reactor sites, significantly further than the $\Delta m^2_{12}$ CRLB. It is not known whether this represents a figment of the approximation since verification was only performed out to 100km.}
\label{crlb_s2t12}
\end{figure}

\begin{figure}[!htbp]
\includegraphics[width=\linewidth]{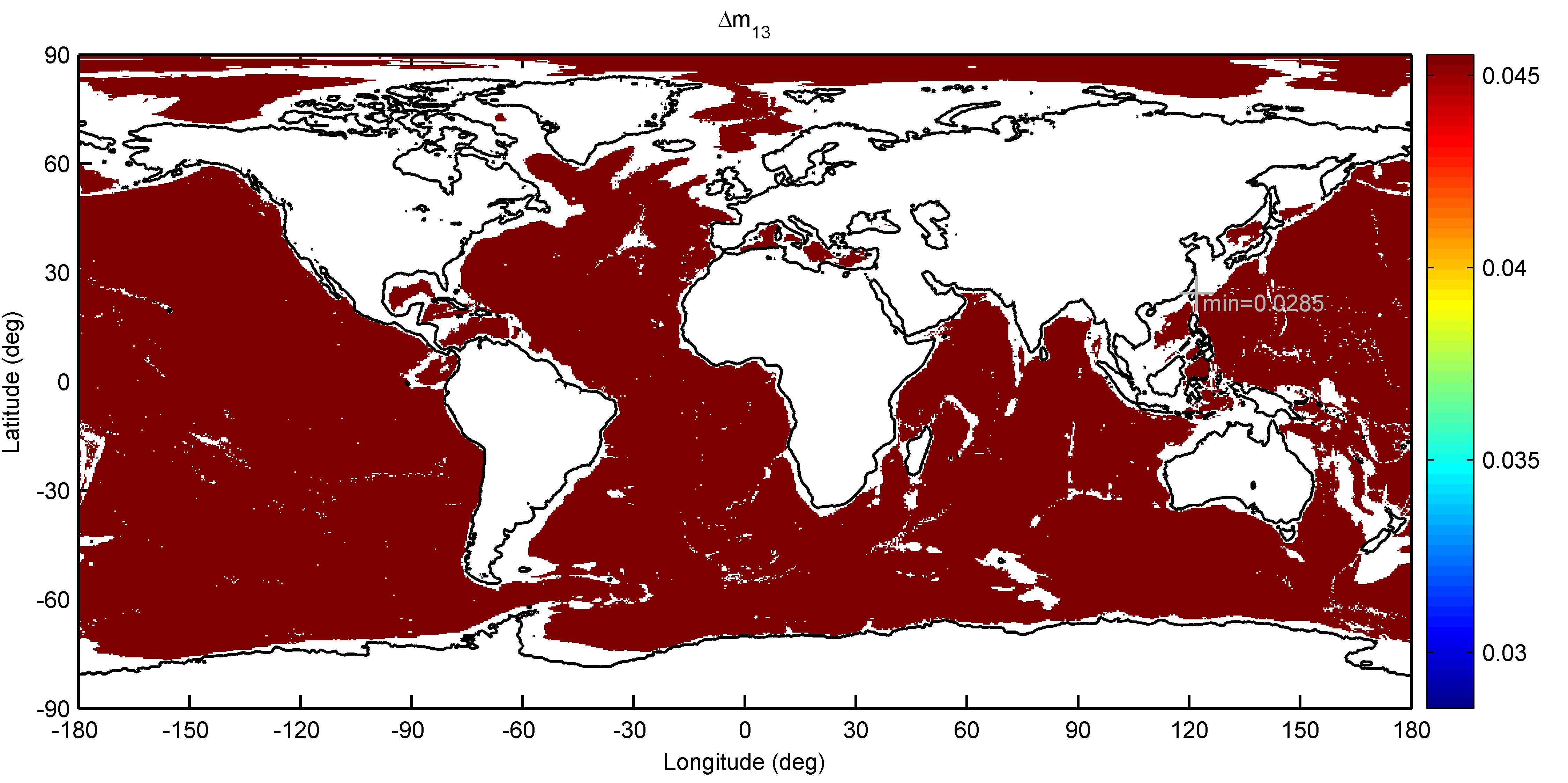}
\centering
\caption{MAP estimator approximation CRLB, $\Delta m^2_{13}$. Units are \textit{a posteriori} fractional uncertainty. $\Delta m^2_{13}$ observability appears to be nearly non-existant across the map, though at these scales it would be hard to spot as $\Delta m^2_{13}$ observability relies heavily on a very closely spaced detector. Nevertheless the absolute minima next to Taiwan is showing about 50\% reduction in uncertainty. This absolute minima region is examined more closely in Figure \ref{crlb_close_taiwan}.}
\label{crlb_dm13}
\end{figure}

\begin{figure}[!htbp]
\includegraphics[width=\linewidth]{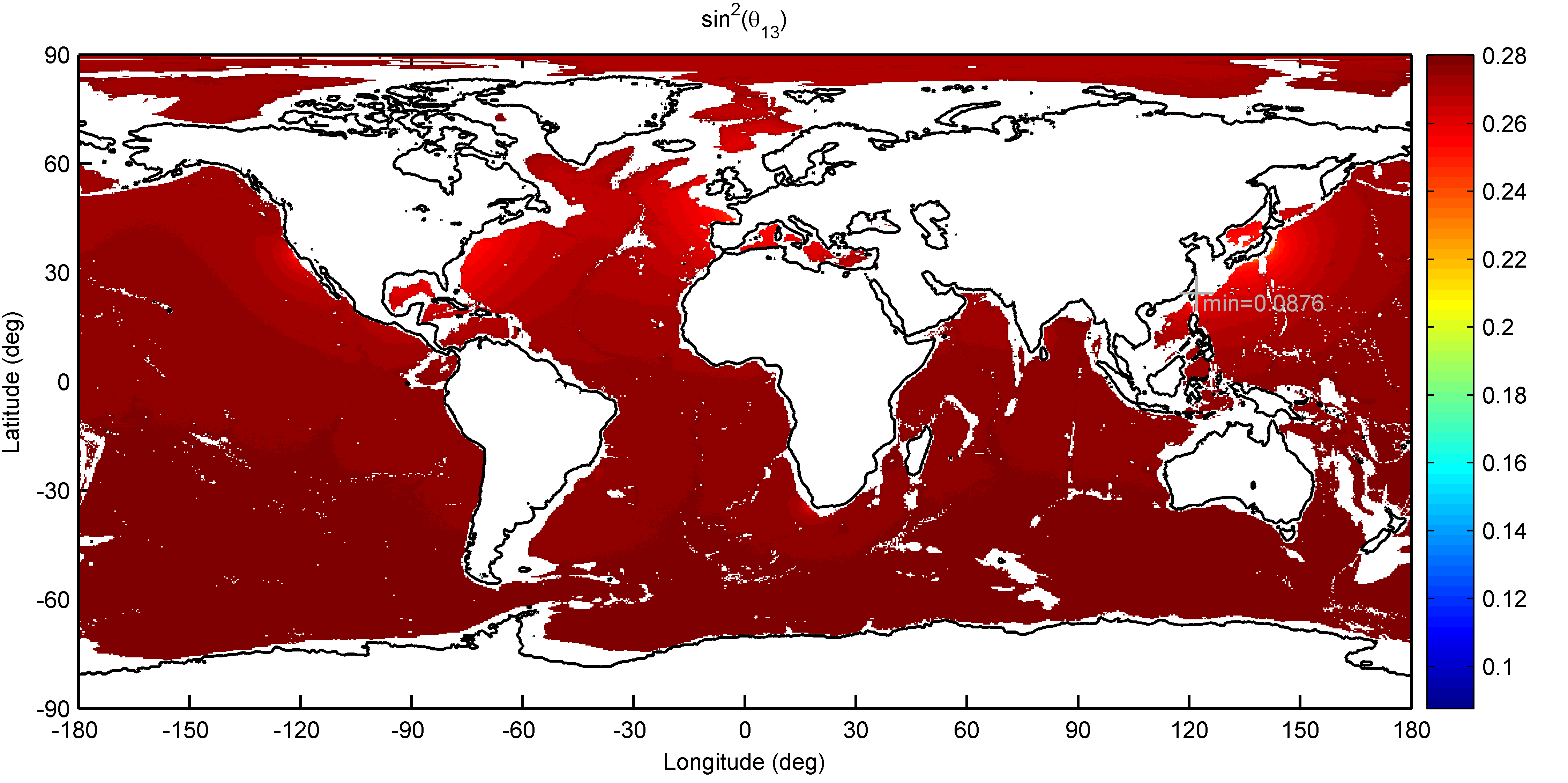}
\centering
\caption{MAP estimator approximation CRLB, $\sin^2\theta_{13}$. Units are \textit{a posteriori} fractional uncertainty. The absolute minima again appears next to Taiwan by the Lung-Mei reactor site, just as in the $\sin^2\theta_{12}$ results in Figure \ref{crlb_s2t12}. This absolute minima is showing about 70\% reduction in \textit{a priori} uncertainty. $\sin^2\theta_{13}$ observability is seen to drop off rapidly at any significant range, however again as in the $\sin^2\theta_{12}$ results in Figure \ref{crlb_s2t12} we see a long ``tail" of slight observability surviving out to 1000's of km from the nearest reactors. Verification plots shown in Figure \ref{crlbverification} do seem to substantiate a long tail of slight $\sin^2\theta_{13}$ observability, at least out to the verification limit of 100km.}
\label{crlb_s2t13}
\end{figure}

\begin{figure}[!htbp]
\includegraphics[width=\linewidth]{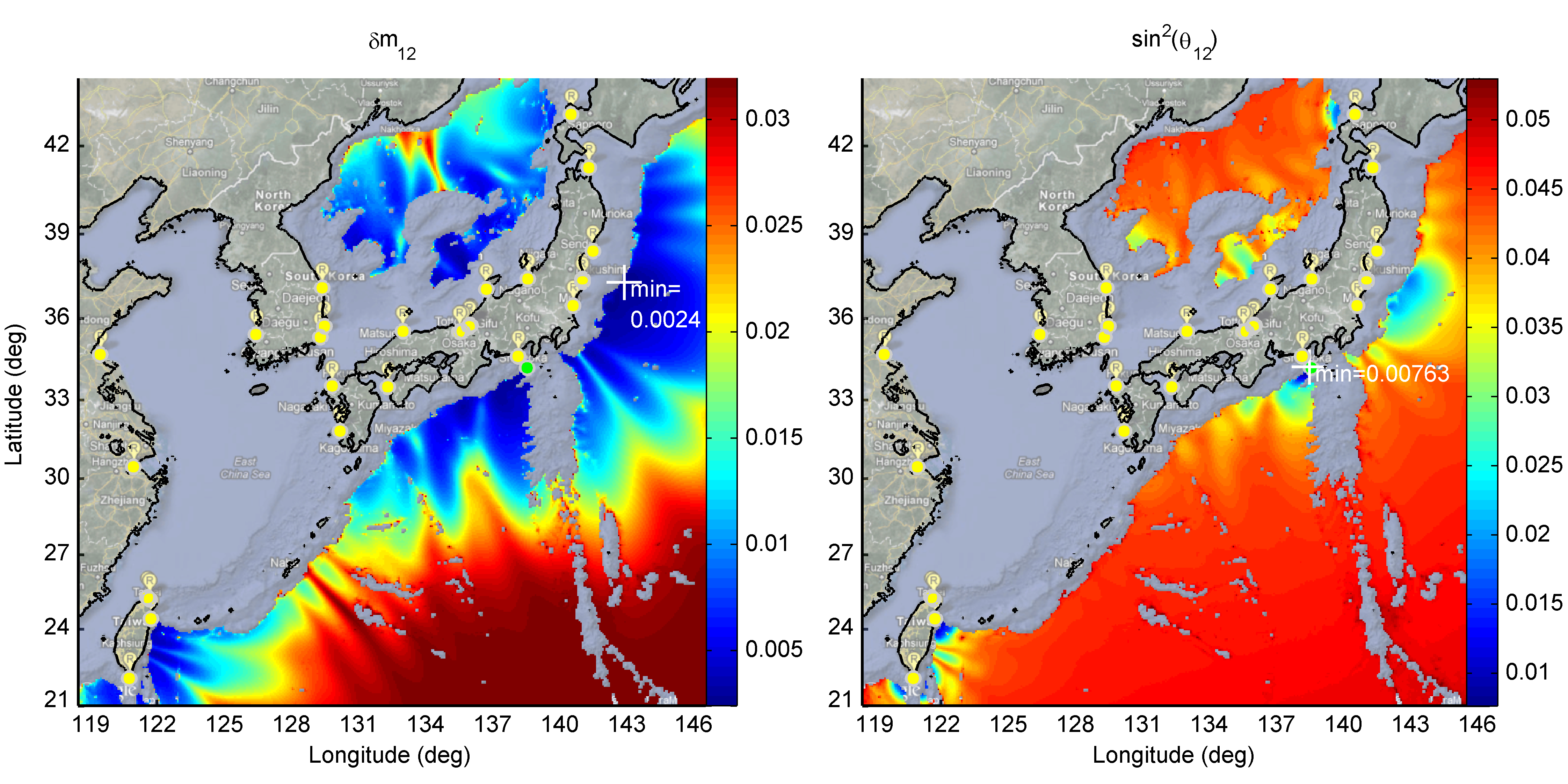}
\centering
\caption{Japan close up, $\Delta m^2_{12}$ \textit{a posteriori} CRLB on the left and $\sin^2\theta_{12}$ CRLB on the right. Units are \textit{a posteriori} fractional uncertainty. \textit{A priori} uncertainties are at the top of the colorbar in dark red. Yellow dots indicate existing nuclear reactor sites. Green dot indicates chosen TREND detector \#1 site, a compromise location selected for its excellent $\Delta m^2_{12}$ and $\sin^2\theta_{12}$ observability from a single detector placed at the green dot. Un-evaluated locations indicate depths too shallow for TREND to operate, generally $<2500$m.}
\label{crlb_close_japan}
\end{figure}

\begin{figure}[!htbp]
\includegraphics[width=\linewidth]{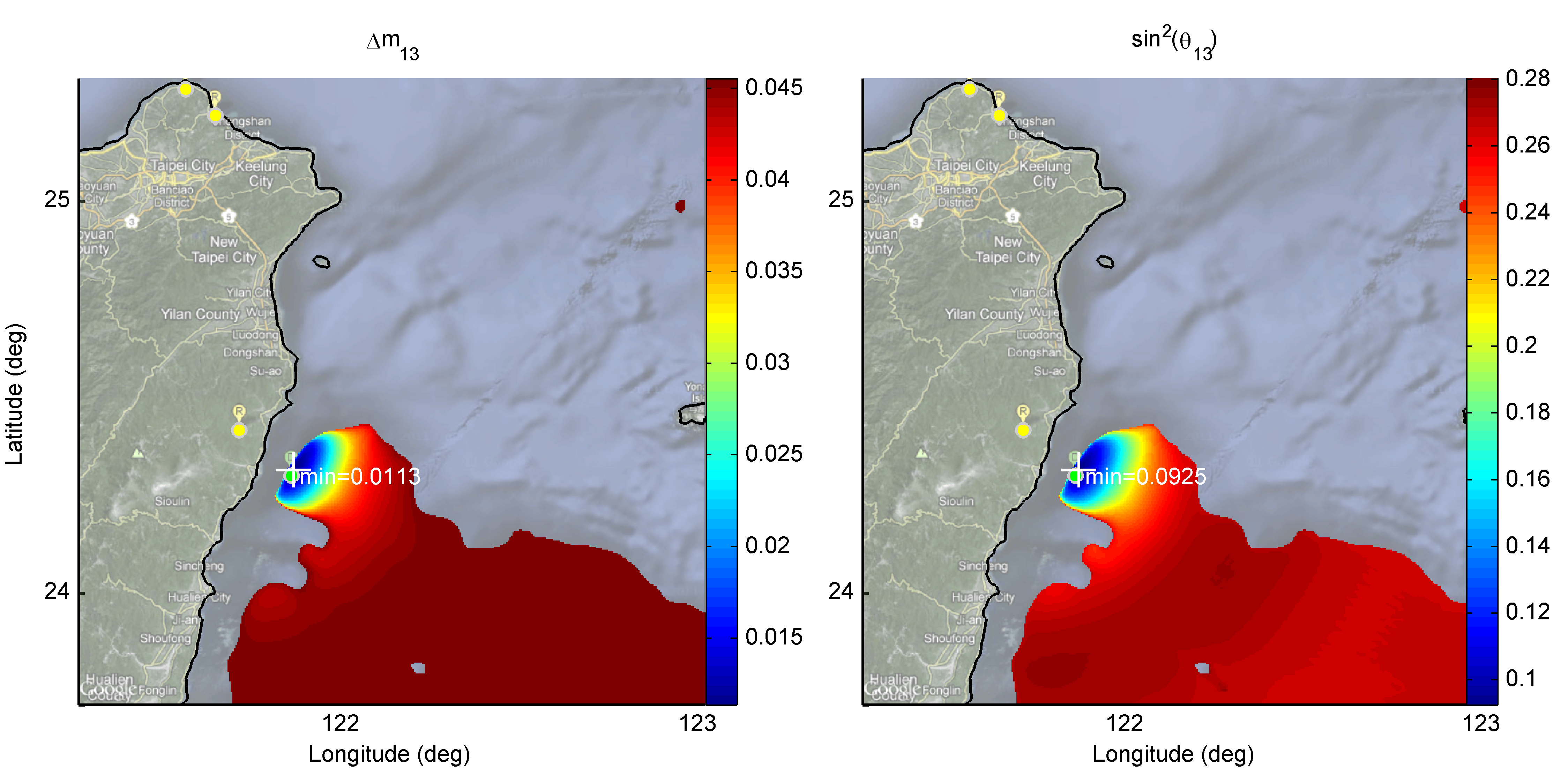}
\centering
\caption{Taiwan close up, $\Delta m^2_{13}$ \textit{a posteriori} CRLB on the left and $\sin^2\theta_{13}$ CRLB on the right. Units are \textit{a posteriori} fractional uncertainty. \textit{A priori} uncertainties are at the top of the colorbar in dark red. Yellow dots indicate existing nuclear reactor sites. The nearest site is the 8.2GW$_\mathrm{th}$ Taiwanese Lung-Mei reactor. Green dot indicates chosen TREND detector \#2 site, selected for its excellent $\Delta m^2_{13}$ and $\sin^2\theta_{13}$ observability from a single detector placed at the green dot. The green dot is about 20km distant from the Lung-Mei reactor yellow dot. Un-evaluated locations indicate depths too shallow for TREND to operate, generally $<2500$m.}
\label{crlb_close_taiwan}
\end{figure}

For $\Delta m^2_{12}$ and $\sin^2\theta_{12}$ terms, the TREND CRLB results returned two very close locations just west of Japan. A close-up of $\sin^2\theta_{12}$ results are shown in Figure \ref{crlb_close_japan}. The two location returned both offer comparable $\Delta m^2_{12}$ observability (0.0026 \textit{a posteriori} fraction uncertainty for the southern site vs. 0.0024 for the northern), but the southernmost location offers markedly better $\sin^2\theta_{12}$ observability (0.008 \textit{a posteriori} fraction uncertainty vs. 0.018). Thus, we reason that a single TREND detector placed at the absolute minima shown in Figure \ref{crlb_s2t12} would best be able to capture both $\Delta m^2_{12}$ and $\sin^2\theta_{12}$ terms simultaneously. We assign this location to TREND detector \#1 as the most optimal site in the world for a single TREND detector to simultaneously measure both $\Delta m^2_{12}$ and $\sin^2\theta_{12}$.

For $\Delta m^2_{13}$ and $\sin^2\theta_{13}$ terms, the CRLB \textit{a posteriori} uncertainties shown in Figures \ref{crlb_dm13} and \ref{crlb_s2t13} were both smallest at the same exact location, about 20km east of the Lung-Mei reactor site in Taiwan. A close-up of this site is shown in Figure \ref{crlb_close_taiwan}. We assign this location to TREND detector \#2 as the most optimal site in the world for a single TREND detector to simultaneously measure both $\Delta m^2_{13}$ and $\sin^2\theta_{13}$.

A single TREND detector does seem able to observe both the $\sin^2\theta_{12}$ and $\sin^2\theta_{13}$ terms simultaneously, but with significant compromise to the observability of both. In the verification plots in Figure \ref{crlbverification} (based on smaller KamLAND-sized detectors), we do see some $\sin^2\theta_{12}$ and $\sin^2\theta_{13}$ observability overlap in the 20km-25km range.  Unfortunately, these ranges are generally not accessible to an unsegmented detector as large as TREND. From Figure \ref{crlb_close_taiwan}, we observe that simultaneous estimation of $\sin^2\theta_{12}$ and $\sin^2\theta_{13}$ ideally requires two separately placed TREND detectors since the $\sin^2\theta_{12}$ and $\sin^2\theta_{13}$ oscillations operate on largely different distance scales.

In Figure \ref{crlbverification} we see that $\Delta m^2_{13}$ and $\sin^2\theta_{13}$ terms are most observable only a few km away from a reactor, yet $\Delta m^2_{12}$ and $\sin^2\theta_{12}$ are nearly unobservable at such short ranges. To approach optimal $\Delta m^2_{12}$ and $\sin^2\theta_{12}$ observability requires longer range detection, around 50km, whereas the $\Delta m^2_{13}$ and $\sin^2\theta_{13}$ observability has largely disappeared. From this assessment, we reason that the best strategy for observing all four oscillation parameters simultaneously would \textit{not} be to compromise on range, but rather to employ two detectors for the task, one near and one far. 

To this pair of detectors tasked with oscillation parameter estimation, we add a second pair of detectors tasked with geo-neutrino detection: one near Hawaii for optimal mantle geo-neutrino flux observability, and one east of Australia for optimal crust geo-neutrino flux observability. The rationale behind the additional pair of detectors is that a better understanding of geo-neutrino flux will help out during later unknown-reactor searches, and the first two TREND detectors placed by Japan and Taiwan would be located so close to major reactor sites that their geo-neutrino signature will be nearly completely obscured.

Hawaii was selected for mantle geo-neutrino observability for its distance from any known reactors (3900km distant from the closest reactor), as well as its isolation from crustal geo-neutrino sources (i.e. continental shelves). Per our Geospatial Model, we found the Earth's mantle to make up about 30\% of the total flux in Hawaii and the crust about 44\%, making the mantle much more prominent in Hawaii than in most locations around the world.

We found Australia to be an excellent place to study the Earth's crust due to its remote isolation from known reactors and the potential proximity to the continent with which a submerged detector may approach. This proximity is greater than in many parts of the world due to the nearby deep Pacific waters and the steep drop seen at the eastern edge of the continental shelf, where water depths rapidly plunge from only $\sim$100m near the coast to $\sim$5000m about 50km offshore. Sydney, Australia is 6500km from the closest nuclear reactor. Per our Geospatial Model, we find the crust contribution to be 68\% of the total flux seen at a detector 100km north-west of Sydney with the mantle contribution limited to 19\%.

\begin{figure}[H]
\includegraphics[width=\linewidth]{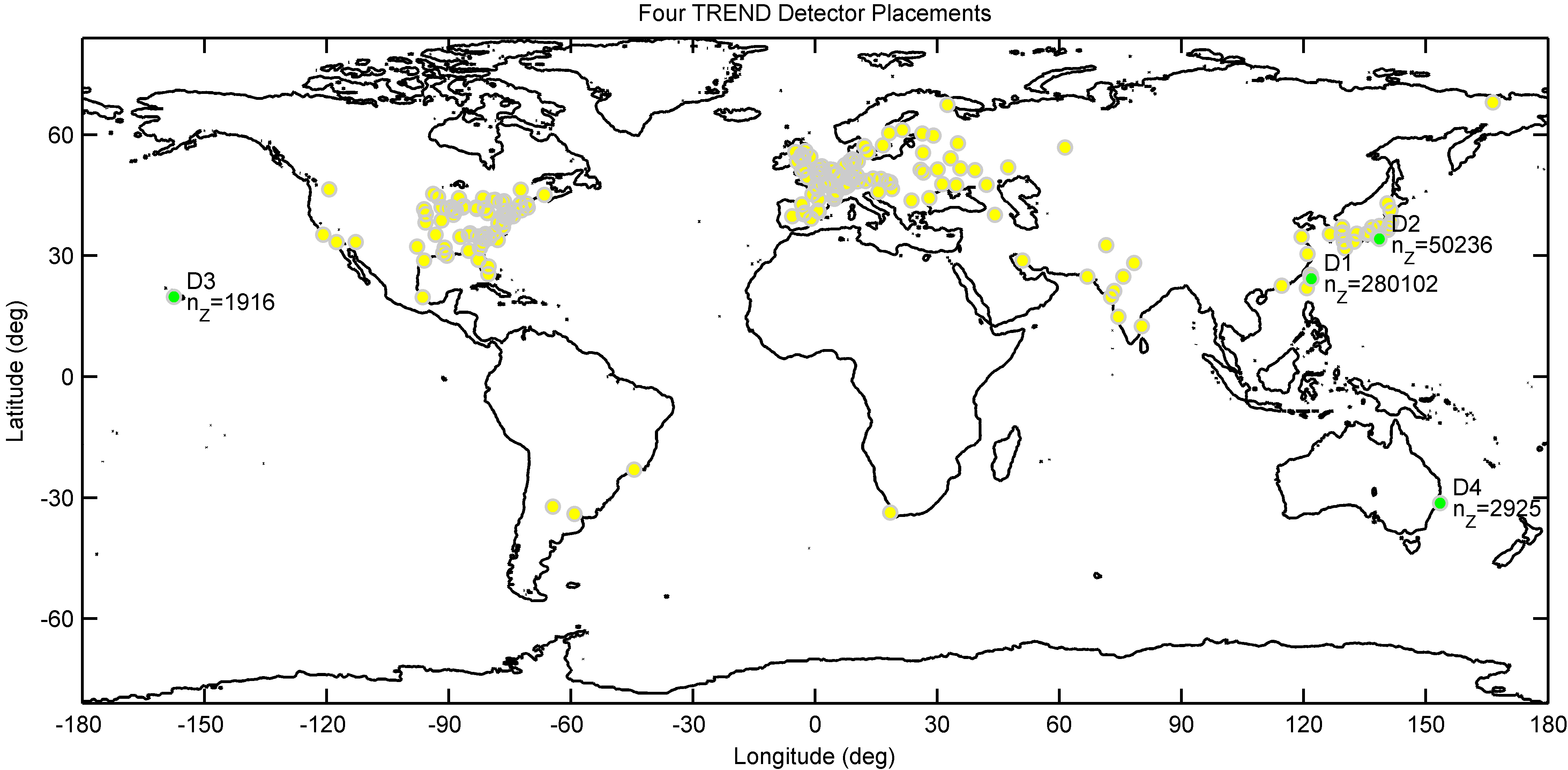}
\centering
\caption{Four TREND detector placements. Detectors are green circles, known reactor sites are yellow circles. Mean event rate per detector per year is displayed next to each TREND detector.}
\label{worldwide_dall}
\end{figure}

Figure \ref{worldwide_dall} shows the four TREND detector sites on a world map, followed by a detailed accounting of each detector site.

\begin{enumerate}
\item Detector \#1, Lung-Mei, Taiwan: One TREND detector, intended to capture $\Delta m^2_{13}$ and $\sin^2\theta_{13}$ terms, is placed in the waters 20km south-east of the 8.2GW$_\mathrm{th}$ Lung-Mei reactor site in Taiwan. This detector is in 2700m of water, and receives about 271,000 events per year after factoring in a 53\% duty cycle, of which 260,000 events originate from the Lung-Mei reactor site itself. Its energy spectrum can be seen in Figure \ref{worldwide_d1epdf}.
\begin{figure}[H]
\includegraphics[width=\linewidth]{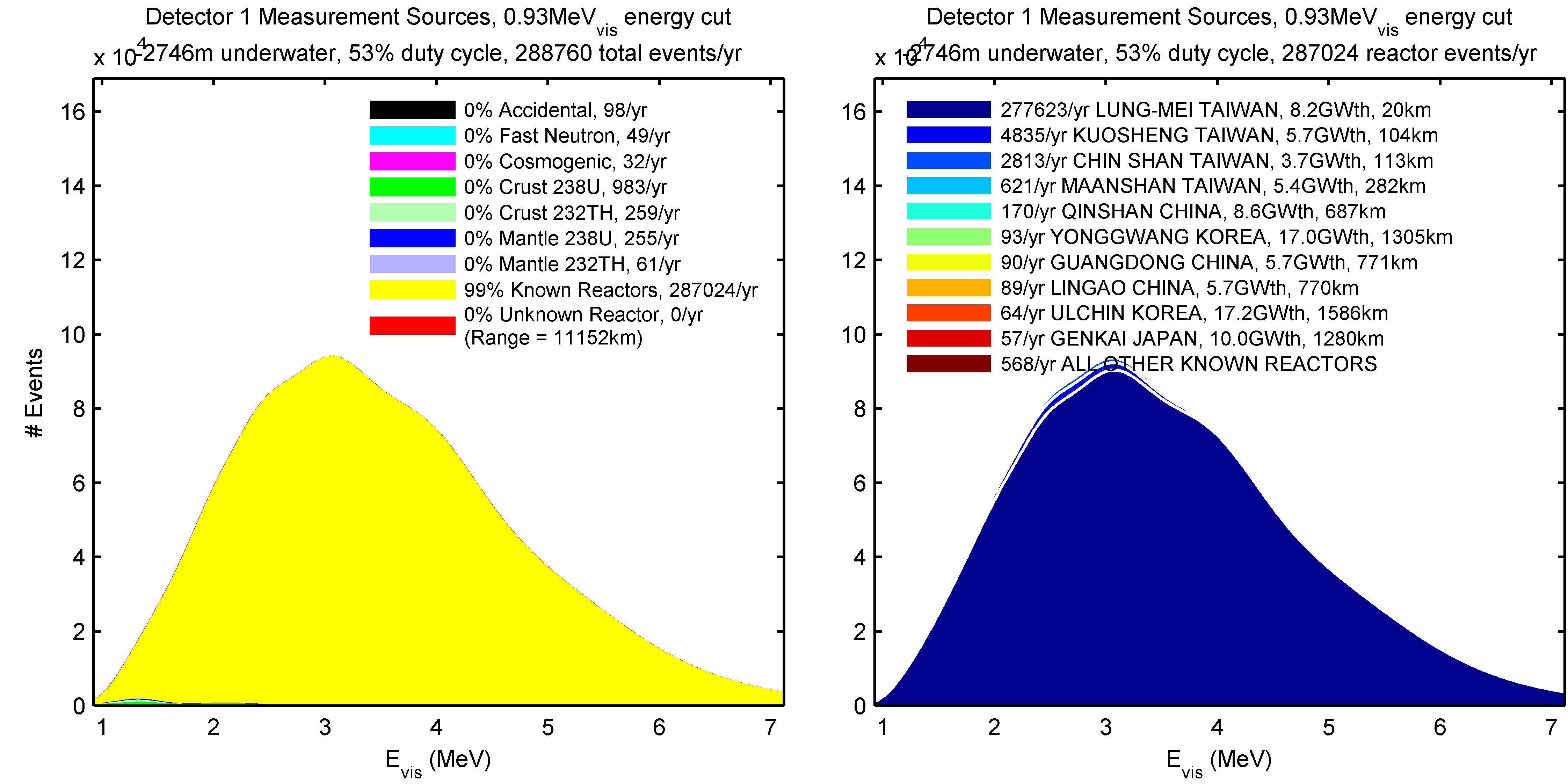}
\centering
\caption{Lung-Mei, Taiwan, TREND detector \#1 (targeted to $\Delta m^2_{13}$ and $\sin^2\theta_{13}$) energy spectra. Left plot represents all sources of flux at detector, right plot represents known-reactor flux only.}
\label{worldwide_d1epdf}
\end{figure}

\item Detector \#2, Hamaoka, Japan: One TREND detector, intended to capture $\Delta m^2_{12}$ and $\sin^2\theta_{12}$ terms, is placed in the waters about 60km south-east of 
the 14.5GW$_{\mathrm{th}}$ Hamaoka reactor site in Japan. This detector is in 3500m of water, and receives about 50,500 events per year after factoring in an 86\% duty cycle, of which 33,500 originate from the Hamaoka reactor site itself. Its energy spectrum can be seen in Figure \ref{worldwide_d2epdf}.
\begin{figure}[H]
\includegraphics[width=\linewidth]{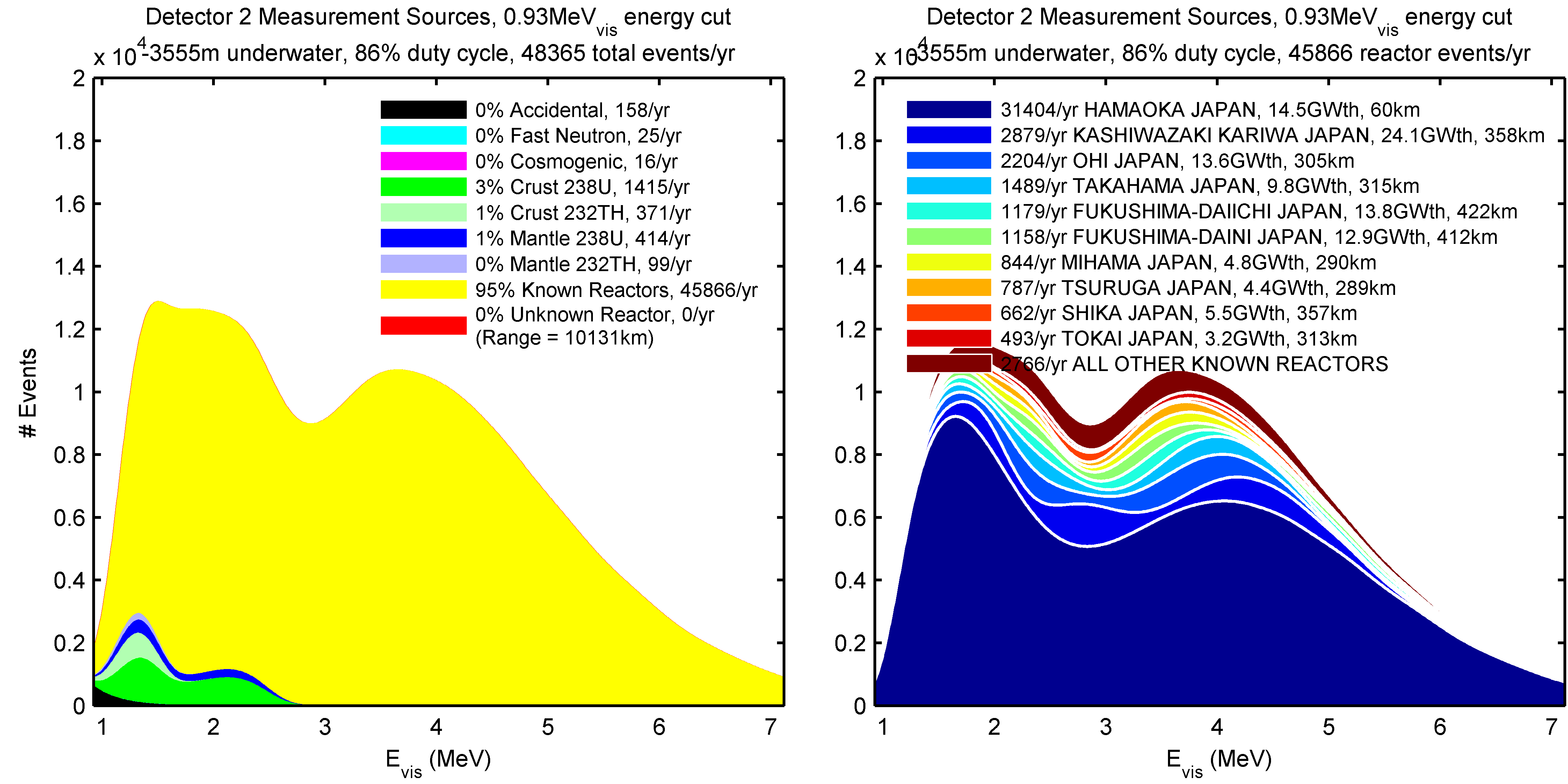}
\centering
\caption{Hamaoka, Japan, TREND detector \#2 (targeted to $\Delta m^2_{12}$ and $\sin^2\theta_{12}$) energy spectra. Left plot represents all sources of flux at detector, right plot represents known-reactor flux only.}
\label{worldwide_d2epdf}
\end{figure}

\item Detector \#3, Hawaii: One TREND detector, intended to capture the geo-neutrino flux from the Earth's mantle, is placed in the waters about 50km south of Honolulu, Hawaii. This detector is in 4200m of water, and receives about 1920 events per year after factoring in a 95\% duty cycle, of which about 560 events (30\% of the total) originate from the Earth's mantle, and 850 events (44\% of the total) originate from the Earth's crust. This detector is 3900km from the closest nuclear reactor site and receives only 314 mean reactor events per year. Its energy spectrum can be seen in Figure \ref{worldwide_d3epdf}.
\begin{figure}[H]
\includegraphics[width=\linewidth]{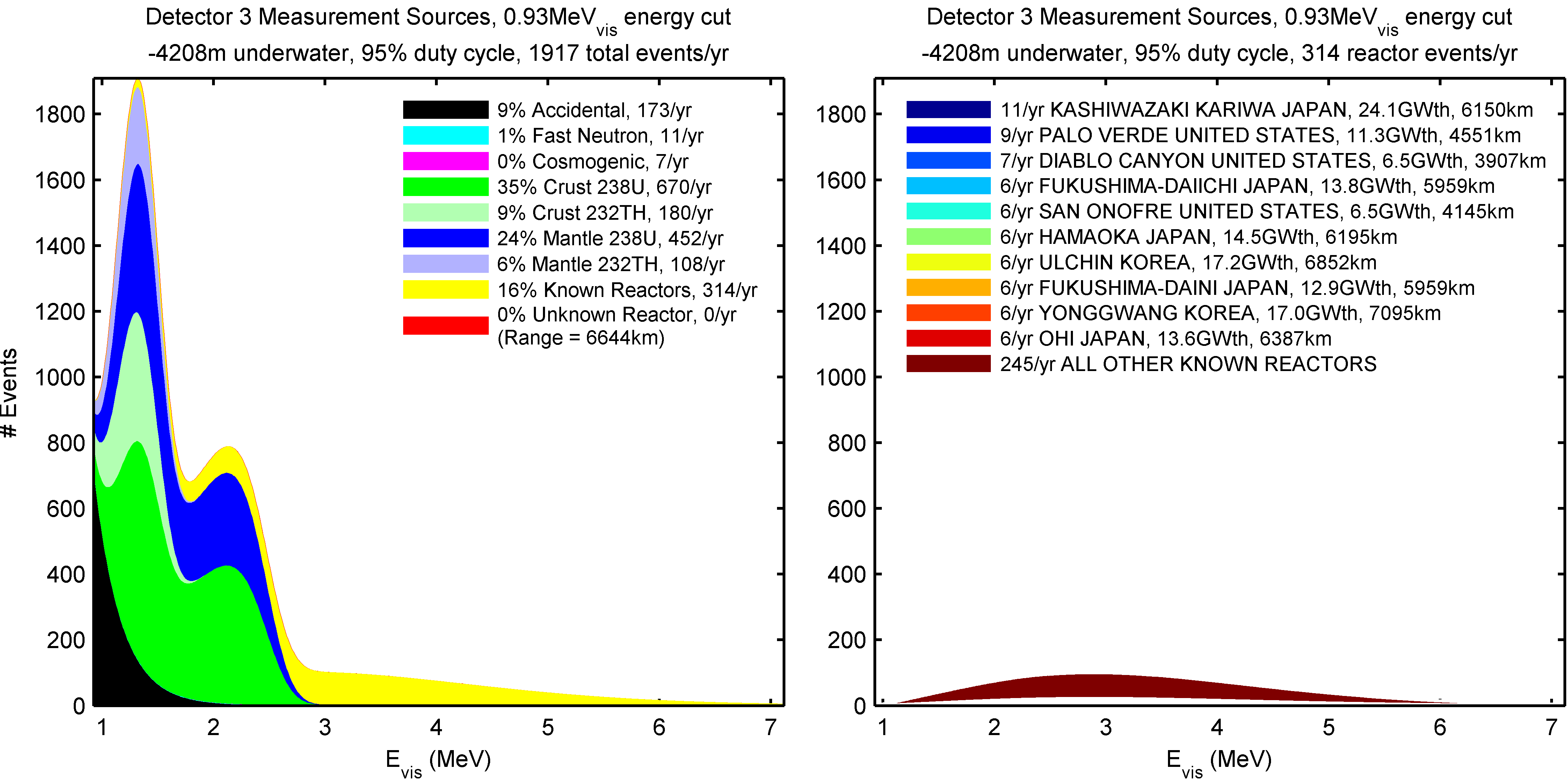}
\centering
\caption{Hawaii, TREND detector \#3 (targeted to observing mantle geo-neutrino flux) energy spectra. Left plot represents all sources of flux at detector, right plot represents known-reactor flux only.}
\label{worldwide_d3epdf}
\end{figure}

\item Detector \#4, Sydney, Australia: One TREND detector, intended to capture the geo-neutrino flux from the Earth's crust, is placed in the waters just north-west of Sydney, Australia. This detector is in 3970m of water, and receives about 2930 events per year after factoring in a 93\% duty cycle, of which about 550 events (19\% of the total) originate from the Earth's mantle, and 2000 events (68\% of the total) originate from the Earth's crust. This detector is 6500km from the closest nuclear reactor site and receives only 184 mean reactor events per year. Its energy spectrum can be seen in Figure \ref{worldwide_d4epdf}.
\begin{figure}[H]
\includegraphics[width=\linewidth]{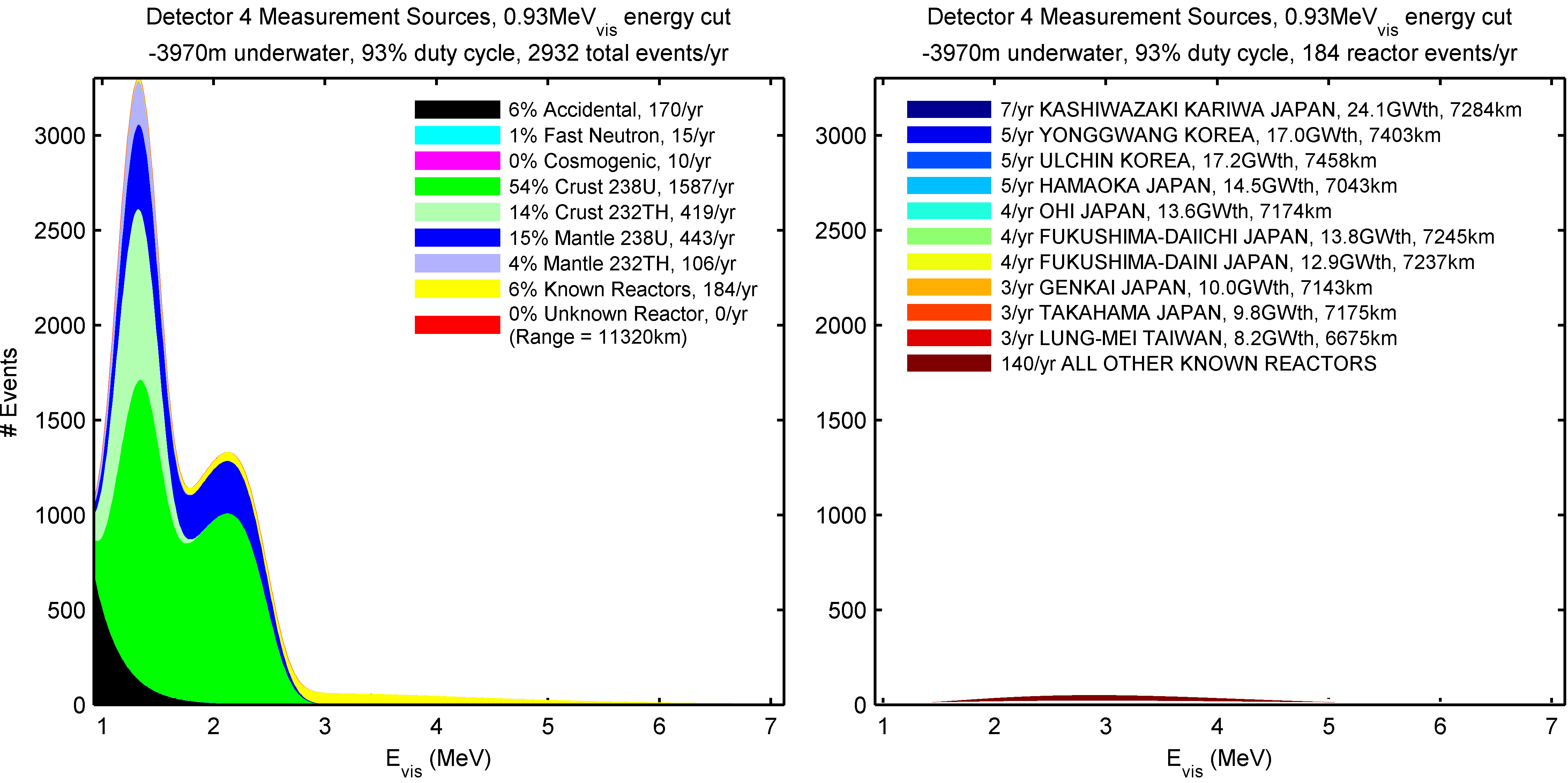}
\centering
\caption{Sydney, Australia, TREND detector \#4 (targeted to observing crust geo-neutrino flux) energy spectra. Left plot represents all sources of flux at detector, right plot represents known-reactor flux only.}
\label{worldwide_d4epdf}
\end{figure}

\end{enumerate}

During oscillation parameter estimation we choose to co-estimate the six aforementioned flux categories (IAEA flux, crust flux, mantle flux, accidental flux, fast neutron flux, and cosmogenic flux) simultaneously with the four oscillation parameters: $\Delta m^2_{13}$, $\sin^2\theta_{13}$, $\Delta m^2_{12}$ and $\sin^2\theta_{12}$. This entails a 10-dimensional optimization performed along four correlated detectors simultaneously. We opt to include the six extra background flux categories in the optimization so that their MC \textit{a posteriori} values could used to set new \textit{a priori} inputs during the later unknown reactor searches discussed in Section \ref{Reactor geolocation results}.

\subsection{Oscillation parameter search MC results}
\label{Oscillation parameter search MC results}

MC simulation results of the worldwide four-detector, 10-parameter estimation problem illustrated in Figure \ref{worldwide_dall} are shown in Tables \ref{table:worldwide_op} and \ref{table:worldwide_flux}. All 10 parameters were optimized across all four detectors simultaneously for each of 1000 MC runs. The four detector scenario was analyzed at five different levels of measurement resolution:

\begin{enumerate}
\item \textbf{Count+Energy.} Here we pass the measurement count (a scalar) as well as the measurement energies (a vector) to the estimator. For this and all our later cases we maintain a measurement energy cut of $E_\mathrm{vis}\geq 0.9$MeV. The MAP estimator co-estimates independent background fluxes from each background source category simultaneously with the four oscillation parameters. The \textit{oscillation parameters are only optimized for known reactor sources} while other sources utilize a smeared \textit{a priori} oscillation spectrum to reduce computational burden. This is a simple choice from a performance perspective since we account for only 206 active reactor sites (471 reactor cores) worldwide, though we have millions of geo-neutrino point sources produced during the ``smart" crust and mantle integration discussed in Section \ref{Estimation theory}. With enough computation resources the geo-neutrino sources can, of course, also contribute to the oscillation parameter estimation. We use our TREND energy resolution, as determined by our TREND MC runs, of $\left. 8.9\% \right|_{E_\mathrm{vis}=1\mathrm{MeV}}$, to add TREND measurement noise to the true neutrino event energies. We pass the TREND energy resolution, the 4x4 count covariance matrix, the oscillation parameters, the background fluxes, the smeared \textit{a priori} oscillations, and the known constants to the MAP estimator, where they all assist in the creation of the smeared measurement space $Z$.

\item \textbf{Count+Energy+Angle.} Here we pass the direction vectors of the measurements (a matrix) to the MAP estimator alongside the (scalar) count and (vector) energy. Our angle resolution for TREND, using the CHOOZ CE-to-CE technique cited in  \cite{chooz_2003}, was about SNR=0.045. Estimation is performed identically to the Count+Energy case except that the measurement space has expanded from 1D (Energy) to 3D (Energy+Elevation+Azimuth). Again the MAP estimator co-estimates independent background fluxes from each background source category simultaneously with the four oscillation parameters.  Again, the \textit{oscillation parameters are only optimized for known reactor sources} while other sources utilize a smeared \textit{a priori} oscillation spectrum to reduce computational burden.

\item \textbf{Count+Energy+Angle$_{\mathrm{(SNRx10)}}$.} This is where we use our MAP estimator to quantify performance gains made possible by future hypothetical hardware advancements coupled with improved algorithms that replace the CHOOZ CE-to-CE technique.  We hypothesize 10 times better direction vector resolution than what we found in our TREND MC runs, or about 0.45 SNR. We believe this level of directional resolution may be possible in the near future for smaller detectors using multi-channel plate technology.

\item \textbf{Count+Energy+Angle$_{\mathrm{(SNRx20)}}$.} Here we hypothesize 20 times better direction vector resolution than what we found in our TREND MC runs, or about 0.90 SNR.

\item \textbf{Count+Energy+Angle$_{\mathrm{(SNRx30)}}$.} Here we hypothesize 30 times better direction vector resolution than what we found in our TREND MC runs, or about 1.35 SNR.
\end{enumerate}

\begin{table} [!htbp]
\centering
\fontsize{8}{10}\selectfont
\begin{tabular}{l|l|l|l|l}
\hline
\textbf{Scenario} & \textbf{$\Delta m^2_{12}$} & \textbf{$\Delta m^2_{13}$} & \textbf{$\sin^2\theta_{12}$} & \textbf{$\sin^2\theta_{13}$}\\
& $(eV^2)$ & $(eV^2)$ & & \\
%\hline \textit{a priori} max. likelihood $\mu$ \cite{fogli_2012}			& $7.54E-5$		& $0.00243$		& $0.307$		& $0.0243$\\
%\hline \textit{a priori} uncertainty \cite{fogli_2012}		& $\mu+2.43\%-2.83\%$		& $\mu+3.37\%-2.61\%$		& $\mu+5.21\%-5.65\%$		& $\mu+9.97\%-9.83\%$\\
%\hline Count							        			& $\mu+1.61\%-1.41\%$		& $\mu+1.54\%-2.08\%$		& $\mu+2.42\%-2.44\%$		& $\mu+6.09\%-6.08\%$\\
%\hline Count+Energy							    			& $\mu+0.27\%-0.27\%$		& $\mu+0.49\%-0.50\%$		& $\mu+0.85\%-0.81\%$		& $\mu+5.62\%-5.88\%$\\
%\hline Count+Energy+Angle					    			& $\mu+0.27\%-0.26\%$		& $\mu+0.51\%-0.50\%$		& $\mu+0.88\%-0.77\%$		& $\mu+5.90\%-5.95\%$\\
%\hline Count+Energy+Angle$_{\mathrm{(SNRx10)}}$ 			& $\mu+0.25\%-0.26\%$		& $\mu+0.51\%-0.47\%$		& $\mu+0.83\%-0.80\%$		& $\mu+6.05\%-5.93\%$\\
%\hline Count+Energy+Angle$_{\mathrm{(SNRx20)}}$ 			& $\mu+0.25\%-0.25\%$		& $\mu+0.49\%-0.47\%$		& $\mu+0.78\%-0.84\%$		& $\mu+5.91\%-5.79\%$\\
%\hline Count+Energy+Angle$_{\mathrm{(SNRx30)}}$ 			& $\mu+0.25\%-0.24\%$		& $\mu+0.54\%-0.49\%$		& $\mu+0.80\%-0.77\%$		& $\mu+5.79\%-5.90\%$\\
\hline \textit{a priori} maximum likelihood $\mu$ \cite{fogli_2012}			& $7.54E-5$		& $0.00243$		& $0.307$		& $0.0243$\\
\hline \textit{a priori} uncertainty \cite{fogli_2012}		& $\pm 2.6\%$		& $\pm 3.0\%$		& $\pm 5.5\%$		& $\pm 9.9\%$\\
\hline Count su							        			& $\pm 1.5\%$		& $\pm 1.8\%$		& $\pm 2.4\%$		& $\pm 6.1\%$\\
\hline Count+Energy							    			& $\pm 0.3\%$		& $\pm 0.5\%$		& $\pm 0.9\%$		& $\pm 5.8\%$\\
\hline Count+Energy+Angle					    			& $\pm 0.3\%$		& $\pm 0.5\%$		& $\pm 0.9\%$		& $\pm 5.9\%$\\
\hline Count+Energy+Angle$_{\mathrm{(SNRx10)}}$ 			& $\pm 0.3\%$		& $\pm 0.5\%$		& $\pm 0.8\%$		& $\pm 6.0\%$\\
\hline Count+Energy+Angle$_{\mathrm{(SNRx20)}}$ 			& $\pm 0.3\%$		& $\pm 0.5\%$		& $\pm 0.8\%$		& $\pm 5.9\%$\\
\hline Count+Energy+Angle$_{\mathrm{(SNRx30)}}$ 			& $\pm 0.3\%$		& $\pm 0.5\%$		& $\pm 0.8\%$		& $\pm 5.8\%$\\
\end{tabular}
\caption{Worldwide oscillation parameter optimization results using 4 TREND detectors. The four oscillation parameters were optimized simultaneously with the background flux, the results of which can be seen in Table \ref{table:worldwide_flux}. 1000 MC runs per scenario were used.}
\label{table:worldwide_op}
\end{table}

\begin{table} [!htbp]
\centering
\fontsize{8}{10}\selectfont
\begin{tabular}{l|l|l|l|l|l|l}
\hline
\textbf{Scenario} & \textbf{p($\theta_{\mathrm{IAEA}}$)} & \textbf{p($\theta_{\mathrm{mantle}}$)} & \textbf{p($\theta_{\mathrm{crust}}$)} & \textbf{p($\theta_{\mathrm{fast}_{n^0}}$)} & \textbf{p($\theta_{\mathrm{accidental}}$)} & \textbf{p($\theta_{\mathrm{cosmogenic}}$)}\\
\hline \textit{a priori} uncertainty			& $\pm 2.0\%$	& $\pm 50.0\%$	& $\pm 20.0\%$	& $\pm 10.0\%$ & $\pm 1.3\%$ & $\pm 3.3\%$ \\
\hline Count    							    & $\pm 1.7\%$	& $\pm 34.2\%$	& $\pm 14.6\%$	& $\pm 9.5\%$  & $\pm 1.3\%$ & $\pm 3.2\%$ \\
\hline Count+Energy								& $\pm 1.5\%$	& $\pm 25.1\%$	& $\pm 9.8\%$	& $\pm 9.4\%$  & $\pm 1.2\%$ & $\pm 3.3\%$ \\
\hline Count+Energy+Angle						& $\pm 1.5\%$	& $\pm 26.0\%$	& $\pm 10.2\%$	& $\pm 9.8\%$  & $\pm 1.2\%$ & $\pm 3.3\%$ \\
\hline Count+Energy+Angle$_{\mathrm{(SNRx10)}}$ & $\pm 1.5\%$	& $\pm 20.3\%$	& $\pm 8.3\%$	& $\pm 9.4\%$  & $\pm 1.2\%$ & $\pm 3.3\%$ \\
\hline Count+Energy+Angle$_{\mathrm{(SNRx20)}}$ & $\pm 1.5\%$	& $\pm 13.9\%$	& $\pm 6.3\%$	& $\pm 9.5\%$  & $\pm 1.2\%$ & $\pm 3.2\%$ \\
\hline Count+Energy+Angle$_{\mathrm{(SNRx30)}}$ & $\pm 1.5\%$	& $\pm 10.2\%$	& $\pm 5.3\%$	& $\pm 9.4\%$  & $\pm 1.2\%$ & $\pm 3.4\%$ \\
\end{tabular}
\caption{Worldwide flux uncertainty optimization results using 4 TREND detectors. The six background flux categories were optimized simultaneously with the oscillation parameters, the results of which can be seen in Table \ref{table:worldwide_op}. 1000 MC runs per scenario were used.}
\label{table:worldwide_flux}
\end{table}

Tables \ref{table:worldwide_op} and \ref{table:worldwide_flux} clearly show tremendous scientific utility of a network of TREND detectors.   With a one year deployment to the sites specified, the bounds on $\Delta m^2_{12}$ and $\sin^2\theta_{12}$ can be tightened by 11X and 6.5X respectively.    Similarly, the bounds on $\Delta m^2_{13}$ and $\sin^2\theta_{13}$ can be tightened by 8X and 3X respectively.  The systematic uncertainties in crust and mantle geo-neutrino flux can be reduced by 2X.  These levels of improvement get better still if hardware and algorithm advances permit better angular resolution in the future.  It is worth noting the steady improvement in the knowledge of the crust and mantle geo-neutrino signal with increasing angular resolution.  This is because the crust and mantle components (which are not separable in energy spectrum) become more separable in direction as the angular resolution improves.  The MC Count+Energy+Angle \textit{a posteriori} uncertainties in Tables \ref{table:worldwide_op} and \ref{table:worldwide_flux} will be used as the \textit{a priori} uncertainties in the search for unknown reactors.

\newpage

\section{Reactor geolocation results}
\label{Reactor geolocation results}

Here we apply the suboptimal MAP estimator presented in Section \ref{Reactor geolocation} to the unknown-reactor search problem at various locations around the world. For each scenario we assume one or more TREND detectors collecting measurements in the vicinity of an unknown 300MW$_\mathrm{th}$ reactor for one calendar year (minus downtime).

The first scenario, shown in Figure \ref{Portugal1_GE}, takes place in the Iberian Peninsula on the Atlantic side. The unknown reactor is located at the south-westernmost tip of the peninsula, and two TREND detectors are each placed about 200km away in the deep waters of the Atlantic. The majority of measurements in this scenario come from nearby known reactors across Europe. The overall background flux is about half as high as in our third scenrio (Europe-Mediterranean), while the unknown-reactor flux remains about the same.

The second scenario takes place in Southern Africa. Here the unknown reactor is placed near a known nuclear power plant, the only nuclear reactor site within thousands of kilometers. The latitude-longitude search area selected for our Southern Africa scenario intentionally encompasses this nearby known reactor to test the robustness of the MAP estimator in the presence of a strong confuser in an otherwise low reactor-flux environment. Two TREND detectors are located about 200km offshore from the unknown reactor, and three MAP estimator approximation results are shown, first using measurements from detector \#1 only, then detector \#2 only, and finally from both detectors together.

Lastly, we present the most challenging scenario we consider, where we move the unknown reactor to the Mediterranean side of the Iberian peninsula. Three TREND detectors are placed in the relatively shallow waters of the southern Mediterranean, shadowing the coastline about 100km out. Since the Mediterranean waters are much shallower than in the Atlantic ocean, the detectors can not be submerged as deeply. Hence these detectors suffer higher muon flux rates and higher veto-related downtime, leading to reduced detector duty cycles. Another challenge in this scenario is the presence of multiple known reactors inside the latitude-longitude parameter space, the closest being only (10km) from the unknown reactor.

Throughout all three scenarios, one common theme is that the unknown reactor supplies very few measurements to the TREND detectors. Due to its low power output (300MW$_\mathrm{th}$) and the significant stand-off distances involved (150km-350km) unknown-reactor measurements always comprises a very small percentage of the total measurements. This unknown-reactor typically accounts for 2\%-3\% of each detector's signal across the scenarios, and as little as 1\%.

For each scenario we test six different measurement resolutions. We begin with the scalar count-only measurement approach detailed in Lasserre, \textit{et al.} \cite{lasserre_2010}, to which we add the energy component of the measurements, and finally the directional (or angle) component of the measurements. After these three approaches, we apply progressively better levels of hypothetical directional resolution to quantify possible future performance gains (which the MAP estimator is capable of exploiting) over current detector technology. These last three scenarios are not applicable today since they posit better directional resolution than today's hardware can produce.  They may be possible in the future, and they provide a fascinating look at the capabilities of the MAP estimator itself (whose potential is largely hardware agnostic).

Throughout all the scenarios, the TREND detectors are assumed submerged underwater to the ocean floor at their respective operational locations to minimize cosmogenic background. Thus each TREND detector operates at a different depth, ranging from about 2500m depth in the Mediterranean to almost 5000m depth in the Atlantic Ocean. Cosmogenic veto times are a function of detector depth, so we end up with different duty cycles for each detector ranging from about 45\% duty cycle at 2500m in the Mediterranean to 99\% duty cycle at 5000m depth in the Atlantic off of the coast of Europe-Atlantic. Naturally these detector-to-detector variations in depth and duty cycle must be properly accounted for when applying the MAP estimator with multiple detectors.

\subsection{No unknown reactor (Reactor Exclusion)}
\label{No reactor}
One basic question relating to the search for an unknown reactor is what the results would look like if there were simply no unknown reactor present within the constrained latitude-longitude search area. To answer this question we remove the unknown reactor from the Geospatial Model entirely, and let four detectors observe for one year.

Due to Poisson statistics, \textit{only the presence of a reactor can be confirmed, not the absence of one}.  This is because a sufficiently small and/or sufficiently distant reactor may not contribute measurements to any of the four detectors, or the number of measurements contributed may be overwhelmed by the statistical noise (associated with the background) for all four detectors. It would, thus, be impossible to assert with 100\% confidence that no reactor of any power is present within the search area.  If we constrain the minimum unknown-reactor power though, as well as the search area, then it \textit{is} possible to rule out its existence of a reactor to a high degree of confidence.

We find that, in lieu of confirming the absence of a reactor, one may express a certain confidence of no reactor existing past a certain power output within the constrained search area. This confidence may be expressed for the search area as a whole (by marginalizing through the latitude and longitude dimensions), or may be more accurately expressed conditional on a given  latitude-longitude location.  Our approach has an distinct advantage in this area over count-only detection theory metrics because it fully considers the information contained in the energy and direction measurements.

\begin{figure}[!htbp]
\includegraphics[width=\linewidth]{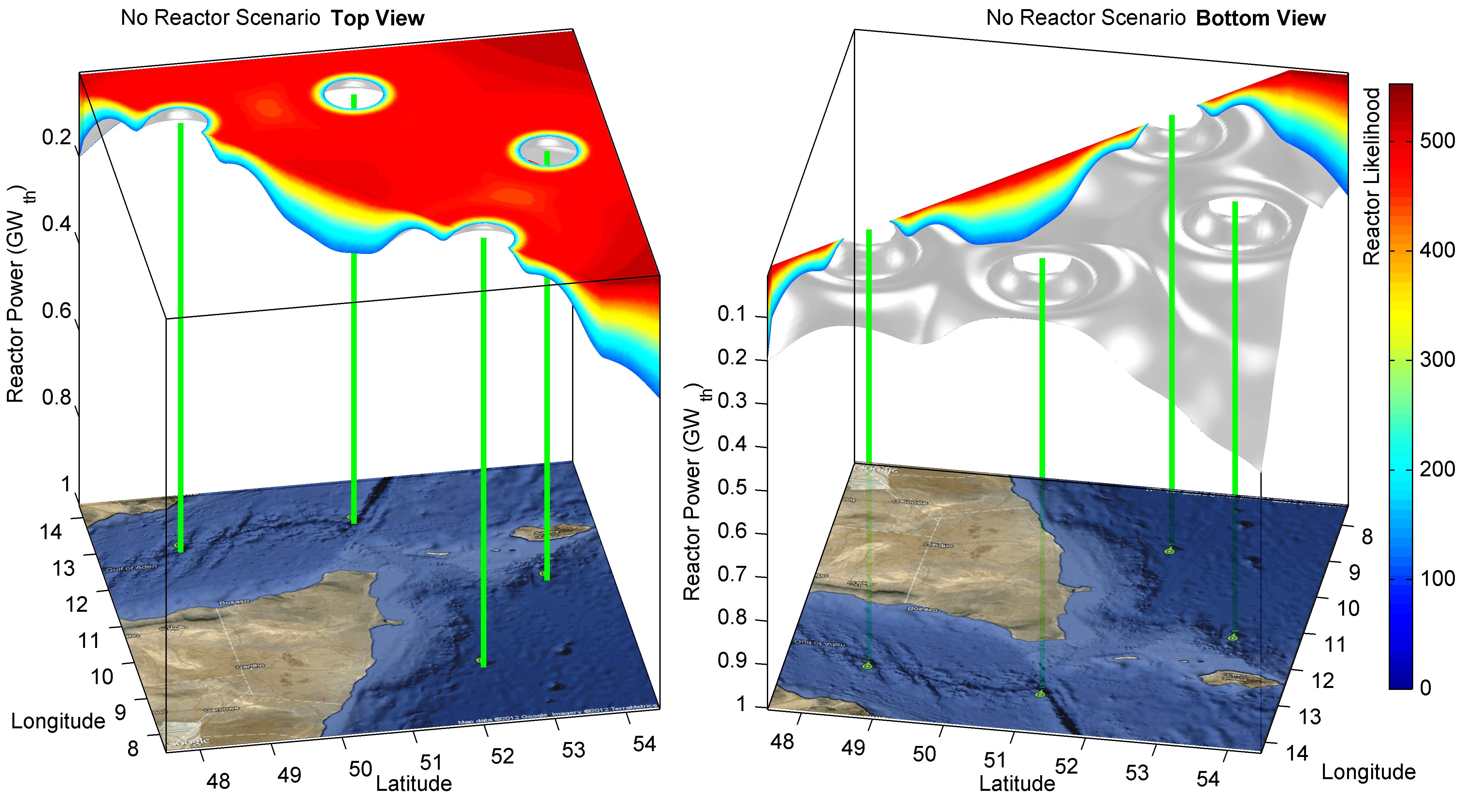}
\centering
\caption{Reactor exclusion results. Silver-colored 90\% confidence isosurface puts a ceiling on possible unknown-reactor thermal power output at each location in the lat-long search area. Isosurface has been sliced diagonally to allow observation of the internal \textit{a posteriori} pdf.}
\label{SomaliaD0conditional}
\end{figure}

\begin{figure}[!htbp]
\includegraphics[width=\linewidth]{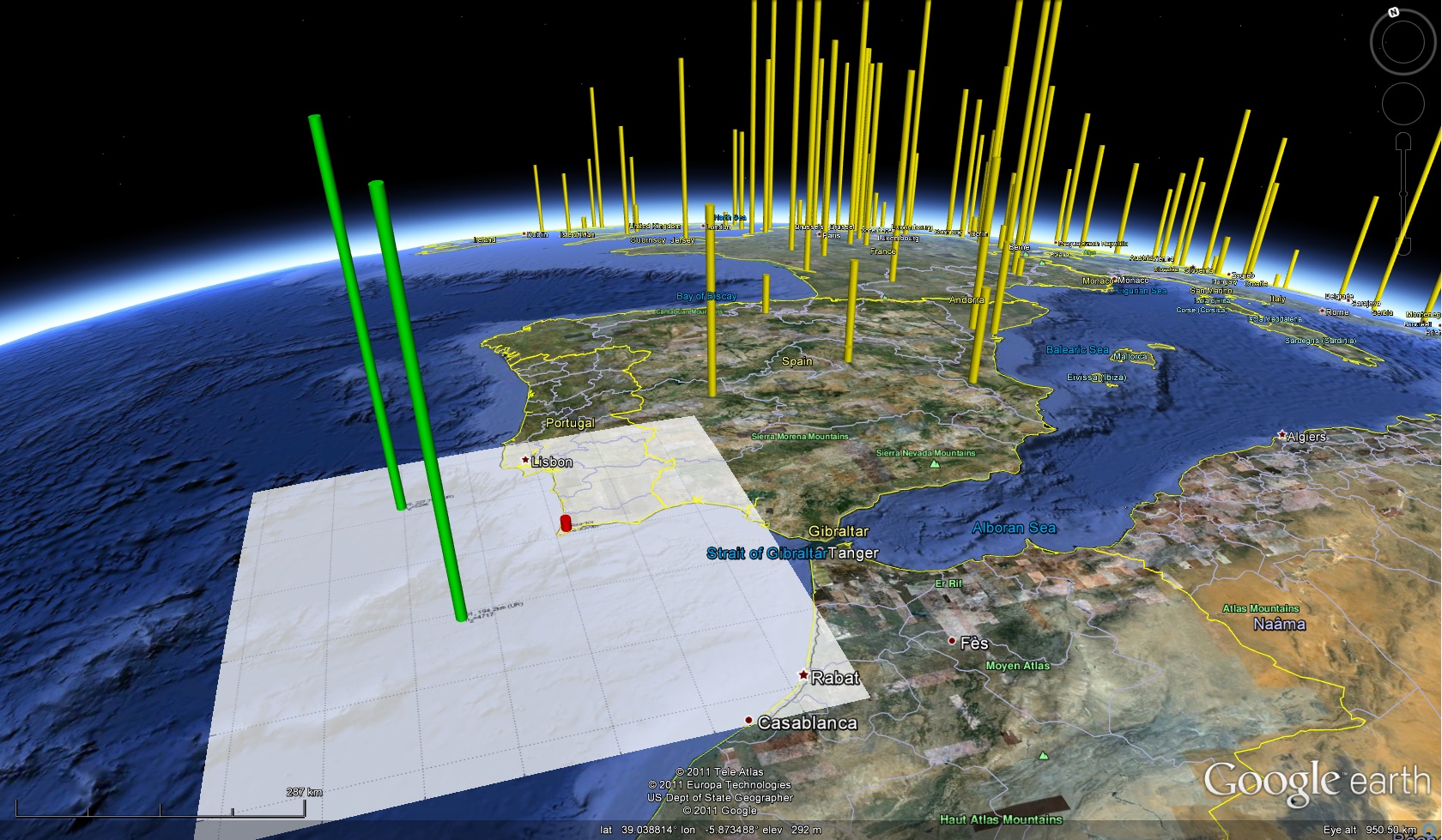}
\centering
\caption{Europe-Atlantic scenario layout. Green cylinders are TREND detectors. The small red cylinder is the unknown reactor. Tall yellow cylinders are known reactors in Europe. Red and yellow cylinders are shown to scale to easily compare thermal power output between known and unknown reactors. The white square represents the latitude-longitude constrained search area.}
\label{Portugal1_GE}
\end{figure}

Figure \ref{SomaliaD0conditional} shows the results of this test. The MAP estimator has returned an \textit{a posteriori} conditional pdf which puts an effective upper confidence bound on possible unknown reactor powers at any location. Using this \textit{a posteriori} pdf, at a given location, one could express 90\% confidence that no reactor exists with a thermal power output greater than that denoted by the silver 90\% confidence isosurface. As better direction vector resolution is employed, we can see that the 90\% confidence ceiling lowers to smaller and smaller power outputs.

Near each of the detectors we can assert with 90\% confidence that no reactor exists greater than 10MW$_\mathrm{th}$ using today's technology. Near the center of our search area we can assert with 90\% confidence that no reactor exists larger than 90MW$_\mathrm{th}$. Over our entire latitude-longitude constrained search area we can assert with 90\% confidence that no reactor exists with a mean thermal power output greater than 190MW$_\mathrm{th}$ over the past year.

\subsection{High reactor background (Europe-Atlantic)}
\label{High reactor background (Europe-Atlantic)}

Here two TREND detectors are submerged all the way down to the Atlantic Ocean floor, about 200km off the coast of the coast of the Iberian Peninsula. Detector 1 is the southernmost detector, while detector 2 is about 200km north of detector 1. The two detectors are 194km and 228km from our reactor, respectively. Here the signal is dominated by neutrinos emitted from nearby known reactors. Both of these detectors are very deep, about 5000m underwater, and consequently enjoy 99\% duty cycles as a virtue of the low muon flux at that depth. The geo-neutrino flux is, of course, nearly identical here as in Europe-Atlantic and Southern Africa.

TREND detector 1 has intentionally been placed along a poor line of sight. Even with good angle measurement resolution, detector 1 should have a difficult time isolating the unknown-reactor signal since the unknown reactor lies along the same line of sight as all the powerful European reactors directly behind it. The unknown-reactor neutrinos will be arriving at detector 1 from almost the same exact direction as the far more numerous neutrinos being emitted from all the known reactors across Europe. For detector 1, the unknown-reactor signal should strongly resemble the background signal both in energy and direction. TREND detector \#2 has no such directional problems, as it has a clean line of sight to the unknown reactor, with no major neutrino sources ahead of, or behind the unknown reactor.

\subsubsection{One detector} 
\begin{figure}[!htbp]
\includegraphics[width=\linewidth]{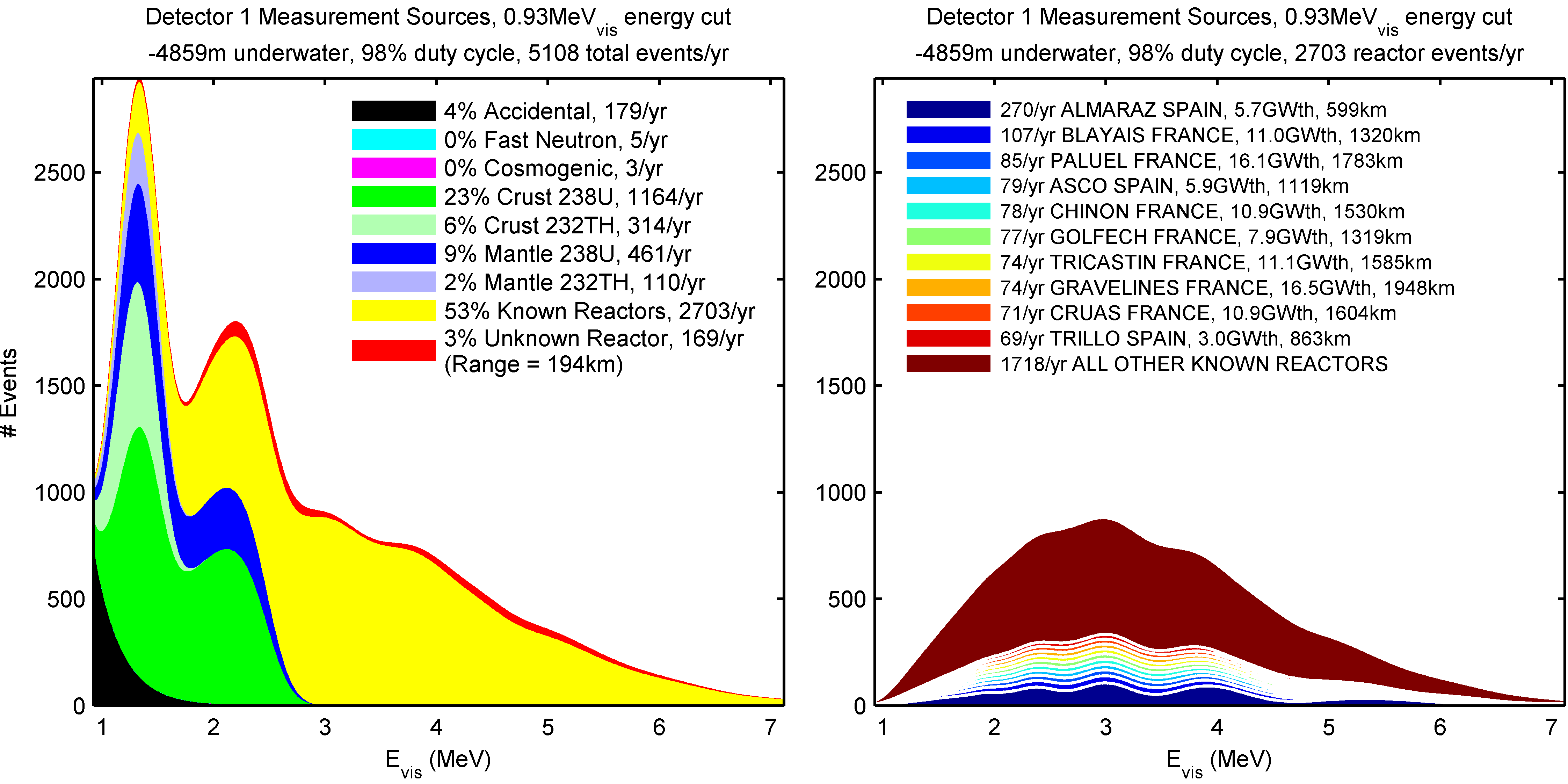}
\centering
\caption{Europe-Atlantic scenario TREND detector \#1 mean smeared energy measurement space. Flux from all sources shown on left plot, known-reactor flux shown isolated in right plot.}
\label{Portugal1_epdf}
\end{figure} 

Figure \ref{Portugal1_epdf} shows the measurement sources as seen by detector 1 in Europe-Atlantic. Only about 3\% of the measurements each detector receives originate from the unknown reactor, while known reactors account for over 50\% of the measurements. Geo-neutrinos comprise about 40\% of the measurement total, with accidentals responsible for most of the rest, about 5\%.

\begin{table} [!htbp]
\centering
\fontsize{8}{10}\selectfont
\begin{tabular}{l|ll|ll|ll|ll}
\hline \textbf{Measurement Resolution} & \multicolumn{2}{c|}{\textbf{Area@300MW$_{\mathrm{th}}$ $90\%$}} & \multicolumn{2}{c|}{\textbf{Volume $90\%$}} & \multicolumn{2}{c|}{\textbf{Power $1\sigma$}} & \multicolumn{2}{c}{\textbf{Location $1\sigma$}}\\
& \multicolumn{2}{c|}{(km$^2$)} & \multicolumn{2}{c|}{(km$^2$GW$_{\mathrm{th}}$)} & \multicolumn{2}{c|}{(MW$_{\mathrm{th}}$)} & \multicolumn{2}{c}{(km)}\\
& $U(\theta)$ & $p(\theta)_\mathrm{H_2O}$& $U(\theta)$ &$p(\theta)_\mathrm{H_2O}$& $U(\theta)$ &$p(\theta)_\mathrm{H_2O}$& $U(\theta)$ &$p(\theta)_\mathrm{H_2O}$ \\
\hline Count-Only							   &         294528& 62143&         268439& 60350&            409& 537&            330& 186\\
\hline Count+Energy							   &         264449& 60908&         226482& 57496&            247& 299&            319& 77\\
\hline Count+Energy+Angle					   &         263985& 60839&         225470& 57543&            247& 329&            299& 118\\
\hline Count+Energy+Angle$_{\mathrm{(SNRx10)}}$&         249736& 60671&         204199& 57418&            247& 500&            194& 161\\
\hline Count+Energy+Angle$_{\mathrm{(SNRx20)}}$&         199344& 60780&         172340& 56286&            249& 482&            145& 150\\
\hline Count+Energy+Angle$_{\mathrm{(SNRx30)}}$&         160506& 59269&         136284& 53902&            249& 511&            133& 146\\
\end{tabular}
\caption{Europe-Atlantic detector \#1 MC results. Left-hand values reflect uniform parameter-space \textit{a priori} pdf $U(\theta)$ corresponding to Figure \ref{PortugalD1mc} results, right-hand values reflect water \textit{a priori} pdf $p(\theta)_\mathrm{H_2O}$ expressing belief that an unknown reactor must necessarily exist on land.}
\label{table:Portugal1table}
\end{table}

\begin{figure}[!htbp]
\includegraphics[width=.879\linewidth]{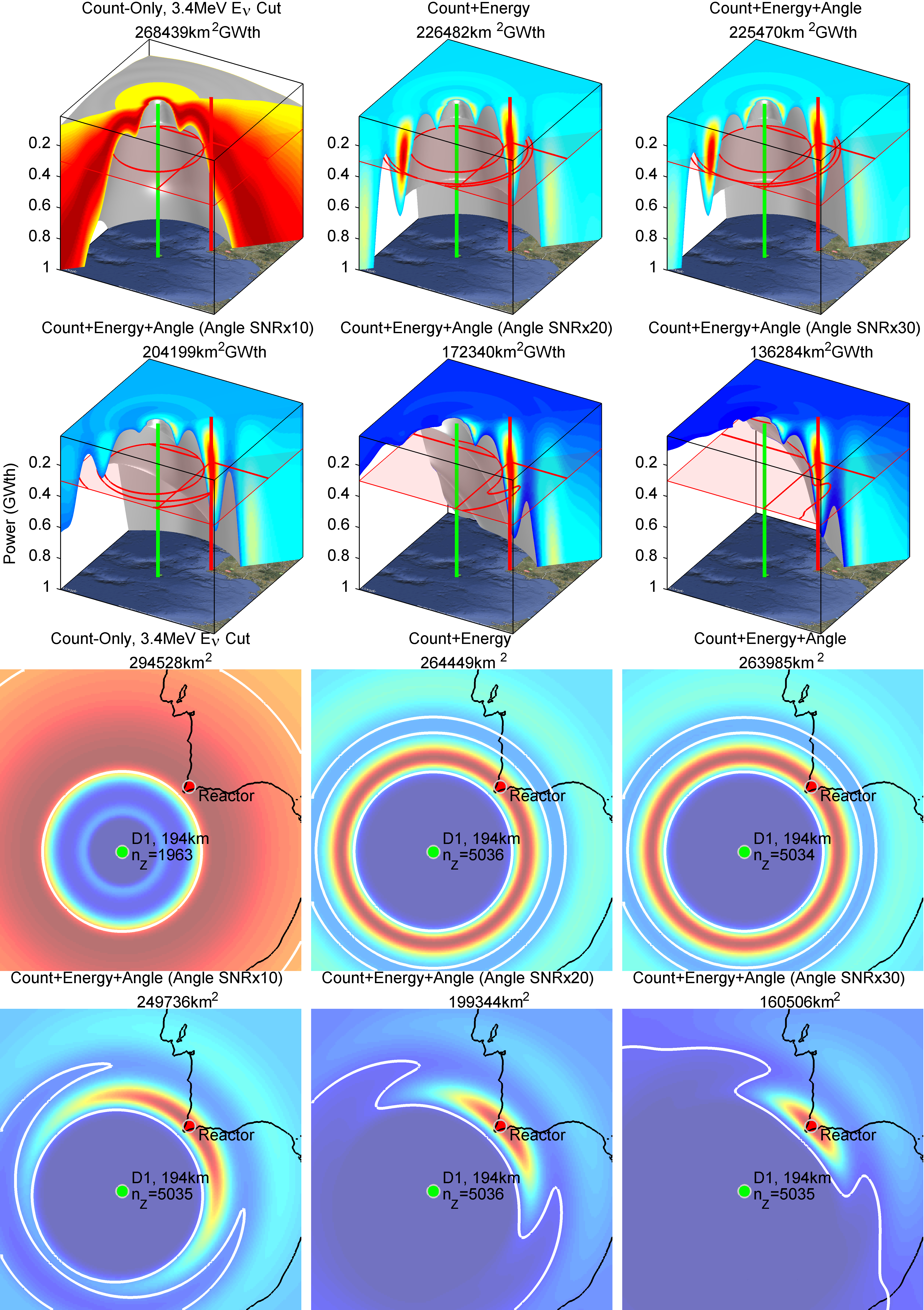}
\centering
\caption{Europe-Atlantic scenario with 1 TREND detector. 300MW$_\mathrm{th}$ unknown reactor shown in red, TREND detectors in green. Full parameter space shown in upper six plots (90\% confidence isosurface shown in silver), 300MW$_\mathrm{th}$ conditional slice shown in lower six plots (90\% 300MW$_\mathrm{th}$ conditional confidence contour shown in white).}
\label{PortugalD1mc}
\end{figure}

%PASTED from Somalia!! ---------------------------------------------------------------------------------------------------------------------------------------------
Table \ref{table:Portugal1table} and Figure \ref{PortugalD1mc} show the Europe-Atlantic single-detector MC simulation results after 1000 MC runs. The results are presented two ways in Figure \ref{PortugalD1mc}: as a 300MW$_\mathrm{th}$ conditional 2D slice through the 3D parameter space, and as the full conditional 3D latitude-longitude-power pdf. The area in km$^2$ (and volume in km$^2$GW$_{\mathrm{th}}$) enclosed by a 90\% confidence 2D contour (and 3D isosurface) are used as the primary performance metrics when comparing the results across scenarios or measurement resolutions.  The area (or volume) shown in Figure \ref{PortugalD1mc} is formed by the careful scaling and combination of parameter-space pdfs from all 1000 MC runs.  This smooths out the impact that statistical (Poisson) noise can have on any single observation run.  The power and location standard deviations in Table \ref{table:Portugal1table} represent the standard error between the MAP estimate (a single best-estimate value for latitude, longitude, and power for each MC run) and the truth used in the simulations\footnote{These MAP estimates are found using a standard optimizer implemented in the Matlab programming language.  The optimizer used was not customized for this particular problem and, hence, optimizer-related issues may add some additional ``noise" to the MAP estimates.  As the parameter space becomes more complicated (e.g. more dimensions, multiple detectors, the use of Bayesian spatial priors with sharp edges), optimizer robustness will become increasingly important.}.
%PASTED from Somalia!! ---------------------------------------------------------------------------------------------------------------------------------------------

%PASTED from Somalia!! ---------------------------------------------------------------------------------------------------------------------------------------------
In Figure \ref{PortugalD1mc} six sub-cases are presented, corresponding to six different levels of detector measurement resolution:

\begin{enumerate}
\item \textbf{Count-Only.} Here we pass only the measurement $count$ to the MAP estimator. No background flux estimation is possible in this case, so we impose an energy cut of $E_\mathrm{vis}\geq 2.6$MeV on the measurements to reject the uncertain geo-neutrino and accidental signatures. This method has previously been proposed in detail by Lasserre \textit{et al.} \cite{lasserre_2010}. This method stills accounts for neutrino oscillations, though the impact is markedly less than it is on the other sub-cases which consider the energy of individual measurements. Count-Only \textit{a priori} oscillation parameter uncertainty and background flux uncertainty are taken to be the \textit{a priori} uncertainties from Tables \ref{table:worldwide_op} and \ref{table:worldwide_flux} because oscillation and flux optimization is not possible in this sub-case.

\item \textbf{Count+Energy.} Here we pass the measurement count (a scalar) as well as the measured energies (a vector) to the estimator. The energy cut drops to $E_\mathrm{vis}\geq 0.9$MeV, enabling the MAP estimator to co-estimate independent background fluxes from each background source category simultaneously with the unknown-reactor power and location. We use the TREND energy resolution, $\left. 8.9\% \right|_{E_\mathrm{vis}=1\mathrm{MeV}}$, to add measurement error to the true energies in the MC runs. The MAP estimator uses the modeled TREND energy resolution, the count covariance matrix, the oscillation parameters, the background fluxes, the smeared \textit{a priori} oscillation survival fraction, and the known constants to create the smeared measurement space $Z$. Count+Energy \textit{a priori} oscillation parameter uncertainty and background flux uncertainty are taken to be the Count+Energy \textit{a posteriori} uncertainties from Tables \ref{table:worldwide_op} and \ref{table:worldwide_flux} in Section \ref{Oscillation parameter estimation results}.

\item \textbf{Count+Energy+Angle.} Here we pass the IBD CE-to-CE direction vectors of the measurements (a matrix) to the MAP estimator alongside the (scalar) count and (vector) energy measurements. For this and all the later sub-cases, we maintain an energy cut of $E_\mathrm{vis}\geq 0.9$MeV and co-estimate the background flux with the unknown reactor. Direction vector resolution for TREND, using the CHOOZ CE-to-CE technique cited in \cite{chooz_2003}, was taken to be about SNR=0.045 (or 21mm/461mm).  This resolution derives from the mean neutron displacement of 21mm, the mean neutron capture point resolution of 243mm and the mean positron vertex point resolution of 374mm (the point resolutions combine with the IBD uncertainty to form a reconstruction vector resolution of 461mm per Cartesian axis). Count+Energy+Angle \textit{a priori} oscillation parameter uncertainty and background flux uncertainty are taken to be the Count+Energy+Angle \textit{a posteriori} uncertainties from Tables \ref{table:worldwide_op} and \ref{table:worldwide_flux} in Section \ref{Oscillation parameter estimation results}.

\item \textbf{Count+Energy+Angle$_{\mathrm{(SNRx10)}}$.} Here the MAP estimator is used to quantify performance gains enabled by future hypothetical hardware advancements. We hypothesize 10 times better direction vector resolution than what was found by the TREND Detector Model MC runs, or about SNR=0.45). We believe this level of directional resolution may be possible in the near future for smaller detectors using multi-channel plate technology. Count+Energy+Angle$_{\mathrm{(SNRx10)}}$ \textit{a priori} oscillation parameter uncertainty and background flux uncertainty are taken to be the corresponding \textit{a posteriori} uncertainties from Tables \ref{table:worldwide_op} and \ref{table:worldwide_flux} in Section \ref{Oscillation parameter estimation results}.

\item \textbf{Count+Energy+Angle$_{\mathrm{(SNRx20)}}$.} Here we hypothesize 20 times better direction vector resolution than what was found by the TREND Detector Model MC runs, or about SNR=0.90. Count+Energy+Angle$_{\mathrm{(SNRx20)}}$ \textit{a priori} oscillation parameter uncertainty and background flux uncertainty are taken to be the corresponding \textit{a posteriori} uncertainties from Tables \ref{table:worldwide_op} and \ref{table:worldwide_flux} in Section \ref{Oscillation parameter estimation results}.

\item \textbf{Count+Energy+Angle$_{\mathrm{(SNRx30)}}$.} Here we hypothesize 30 times better direction vector resolution than what was found by the TREND Detector Model MC runs, or about SNR=1.35. Count+Energy+Angle$_{\mathrm{(SNRx30)}}$ \textit{a priori} oscillation parameter uncertainty and background flux uncertainty are taken to be the corresponding \textit{a posteriori} uncertainties from Tables \ref{table:worldwide_op} and \ref{table:worldwide_flux} in Section \ref{Oscillation parameter estimation results}.
\end{enumerate}
%PASTED from Somalia!! ---------------------------------------------------------------------------------------------------------------------------------------------

%PASTED from Somalia!! ---------------------------------------------------------------------------------------------------------------------------------------------
Note that directional SNR is a function of neutrino energy as shown in Figure \ref{SNR1}. Figure \ref{SNR1} indicates a tendency for higher energy neutrinos to have higher angular SNRs. Detectors are not co-located and, as a consequence, each detector will see a slightly different energy spectrum. This means no two detectors will have identical spectrum-weighted mean directional SNRs. The TREND MC value of SNR=0.045 is obtained by taking a weighted mean across a noiseless 1km-range reactor spectrum from 2MeV to 11MeV. If we weight across the background energy spectrum rather than a 1km reactor spectrum, the lower energies influence the mean to a greater degree, resulting in slightly lower SNR of about 0.035 to 0.040.

Looking at Figure \ref{PortugalD1mc} (lower panel), the most obvious conclusion that can be drawn is that the inclusion of energy measurements produces a great increase in parameter observability. The Count+Energy subplot (Subplot 2) shows much smaller 90\% confidence contours than the Count-Only subplot (Subplot 1), and the pdf is much tighter about the correct range. The information the measurements contain regarding the parameter space (i.e. Fisher information) is commonly quantified as the inverse of the parameter variance, so in this case we can conclude that the energy component of the measurements contain much information about the parameter space, even with modest energy measurement resolution of $\left. 8.9\% \right|_{E_\mathrm{vis}=1\mathrm{MeV}}$.

A more interesting conclusion can be drawn from examination of the full 3D parameter space. A look at Subplot 1 (upper panel) shows very poor reactor power observability from the Count-Only measurements. Using Count-Only measurements from one single detector, the unknown-reactor search problem seems to be largely under-determined. We see a cone of nearly constant probability density extending down to larger reactor powers at further ranges. This shows that a more powerful unknown-reactor may exist at further ranges, with equal probability to a smaller unknown-reactor at closer ranges. A single Count-Only detector is thus incapable of resolving the range-power ambiguity.

Though the range and power are obviously both constrained to a cone of probability, there are a wide range of equally probable range-power combinations. The 3D parameter space is azimuthally symmetric about a single Count-Only detector and, at long ranges, a degeneracy between range and power forms that reduces the parameter space to 1D ($\frac{range^2[\mathrm{km^2}]}{power[\mathrm{GW_{th}}]}$). This is highly suboptimal, and a quick look at the rest of the subplots in Figure \ref{PortugalD1mc} confirms that we are not extracting all of the available information from the measurements using this Count-Only approach.

In Subplot 2, we introduce the Count+Energy results. Unknown-reactor range observability in the 300MW$_\mathrm{th}$ power slice (lower panel) $and$ unknown-reactor power observability (upper panel) are both greatly increased relative to the Count-Only results. Subplot 2 (upper panel) shows a well-defined absolute maximum forming at the true power; the problem is no longer under-determined. There now exists a unique range and a unique power estimate along which the probability density is highest, though there are also several local minima appearing. These local minima reflect other likely reactor locations and are caused by the oscillations in the neutrino energy spectra over range. Not surprisingly, the local minima tend to be occur along the range-power degeneracy seen in the Count-Only results ($\frac{range^2[\mathrm{km^2}]}{power[\mathrm{GW_{th}}]}$).

The 3D latitude-longitude-power parameter space in Subplot 2 (upper panel) can be reduced to two degrees of freedom: range and power. A single Count+Energy detector is, thus, incapable of determining the direction towards an unknown reactor even though it may have good range resolution. This limitation can be mitigated in a straightforward manner by using multiple detectors, but it requires a multiple of the cost.  In the future, advances in hardware may allow for detectors with better intrinsic directionality.   This would allow for improvement with even a single detector, and that improvement is further leveraged by multiple detectors.

In Subplot 3, the Europe-Atlantic single-detector Count+Energy+Angle results are presented.  These results include the angle information inherent in the measured IBD direction vectors. The Subplot 3 results are difficult to distinguish from the Subplot 2 Count+Energy results. The metrics in Table \ref{table:Portugal1table} quantify the very small (by finite) improvement that results from including direction. It is not surprising that the improvement is very small since the current hardware approaches (combined with the CHOOZ CE-to-CE technique) result in a nearly isotropic/uninformative directional resolution.  It is significant, however, that the estimator performed as designed.   It accepted the direction measurements, weighted the measurements properly according to the information that they contained, and produced results at least as good as the Count+Energy results (or within MC noise).

Subplot 3 (upper panel) does, however, show a truly irreducible 3D parameter space for the first time.  This is because the likelihood of the range ``ring" in Subplot 3 (lower panel) varies ever so slightly with azimuth. The ``ring" starts to turn into a ``crescent moon" shape as greater angular measurement information is introduced. This is made possible through the mixture distribution nature of the MAP estimator even though only about 2\% of the measurements are from the unknown reactor.

Despite the fact that the improvement between Subplot 2 and 3 is not readily apparent visually, the MAP estimator is extracting all of available Fisher information for the first time. The estimator is showing its capacity to exploit direction vector information no matter how poor the angular resolution. Subplot 3 presents the first ``optimal" result in the context of the MAP estimator extracting $all$ the available information from the measurements (Count+Energy+Angle), not just extracting subsets of the total information (i.e. Count+Angle or Count-Only). There is no information left in the measurements that the MAP estimator has not exploited\footnote{If the detector design results in the non-neutrino background being a strong function of vertex location within the detector, then the estimator could be modified to utilize the CE vertices themselves.  The estimator in this paper currently only uses the CE-to-CE direction vector.}, thus Subplot 3 presents the best possible result given the detector hardware and the scenario specifics.

If one seeks performance gains above and beyond the results shown in Subplot 3 using today's CE-to-CE algorithms and detector hardware, we believe that any such gains must be sought after some other place than in the MAP estimator, such as in the detector hardware or the detector placement strategy. We examine the potential impact that detector hardware improvements may have on estimation results in the remaining subplots.

Subplots 4, 5 and 6 show the same ``optimal" MAP estimates as Subplot 3, but with hypothetical future angle SNR performance improvements of 10X, 20X, and 30X corresponding to an SNR of 0.37, 0.74, and 1.12 respectively. As angle measurement resolution increases we see increased unknown-reactor observability, not just in direction (azimuth angle), \textit{but also in range and thermal power}. This is indicating increased observability across the entire parameter space due to increased angular measurement resolution.

Throughout all subplots, the energy resolution remains constant at $\left. 8.9\% \right|_{E_\mathrm{vis}=1\mathrm{MeV}}$ while the angle resolution varies. We opted not to study improvements in energy resolution due to the diminishing returns imposed by Poisson noise, and due to the fact that the current state of the art is already well placed to exploit increased energy resolution.
%PASTED from Somalia!! ---------------------------------------------------------------------------------------------------------------------------------------------

We conclude from Figure \ref{PortugalD1mc} that our estimator is robust in a low signal-to-background environments. The estimator is stressed here (considering the 1:15 signal-to-background ratio), yet it still provides for robust unbiased estimates across all six measurement resolutions even with \textit{a priori} uncertainties assumed in this paper. This verifies the utility of the MAP estimator presented in Section \ref{Reactor geolocation}, and also indicates that increased angular resolution helps \textit{even if the unknown neutrino source is not isolated in angle space}.  The increased angular resolution reduces the confusion between different background sources, and between much of the background and the unknown reactor.

\subsubsection{Two detectors}
\begin{figure}[!htbp]
\includegraphics[width=.879\linewidth]{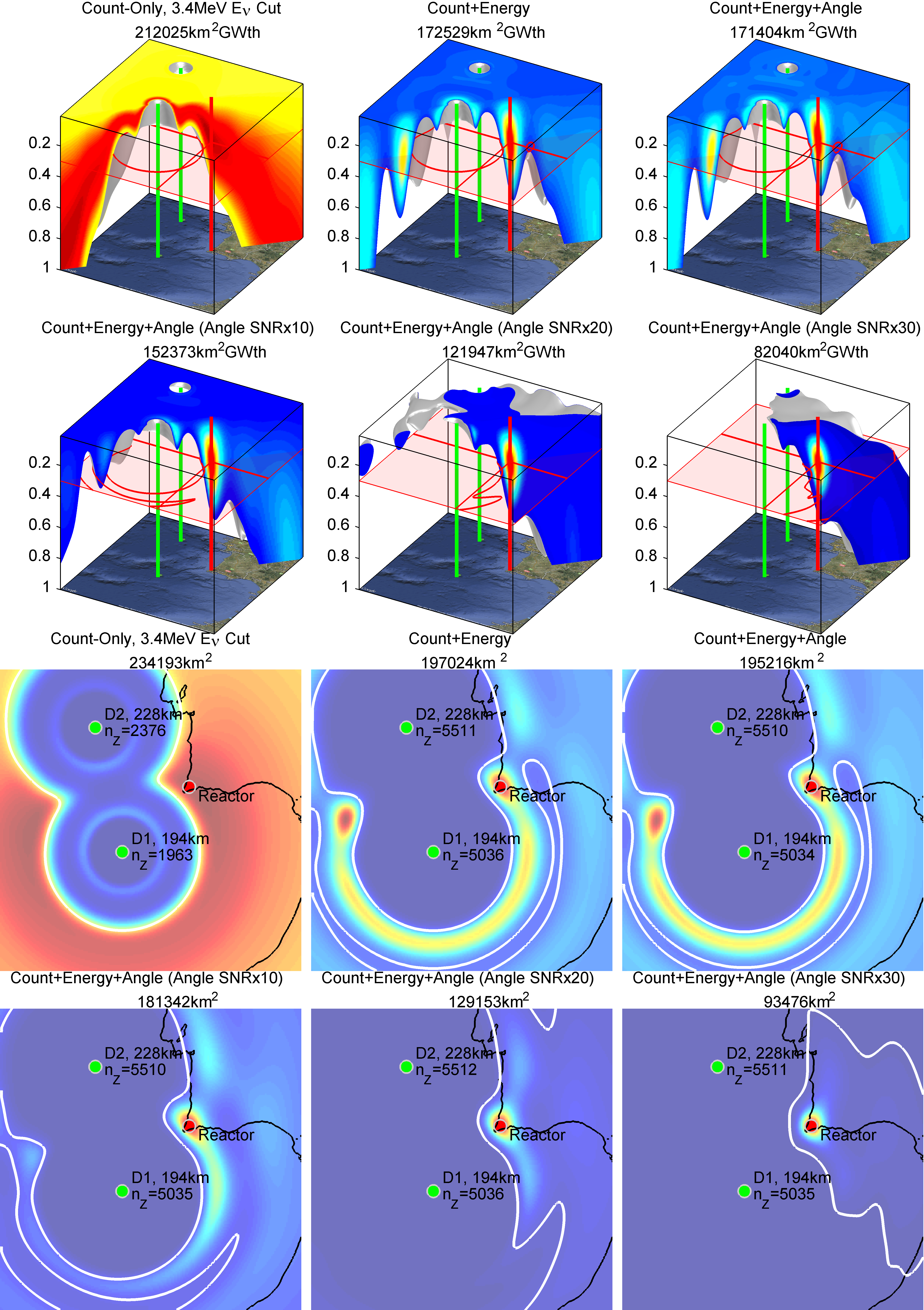}
\centering
\caption{Europe-Atlantic scenario with 2 TREND detectors. 300MW$_\mathrm{th}$ unknown reactor shown in red, TREND detectors in green. Full parameter space shown in upper six plots (90\% confidence isosurface shown in silver), 300MW$_\mathrm{th}$ conditional slice shown in lower six plots (90\% 300MW$_\mathrm{th}$ conditional confidence contour shown in white).}
\label{PortugalD2mc}
\end{figure}

Table \ref{table:Portugal2table} and Figure \ref{PortugalD2mc} show the Europe-Atlantic single-detector MC simulation results after 1000 runs. Here, as in our other scenarios, we see that using multiple detectors (just two in this case) improves performance relative to a single detector, independent of measurement resolution. With low (or no) direction vector resolution, two detectors tend to produce two local minima which are symmetric about the great circle connecting the two detector sites. The addition of a third detector, or the use of Bayesian spatial priors, should remove this ambiguity in practice.

\begin{table} [!htbp]
\centering
\fontsize{8}{10}\selectfont
\begin{tabular}{l|ll|ll|ll|ll}
\hline \textbf{Measurement Resolution} & \multicolumn{2}{c|}{\textbf{Area@300MW$_{\mathrm{th}}$ $90\%$}} & \multicolumn{2}{c|}{\textbf{Volume $90\%$}} & \multicolumn{2}{c|}{\textbf{Power $1\sigma$}} & \multicolumn{2}{c}{\textbf{Location $1\sigma$}}\\
& \multicolumn{2}{c|}{(km$^2$)} & \multicolumn{2}{c|}{(km$^2$GW$_{\mathrm{th}}$)} & \multicolumn{2}{c|}{(MW$_{\mathrm{th}}$)} & \multicolumn{2}{c}{(km)}\\
& $U(\theta)$ & $p(\theta)_\mathrm{H_2O}$& $U(\theta)$ &$p(\theta)_\mathrm{H_2O}$& $U(\theta)$ &$p(\theta)_\mathrm{H_2O}$& $U(\theta)$ &$p(\theta)_\mathrm{H_2O}$ \\
\hline Count-Only							   &         234193& 61343&         212025& 57382&            345& 523&            264& 262\\
\hline Count+Energy							   &         197024& 58736&         172529& 51153&            230& 275&            261& 142\\
\hline Count+Energy+Angle					   &         195216& 58440&         171404& 51245&            230& 322&            266& 152\\
\hline Count+Energy+Angle$_{\mathrm{(SNRx10)}}$&         181342& 57946&         152373& 51427&            207& 286&            180& 141\\
\hline Count+Energy+Angle$_{\mathrm{(SNRx20)}}$&         129153& 56168&         121947& 49568&            207& 228&            138& 124\\
\hline Count+Energy+Angle$_{\mathrm{(SNRx30)}}$&          93476& 44378&          82040& 39386&            152& 117&             60& 27\\
\end{tabular}
\caption{Europe-Atlantic scenario MC results from both detectors combined. Left-hand values reflect uniform parameter-space \textit{a priori} pdf $U(\theta)$ corresponding to Figure \ref{PortugalD2mc} results, right-hand values reflect water \textit{a priori} pdf $p(\theta)_\mathrm{H_2O}$ expressing belief that an unknown reactor must necessarily exist on land.}
\label{table:Portugal2table}
\end{table}

With high enough angular resolution, such as those hypothesized in Figure \ref{PortugalD2mc}, it seems that much of the benefit of multiple detectors can be attained with just two detectors. At angular resolutions of SNR=2.0 and better, we can extrapolate to say that one TREND detector might be sufficient.  

With today's low directional resolution we see the absolute minima in Europe-Atlantic reflected in the Atlantic Ocean as a local minima. As the angular resolution of the detectors increases we see the reflected local minima begin to disappear and the confidence contour shrink around the absolute minima. The shape of the two local minima, at today's low angular resolution, is nearly circular. The Europe-Atlantic scenario was slightly ``lucky" in its detector placement, as the two detectors form vectors to the unknown reactor at nearly right angles to each other, resulting in near-optimal geometry for the discovery of an unknown reactor in the exact area where it actually exists.
 
If the two detectors had been placed much closer to each other, and the angle between the reactor and the two detectors were to shrink, we can surmise that we would begin to lose reactor location observability (as well as observability across the rest of the parameter space). The circular local minima would begin to elongate into a crescent shape and the 90\% confidence contour would increase to encompass a more area, indicating lower information content within the measurements \textit{even at constant measurement count}. As the two detectors become co-located, the two-detector result would approach the single-detector result, albeit with twice the number of measurements.

This indicates that information content contained within a single measurement is a function of multiple factors. Whereas measurement resolution is typically thought of in isolation (i.e. it can be expressed without knowing the application of said measurement), measurement information content (i.e. Fisher information) can only be expressed \textit{as it pertains to the parameter space}. Measurement \textit{resolution} plays but a small role in defining measurement \textit{information}. For the unknown-reactor problem, it seems that the Fisher information contained in the measurements (bounded from above in Equation \ref{crlb2} by the inverse of the variances of the parameter estimates) is a function of at least five factors:

\begin{enumerate}
\item Detector measurement resolution. Naturally, more accurate measurements produce more accurate estimates. Due to the mixture distribution nature of the MAP estimator, this is true \textit{even for background measurements which do not come from the unknown reactor itself}.

\item Unknown reactor power and location, a subset of SBR. As expected, a more powerful reactor will be more observable due to its higher measurement count and increased fraction of the total flux. The range to a reactor is more important than its power due to the inverse squared flux-range relationship versus the linear flux-power relationship.

\item Background flux, another subset of SBR. An unknown reactor in a low background flux area (i.e. Australia) will be more observable than the same unknown reactor placed in a high background flux area (i.e. Europe). The type of background flux also effects the reactor observability, since it will dictate its incoming directions and neutrino energies.

\item \textit{A priori} ``nuisance" parameter uncertainty. The MAP estimator depends on accurate \textit{a priori} knowledge of background flux and oscillation parameter uncertainty. Uncertain nuisance parameters may be estimated or marginalized away, however accurate characterization of the \textit{a priori} uncertainty about a nuisance parameter is (assumed) critical for this to work properly in a Bayesian estimator. This is an interesting topic, but a complete characterization of the robustness of the MAP estimator to poor \textit{a priori} assumptions is beyond the scope of this paper.

\item Detector placement geometry (if multiple detectors are used). The observability of the unknown reactor will be increased by virtue of multiple detectors, however not all multiple detector placements will return the same dividends. Some detector-reactor geometries produce much better observability when dealing with low angular resolution detectors. Given an accurate \textit{a priori} parameter space pdf, there should be a single optimal placement which would statistically provide the best performance for the least cost. This is another interesting topic, though a full discussion of this is beyond the scope of this paper.

\end{enumerate}

%PASTED from Somalia!! ---------------------------------------------------------------------------------------------------------------------------------------------
Table \ref{table:Portugal2table} and Figure \ref{PortugalD2mc} show the Europe-Atlantic two-detector MC simulation results after 1000 runs. Similar to the single detector case, the results are presented two ways, as a 300MW$_\mathrm{th}$ conditional 2D slice through the 3D parameter space in the lower subplots of Figure \ref{PortugalD2mc}, and over the full 3D latitude-longitude-power space in the upper subplots of Figure \ref{PortugalD2mc}. Again we use the area (and volume) enclosed by a 90\% confidence 2D contour (and 3D isosurface) as the primary metrics of performance.

The multi-detector multivariate normal distribution presented in the Section \ref{Reactor geolocation} MAP estimator approximation is used here for the first time. The multivariate distribution exploits the correlations in count rate uncertainty across multiple detectors. Not surprisingly, the main conclusion comparing Figure \ref{PortugalD2mc} to Figure \ref{PortugalD1mc} is that the addition of multiple detectors greatly increases the unknown reactor's observability, independent of measurement resolution. Significant improvements in background flux estimates (not shown) are also seen.

Observability is improved not only in the 300MW$_\mathrm{th}$ conditional latitude-longitude plane but also in the reactor power dimension. Comparing Tables \ref{table:Portugal1table} and \ref{table:Portugal2table}, the uncertainty in the reactor power dimension is reduced by more than half. Likewise, location uncertainty is also drastically diminished. The four TREND detectors effectively exploit the dispersed multiple-detector geometry presented in this scenario to intersect range ``rings" to great effect.  It is also clear from the results that improving directional resolution will have tangible benefits even when multiple detectors are employed.
%PASTED from Somalia!! ---------------------------------------------------------------------------------------------------------------------------------------------

\subsection{Low reactor background with nearby reactor (Southern Africa)}

\begin{figure}[!htbp]
\includegraphics[width=\linewidth]{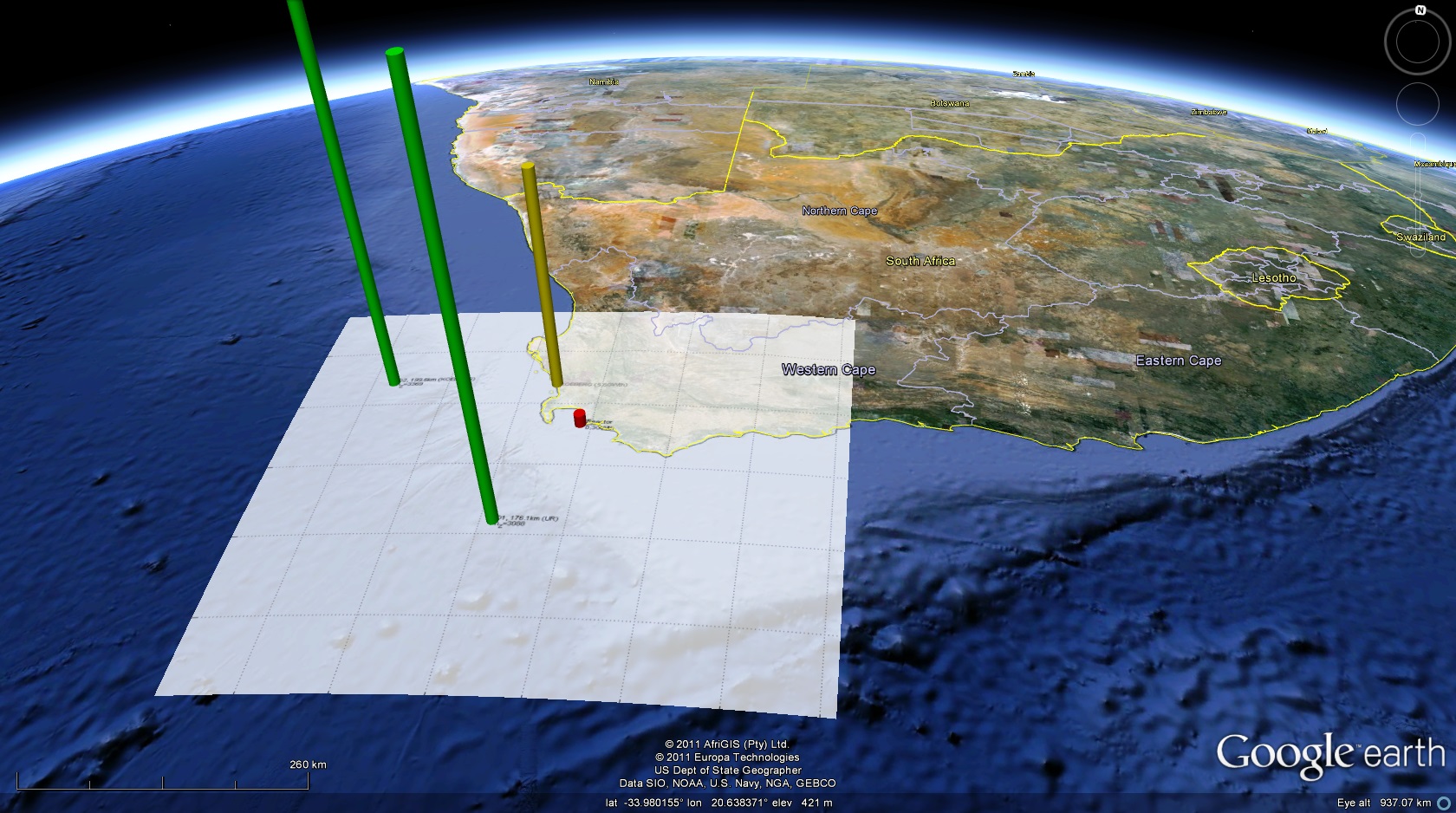}
\centering
\caption{Southern Africa scenario layout. Green cylinders are TREND detectors. The small red cylinder is the unknown reactor. Tall yellow cylinders are known reactors. Red and yellow cylinders are shown to scale to easily compare thermal power output between known and unknown reactors. The white square represents the latitude-longitude constrained search area.}
\label{southafrica_GE}
\end{figure}

The African continental shelf extends out around Southern Africa to nearly 200km, making approach by a large submerged neutrino detector difficult. A single large nuclear reactor site in Southern Africa, the 5.6GW$_{\mathrm{th}}$ Koeberg site, supplies nearly all the man-made neutrino flux in the region. The unknown reactor was placed about 100km away from the Koeberg reactor for this scenario, and two TREND detectors were dropped about 200km offshore. This scenario differs from the Europe-Atlantic scenario in that, though they both share high known-reactor flux, 99\% of the known-reactor flux originates from the lone Koeberg reactor which is located within the latitude-longitude search area (see Figure \ref{southafrica_GE}).

\subsubsection{One detector, detector \#1}
Here only the measurements from Southern Africa detector \#1 are passed to the MAP estimator. Evaluation of single-detector measurements provides insight into the unique topology of a single-detector parameter space that may be obscured by a two, three or four detector solution. It enables  better understanding of the dynamics at play in a specific scenario for a single detector; which is a pre-requisite for the successful understanding (and exploitation) of multiple-detector measurements. 

It is important to note that a dual-detector solution (as seen in Figure \ref{SAD12mc}) is not simply the product of two single-detector solutions (such as Figures \ref{SAD1mc} and \ref{SAD2mc}), but rather also exploits expected correlations in flux across detectors, making a correct dual-detector solution ``greater than the sum of its parts".

\begin{figure}[!htbp]
\includegraphics[width=\linewidth]{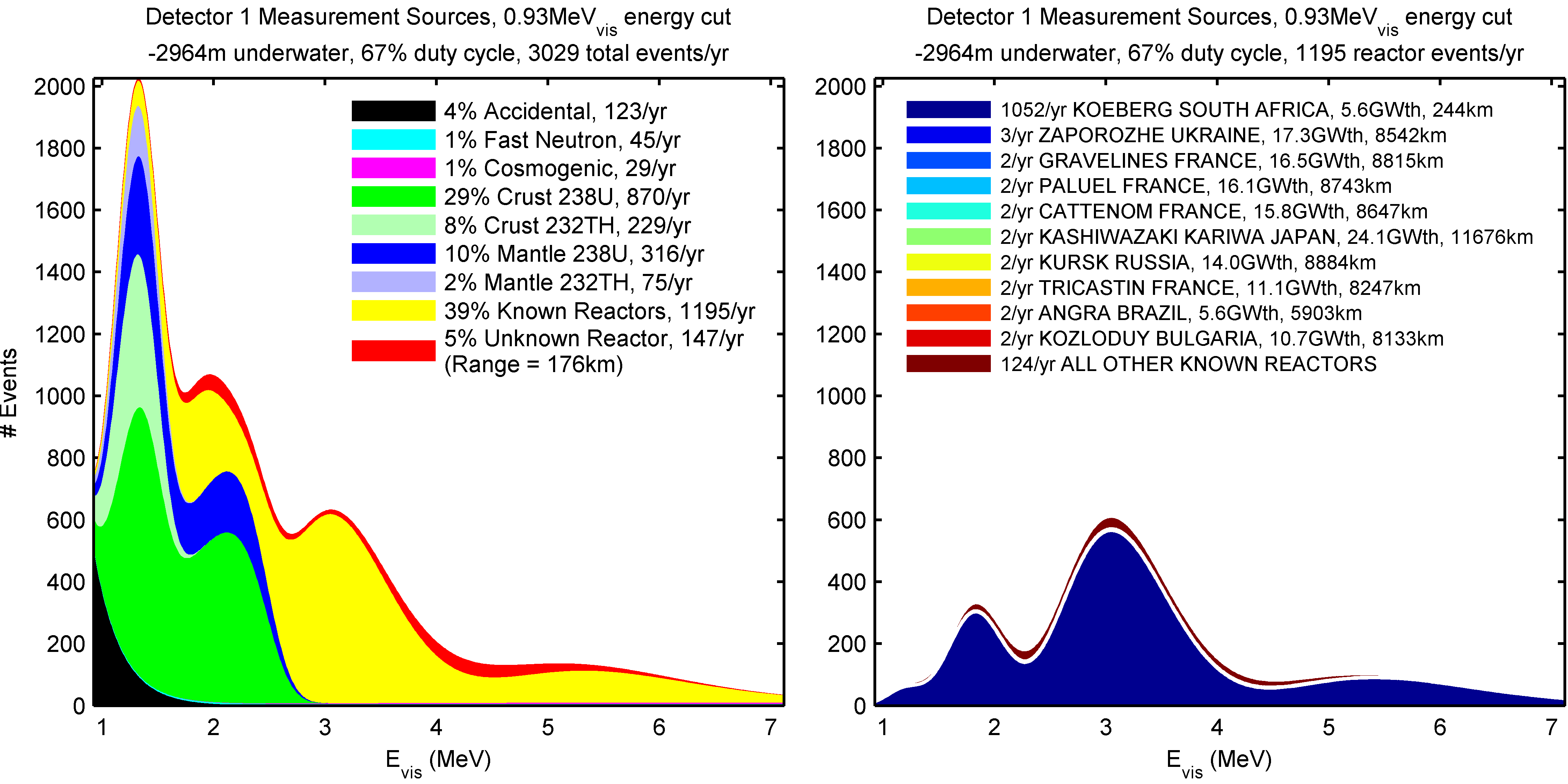}
\centering
\caption{Southern Africa scenario TREND detector \#1 mean smeared energy measurement space. Flux from all sources shown on left plot, known-reactor flux shown isolated in right plot.}
\label{southafrica1_epdf}
\end{figure} 

Figure \ref{southafrica1_epdf} shows the measurement sources as seen by detector 1 (southernmost detector) in the Southern Africa scenario. Note that nearly all the known-reactor flux originates from the Koeberg reactor site, and the second closest reactor site is over 5000km away. This single-source reactor flux property makes this scenario quite different than the Europe-Atlantic scenario, where many reactors contribute significantly to the reactor background.

\begin{table} [!htbp]
\centering
\fontsize{8}{10}\selectfont
\begin{tabular}{l|ll|ll|ll|ll}
\hline \textbf{Measurement Resolution} & \multicolumn{2}{c|}{\textbf{Area@300MW$_{\mathrm{th}}$ $90\%$}} & \multicolumn{2}{c|}{\textbf{Volume $90\%$}} & \multicolumn{2}{c|}{\textbf{Power $1\sigma$}} & \multicolumn{2}{c}{\textbf{Location $1\sigma$}}\\
& \multicolumn{2}{c|}{(km$^2$)} & \multicolumn{2}{c|}{(km$^2$GW$_{\mathrm{th}}$)} & \multicolumn{2}{c|}{(MW$_{\mathrm{th}}$)} & \multicolumn{2}{c}{(km)}\\
& $U(\theta)$ & $p(\theta)_\mathrm{H_2O}$& $U(\theta)$ &$p(\theta)_\mathrm{H_2O}$& $U(\theta)$ &$p(\theta)_\mathrm{H_2O}$& $U(\theta)$ &$p(\theta)_\mathrm{H_2O}$ \\
\hline Count-Only							   &         315729& 79795&         273054& 74529&            471& 571&            335& 144\\
\hline Count+Energy							   &         201963& 72481&         201692& 60433&            216& 217&            269& 79\\
\hline Count+Energy+Angle					   &         201102& 70207&         199756& 58082&            218& 235&            259& 95\\
\hline Count+Energy+Angle$_{\mathrm{(SNRx10)}}$&         218651& 67635&         183519& 55775&            207& 235&            140& 88\\
\hline Count+Energy+Angle$_{\mathrm{(SNRx20)}}$&         189720& 68526&         155126& 55678&            210& 220&            115& 69\\
\hline Count+Energy+Angle$_{\mathrm{(SNRx30)}}$&         133047& 62687&         106918& 50491&            217& 221&            112& 57\\
\end{tabular}
\caption{Southern Africa detector \#1 MC results. Left-hand values reflect uniform parameter-space \textit{a priori} pdf $U(\theta)$ corresponding to Figure \ref{SAD1mc} results, right-hand values reflect water \textit{a priori} pdf $p(\theta)_\mathrm{H_2O}$ expressing belief that an unknown reactor must necessarily exist on land.}
\label{table:southafrica1table}
\end{table}

Table \ref{table:southafrica1table} displays Southern Africa scenario MC results employing TREND detector \#1 only after 1000 MC runs, corresponding to Figure \ref{SAD1mc} subplots containing the conditional and marginal results of each of the six measurement resolution sub-cases.

\begin{figure}[!htbp]
\includegraphics[width=.879\linewidth]{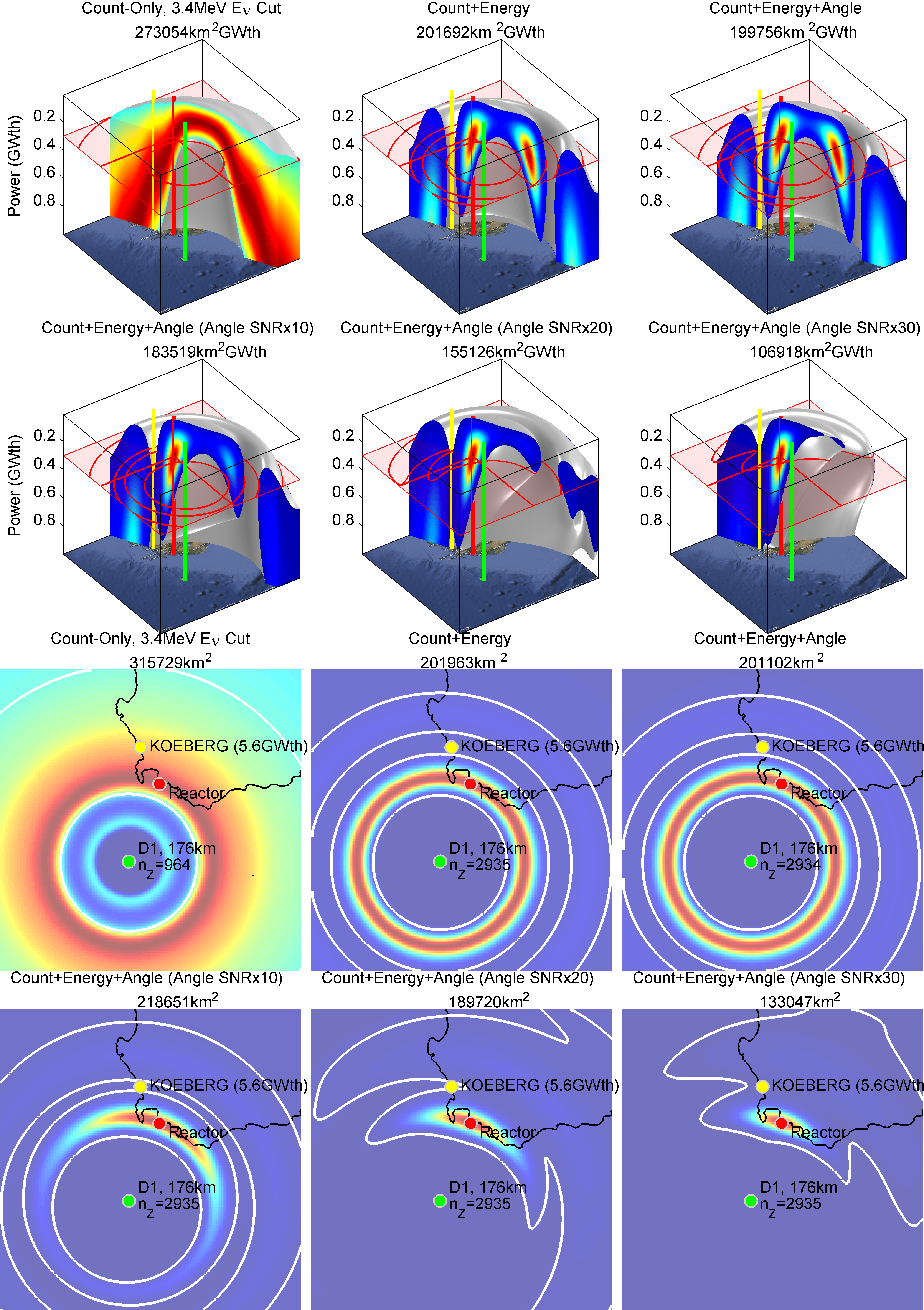}
\centering
\caption{Southern Africa scenario for TREND detector \#1 only. 300MW$_\mathrm{th}$ unknown reactor shown in red, TREND detectors in green, known reactors in yellow. Full parameter space shown in upper six plots (90\% confidence isosurface shown in silver), 300MW$_\mathrm{th}$ conditional slice shown in lower six plots (90\% 300MW$_\mathrm{th}$ conditional confidence contour shown in white).}
\label{SAD1mc}
\end{figure}
 
\subsubsection{One detector, detector \#2}
\begin{figure}[!htbp]
\includegraphics[width=\linewidth]{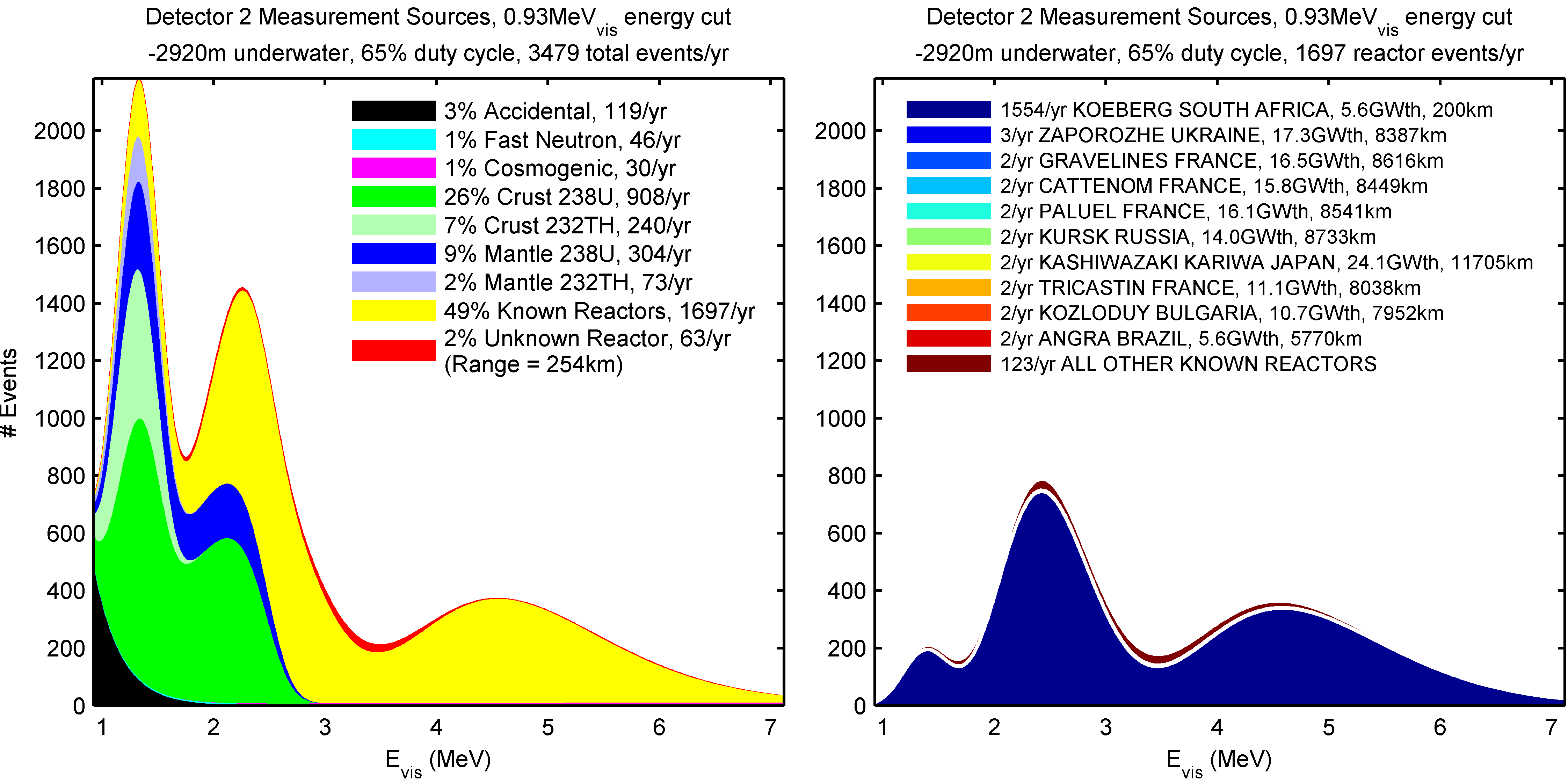}
\centering
\caption{Southern Africa scenario TREND detector \#2 mean smeared energy measurement space. Flux from all sources shown on left plot, known-reactor flux shown isolated in right plot.}
\label{southafrica2_epdf}
\end{figure}

Here only the measurements from Southern Africa detector \#2 (northernmost detector) are passed to the MAP estimator. Figure \ref{southafrica2_epdf} shows the measurement sources as seen by this detector in the Southern Africa scenario. Again, nearly all the known-reactor flux originates from the Koeberg reactor site, though the Koeberg 200km range spectrum observed by detector 2 is significantly different from the Koeberg 244km range spectrum observed at detector 1 in Figure \ref{southafrica1_epdf}. This shifting of the peak is, of course, primarily caused by $\sin^2\theta_{12}$ oscillation.

\begin{table} [!htbp]
\centering
\fontsize{8}{10}\selectfont
\begin{tabular}{l|ll|ll|ll|ll}
\hline \textbf{Measurement Resolution} & \multicolumn{2}{c|}{\textbf{Area@300MW$_{\mathrm{th}}$ $90\%$}} & \multicolumn{2}{c|}{\textbf{Volume $90\%$}} & \multicolumn{2}{c|}{\textbf{Power $1\sigma$}} & \multicolumn{2}{c}{\textbf{Location $1\sigma$}}\\
& \multicolumn{2}{c|}{(km$^2$)} & \multicolumn{2}{c|}{(km$^2$GW$_{\mathrm{th}}$)} & \multicolumn{2}{c|}{(MW$_{\mathrm{th}}$)} & \multicolumn{2}{c}{(km)}\\
& $U(\theta)$ & $p(\theta)_\mathrm{H_2O}$& $U(\theta)$ &$p(\theta)_\mathrm{H_2O}$& $U(\theta)$ &$p(\theta)_\mathrm{H_2O}$& $U(\theta)$ &$p(\theta)_\mathrm{H_2O}$ \\
\hline Count-Only							   &         310505& 81547&         294895& 77518&            353& 446&            362& 251\\
\hline Count+Energy							   &         291122& 79632&         276735& 73575&            349& 475&            291& 192\\
\hline Count+Energy+Angle					   &         289893& 79457&         274153& 73012&            309& 411&            332& 235\\
\hline Count+Energy+Angle$_{\mathrm{(SNRx10)}}$&         286747& 79642&         266399& 72731&            286& 432&            268& 200\\
\hline Count+Energy+Angle$_{\mathrm{(SNRx20)}}$&         275806& 79580&         250053& 71848&            266& 433&            220& 156\\
\hline Count+Energy+Angle$_{\mathrm{(SNRx30)}}$&         260644& 78443&         232287& 70984&            267& 391&            199& 125\\

\end{tabular}
\caption{Southern Africa detector \#2 MC results. Left-hand values reflect uniform parameter-space \textit{a priori} pdf $U(\theta)$ corresponding to Figure \ref{SAD2mc} results, right-hand values reflect water \textit{a priori} pdf $p(\theta)_\mathrm{H_2O}$ expressing belief that an unknown reactor must necessarily exist on land.}
\label{table:southafrica2table}
\end{table}

\begin{figure}[!htbp]
\includegraphics[width=.879\linewidth]{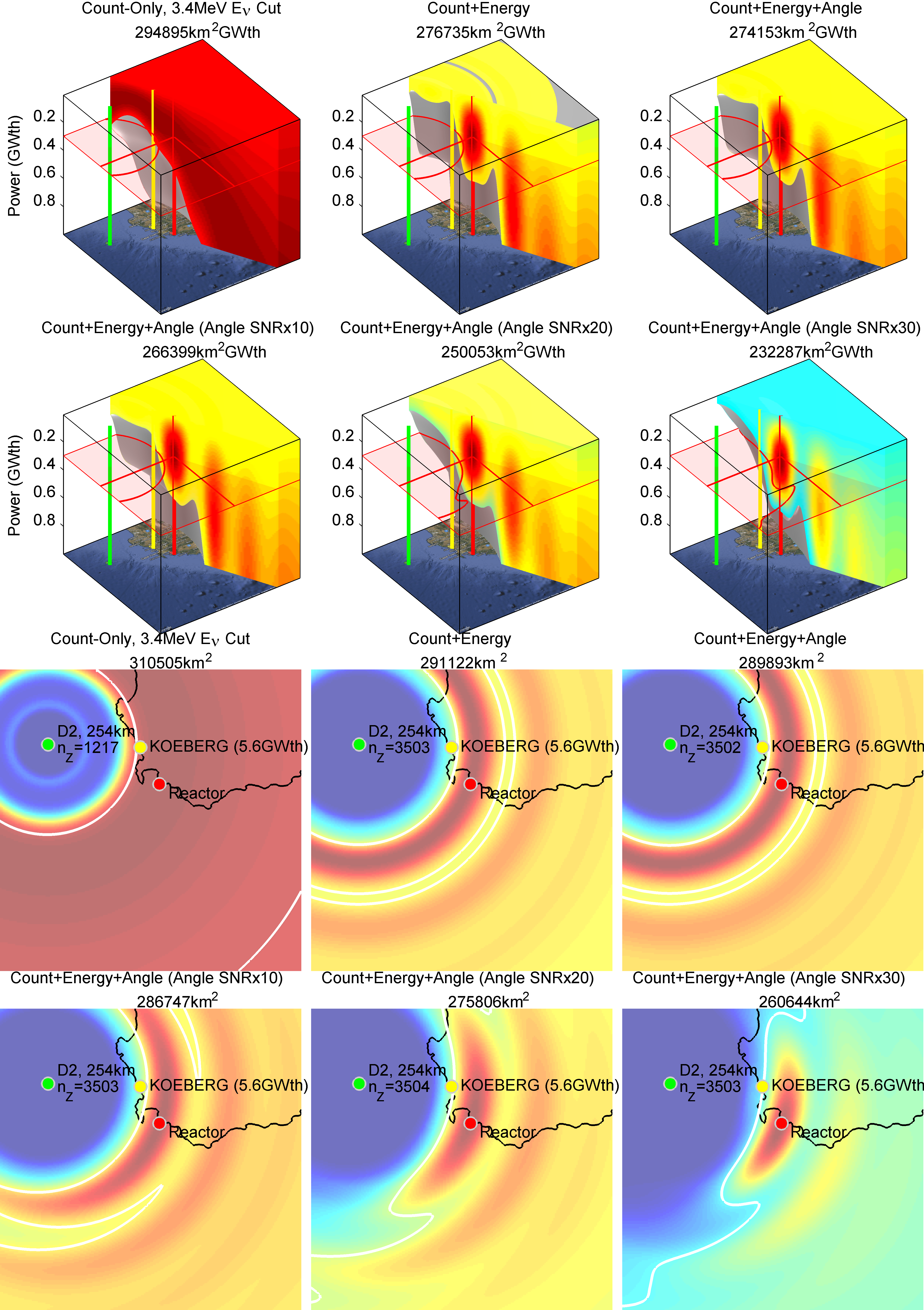}
\centering
\caption{Southern Africa scenario for TREND detector \#2 only. 300MW$_\mathrm{th}$ unknown reactor shown in red, TREND detectors in green, known reactors in yellow. Full parameter space shown in upper six plots (90\% confidence isosurface shown in silver), 300MW$_\mathrm{th}$ conditional slice shown in lower six plots (90\% 300MW$_\mathrm{th}$ conditional confidence contour shown in white).}
\label{SAD2mc}
\end{figure}

Table \ref{table:southafrica2table} displays Southern Africa scenario MC results employing TREND detector \#2 only after 1000 MC runs, corresponding to Figure \ref{SAD2mc} subplots containing the conditional and marginal results of each of the six measurement resolution sub-cases.

The detector \#1-only scenario out-performed the detector \#2-only scenario primarily due to the different geometries inherent in each detector placement. Detector \#1 is closer to the unknown reactor and thus receives more signal than detector \#2: 149 events/year (5\% of the signal) versus 54 events/year (2\% of the signal).  More importantly, the detector \#1 placement results in a greater mismatch between the observed energy spectra of the Koeberg reactor and the unknown reactor at the detector.  This mismatch in observed energy spectra from these two point sources provides the estimator with more information to leverage, and results in better estimates of the unknown reactor's power and location.  This is confirmed by the observation that the performance is comparable for the Count-Only case, but it begins to strongly favor detector \#1 when energy measurements are included. We conclude that unknown-reactor observability is very much a function of detector placement. Note that these detector placements are not optimized at all; the process of optimized detector placement is an interesting study on its own, though unfortunately outside of the scope of this work.

\subsubsection{Two detectors}
\begin{figure}[!htbp]
\includegraphics[width=.879\linewidth]{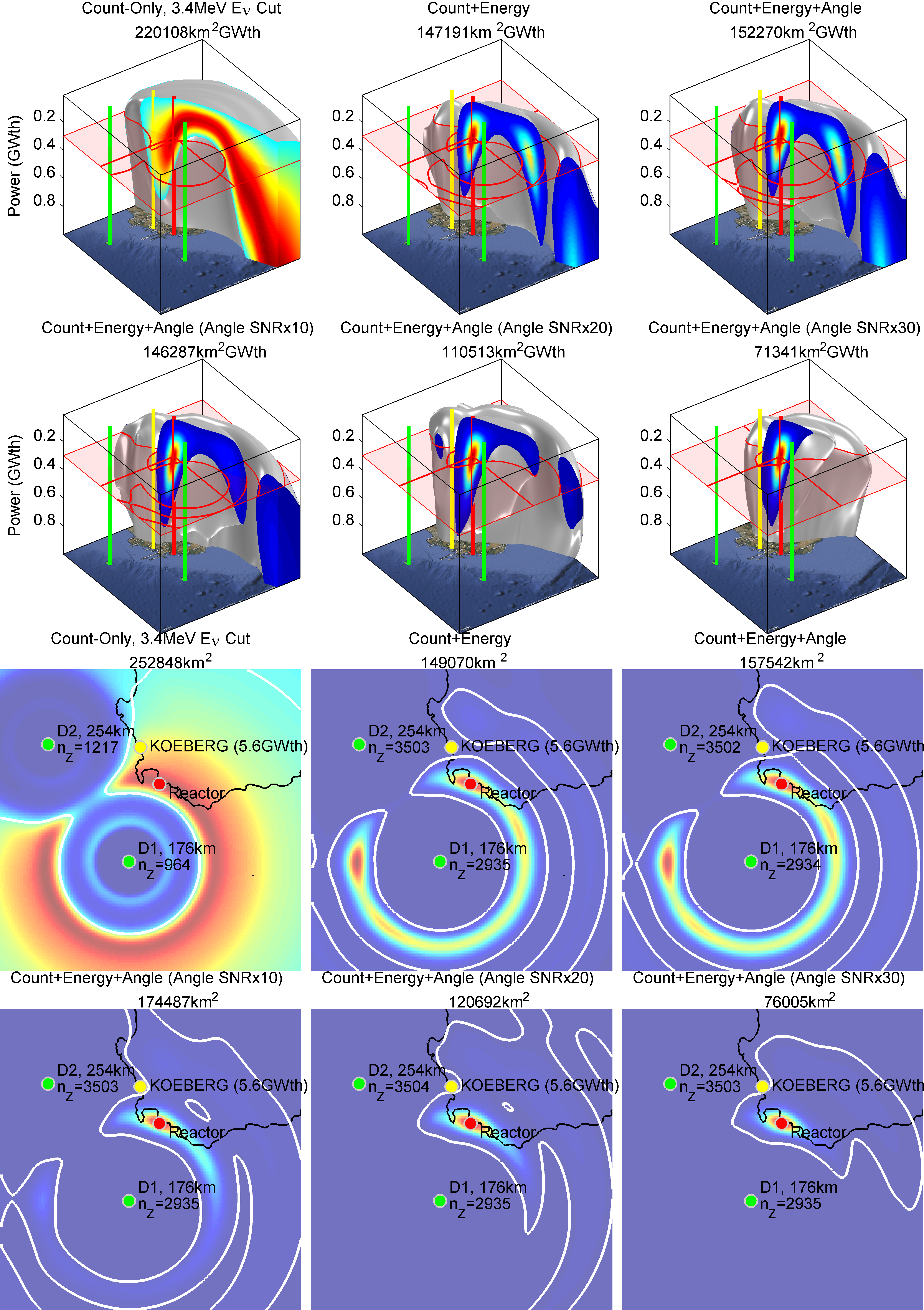}
\centering
\caption{Southern Africa scenario using both TREND detectors. 300MW$_\mathrm{th}$ unknown reactor shown in red, TREND detectors in green, known reactors in yellow. Full parameter space shown in upper six plots (90\% confidence isosurface shown in silver), 300MW$_\mathrm{th}$ conditional slice shown in lower six plots (90\% 300MW$_\mathrm{th}$ conditional confidence contour shown in white).}
\label{SAD12mc}
\end{figure}

Here all the measurements from both of Southern Africa's detectors are passed to the MAP estimator. Table \ref{table:southafrica12table} displays Southern Africa scenario MC results employing both TREND detectors after 1000 MC runs, corresponding to the Figure \ref{SAD12mc} subplots containing the conditional and marginal results of each of the six measurement resolution sub-cases

\begin{table} [!htbp]
\centering
\fontsize{8}{10}\selectfont
\begin{tabular}{l|ll|ll|ll|ll}
\hline \textbf{Measurement Resolution} & \multicolumn{2}{c|}{\textbf{Area@300MW$_{\mathrm{th}}$ $90\%$}} & \multicolumn{2}{c|}{\textbf{Volume $90\%$}} & \multicolumn{2}{c|}{\textbf{Power $1\sigma$}} & \multicolumn{2}{c}{\textbf{Location $1\sigma$}}\\
& \multicolumn{2}{c|}{(km$^2$)} & \multicolumn{2}{c|}{(km$^2$GW$_{\mathrm{th}}$)} & \multicolumn{2}{c|}{(MW$_{\mathrm{th}}$)} & \multicolumn{2}{c}{(km)}\\
& $U(\theta)$ & $p(\theta)_\mathrm{H_2O}$& $U(\theta)$ &$p(\theta)_\mathrm{H_2O}$& $U(\theta)$ &$p(\theta)_\mathrm{H_2O}$& $U(\theta)$ &$p(\theta)_\mathrm{H_2O}$ \\
\hline Count-Only							   &         252848& 79673&         220108& 74380&            399& 421&            262& 105\\
\hline Count+Energy							   &         149070& 65392&         147191& 54947&            221& 302&            264& 102\\
\hline Count+Energy+Angle					   &         157542& 68547&         152270& 56590&            202& 261&            252& 96\\
\hline Count+Energy+Angle$_{\mathrm{(SNRx10)}}$&         174487& 65197&         146287& 54662&            198& 241&            130& 86\\
\hline Count+Energy+Angle$_{\mathrm{(SNRx20)}}$&         120692& 64101&         110513& 53623&            196& 203&            110& 59\\
\hline Count+Energy+Angle$_{\mathrm{(SNRx30)}}$&          76005& 52791&          71341& 46872&            168& 146&             68& 44\\
\end{tabular}
\caption{Southern Africa scenario MC results from both detectors combined. Left-hand values reflect uniform parameter-space \textit{a priori} pdf $U(\theta)$ corresponding to Figure \ref{SAD12mc} results, right-hand values reflect water \textit{a priori} pdf $p(\theta)_\mathrm{H_2O}$ expressing belief that an unknown reactor must necessarily exist on land.}
\label{table:southafrica12table}
\end{table}

\subsection{High reactor background with nearby reactors (Europe-Mediterranean)}
\label{High reactor background with nearby reactors (Europe-Mediterranean)}

\begin{figure}[!htbp]
\includegraphics[width=\linewidth]{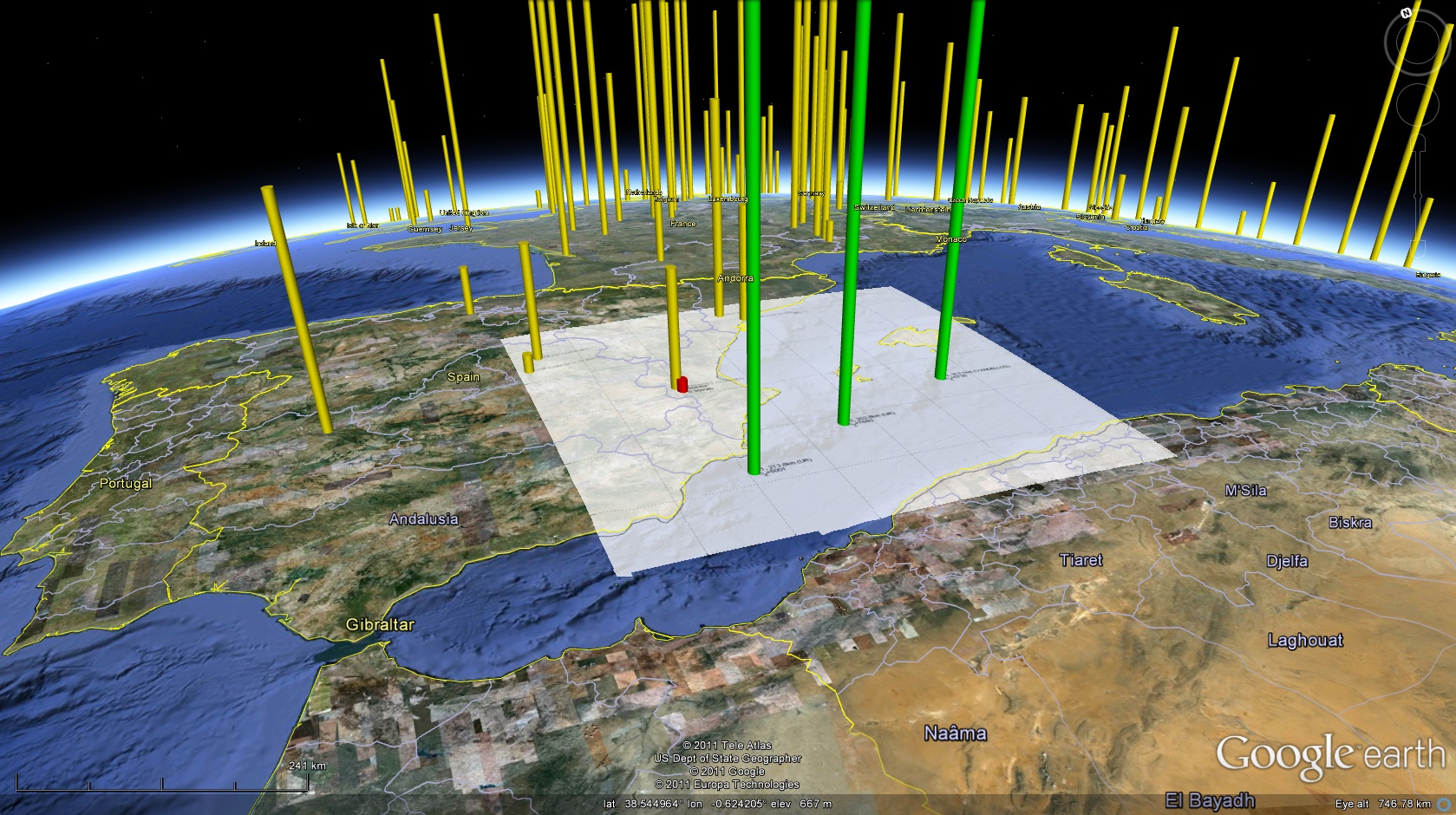}
\centering
\caption{Europe-Mediterranean scenario layout. Green cylinders are TREND detectors. The small red cylinder is the unknown reactor. Tall yellow cylinders are known reactors. Red and yellow cylinders are shown to scale to easily compare thermal power output between known and unknown reactors. The white square represents the latitude-longitude constrained search area.}
\label{spain_GE}
\end{figure}

Figure \ref{spain_GE} presents the most challenging scenario, where we move the unknown reactor to the Spain's Costa Blanca region near Valencia. Three TREND detectors are placed in the relatively shallow waters of the southern Mediterranean, shadowing the Spanish coastline about 100km out. Since the Mediterranean waters are much shallower than in the Atlantic ocean, the detectors can not be submerged as deeply. Hence these detectors suffer higher muon flux rates and higher veto times, leading to reduced detector duty cycles. One challenge in this scenario is the presence of multiple known reactors inside the latitude-longitude parameter space, the closest being extremely near (10km) to the unknown reactor. 

\subsubsection{Three detectors}
\begin{figure}[!htbp]
\includegraphics[width=\linewidth]{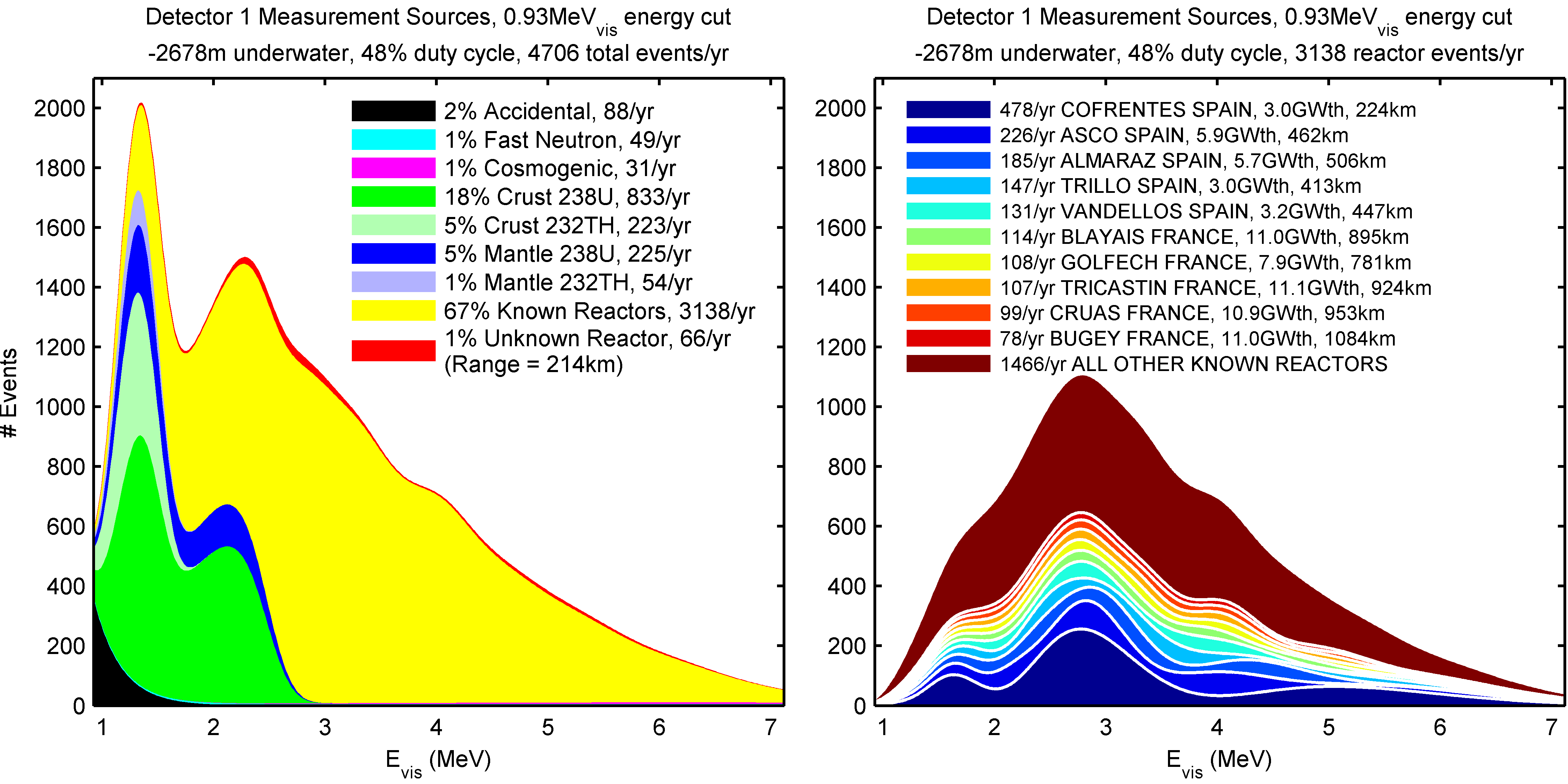}
\centering
\caption{Europe-Mediterranean scenario TREND detector \#1 mean smeared energy measurement space. Flux from all sources shown on left plot, known-reactor flux shown isolated in right plot.}
\label{spain1_epdf}
\end{figure}

Figure \ref{spain1_epdf} shows the measurement sources as seen by detector 1 in Europe-Mediterranean. The unknown-reactor flux at this detector is extremely low; only 67 out of 4803 mean events per year are projected to originate from the unknown reactor, creating a SBR of only 0.014. The corresponding rates for detectors 2 and 3 are 107/7497 events (also 0.014 SBR) and 35/5999 events (0.006 SBR). The ranges from the detectors to the unknown-reactor are not dissimilar from the other scenarios (from 203km to 313km in the Europe-Mediterranean scenario), but the known-reactor flux has increased substantially due to the proximity to the mainland European (especially French) reactors.

\begin{table} [!htbp]
\centering
\fontsize{8}{10}\selectfont
\begin{tabular}{l|ll|ll|ll|ll}
\hline \textbf{Measurement Resolution} & \multicolumn{2}{c|}{\textbf{Area@300MW$_{\mathrm{th}}$ $90\%$}} & \multicolumn{2}{c|}{\textbf{Volume $90\%$}} & \multicolumn{2}{c|}{\textbf{Power $1\sigma$}} & \multicolumn{2}{c}{\textbf{Location $1\sigma$}}\\
& \multicolumn{2}{c|}{(km$^2$)} & \multicolumn{2}{c|}{(km$^2$GW$_{\mathrm{th}}$)} & \multicolumn{2}{c|}{(MW$_{\mathrm{th}}$)} & \multicolumn{2}{c}{(km)}\\
& $U(\theta)$ & $p(\theta)_\mathrm{H_2O}$& $U(\theta)$ &$p(\theta)_\mathrm{H_2O}$& $U(\theta)$ &$p(\theta)_\mathrm{H_2O}$& $U(\theta)$ &$p(\theta)_\mathrm{H_2O}$ \\
\hline Count-Only							   &         274971& 129004&         216117& 115213&            287& 277&            295& 303\\
\hline Count+Energy							   &         212273& 117262&         184379& 103939&            367& 477&            274& 267\\
\hline Count+Energy+Angle					   &         207849& 117151&         183536& 104479&            394& 507&            280& 289\\
\hline Count+Energy+Angle$_{\mathrm{(SNRx10)}}$&         190830& 112819&         171696& 99029&             289& 417&            253& 223\\
\hline Count+Energy+Angle$_{\mathrm{(SNRx20)}}$&         162894& 98592&          153534& 91686&             282& 370&            217& 183\\
\hline Count+Energy+Angle$_{\mathrm{(SNRx30)}}$&         146691& 90059&          137752& 85163&             280& 333&            193& 170\\
\end{tabular}
\caption{Europe-Mediterranean scenario MC results from all 3 detectors combined. Left-hand values reflect uniform parameter-space \textit{a priori} pdf $U(\theta)$ corresponding to Figure \ref{SpainD3mc} results, right-hand values reflect water \textit{a priori} pdf $p(\theta)_\mathrm{H_2O}$ expressing belief that an unknown reactor must necessarily exist on land.}
\label{table:spain3table}
\end{table}

\begin{figure}[!htbp]
\includegraphics[width=.879\linewidth]{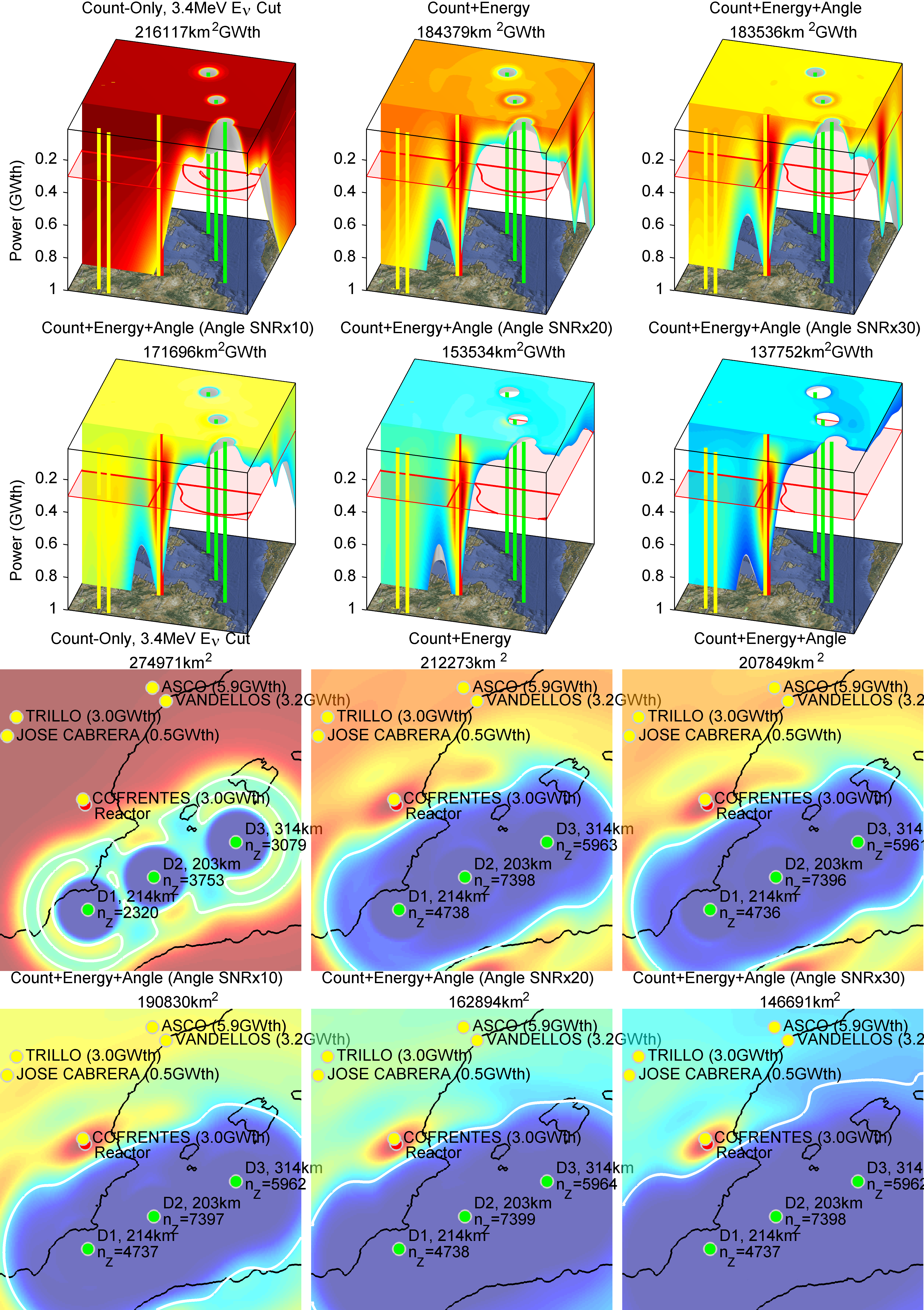}
\centering
\caption{Europe-Mediterranean scenario with 3 TREND detectors. 300MW$_\mathrm{th}$ unknown reactor shown in red, TREND detectors in green, known reactors in yellow. Full parameter space shown in upper six plots (90\% confidence isosurface shown in silver), 300MW$_\mathrm{th}$ conditional slice shown in lower six plots (90\% 300MW$_\mathrm{th}$ conditional confidence contour shown in white).}
\label{SpainD3mc}
\end{figure}

Table \ref{table:spain3table} displays Europe-Mediterranean scenario MC results employing all three TREND detectors after 1000 MC runs, corresponding to Figure \ref{SpainD3mc} subplots containing the conditional and marginal results of each of the six measurement resolution sub-cases.

The Southern Africa and Europe-Mediterranean scenarios collectively demonstrate the estimator's ability to provide robust results even when one or more strong point sources are located within the latitude-longitude search area.  Note that there are no concentrations of elevated probability near isolated known reactors.  This is strong evidence that the estimator is correctly accounting for the presence of background sources.  The Europe-Mediterranean scenario shows that the estimator can address the stressing case of a small unknown reactor operating in very close proximity to a much more powerful existing reactor site.

\section{Conclusions and Recommendations}
\label{Conclusion and Recommendations}

In this paper we have introduced the concept of what we call ``NUDAR" (NeUtrino Direction and Ranging), making the point that measurement of the observed energy spectrum can be employed to passively deduce the range of a neutrino source (with a known at-source energy spectrum) from even a single detector.  The inherent range resolution is due the unique energy-dependent character of neutrino oscillations and can be improved still further by the use of multiple detectors.  Moreover we have shown that with even modest resolution of incoming neutrino directions, NUDAR can be greatly improved.  We believe these points have not been fully appreciated in the neutrino community.

Because this study involves long-range NUDAR, where the signal-to-noise and signal-to-background ratios are very small, we designed the process to incorporate all available information.  We used a Bayesian estimation approach whereby prior constraints on the likely location and power of a reactor could be easily incorporated.  We constrained the reactor location to the Earth's surface and explored the benefits of excluding reactor locations over water.  This is the simplest spatial prior information available, yet it significantly reduced the uncertainty in the estimated location and power of a reactor.  We included all measurement dimensions in our estimator (count rate, energy, and direction) and avoided the use of an energy threshold to exclude the geo-neutrino contributions.  Thus, we do not discard any of the signal from the neutrino source which happens to be below the tail of the geo-neutrino flux.

It is difficult to simulate a general NUDAR scenario, since an experiment could have various objectives.  There are also a variety of geographical cases, some of which we have explored, involving reactors near a straight coast, on a peninsula, or on an island.  For the sort of mobile detectors that we envision, we have considered herein only deep ocean deployment where we can choose the detector location.  Of course land-based detector placements are possible too, but these would likely rely upon locally available mines and tunnels, at least for long range monitoring.  We have also not fully explored the range of detector sizes and depths which are possible.  Fortunately, many reactors are located near shore due to their requirements for cooling water.  Experiments involving small reactors collocated near more powerful reactors would benefit from smaller detectors which can operate at shallower depths closer to shore.

We have carried out the most detailed study to date of the general background for low energy scintillation-based detectors, and have fully parameterized the background due to cosmic rays and other sources.  Note that there is coupling between detector size and depth as well which places a strong limitation on volumetric detectors (such as we consider) due to the cosmic ray muon rate which can overwhelm the search for neutrinos.  This is not to say that one cannot work nearer to the surface, but the price of doing so is the segmentation of the detector into numerous cells with lower individual cosmic ray limitations.  The sort of segmentation needed for detectors operating on the surface seems (at present) only economical for instruments out to a few kilometers, but we have not investigated that option in detail.

In this study we have focused on single-volume configurations.  There is an intermediate case involving the segmentation of a large volume into a small number of more easily manageable volumes.  This could have economic impacts as well as detector advantages.  However once one allows for the consideration of segmentation schemes, one should also consider splitting a fixed scintillating volume into a larger number of more geographically dispersed instruments. Clearly information is gained by having more operating locations, but the detailed tradeoff between these two options can be very complex.  Hence we reserve such considerations for more specific applications, though this sort of optimization is surely an avenue for exploration.

We have clearly demonstrated that detectors in the class of $10^{34}$ protons have the remarkable ability to refine the current knowledge of significant neutrino background sources (e.g. geo-neutrinos), to constrain key parameters describing neutrino oscillations, and to geo-locate and characterize neutrino sources (with a known energy spectrum) from distances of hundreds of kilometers.  Previous studies have suggested this last point, but have not employed the full information available, particularly the energy spectrum.  We realize that detectors of this size are currently very expensive, but such instruments are under present consideration (for example the 50-100 kiloton LENA detector in Europe).  The masses of such instruments are also not outside engineering experience: oil supertankers have been built with capacities of 500,000 tons for example.

We therefore recommend that the neutrino community should work to develop new (and less expensive) photodetectors, doping of water to improve detection without having the cost of oil for large volumes, appropriate low power electronics, and designs for large detectors (particularly mobile ocean-going detectors).  We have identified the great benefit held by direction determination of reactor neutrinos and, while recognizing the difficulty of this problem in large detectors, we want to emphasize its importance for the fields of geology, particle physics and astrophysics.

We would like to conclude by emphasizing that the study and eventual construction of such long range detectors will have scientific benefits far beyond the immediate practical reactor-related applications.  The study of geo-neutrinos, needed to fully support reactor detection, is itself a gateway to meaningful geologic research into the Earth's heat sources and geodynamics.  Moreover, given the general physics interest in such projects, one has the opportunity to engage a larger segment of the scientific community in a long term program with many potential scientific payoffs.

\newpage
\section{Acronyms}
\begin{table} [!htbp]
\fontsize{8}{10}\selectfont
\begin{center}
\begin{tabular}{l|l}
\hline
\textbf{Acronym} & \\
\hline 3D		& 3-dimensional\\
\hline 4D		& 4-dimensional\\
\hline cdf		& cumulative distribution function\\
\hline CE		& Center of Energy\\
\hline CRLB		& Cramer-Rao Lower Bound\\
\hline DCM		& Direction Cosine Matrix\\
\hline DT		& Down Time\\
\hline ECE		& Energy Capture Efficiency\\
\hline ECEF		& Earth-Centered Earth-Fixed\\
\hline EGM		& Earth Gravity Model\\
\hline ETOPO	& Earth TOPOgraphical\\
\hline FIM		& Fisher Information Matrix\\
\hline IAEA		& International Atomic Energy Agency\\
\hline IBD		& Inverse Beta Decay\\
\hline IMB      & Irvine-Michigan-Brookhaven\\
\hline KamLAND  & Kamioka Liquid-scintillator Anti-Neutrino Detector\\
\hline LAB 		& linear alkyl benzene\\
\hline LLH 		& Latitude-Longitude-Height\\
\hline LS 		& Liquid Scintillator\\
\hline MAP  	& Maximum \textit{A Posteriori}\\
\hline MC  		& Monte Carlo\\
\hline ML  		& Maximum Likelihood\\
\hline MLE  	& Maximum Likelihood Estimator\\
\hline MSW  	& Mikheyev\textendash Smirnov\textendash Wolfenstein\\
\hline MWE      & Meters of Water Equivalent\\
\hline NED  	& North-East-Down\\
\hline NGA  	& National Geospatial-Intelligence Agency\\
\hline NOAA  	& National Oceanic and Atmospheric Administration\\
\hline NUDAR    & NeUtrino Direction and Ranging\\
\hline pdf		& probability density function\\
\hline PE		& PhotoElectron\\
\hline PMT		& Photo-Multiplier Tube\\
\hline PREM		& Preliminary Reference Earth Model\\
\hline SNIF		& Secret Neutrino Interactions Finder\\
\hline SBR		& Signal-to-Background Ratio\\
\hline SNR		& Signal-to-Noise Ratio\\
\hline TREND	& Test REactor Neutrino Detector\\
\hline WGS		& World Geodetic System\\
\hline WGS84	& World Geodetic System 1984\\
\hline QE		& Quantum Efficiency\\

\end{tabular}
\caption{All acronyms employed in this document are defined here.}
\label{table:acronyms}
\end{center}
\end{table}

\section{Acknowledgements}
\label{Acknowledgements}
%We would like to thank Mikhail Batygov for generating the GEANT simulations that were used in this research. We would also like to thank Michael J. Lenihan, Todd E. %Johanesen, Stephen Malys, Robert C. Anderson, Dr. Peter F. Bythrow, The Defense Intelligence Agency, the National Consortium for Measures and Signatures Intelligence %(MASINT) Research, the Hawaii Pacific University Trustees Scholarly Development Program, The University of Hawaii and the U.S. Department of Energy for their unwavering %support of this effort.
We would like to thank Mikhail Batygov for generating the GEANT simulations that were used in this research. We would like to thank Dr. Peter F. Bythrow of the Defense Intelligence Agency and the National Consortium for Measures and Signatures Intelligence (MASINT) Research. We would also like to thank the Hawaii Pacific University Trustees Scholarly Development Program, the University of Hawaii and the U.S. Department of Energy for their unwavering support of this effort.

\appendix

\section{Geo-reactor results}

High concentrations of actinide elements inside the earth can lead to sustained nuclear fission reactions \cite{Kuroda_1956}. The fragments of these fission reactions decay with the emission of electron antineutrinos, which suffer negligible attenuation by the earth. Isotopic analysis of uranium deposits confirms the operation, about 2 billion years ago, of natural nuclear reactors \cite{Cowan_1976}, now called geo-reactors. Several models propose currently operating geo-reactors at the center of the core \cite{Herndon_1996}, the inner core boundary \cite{Rusov_2007}, and the core-mantle boundary \cite{deMeijer_2008}. A geo-reactor would contribute to terrestrial heat flow \cite{Davies_2010} and the high content of $^3$He in oceanic basalts \cite{Farley_1998}. The measurement of electron antineutrinos from commercial nuclear reactors leads to an experimental upper limit to the power of an earth-centered geo-reactor of 3 TW$_\mathrm{th}$ (95\% C.L.) \cite{Borexino_2010}.

In this appendix, we explore the observability of a putative earth-centered geo-reactor based on a year of observation with all four (160,000 m$^3$ each) TREND detectors.  We attempt to quantify the uncertainty of the estimated geo-reactor power, as a function of its postulated power.  This analysis is motivated by the fact that detectors as large as TREND would have an unprecedented capability to confirm the existence of a geo-reactor (if present), or to place the tight observational upper bounds on its power (if not present). 

\begin{figure}[!htbp]
\centering
\includegraphics[width=\linewidth]{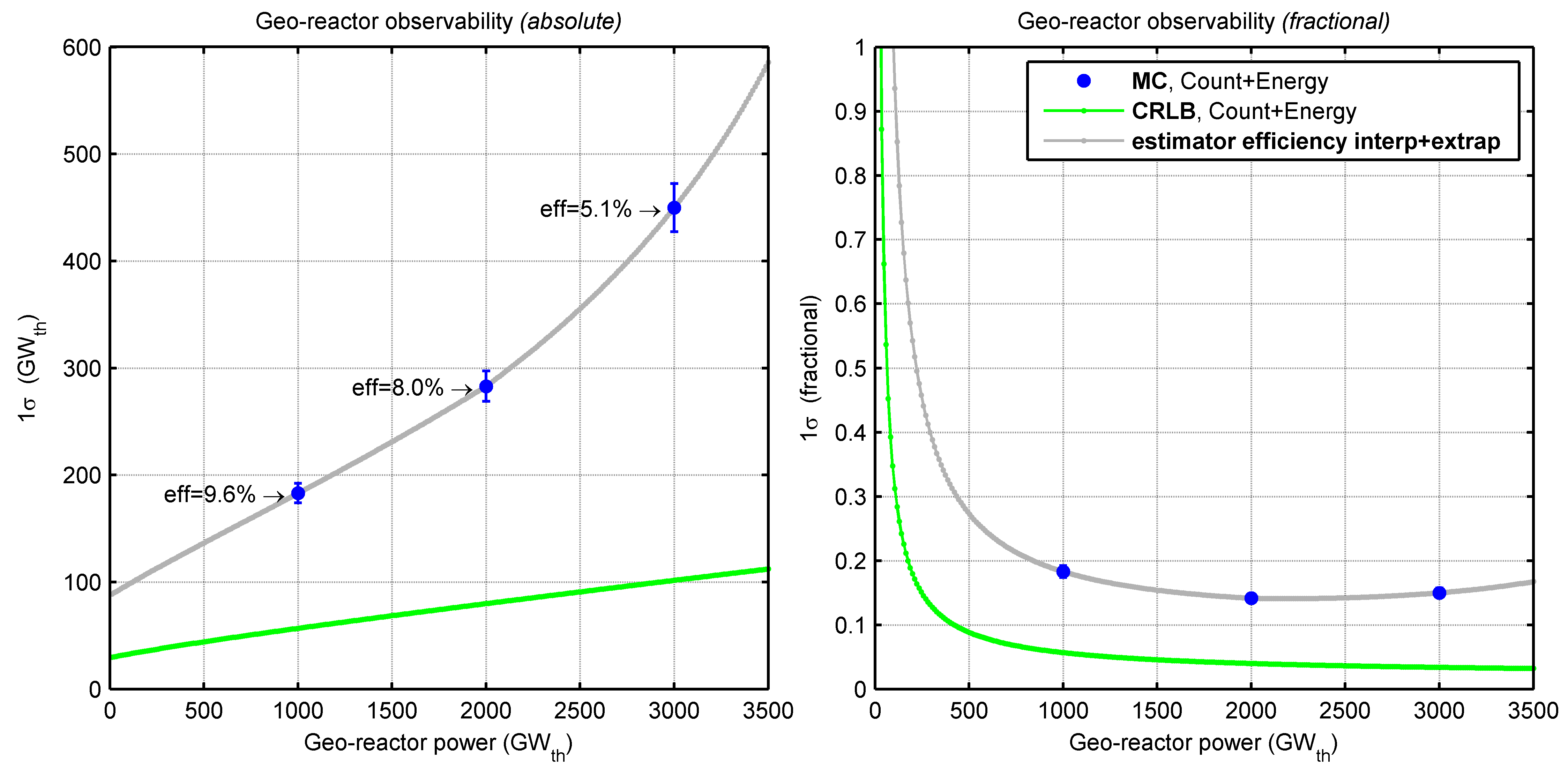}
\caption{Geo-reactor observability as a function of geo-reactor power. MC and CRLB results were used to calculate estimator efficiencies at several geo-reactor powers. These efficiencies were extrapolated across the geo-reactor power axis and used to predict MC observability from 0TW$_\mathrm{th}$ to 3.5TW$_\mathrm{th}$ using the four-detector TREND scenario shown in Figure \ref{worldwide_dall}.}
\label{georeactor_observability}
\end{figure}

In order to explore the observability of a geo-reactor, we computed the Count+Energy CRLB for postulated powers of 0 - 3.5 TW$_\mathrm{th}$.  In order to anchor the CRLB results, we performed 1000 Count+Energy MC runs at assumed geo-reactor powers of 1 TW$_\mathrm{th}$, 2 TW$_\mathrm{th}$, and 3 TW$_\mathrm{th}$.  For each of these geo-reactor powers, the estimator efficiency was computed.   The efficiencies were then linearly interpolated / extrapolated to all powers, and used to scale the CRLB results to absolute uncertainties for the geo-reactor power.   The results are displayed in Figure \ref{georeactor_observability}.  For geo-reactors approaching 0 TW$_{th}$, the uncertainty in the estimated geo-reactor power approaches $\sim$90 GW$_\mathrm{th}$ (1$\sigma$). For geo-reactors approaching 3 TW$_{th}$ (the observational upper limit derived by Borexino), the uncertainty in the estimated geo-reactor power approaches $\sim$440 GW$_\mathrm{th}$ (1$\sigma$).  

The relatively low efficiency of the CRLB results is interesting to note.  The CRLB is evaluated numerically at the true value of the 11x1 parameter vector (6 dimensions for background flux corrections, 4 dimensions for oscillation parameter corrections, and the geo-reactor power).  The MC-computed \textit{a posteriori} variance, on the other hand, is a weighted average of the power uncertainty derived from 1000 different 11x1 parameter vectors, each found by the use of an optimizer.  The low efficiency observed might be an indication that the optimizer used (which uses a simple 1-start point optimization strategy) is not robust enough to avoid local minima and arrive at the true MAP value.  This certainly deserves more study, but we believe that the results presented here are trustworthy if not somewhat conservative (due to optimizer-related robustness issues).

The presence of a geo-reactor, especially at the upper limit power of 3 TW$_\mathrm{th}$, is expected to impact the uncertainty in the estimated correction gains associated with background flux and oscillation parameters.  This is expected because for at least two reasons.  First, the estimator must search for the MAP value within an 11D parameter space instead of a 10D parameter space.   Second, a 3 TW$_\mathrm{th}$ geo-reactor is expected to contribute up to the order of 1500 events per year in a TREND detector, and there is additional statistical (Poisson) noise associated with this additional signal background component.  Tables \ref{table:worldwide_op_gr} and \ref{table:worldwide_flux_gr} displays the results.  The \textit{a priori} probability of the geo-reactor power was assumed to be unstrained (other than a non-negative constraint).

\begin{table} [!htbp]
\centering
\fontsize{8}{10}\selectfont
\begin{tabular}{l|l|l|l|l}
\hline
\textbf{Scenario} & \textbf{$\Delta m^2_{12}$} & \textbf{$\Delta m^2_{13}$} & \textbf{$\sin^2\theta_{12}$} & \textbf{$\sin^2\theta_{13}$}\\
& $(eV^2)$ & $(eV^2)$ & & \\
\hline \textit{a priori} maximum likelihood $\mu$ \cite{fogli_2012}	& $7.58E-5$					& $0.00235$					& $0.312$		& $0.025$\\
\hline \textit{a priori} uncertainty \cite{fogli_2012}	& $\mu^{+2.2E-6}_{-2.6E-6}$	& $\mu^{+1.2E-4}_{-9.0E-5}$	& $\mu^{+0.0170}_{-0.0160}$	& $\mu^{+0.0070}_{-0.0070}$\\
\hline Count+Energy							    	  	& $\mu^{+2.2E-7}_{-2.3E-7}$	& $\mu^{+9.4E-5}_{-9.6E-5}$	& $\mu^{+0.0027}_{-0.0026}$	& $\mu^{+0.0032}_{-0.0078}$\\ 
\hline Count+Energy+Angle					    		& $\mu^{+2.0E-7}_{-2.3E-7}$	& $\mu^{+8.8E-5}_{-9.6E-5}$	& $\mu^{+0.0028}_{-0.0028}$	& $\mu^{+0.0029}_{-0.0074}$	\\ 
\hline Count+Energy+Angle$_{\mathrm{(SNRx10)}}$ 		& $\mu^{+2.0E-7}_{-2.1E-7}$	& $\mu^{+8.7E-5}_{-9.4E-5}$	& $\mu^{+0.0025}_{-0.0027}$	& $\mu^{+0.0032}_{-0.0075}$	\\ 
\hline Count+Energy+Angle$_{\mathrm{(SNRx20)}}$ 		& $\mu^{+1.9E-7}_{-2.0E-7}$	& $\mu^{+9.1E-5}_{-9.1E-5}$	& $\mu^{+0.0027}_{-0.0027}$	& $\mu^{+0.0035}_{-0.0075}$	\\ 
\hline Count+Energy+Angle$_{\mathrm{(SNRx30)}}$ 		& $\mu^{+1.9E-7}_{-1.9E-7}$	& $\mu^{+9.7E-5}_{-9.4E-5}$	& $\mu^{+0.0026}_{-0.0025}$	& $\mu^{+0.0033}_{-0.0077}$	\\ 
\end{tabular}
\caption{Worldwide oscillation parameter optimization results using 4 TREND detectors. The four oscillation parameters were optimized simultaneously with the background flux, including a 3 TW$_\mathrm{th}$ geo-reactor lying at the earth's center, the results of which can be seen in Table \ref{table:worldwide_flux}. 1000 MC runs per scenario were used.}
\label{table:worldwide_op_gr}
\end{table}

A comparison of Table \ref{table:worldwide_op_gr} to Table \ref{table:worldwide_op} reveals that the presence of a 3 TW$_\mathrm{th}$ geo-reactor is expected to increase the uncertainty in the estimated $\theta_{13}$-related parameters, but not in the estimated $\theta_{12}$-related parameters. A comparison of Table \ref{table:worldwide_flux_gr} to Table \ref{table:worldwide_flux}, likewise, shows that a 3 TW$_\mathrm{th}$ geo-reactor is expected to increase the uncertainty in the flux correction to mantle (and to a lesser extend crust) geo-neutrinos, especially in at high angular resolutions.   This is not surprising since, at high angular resolutions, the mantle and geo-reactor backgrounds become somewhat confused in their measured direction.  It is also interesting to note that the 3 TW$_\mathrm{th}$ geo-reactor results in a slightly larger uncertainty in the flux correction for known reactors.   This is also not surprising since the known reactor and geo-reactor backgrounds are somewhat confused in their measured energy spectra.

\begin{table} [!htbp]
\centering
\fontsize{8}{10}\selectfont
\begin{tabular}{l|l|l|l|l|l|l|l}
\hline
\textbf{Scenario} & \textbf{p($\theta_{\mathrm{IAEA}}$)} & \textbf{p($\theta_{\mathrm{mantle}}$)} & \textbf{p($\theta_{\mathrm{crust}}$)} & \textbf{p($\theta_{\mathrm{fast}_{n^0}}$)} & \textbf{p($\theta_{\mathrm{acc}}$)} & \textbf{p($\theta_{\mathrm{cosm}}$)} & \textbf{p($\theta_{\mathrm{geo-reactor}}$)}\\
\hline \textit{a priori} uncertainty			& $\pm 2.0\%$	& $\pm 50.0\%$	 & $\pm 20.0\%$	 & $\pm 10.0\%$  & $\pm 1.3\%$   & $\pm 3.3\%$  & $\pm 100.0\%$ \\
\hline Count+Energy								& $\pm 1.8\%$   & $\pm 32.4\%$   & $\pm 10.6\%$  & $\pm 10.3\%$  & $\pm 1.2\%$   & $\pm 3.3\%$  & $\pm 14.7\%$ \\
\hline Count+Energy+Angle						& $\pm 1.8\%$   & $\pm 31.5\%$   & $\pm 11.1\%$  & $\pm 10.3\%$  & $\pm 1.2\%$   & $\pm 3.2\%$  & $\pm 13.2\%$ \\
\hline Count+Energy+Angle$_{\mathrm{(SNRx10)}}$ & $\pm 1.8\%$   & $\pm 26.0\%$   & $\pm 10.1\%$  & $\pm 9.7\%$   & $\pm 1.2\%$   & $\pm 3.2\%$  & $\pm 13.4\%$ \\
\hline Count+Energy+Angle$_{\mathrm{(SNRx20)}}$ & $\pm 1.8\%$   & $\pm 19.7\%$   & $\pm 8.8\%$   & $\pm 10.5\%$  & $\pm 1.3\%$   & $\pm 3.4\%$  & $\pm 11.3\%$ \\
\hline Count+Energy+Angle$_{\mathrm{(SNRx30)}}$ & $\pm 1.8\%$   & $\pm 15.7\%$   & $\pm 8.2\%$   & $\pm 10.8\%$  & $\pm 1.3\%$   & $\pm 3.3\%$  & $\pm 9.6\%$  \\
\end{tabular}
\caption{Worldwide flux uncertainty optimization results using 4 TREND detectors, assuming a 3TW$_\mathrm{th}$ geo-reactor lying at the earth's center (a point source). The seven background flux categories were optimized simultaneously with the oscillation parameters, the results of which can be seen in Table \ref{table:worldwide_op}. 1000 MC runs per scenario were used.}
\label{table:worldwide_flux_gr}
\end{table}

We conclude by noting that the presence of a 3 TW$_\mathrm{th}$ geo-reactor is also expected to impact the performance of the TREND detectors in applied experiments such as the search for unknown nuclear reactors.   First, the \textit{a priori} uncertainties associated with background flux and oscillation parameters will be higher.  Second, the additional statistical (Poisson) noise associated with the geo-reactor signal will also increase the \textit{a posteriori} uncertainties in the estimated parameters.   We did not attempt to quantify the impact of a geo-reactor on the (man-made) reactor search results in this work, but the capabilities exist to perform such an analysis if stronger observational evidence for a geo-reactor is uncovered in the future.

%% References
%%\cite{key}     ==>>  [#]
%%\cite[chap. 2]{key} ==>>  [#, chap. 2]
%%\citet{key}    ==>>  Author [#]


\bibliographystyle{unsrt}
\bibliography{<your-bib-database>}



\end{document}